\documentclass[10pt,twocolumn,letterpaper]{article}

\usepackage{iccv}

\usepackage{enumitem}
\usepackage{times}
\usepackage{epsfig}
\usepackage{graphicx}
\usepackage{amsmath, bm}
\usepackage{amssymb}
\usepackage{booktabs}
\usepackage{algorithm,algorithmicx,algpseudocode}
\usepackage{subfig}
\usepackage{multirow}
\usepackage{xspace}
\usepackage{url}
\makeatletter
\@namedef{ver@everyshi.sty}{}
\makeatother
\usepackage{tikz}
\usetikzlibrary{decorations.text}
\usetikzlibrary{positioning}
\usepackage{printlen}
\usepackage[font=small,labelfont=bf]{caption}

\usepackage{amsthm,amsmath,amsfonts,bm,xspace}
\usepackage{bbm}
\usepackage{color}
\usepackage{enumitem}

\def\eg{{\em e.g.,}\xspace}
\def\ie{{\em i.e.,}\xspace}

\def\Figref#1{Figure~\ref{#1}}

\def\tabref#1{Table~\ref{#1}}

\def\Secref#1{Section~\ref{#1}}

\def\eqref#1{(\ref{#1})}

\def\1{\bm{1}}

\def\vx{{\bm{x}}}
\def\veps{{\bm{\epsilon}}}

\def\vty{\widetilde{\bm{y}}}
\def\vy{{\bm{y}}}
\def\vz{{\bm{z}}}

\def\stdnormal{\mathcal{N}(\bm{0}, \bm{I})}

\DeclareMathAlphabet{\mathsfit}{\encodingdefault}{\sfdefault}{m}{sl}
\SetMathAlphabet{\mathsfit}{bold}{\encodingdefault}{\sfdefault}{bx}{n}

\def\R{{\mathbb{R}}}

\newcommand{\E}{\mathbb{E}}

\renewcommand{\R}{\mathbb{R}}

\usepackage[breaklinks=true,bookmarks=false]{hyperref}
\hypersetup{
    colorlinks=true,
    urlcolor=magenta,
}
\iccvfinalcopy 

\def\iccvPaperID{11175} 

\newcommand{\modelname}{SR3\xspace}

\ificcvfinal\fi

\begin{document}

\title{Image Super-Resolution via Iterative Refinement}
\author{
\hspace{-.7cm}\stepcounter{footnote}Chitwan Saharia\thanks{Work done as part of the Google AI Residency.},
Jonathan Ho,
William Chan,
Tim Salimans,
David J. Fleet,
Mohammad Norouzi\\[.1cm]
\hspace{-.7cm}{\large \texttt{\{sahariac,jonathanho,williamchan,salimans,davidfleet,mnorouzi\}@google.com}}\\[.2cm]
Google Research, Brain Team
}

\maketitle
\ificcvfinal\thispagestyle{empty}\fi


\begin{abstract}
\vspace*{-0.2cm}
We present \modelname, an approach to image Super-Resolution via Repeated Refinement.
\modelname adapts denoising diffusion probabilistic models~\cite{ho2020denoising,sohldickstein-icml-2015} to conditional image generation and 
performs super-resolution through a stochastic iterative denoising process.
Output generation starts with pure Gaussian noise and iteratively refines the noisy output using a U-Net model trained on denoising at various noise levels.
\modelname exhibits strong performance on super-resolution tasks at different magnification factors, on faces and natural images.
We conduct human evaluation on a standard 8$\times$ face super-resolution task on CelebA-HQ, comparing with SOTA GAN methods.
\modelname achieves a fool rate close to $50\%$, suggesting photo-realistic outputs, while GANs do not exceed a fool rate of 34\%. 
We further show the effectiveness of \modelname in cascaded image generation, where generative models are chained with super-resolution models,
yielding a competitive FID score of 11.3 on ImageNet.
\vspace*{-0.2cm}
\end{abstract}


\section{Introduction}

Single-image super-resolution is the process of generating a high-resolution image that is consistent with an input low-resolution image.
It falls under the broad family of image-to-image translation tasks, including colorization, in-painting, and de-blurring. 
Like many such inverse problems, image super-resolution is challenging because multiple output images may be consistent with a single input image, and
the conditional distribution of output images given the input typically does not conform well to simple parametric distributions, \eg~a multivariate Gaussian. 
Accordingly, while simple regression-based methods with feedforward convolutional nets may work for super-resolution at low magnification ratios, they often lack the high-fidelity details
needed for high magnification ratios.


\begin{figure}
\begin{center}
\small
\setlength{\tabcolsep}{2pt}
\begin{tabular}{ccc}
{\small Input} & {\small \modelname output} & {\small Reference} \\
{\includegraphics[width=0.15\textwidth]{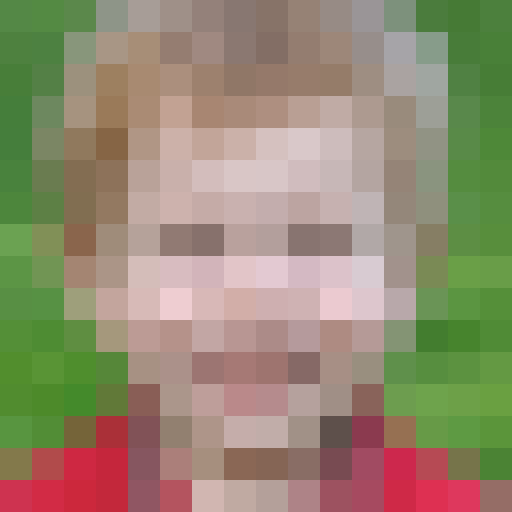}} &
{\includegraphics[width=0.15\textwidth]{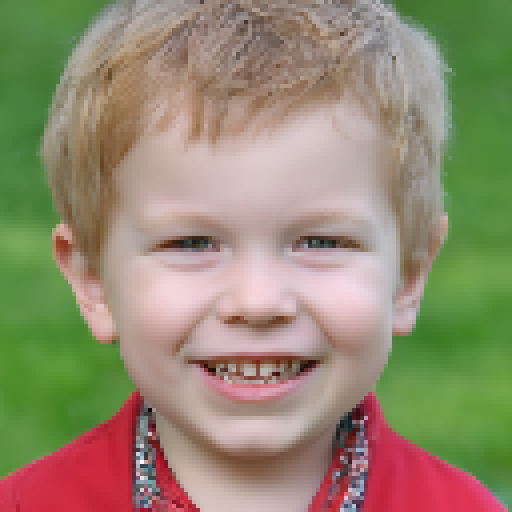}} &
{\includegraphics[width=0.15\textwidth]{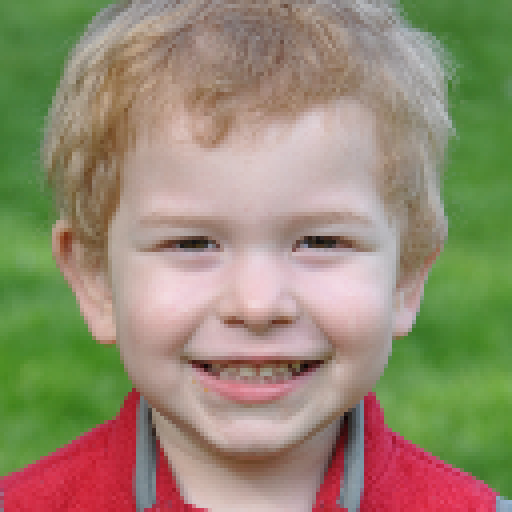}} \\


{\includegraphics[width=0.15\textwidth]{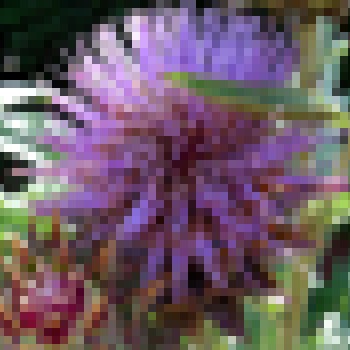}} &
{\includegraphics[width=0.15\textwidth]{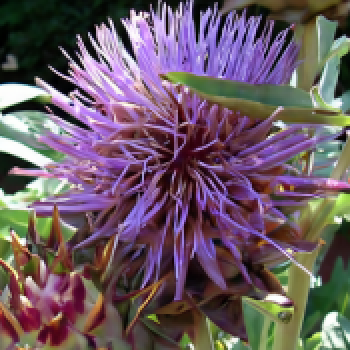}} &
{\includegraphics[width=0.15\textwidth]{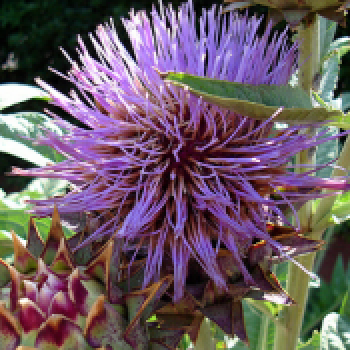}} \\

\end{tabular}
\vspace*{-0.2cm}
\caption{Two representative \modelname outputs: (top) 8$\times$ face super-resolution at
16$\times$16 $\!\rightarrow\!$ 128$\times$128 pixels (bottom)
 4$\times$ natural image super-resolution at 64$\times$64 $\!\rightarrow\!$ 256$\times$256 pixels. 
 \label{fig1}
}
\end{center}
\vspace*{-0.5cm}
\end{figure}

Deep generative models have seen success in learning complex empirical distributions of images (\eg~\cite{sutskever-nips-2014,vaswani-nips-2017}). 
Autoregressive models \cite{oord-arxiv-2016,oord-nips-2016}, variational autoencoders (VAEs) \cite{Kingma2013,vahdat2021nvae}, Normalizing Flows (NFs) \cite{dinh2016density,Kingma2018}, and GANs \cite{goodfellow2014generative,karras2018ProGAN,radford2015unsupervised} have shown convincing image generation results and have been applied to conditional tasks such as image super-resolution \cite{chen2018fsrnet,dahl2017pixel,ledig2017photo,menon2020pulse,parmar2018image}.
However, these approaches often suffer from various limitations; \eg~autoregressive models are prohibitively expensive for high-resolution image generation, NFs and VAEs often yield sub-optimal sample quality, and GANs require
carefully designed regularization and optimization tricks to tame optimization instability~\cite{arjovsky-arxiv-2017,gulrajani2017improved} and mode collapse~\cite{metz2016unrolled,ravuri2019classification}.

We propose \modelname (Super-Resolution via Repeated Refinement), a new approach to conditional image generation, inspired by recent work on Denoising Diffusion Probabilistic Models (DDPM) \cite{ho2020denoising,sohl2015deep}, 
and denoising score matching \cite{ho2020denoising,song2019generative}.
\modelname works by learning to transform a standard normal distribution into an empirical data distribution through a sequence of refinement steps, resembling Langevin dynamics.
The key is a U-Net architecture~\cite{ronneberger2015u} that is trained with a denoising objective to iteratively remove various levels of noise from the output.
We adapt DDPMs to {\em conditional} image generation by proposing a simple and effective modification to the U-Net architecture.
In contrast to GANs that require inner-loop maximization,
we minimize a well-defined loss function.
Unlike autoregressive models, \modelname uses a constant number of inference steps regardless of output resolution.

\modelname works well across a range of magnification factors and input resolutions.
\modelname models can also be cascaded, \eg~going from $64\!\times\!64$ to $256\!\times\!256$, and then 
to $1024\!\times\!1024$.
Cascading models allows one to independently train 
a few small models rather than a single large model with a high magnification factor.
We find that chained models enable more efficient inference, since directly generating a high-resolution image requires more iterative refinement steps for the same quality.
We also find that one can chain an unconditional generative model 
with \modelname models to unconditionally generate high-fidelity images.
Unlike existing work that focuses on specific domains (\eg~faces),
we show that \modelname is effective on both faces and natural images.

Automated image quality scores like PSNR and SSIM do not reflect human preference well when the input resolution is low and the magnification ratio is large~(\eg~\cite{berthelot2020creating,chen2018fsrnet,dahl2017pixel,menon2020pulse}).
These quality scores often penalize synthetic high-frequency details, such as hair texture, because synthetic details do not perfectly align with the reference details.
We resort to human evaluation to compare the quality of super-resolution methods.
We adopt a 2-alternative forced-choice (2AFC) paradigm in which human subjects are shown a low-resolution input and are required to
select between a model output and a ground truth image ({\em cf.}~\cite{zhang2016colorful}).
Based on this study, we calculate {\em fool rate} scores that capture both image quality and 
the consistency of model outputs with low-resolution inputs.
Experiments demonstrate that \modelname achieves a significantly higher fool rate than SOTA GAN methods~\cite{chen2018fsrnet,menon2020pulse} and a strong regression baseline.

\noindent
Our key contributions are summarized as:
\begin{itemize}[topsep=0pt, partopsep=0pt, leftmargin=10pt, parsep=0pt, itemsep=1.75pt]
\item We adapt denoising diffusion models to conditional image generation. Our method, \textit{\modelname}, is an approach to image super-resolution via iterative refinement.
\item \modelname proves effective on face and natural image super-resolution at different magnification factors.
On a standard 8$\times$ face super-resolution task, \modelname achieves a human fool rate close to $50\%$, outperforming FSRGAN~\cite{chen2018fsrnet} and PULSE~\cite{menon2020pulse} that achieve fool rates of at most 34\%. 
\item We demonstrate unconditional and class-conditional generation by cascading a $64\!\times\!64$ image synthesis model with
\modelname models to progressively generate $1024 \!\times\! 1024$ unconditional faces in 3 stages, and $256 \!\times\! 256$ class-conditional ImageNet samples in 2 stages.
Our class conditional ImageNet samples attain competitive FID scores.


\end{itemize}

\vspace{-.1cm}
\section{Conditional Denoising Diffusion Model}
\vspace{-.1cm}

We are given a dataset of input-output image pairs, denoted $\mathcal{D} = \{\vx_i, \vy_i\}_{i=1}^N$, which represent samples drawn from an unknown conditional distribution $p(\vy \,|\, \vx)$.
This is a one-to-many mapping in which many target images may be consistent with a single source image.
We are interested in learning a parametric approximation to $p(\vy \,|\, \vx)$  through a {\em stochastic} iterative refinement process that maps a source image $\vx$ to a target image $\vy \in \R^d$. 
We approach this problem by adapting the denoising diffusion probabilistic (DDPM) model of~\cite{ho2020denoising,sohl2015deep} to \textit{conditional} image generation.

\begin{figure}[t]
\begin{center}
\input{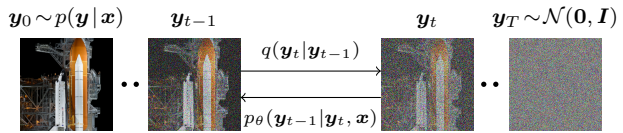}
\end{center}
\vspace*{-0.45cm}
\caption{The forward  diffusion process $q$ (left to right) gradually adds Gaussian noise to the target image.
The reverse inference process $p$ (right to left) iteratively denoises the target image conditioned on a source image $\vx$. 
Source image $\vx$ is not shown here.}
\vspace*{-0.25cm}
\label{fig:DiffusionChain}
\end{figure}

The conditional DDPM model generates a target image $\vy_0$ in $T$ refinement steps. 
Starting with a pure noise image $\vy_T \sim \mathcal{N}(\bm{0}, \bm{I})$,
the model iteratively refines the image through successive iterations $(\vy_{T-1}, \vy_{T-2}, \dotsc, \vy_{0}$) according to learned conditional transition distributions $p_\theta(\vy_{t-1} \,|\, \vy_t, \vx)$ such that $\vy_{0} \sim p(\vy \, |\, \vx)$ (see \Figref{fig:DiffusionChain}).

The distributions of intermediate images in the inference chain are defined in terms of a {\em forward} diffusion process that gradually adds
Gaussian noise to the signal via a fixed Markov chain, denoted $q(\vy_t \,|\, \vy_{t-1})$. 
The goal of our model is to reverse the Gaussian diffusion process by iteratively recovering signal from noise through
a reverse Markov chain conditioned on $\vx$. In principle, each forward process step can be conditioned on $\vx$ too, but we leave that to future work.
We learn the reverse chain using a neural denoising model $f_\theta$
that takes as input a source image and a noisy target image and estimates the noise.
We first give an overview of the forward diffusion process,
and then discuss how our denoising model $f_\theta$ is trained and used for inference.


\subsection{Gaussian Diffusion Process}
\label{sec:formalization}

Following
\cite{ho2020denoising,sohl2015deep}, we first define a \textit{forward} Markovian diffusion process $q$ that gradually adds Gaussian noise to a high-resolution image~$\vy_0$ over $T$ iterations:
\begin{eqnarray}
  q(\vy_{1:T} \mid \vy_0) &=& \prod\nolimits_{t=1}^{T} q(\vy_{t} \mid \vy_{t-1})~, \\
  q(\vy_{t} \mid \vy_{t-1}) &=& \mathcal{N}(\vy_{t} \mid \sqrt{\alpha_t}\, \vy_{t-1}, (1 - \alpha_t) \bm{I} )~, 
\end{eqnarray}
where the scalar parameters $\alpha_{1:T}$ are 
hyper-parameters, subject to $0 < \alpha_t < 1$, which determine the variance of the noise added at each iteration.
Note that $\vy_{t-1}$ is attenuated by $\sqrt{\alpha_t}$ to ensure that the variance of the random variables remains bounded as $t \to \infty$.
For instance, if the variance of $\vy_{t-1}$ is $1$, then the variance of $\vy_{t}$ is also $1$. 


Importantly, one can characterize the distribution of $\vy_t$ given $\vy_0$ 
by marginalizing out the
intermediate steps as
\begin{equation}
    q(\vy_t \mid \vy_0) ~=~ \mathcal{N}(\vy_t \mid \sqrt{\gamma_t}\, \vy_0, (1-\gamma_t) \bm{I})~,
\label{eq:diffusion-marginalized}
\end{equation}
where $\gamma_t = \prod_{i=1}^t \alpha_i$. Furthermore, with some algebraic manipulation and completing the square, 
one can derive the posterior distribution of $\vy_{t-1}$ given $(\vy_0, \vy_t)$ as
\begin{equation}
\begin{aligned}
    q&(\vy_{t-1} \mid \vy_0, \vy_t) = \mathcal{N}(\vy_{t-1} \mid \bm{\mu}, \sigma^2 \bm{I})\\
    &\bm{\mu} =  \frac{\sqrt{\gamma_{t-1}}\,(1-\alpha_t)}{1-\gamma_t}\, \vy_0 \!+\! \frac{\sqrt{\alpha_t}\,(1-\gamma_{t-1})}{1-\gamma_t}\vy_t\\
    &\sigma^2 = \frac{(1-\gamma_{t-1})(1-\alpha_t)}{1-\gamma_t}~.
\end{aligned}
\label{eq:posteriror-ytm1}
\end{equation}
This posterior distribution is helpful when parameterizing the reverse chain and formulating a variational lower bound on the log-likelihood of the reverse chain.
We next discuss how one can learn a neural network to reverse this Gaussian diffusion process.

\subsection{Optimizing the Denoising Model}
\label{sec:training}

\algrenewcommand\algorithmicindent{0.5em}%
\begin{figure}[t]
\vspace*{-.5cm}
\small
\begin{minipage}[t]{0.495\textwidth}
\begin{algorithm}[H]
  \caption{Training a denoising model $f_\theta$} \label{alg:training}
  \small
  \begin{algorithmic}[1]
    \Repeat
      \State $(\vx, \vy_0) \sim p(\vx, \vy)$
      \State $\gamma \sim p(\gamma)$
      \State $\bm{\epsilon}\sim\mathcal{N}(\mathbf{0},\mathbf{I})$
      \State Take a gradient descent step on
      \Statex $\qquad \nabla_\theta \left\lVert f_\theta(\vx, \sqrt{\gamma} \vy_0 + \sqrt{1-\gamma} \bm{\epsilon}, \gamma) - \veps \right\rVert_p^p$ 
    \Until{converged}
  \end{algorithmic}
\end{algorithm}
\end{minipage}
\vspace*{-.2cm}
\end{figure}

To help reverse the diffusion process, we take advantage of additional side information in the form of a source image $\vx$ and
optimize a neural denoising model $f_{\theta}$ that takes as input this source image $\vx$ and a noisy target image $\vty$,
\begin{equation}
    \vty = \sqrt{\gamma}\, \vy_0 + \sqrt{1-\gamma} \,\veps~,
    ~~~~~~ \veps \sim \mathcal{N}(\bm{0},\bm{I})~,
\label{eq:noisy-y}
\end{equation}
and aims to recover the 
noiseless target image $\vy_0$.
This definition of a noisy target image $\vty$ is compatible with the marginal distribution of noisy images at different steps of the forward diffusion process in \eqref{eq:diffusion-marginalized}.



In addition to a source image $\vx$ and a noisy target image~$\vty$,
the denoising model $f_\theta(\vx, \vty, \gamma)$ takes as input 
the sufficient statistics for the variance of the noise $\gamma$,
and is trained to predict the noise vector $\veps$.
We make the denoising model aware of the level of noise through conditioning on a scalar $\gamma$, similar to~\cite{song2019generative,chen-iclr-2021}.
The proposed objective function for training $f_{\theta}$ is
\begin{equation}
    \E_{(\vx, \vy)} \E_{\veps, \gamma} \bigg\lVert f_\theta(\vx, \underbrace{\sqrt{\gamma} \,\vy_0 + \sqrt{1-\gamma}\, \veps}_{\vty}, \gamma) - \veps\, \bigg\rVert^{p}_p~,
\label{eq:loss}
\end{equation}
where $\veps \sim \mathcal{N}(\bm{0}, \bm{I})$, $(\vx, \vy)$ is sampled from the training dataset, $p \in \{1, 2\}$, and $\gamma \sim p(\gamma)$. 
The distribution of $\gamma$ has a big impact on the quality of the model and the generated outputs. We discuss our choice of $p(\gamma)$ in \Secref{sec:details}.

Instead of regressing the output of $f_{\theta}$ to $\veps$, as in \eqref{eq:loss}, one can also regress the output of $f_{\theta}$ to $\vy_0$.
Given $\gamma$ and $\vty$, the values of $\veps$ and $\vy_0$ can be derived from each other deterministically,
but changing the regression target has an impact on the scale of the loss function.
We expect both of these variants to work reasonably well if $p(\gamma)$ is modified to account for the scale of the loss function.
Further investigation of the loss function used for training the denoising model is an interesting avenue for future research in this area.

\subsection{Inference via Iterative Refinement}
\label{sec:inference}

\begin{figure}
\vspace*{-.5cm}
\small
\begin{minipage}[t]{0.495\textwidth}
\begin{algorithm}[H]
  \caption{Inference in $T$ iterative refinement steps} \label{alg:sampling}
  \small
  \begin{algorithmic}[1]
    \vspace{.04in}
    \State $\vy_T \sim \mathcal{N}(\mathbf{0}, \mathbf{I})$
    \For{$t=T, \dotsc, 1$}
      \State $\vz \sim \mathcal{N}(\mathbf{0}, \mathbf{I})$ if $t > 1$, else $\vz = \mathbf{0}$
      \State $\vy_{t-1} = \frac{1}{\sqrt{\alpha_t}}\left( \vy_t - \frac{1-\alpha_t}{\sqrt{1-\gamma_t}} f_\theta(\vx, \vy_t, \gamma_t) \right) + \sqrt{1 - \alpha_t} \vz$
    \EndFor
    \State \textbf{return} $\vy_0$
    \vspace{.04in}
  \end{algorithmic}
\end{algorithm}
\end{minipage}
\vspace*{-0.2cm}
\end{figure}

Inference under our model is defined as a \emph{reverse} Markovian process, which goes
in the reverse direction of the forward diffusion process, starting from Gaussian noise $\vy_T$:
\begin{eqnarray}
    p_\theta(\vy_{0:T} | \vx) &=& p(\vy_T) \prod\nolimits_{t=1}^T p_\theta(\vy_{t-1} | \vy_t, \vx) \\
    p(\vy_T) &=& \mathcal{N}(\vy_T \mid \bm{0}, \bm{I}) \label{eq:mutheta}\\
    p_\theta(\vy_{t-1} | \vy_{t}, \vx) &=& \mathcal{N}(\vy_{t-1} \mid \mu_{\theta}(\vx, {\vy}_{t}, \gamma_t), \sigma_t^2\bm{I})~. \label{eq:reverse_process}
\end{eqnarray}
We define the inference process in terms of isotropic Gaussian conditional distributions, $p_\theta(\vy_{t-1} | \vy_{t}, \vx)$, which are learned.
If the noise variance of the forward process steps are set as small as possible, \ie~$\alpha_{1:T} \approx 1$, the optimal reverse process $p(\vy_{t-1} | \vy_{t}, \vx)$ will be approximately Gaussian~\cite{sohl2015deep}.
Accordingly, our choice of Gaussian conditionals in the inference process \eqref{eq:reverse_process} can provide a reasonable fit to the true reverse process.
Meanwhile, $1 - \gamma_T$ should be large enough so that $\vy_T$ is approximately distributed according to the prior $p(\vy_T) = \mathcal{N}(\vy_T | \bm{0}, \bm{I})$, allowing the sampling process to start at pure Gaussian noise.

Recall that the denoising model $f_{\theta}$ is trained to estimate $\veps$, given any noisy image $\vty$ including $\vy_t$. Thus, given $\vy_t$, we approximate $\vy_0$ by rearranging the terms in \eqref{eq:noisy-y} as
\begin{equation}
    \hat{\vy}_0 = \frac{1}{\sqrt{\gamma_t}} \left( \vy_t - \sqrt{1 - \gamma_t}\, f_{\theta}(\vx, \vy_{t}, \gamma_t) \right)~.
\end{equation}
Following the formulation of \cite{ho2020denoising}, we substitute our estimate $\hat{\vy}_0$ into the posterior distribution of $q(\vy_{t-1} | \vy_0, \vy_t)$ in \eqref{eq:posteriror-ytm1} to parameterize the mean of $p_\theta(\vy_{t-1} | \vy_t, \vx)$ as
\begin{equation}
    \mu_{\theta}(\vx, {\vy}_{t}, \gamma_t) = \frac{1}{\sqrt{\alpha_t}} \left( \vy_t - \frac{1-\alpha_t}{ \sqrt{1 - \gamma_t}} f_{\theta}(\vx, \vy_{t}, \gamma_t) \right)~,
\end{equation}
and we set the variance of $p_\theta(\vy_{t-1}|\vy_t, \vx)$ to $(1 - \alpha_t)$, a default given by the variance of the forward process \cite{ho2020denoising}.

Following this parameterization, each iteration of iterative refinement under our model takes the form,
\begin{equation*}
\vy_{t-1} \leftarrow \frac{1}{\sqrt{\alpha_t}} \left( \vy_t - \frac{1-\alpha_t}{ \sqrt{1 - \gamma_t}} f_{\theta}(\vx, \vy_{t}, \gamma_t) \right) + \sqrt{1 - \alpha_t}\veps_t~,
\end{equation*}
where $\veps_t \sim \stdnormal$. This resembles one step of Langevin dynamics with $f_{\theta}$ providing an estimate of the
gradient of the data log-density.
We justify the choice of the training objective in \eqref{eq:loss} for the probabilistic model outlined in \eqref{eq:reverse_process} from a variational lower bound perspective
and a denoising score-matching perspective in Appendix~\ref{sec:justification}.


\begin{figure*}[t]
\vspace*{-0.3cm}
\setlength{\tabcolsep}{2pt}
\begin{center}
\begin{tabular}{cccc}
{\small Bicubic} & 
{\small Regression} & {\small SR3 (ours)} & {\small Reference} \\
{\includegraphics[width=0.175\textwidth]{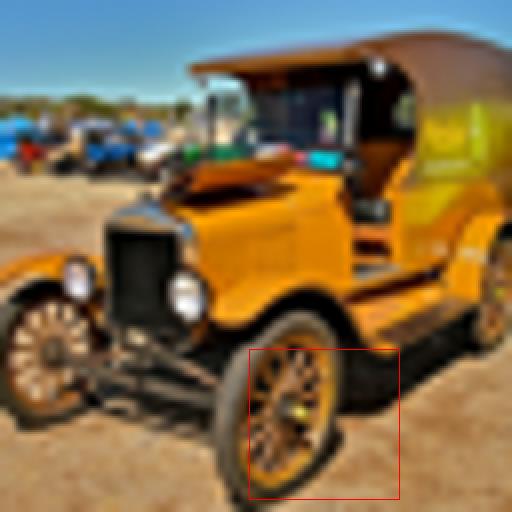}} &
{\includegraphics[width=0.175\textwidth]{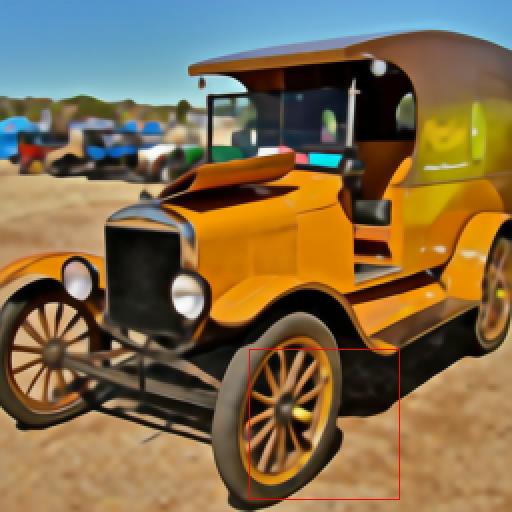}} &
{\includegraphics[width=0.175\textwidth]{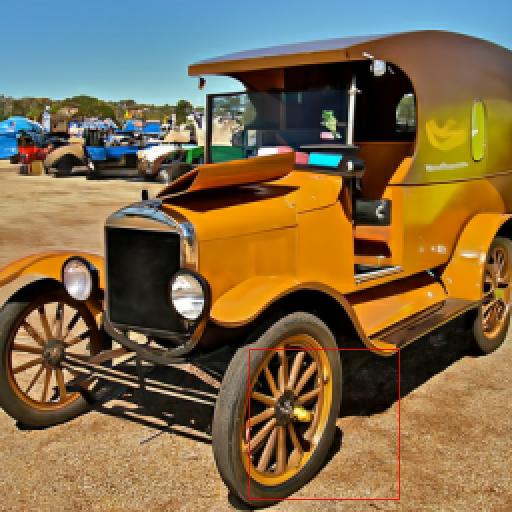}} &
{\includegraphics[width=0.175\textwidth]{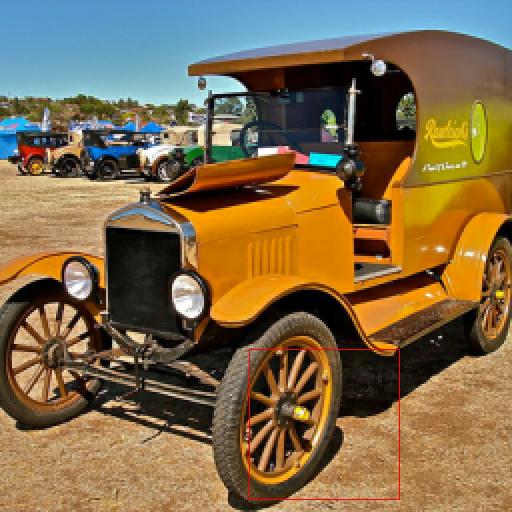}} \\

{\includegraphics[width=0.175\textwidth]{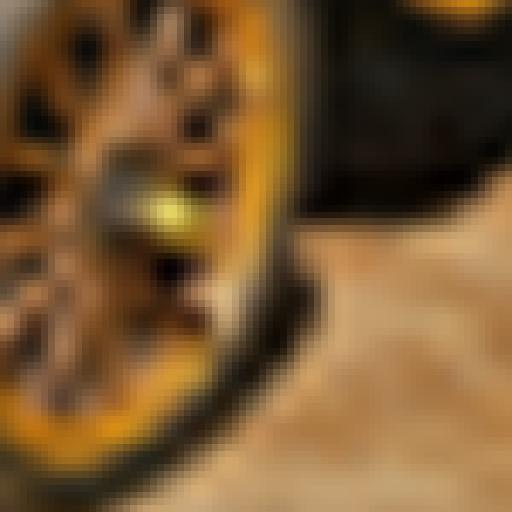}} &
{\includegraphics[width=0.175\textwidth]{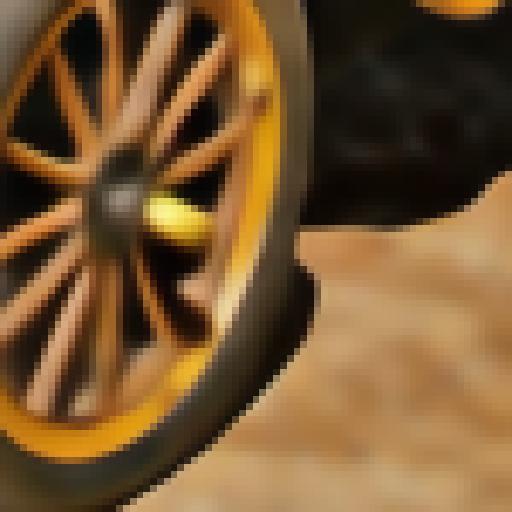}} &
{\includegraphics[width=0.175\textwidth]{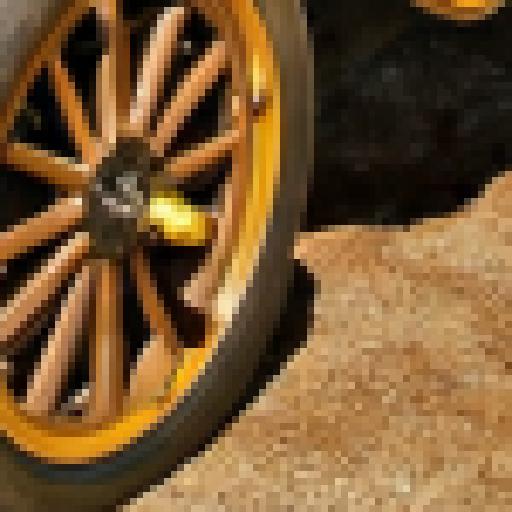}} &
{\includegraphics[width=0.175\textwidth]{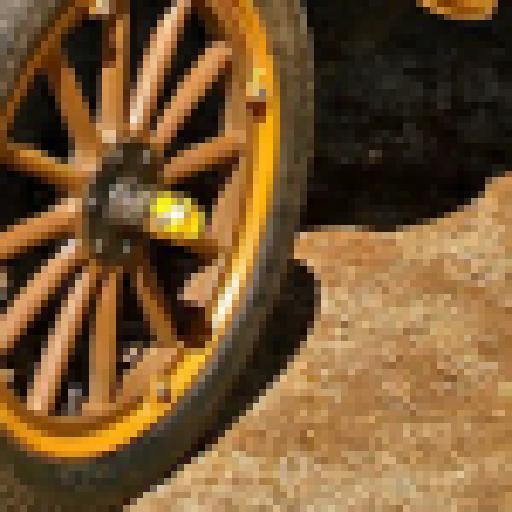}} \\

{\includegraphics[width=0.175\textwidth]{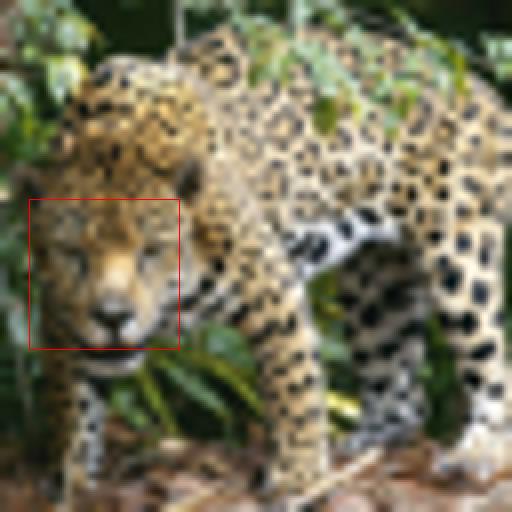}} &
{\includegraphics[width=0.175\textwidth]{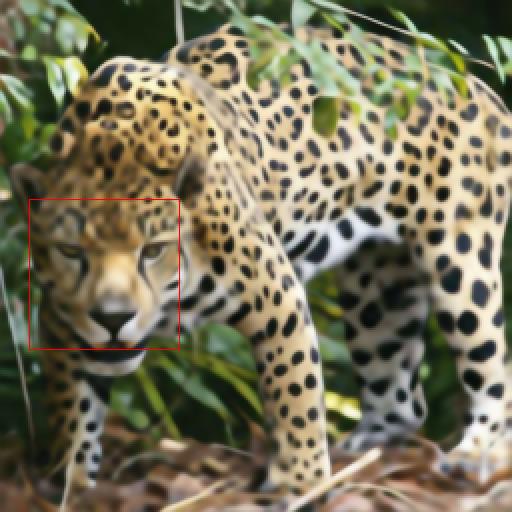}} &
{\includegraphics[width=0.175\textwidth]{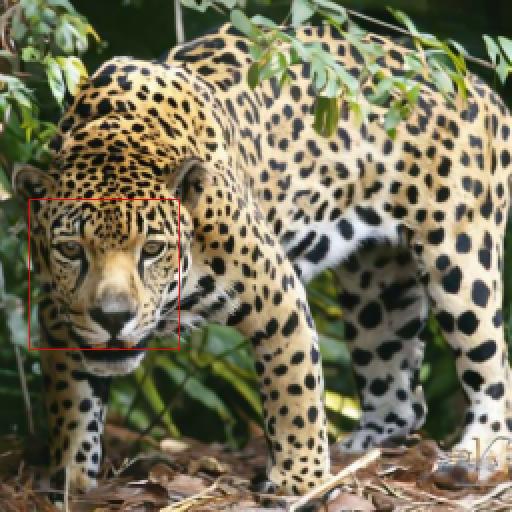}} &
{\includegraphics[width=0.175\textwidth]{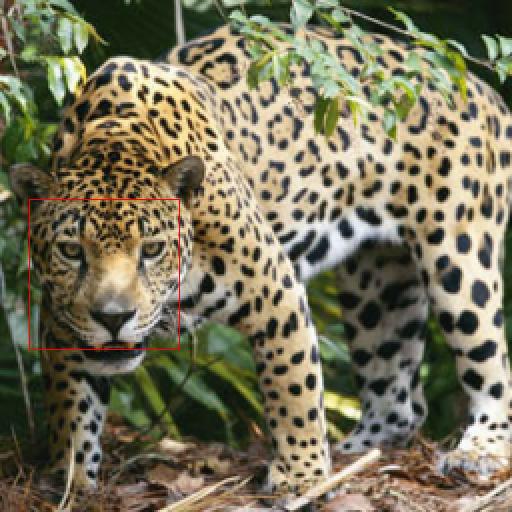}} \\

{\includegraphics[width=0.175\textwidth]{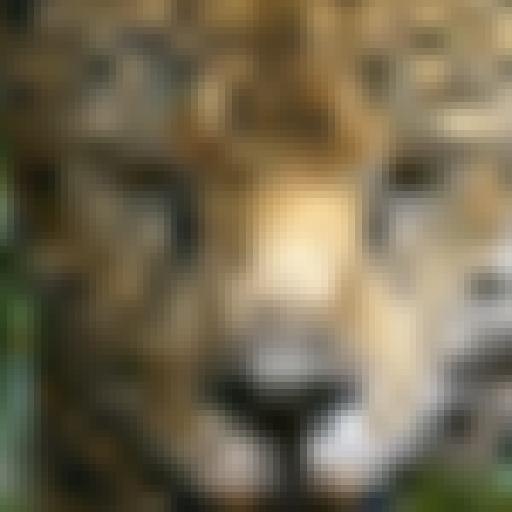}} &
{\includegraphics[width=0.175\textwidth]{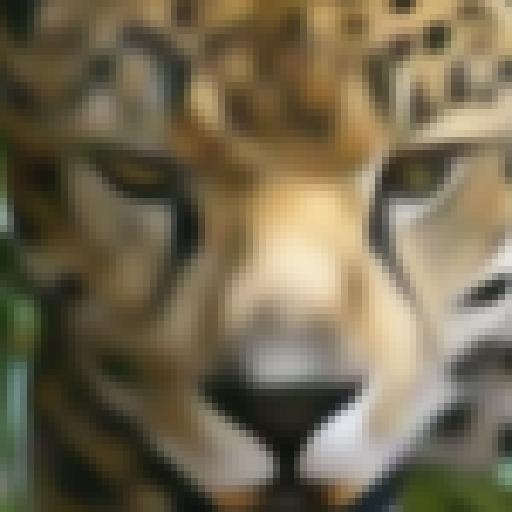}} &
{\includegraphics[width=0.175\textwidth]{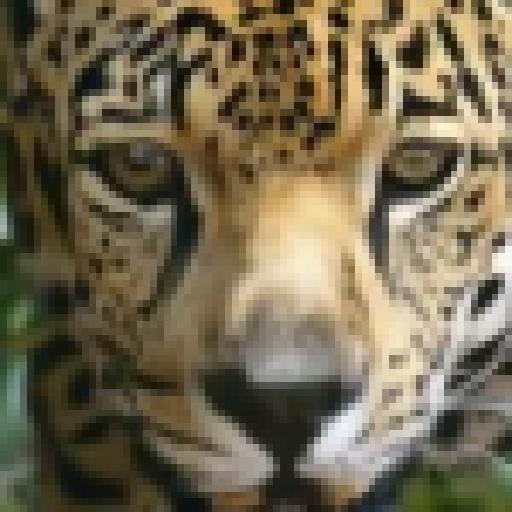}} &
{\includegraphics[width=0.175\textwidth]{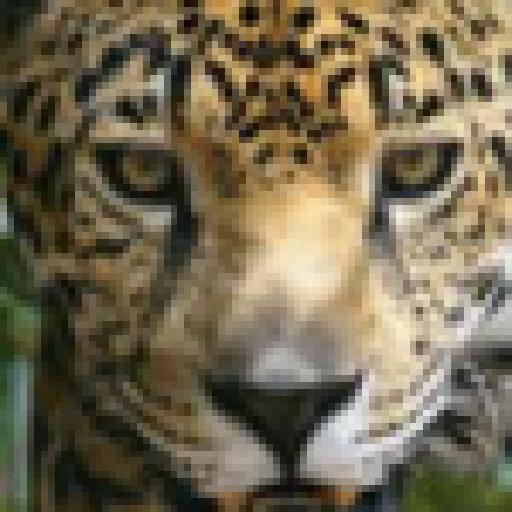}} \\

\end{tabular}
\end{center}
\vspace*{-0.55cm}
\caption{Results of a \modelname model (64$\times$64 $\rightarrow$ 256$\times$256), trained on ImageNet and evaluated on two ImageNet test images.  For each we also show an enlarged patch in which finer details are more apparent.  Additional samples are shown  in Appendix \ref{fig:64x_256x_natural_images2} and \ref{fig:64x_256x_natural_images3}.
\vspace*{-0.25cm}}
\label{fig:64x_256x_natural_images}
\end{figure*}

\subsection{\modelname Model Architecture and  Noise Schedule}
\label{sec:details}

The SR3 architecture is similar to the U-Net found in DDPM \cite{ho2020denoising},
with modifications adapted from \cite{song-iclr-2021}; we replace the original DDPM residual 
blocks with residual blocks from BigGAN \cite{brock2018large}, and we re-scale skip 
connections by $\frac{1}{\sqrt{2}}$. 
We also increase the number of residual blocks, and the channel multipliers at
different resolutions (see Appendix~\ref{sec:task-arch-details} for details). 
To condition the model on the input $\vx$, we up-sample the low-resolution
image to the target resolution using bicubic interpolation.
The result is concatenated with $\vy_t$ along the channel dimension. 
We experimented with more sophisticated methods of conditioning, such as using FiLM \cite{perez2018film}, but we found that the simple concatenation yielded similar generation quality.




For our training noise schedule, we follow \cite{chen-iclr-2021}, and use a piece wise distribution for $\gamma$,
$p(\gamma) = \sum_{t=1}^T \frac{1}{T} U(\gamma_{t-1}, \gamma_t)$. Specifically, during training, we first uniformly sample a time step $t \sim \{0, ..., T\}$ followed by sampling $\gamma \sim U(\gamma_{t-1}, \gamma_t)$. We set $T = 2000$ in all our experiments.

Prior work of diffusion models \cite{ho2020denoising, song-iclr-2021} require 1-2k diffusion steps during inference, making generation slow for large target resolution tasks. 
We adapt techniques from \cite{chen-iclr-2021} to enable more efficient inference. 
Our model conditions on $\gamma$ directly (vs $t$ as in \cite{ho2020denoising}), which allows us flexibility in choosing number of diffusion steps, and the noise schedule during inference. 
This has been demonstrated to work well for speech synthesis \cite{chen-iclr-2021}, but has not been explored for images. For efficient inference we set the maximum inference budget to 100 diffusion steps, and hyper-parameter search over the inference noise schedule. This search is inexpensive as we only need to train the model once \cite{chen-iclr-2021}. 
We use FID on held out data
to choose the best noise schedule, as we found PSNR did not correlate well with image quality.

\section{Related Work}
\vspace*{-0.1cm}

\noindent
\modelname is inspired by recent work on deep generative models and recent learning-based approaches to super-resolution.

\vspace*{0.1cm}
\noindent
\textbf{Generative Models.}~Autoregressive models (ARs) \cite{oord2016pixel, salimans2017pixelcnn++} can model exact data log likelihood, capturing rich distributions. 
However, their sequential generation of pixels is expensive, limiting
application to low-resolution images.
Normalizing flows \cite{rezende2015variational,dinh2016density,Kingma2018} improve on sampling speed while modelling the exact data likelihood, but the need for invertible parameterized transformations with a tractable Jacobian determinant limits their expressiveness. 
VAEs~\cite{Kingma2013,rezende2014stochastic} offer fast sampling, but tend to underperform GANs and ARs in image quality \cite{vahdat2021nvae}. 
Generative Adversarial Networks (GANs) \cite{goodfellow2014generative} are
popular for class conditional image generation and super-resolution. 
Nevertheless, the inner-outer loop optimization often requires tricks to 
stabilize training \cite{arjovsky-arxiv-2017,gulrajani2017improved}, and conditional tasks like super-resolution usually require an auxiliary consistency-based loss to avoid mode collapse \cite{ledig2017photo}. Cascades of GAN models have been used to generate higher resolution images \cite{denton-nips-2015}.

Score matching \cite{hyvarinen2005estimation} models the gradient of the data 
log-density with respect to the image. Score matching on noisy data, called denoising score matching \cite{vincent2011connection}, is equivalent to training a denoising autoencoder, and 
to DDPMs \cite{ho2020denoising}. 
Denoising score matching over multiple noise scales with Langevin dynamics sampling from the learned score functions has recently been shown to be effective for high quality unconditional image generation \cite{song2019generative,ho2020denoising}. 
These models have also been generalized to continuous time \cite{song-iclr-2021}. 
Denoising score matching and diffusion models have also found success in shape generation \cite{cai-eccv-2020}, and speech synthesis \cite{chen-iclr-2021}. 
We extend this method to super-resolution, with a simple learning objective, a constant number of inference generation steps, and high quality generation.

\vspace*{0.1cm}\noindent
\textbf{Super-Resolution.}~ Numerous super-resolution methods have been proposed in the computer vision community \cite{dong2014learning, ahn-cvpr-2018, kim2016deeply, tai2017image, ledig2017photo, sajjadi2017enhancenet}. Much of the early work on super-resolution is regression based and trained with an MSE loss \cite{dong2014learning, ahn-cvpr-2018, wang2015deep, dong-pami-2016, KimCVPR2016}. 
As such, they effectively estimate the posterior mean, yielding blurry images when the posterior is multi-modal \cite{ledig2017photo,sajjadi2017enhancenet,menon2020pulse}. 
Our regression baseline defined below is also a one-step regression model trained with MSE (cf.\  \cite{ahn-cvpr-2018,KimCVPR2016}), but with a large U-Net architecture.
\modelname, by comparison, relies on a series of iterative refinement steps, each of which is trained with a regression loss. This difference permits our iterative  approach to capture richer distributions.
Further, rather than estimating the posterior mean, \modelname generates samples from the target posterior.

Autoregressive models have been used successfully for super-resolution and cascaded
up-sampling \cite{dahl2017pixel,menick-iclr-2019,oord2016conditional,parmar2018image}. Nevertheless, the expensive of inference limits their applicability to low-resolution images. \modelname can generate high-resolution images, e.g., 1024$\times$1024, but with a constant number of refinement steps (often no more than 100).

Normalizing flows have been used for super-resolution with a multi-scale approach \cite{yu-arxiv-2020}. They are capable of generating 1024$\times$1024 images due in part to their efficient inference process.
But \modelname uses a series of reverse diffusion steps to transform a Gaussian distribution to an image distribution while flows require a deep and invertible network.

GAN-based super-resolution methods have also found considerable success  \cite{karras2018ProGAN,ledig2017photo,menon2020pulse,yang-arxiv-2020-hifacegan,sajjadi2017enhancenet}. FSRGAN \cite{chen2018fsrnet} and PULSE \cite{menon2020pulse} in particular have demonstrated high quality face super-resolution results. However, many such GAN based methods are generally difficult to optimize, and often require auxiliary objective functions to ensure consistency with the low resolution inputs. 



\section{Experiments}
\vspace*{-0.1cm}

\begin{figure*}[t]
\vspace*{-0.3cm}
\setlength{\tabcolsep}{2pt}
\begin{center}
\begin{tabular}{cccc}
{\small Bicubic} & {\small Regression} & {\small \modelname (ours)} & {\small Reference} \\
{\includegraphics[width=000000.185\textwidth]{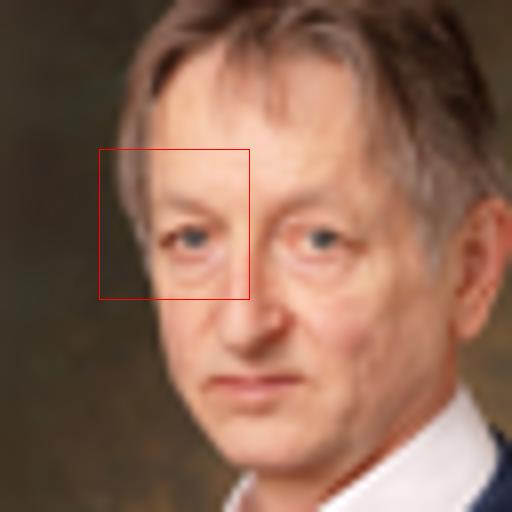}}&
{\includegraphics[width=000000.185\textwidth]{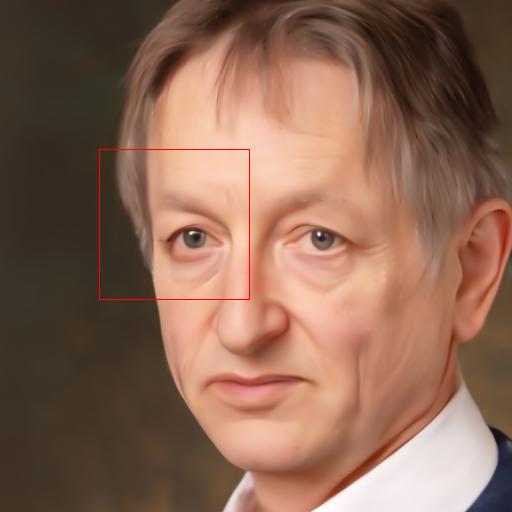}}&
{\includegraphics[width=000000.185\textwidth]{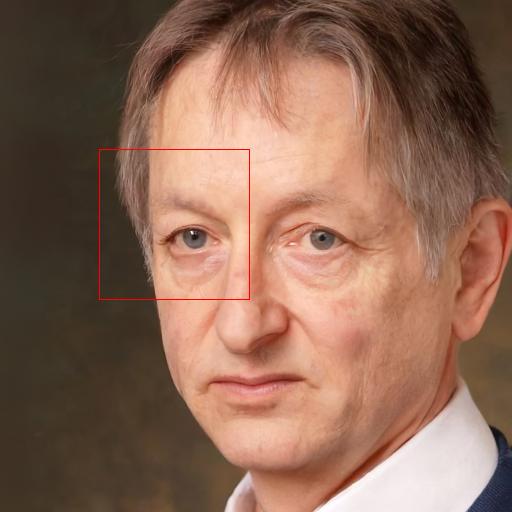}} &
{\includegraphics[width=000000.185\textwidth]{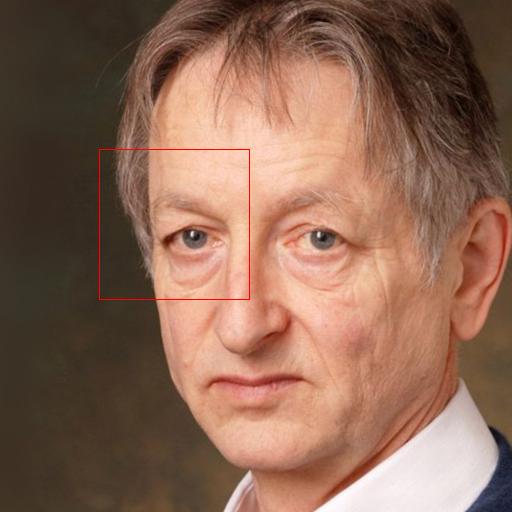}} \\
{\includegraphics[width=000000.185\textwidth]{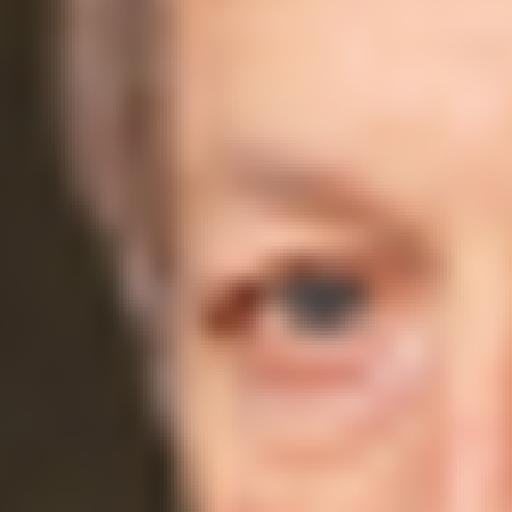}}&
{\includegraphics[width=000000.185\textwidth]{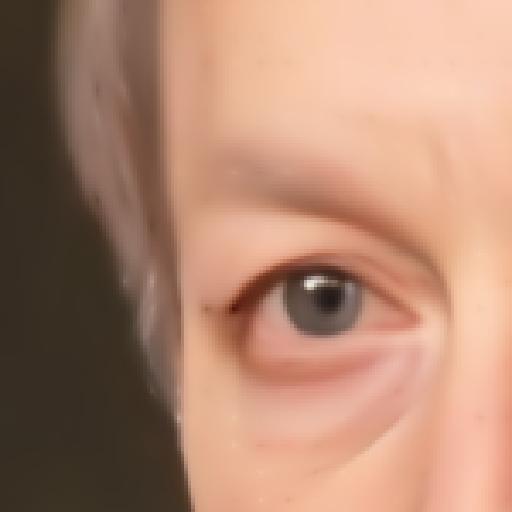}}&
{\includegraphics[width=000000.185\textwidth]{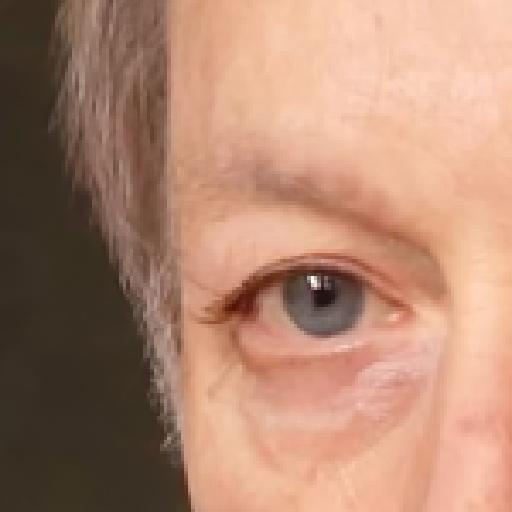}} &
{\includegraphics[width=000000.185\textwidth]{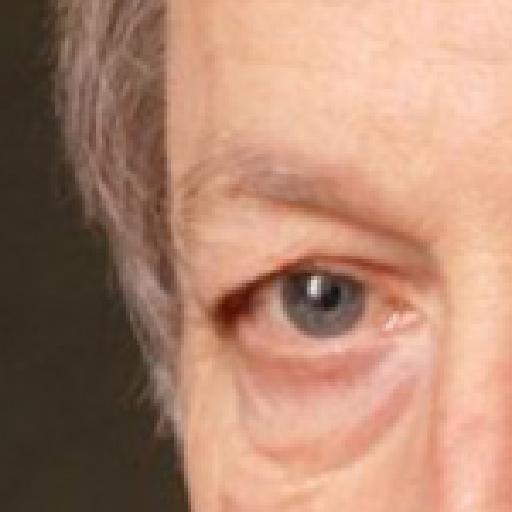}} \\

{\includegraphics[width=000000.185\textwidth]{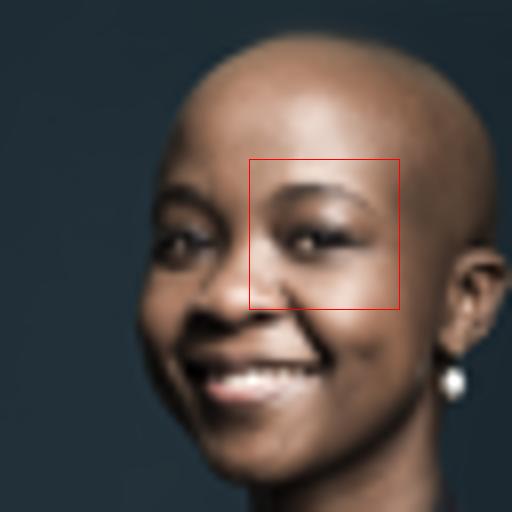}}&
{\includegraphics[width=000000.185\textwidth]{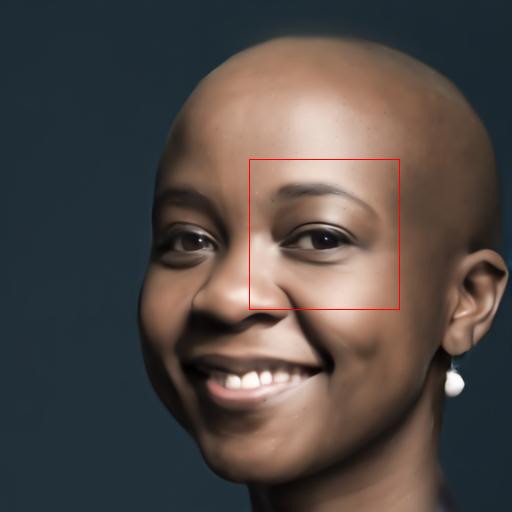}}&
{\includegraphics[width=000000.185\textwidth]{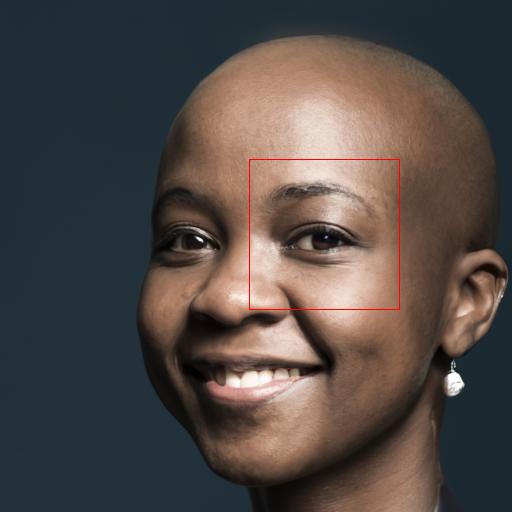}} &
{\includegraphics[width=000000.185\textwidth]{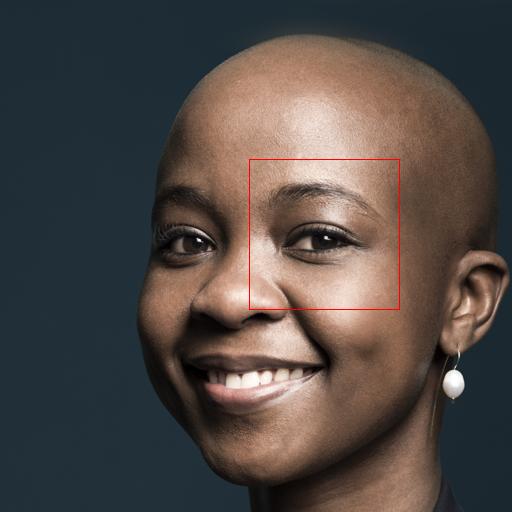}} \\
{\includegraphics[width=000000.185\textwidth]{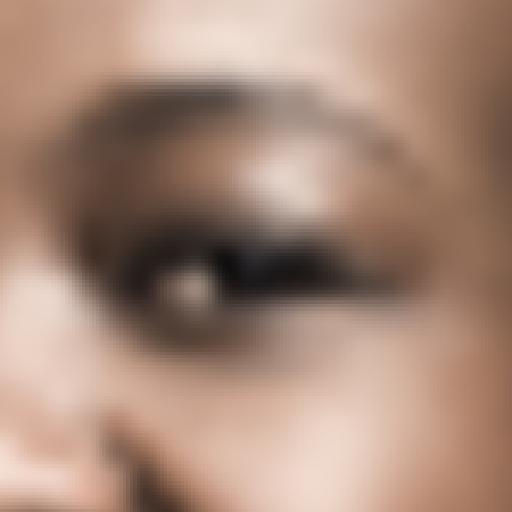}}&
{\includegraphics[width=000000.185\textwidth]{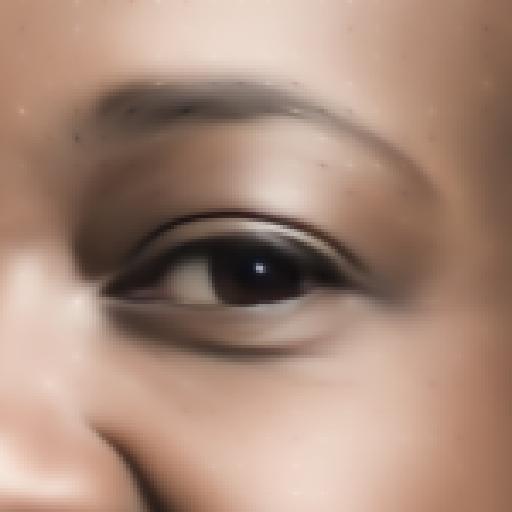}}&
{\includegraphics[width=000000.185\textwidth]{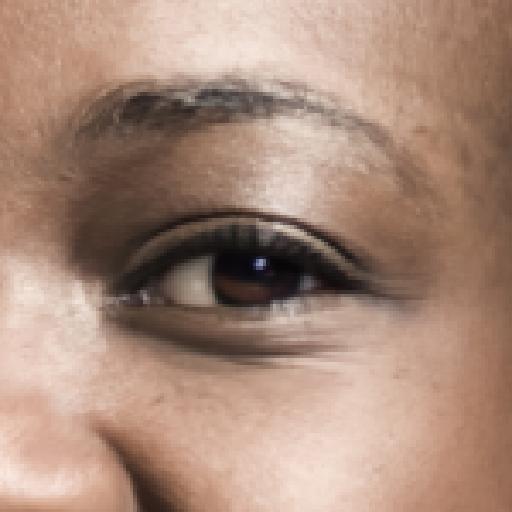}} &
{\includegraphics[width=000000.185\textwidth]{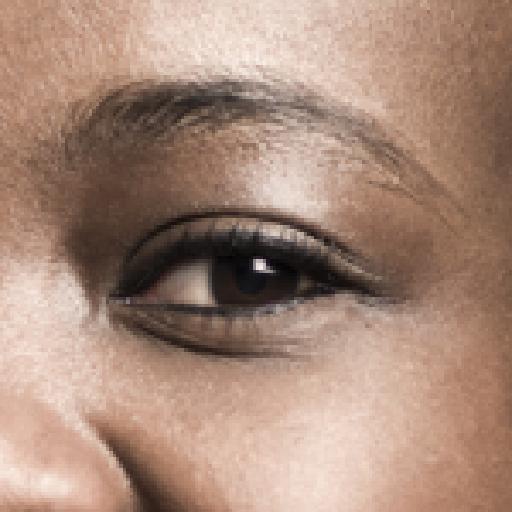}} 
\end{tabular}
\end{center}
\vspace*{-0.55cm}
\caption{Results of a \modelname model (64$\times$64 $\rightarrow$ 512$\times$512), trained on FFHQ, and applied to images outside of the training set, along with enlarged patches to show finer details. 
Additional results are shown in Appendix \ref{fig:64x_512x_faces2_arxiv} and \ref{fig:64x_512x_faces3_arxiv}. \vspace*{-.4cm}
}
\label{fig:64x_512x_faces}
\end{figure*}

We assess the effectiveness of \modelname models in super-resolution on faces, natural images, and synthetic images
obtained from a low-resolution generative model. The latter enables high-resolution image synthesis using model cascades.
We compare \modelname with recent methods such as FSRGAN~\cite{chen2018fsrnet} and PULSE~\cite{menon2020pulse} using human evaluation\footnote{Samples generously provided by the authors of \cite{menon2020pulse}},
and report FID for various tasks.
We also compare to a regression baseline model that shares the same architecture as \modelname, but is trained with a MSE loss.
Our experiments include:
\begin{itemize}[topsep=0pt, partopsep=0pt, leftmargin=10pt, parsep=0pt, itemsep=1.75pt]
\item Face super-resolution at $16\!\times\!16 \!\rightarrow\! 128\!\times\!128$ and $64\!\times\!64 \!\rightarrow\! 512\!\times\!1512$
trained on FFHQ and evaluated on CelebA-HQ. 
\item Natural image super-resolution at $64\!\times\!64 \rightarrow 256\!\times\!256$ pixels on ImageNet~\cite{russakovsky2015imagenet}.
\item Unconditional $1024\!\times\!1024$ face generation by a cascade of 3 models,
and class-conditional $256\!\times\!256$ ImageNet image generation by a cascade of 2 models.
\end{itemize}

\vspace*{0.1cm}
\noindent
{\bf Datasets:}~We follow previous work \cite{menon2020pulse}, training
face super-resolution models on Flickr-Faces-HQ (FFHQ) \cite{karras2019style} and evaluating 
on CelebA-HQ~\cite{karras2018ProGAN}. For natural image super-resolution, we train on ImageNet 1K~\cite{russakovsky2015imagenet} and use the dev split for evaluation.
We train unconditional face and class-conditional ImageNet generative models using DDPM on the same datasets discussed above.
For training and testing, we use low-resolution images that are down-sampled using bicubic interpolation with anti-aliasing enabled. For ImageNet, we discard images where the shorter side is less than the target resolution. We use the largest central crop like \cite{brock2018large}, which is then resized to the target resolution using area resampling as our high resolution image.

\vspace*{0.1cm}
\noindent {\bf Training Details: } We train all of our \modelname and regression models for 1M training steps with a batch size of 256.
We choose a checkpoint for the regression baseline based on peak-PSNR on the held out set.
We do not perform any checkpoint selection on \modelname models and simply select the latest checkpoint.
Consistent with \cite{ho2020denoising}, we use the Adam optimizer with a linear warmup schedule over 10k training steps, followed by a fixed learning rate of 1e-4 for \modelname models and 1e-5 for regression models. We use 625M parameters for our $64\!\times\!64 \rightarrow \{ 256\!\times\!256, 512\!\times\!512 \}$ models, 550M parameters for the 16$\times$16 $\rightarrow$ 128$\times$128 models, and 150M parameters for 256$\times$256 $\rightarrow$ 1024$\times$1024 model. We use a dropout rate of 0.2 for 16$\times$16 $\rightarrow$ 128$\times$128 models super-resolution, but otherwise, we do not use dropout. 
(See Appendix~\ref{sec:task-arch-details} for task specific architectural details.)


\subsection{Qualitative Results}

\noindent

\textbf{Natural Images: } Figure \ref{fig:64x_256x_natural_images}
gives examples of super-resolution natural images for 64$\times$64 $\rightarrow$ 256$\times$256 on the ImageNet dev set,
along with enlarged patches for finer inspection.
The baseline Regression model generates images that are faithful to the inputs, but are blurry and lack detail.
By comparison, \modelname produces sharp images with more detail; this is most evident in the enlarged patches. 
For more samples see Appendix \ref{fig:64x_256x_natural_images2} and \ref{fig:64x_256x_natural_images3}.

\textbf{Face Images: } \Figref{fig:64x_512x_faces} shows outputs of a face super-resolution model ($64\!\times\!64\rightarrow 512\!\times\!512$) on 
two test images, again with selected patches enlarged.
With the 8$\times$ magnification factor one can clearly see the detailed structure inferred. 
Note that, because of the large magnification factor, there are many plausible outputs, so we do not expect the output to exactly match the reference image.
This is evident in the regions highlighted in the faces. For more samples see Appendix \ref{fig:64x_512x_faces2_arxiv} and \ref{fig:64x_512x_faces3_arxiv}.


\subsection{Benchmark Comparison}
\label{sec:benchmark}
\subsubsection{Automated metrics}


\noindent Table \ref{tab:psnr_ssim_faces} shows the PSNR, SSIM  \cite{SSIM2004}  and Consistency scores for 16$\times$16 $\rightarrow$ 128$\times$128 face super-resolution. \modelname outperforms PULSE and FSRGAN on PSNR and SSIM while underperforming the regression baseline.  Previous work~\cite{chen2018fsrnet,dahl2017pixel,menon2020pulse} observed that these conventional automated evaluation measures do not correlate well with human perception when the input resolution is low and the magnification factor is large. This is not surprising because these metrics tend to penalize any synthetic high-frequency detail that is not perfectly aligned with the target image.
Since generating perfectly aligned high-frequency details, \eg~the exact same hair strands in \Figref{fig:64x_512x_faces} and identical leopard spots in
\Figref{fig:64x_256x_natural_images}, is almost impossible, PSNR and SSIM tend to prefer MSE regression-based techniques that are extremely conservative with high-frequency details.
This is further confirmed in \tabref{tab:imagenet_quantitative} for ImageNet super-resolution ($64\!\times\!64 \to 256\!\times\!256$) where the outputs of \modelname achieve higher sample quality scores (FID and IS), but worse PSNR and SSIM than regression. 

\vspace{0.1cm}

\noindent {\bf Consistency:} As a measure of the consistentcy of the super-resolution outputs, we compute MSE between the downsampled outputs and the low resolution inputs. Table \ref{tab:psnr_ssim_faces} shows that \modelname achieves the best consistency error beating PULSE and FSRGAN by a significant margin slightly outperforming even the regression baseline. This result demonstrates the key advantage of \modelname over state of the art GAN based methods as they do not require any auxiliary objective function in order to ensure consistency with the low resolution inputs.

\vspace{0.1cm}

\noindent{\bf Classification Accuracy:} Table \ref{tab:cas_scores} compares our 4$\times$ natural image super-resolution models with previous work in terms of object classification on low-resolution images. We mirror the evaluation setup of \cite{sajjadi2017enhancenet,zhang2018image} and apply 4$\times$ super-resolution models to 56$\times$56
center crops from the validation set of ImageNet. Then, we report classification error based on a pre-trained ResNet-50 \cite{he2016deep}. Since, our super-resolution models are trained on the task of 64$\times$64 $\rightarrow$ 256$\times$256, we use bicubic interpolation to resize the input 56$\times$56 to 64$\times$64, then we apply 4$\times$ super-resolution, followed by resizing back to 224$\times$224. SR3 outperforms existing methods by a large margin on top-1 and top-5 classification errors, demonstrating high perceptual quality of SR3 outputs. The Regression model achieves strong performance compared to existing methods demonstrating the strength of our baseline model. However, SR3 significantly outperforms Regression re-affirming the limitation of conventional metrics such as PSNR and SSIM.  

\vspace*{-0.08cm}
\begin{table}[h]
    \centering
    \small
\scalebox{.95}{    
    \begin{tabular}{l@{\hspace{.2cm}}c@{\hspace{.2cm}}c@{\hspace{.2cm}}c@{\hspace{.2cm}}c}
    \toprule
    \bfseries Metric & \bfseries PULSE \cite{menon2020pulse} & \bfseries FSRGAN \cite{chen2018fsrnet} & \bfseries Regression & \bfseries SR3 \\
    \midrule
    \textbf{PSNR} $\uparrow$ & 16.88 & 23.01 & {\bf 23.96} & 23.04 \\
    \textbf{SSIM} $\uparrow$ & 0.44  & 0.62  & {\bf 0.69} & 0.65 \\
    \midrule
    \textbf{Consistency} $\downarrow$ & 161.1 & 33.8 & 2.71 & {\bf 2.68} \\
    \bottomrule
    \end{tabular}
    }
    \vspace*{0.1cm}
    \caption{PSNR \& SSIM on  16$\times$16 $\rightarrow$ 128$\times$128 face super-resolution.
    Consistency measures MSE ($\times 10^{-5}$) between the low-resolution inputs and the down-sampled super-resolution outputs.}
    \vspace*{-0.1cm}
    \label{tab:psnr_ssim_faces}
\end{table}

\begin{table}[h]
    \centering
    \begin{small}
    \begin{tabular}{lcccc}
    \toprule
    \bfseries Model &  \bfseries FID $\downarrow$ & \bfseries IS $\uparrow$  & \bfseries PSNR $\uparrow$  & \bfseries SSIM $\uparrow$  \\
    \midrule
    Reference & 1.9 & 240.8 & - & - \\
    \midrule
    Regression & 15.2 & 121.1 & \textbf{27.9} & \textbf{0.801} \\
    SR3 &  \textbf{5.2} & \textbf{180.1} & 26.4 & 0.762 \\
    \bottomrule
    \end{tabular}
    \end{small}
    \caption{Performance comparison between SR3 and Regression baseline on natural image super-resolution using standard metrics computed on the ImageNet validation set.}
\vspace*{-0.1cm}
\label{tab:imagenet_quantitative}
\end{table}

\begin{table}[h]
    \centering
    \small
\scalebox{.89}{    
    \begin{tabular}{lcccccccccc}
    \toprule
    \bfseries Method & \bfseries Top-1 Error & \bfseries Top-5 Error \\
    \midrule
    Baseline & 0.252 & 0.080  \\
    \midrule
    DRCN \cite{kim2016deeply} & 0.477 & 0.242  \\
    FSRCNN \cite{dong2016accelerating} & 0.437 & 0.196  \\
    PsyCo \cite{PrezPellitero2016PSyCoMS} & 0.454 & 0.224  \\
    ENet-E \cite{sajjadi2017enhancenet} & 0.449 & 0.214  \\
    RCAN \cite{zhang2018image} & 0.393 & 0.167  \\
    \midrule
    Regression & 0.383 & 0.173 \\
    SR3 & \textbf{0.317} &  \textbf{0.120} \\
    \bottomrule
    \end{tabular}
    }
    \vspace*{0.08cm}
    \caption{Comparison of classification accuracy scores for 4$\times$ natural image super-resolution on the first 1K images from the ImageNet Validation set.}
    \vspace*{-0.4cm}
    \label{tab:cas_scores}
\end{table}


\begin{figure*}[t]
\vspace*{-0.25cm}
\setlength{\tabcolsep}{2pt}
\begin{center}
\begin{tabular}{cccccc}
{\small Bicubic} & {\small FSRGAN \cite{chen2018fsrnet}}  & {\small PULSE \cite{menon2020pulse}} &  {\small Regression} & {\small \modelname} 
\\
{\includegraphics[width=0.12\textwidth]{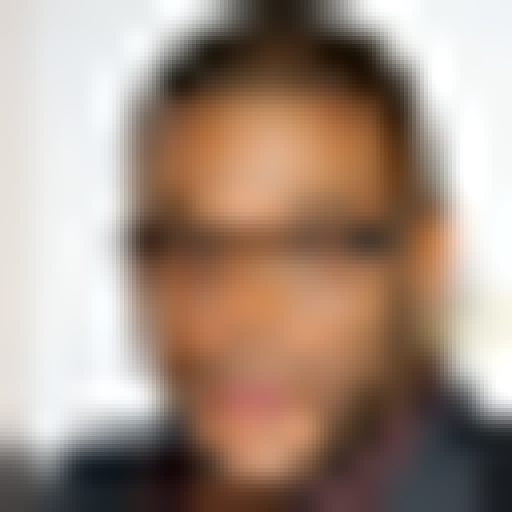}} &
{\includegraphics[width=0.12\textwidth]{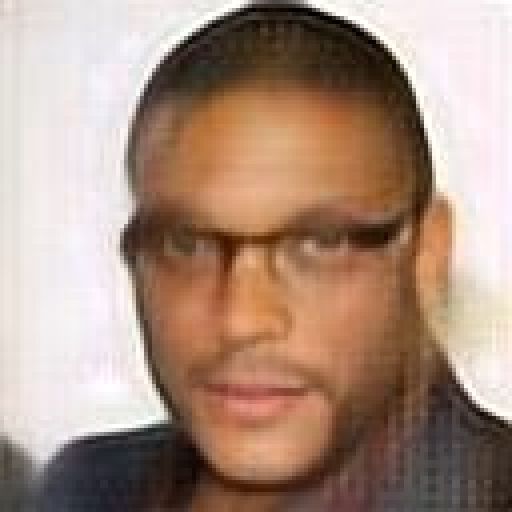}} &
{\includegraphics[width=0.12\textwidth]{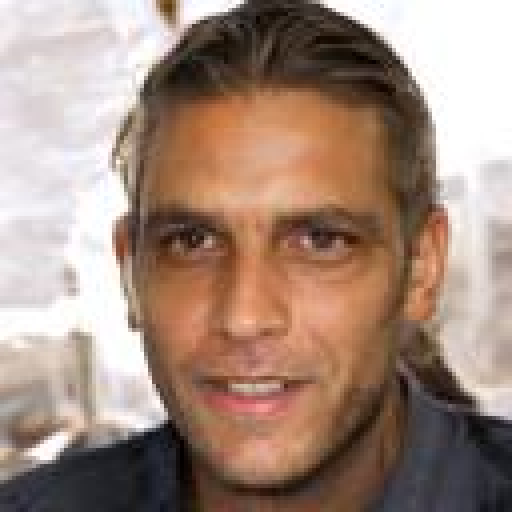}} &
{\includegraphics[width=0.12\textwidth]{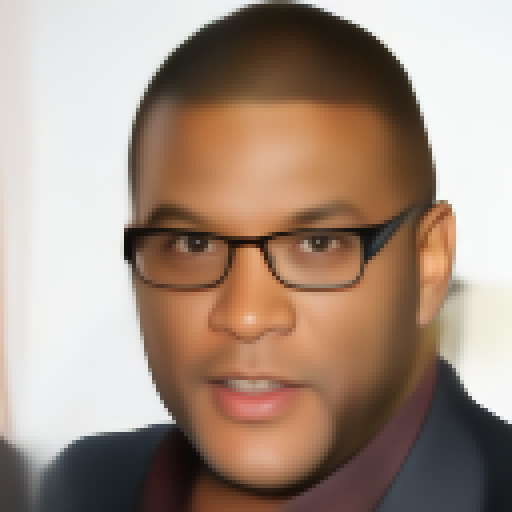}} &
{\includegraphics[width=0.12\textwidth]{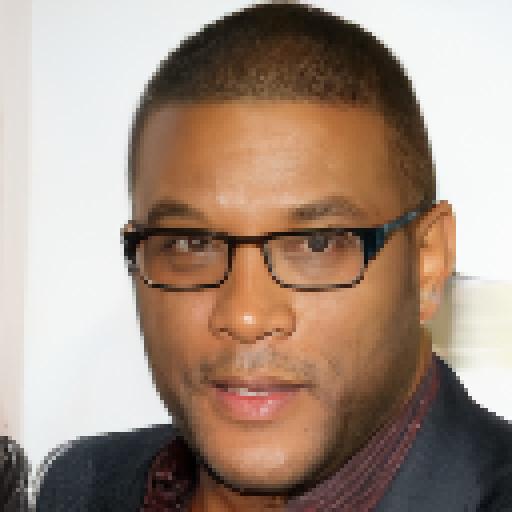}} &
\\

{\includegraphics[width=0.12\textwidth]{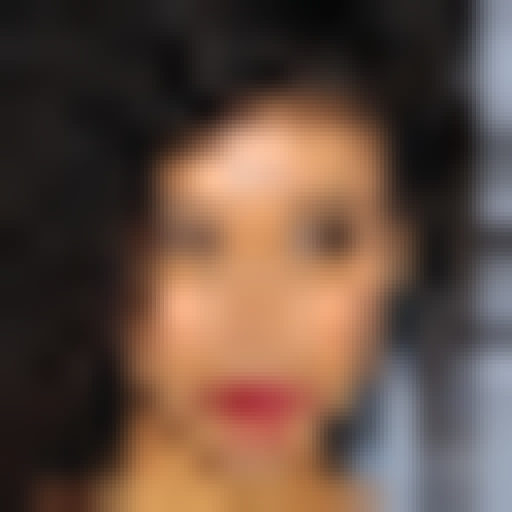}} &
{\includegraphics[width=0.12\textwidth]{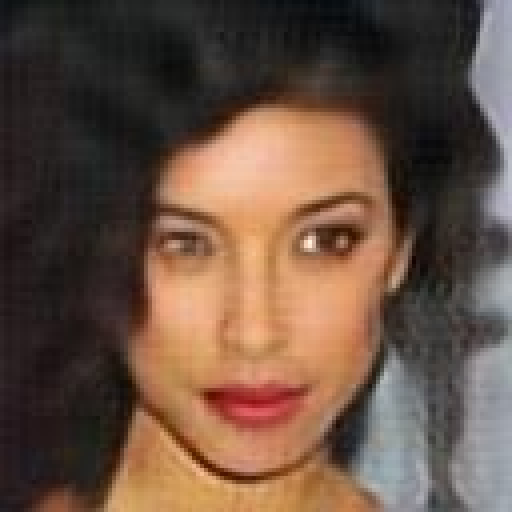}} &
{\includegraphics[width=0.12\textwidth]{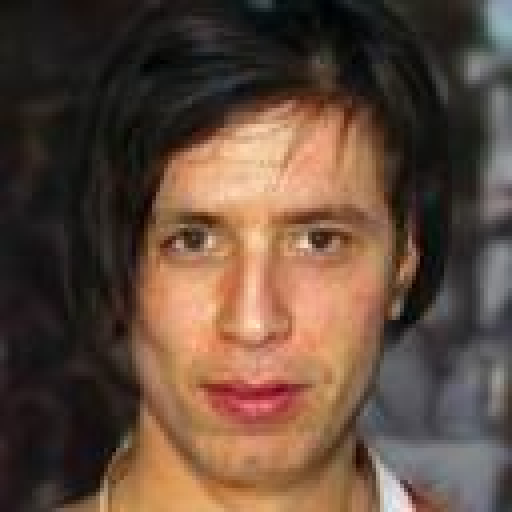}} &
{\includegraphics[width=0.12\textwidth]{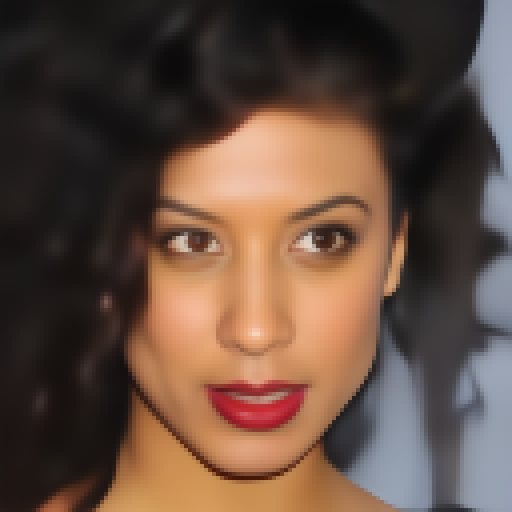}} &
{\includegraphics[width=0.12\textwidth]{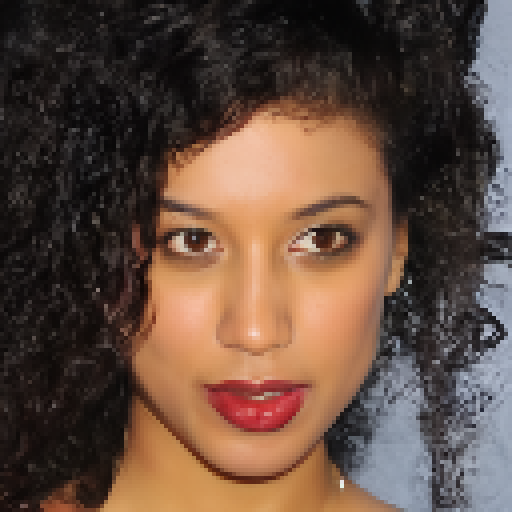}} &
\end{tabular}
\end{center}
\vspace*{-0.5cm}
\caption{Comparison of different methods on the 
16$\times$16 $\rightarrow$ 128$\times$128
face super-resolution task. Reference image has not been included because of privacy concerns. Additional results in Appendix \ref{fig:16x_128x_faces_with_baselines2}. 
\vspace*{-0.3cm}}
\label{fig:16x_128x_faces_with_baselines}
\end{figure*}


% ...

\begin{figure}[t]
\small
\begin{center}
{Fool rates ($3$ sec display w/ inputs, $16\!\times\!16 \to 128\!\times\!128$)} \\
\includegraphics[width=0.44\textwidth]{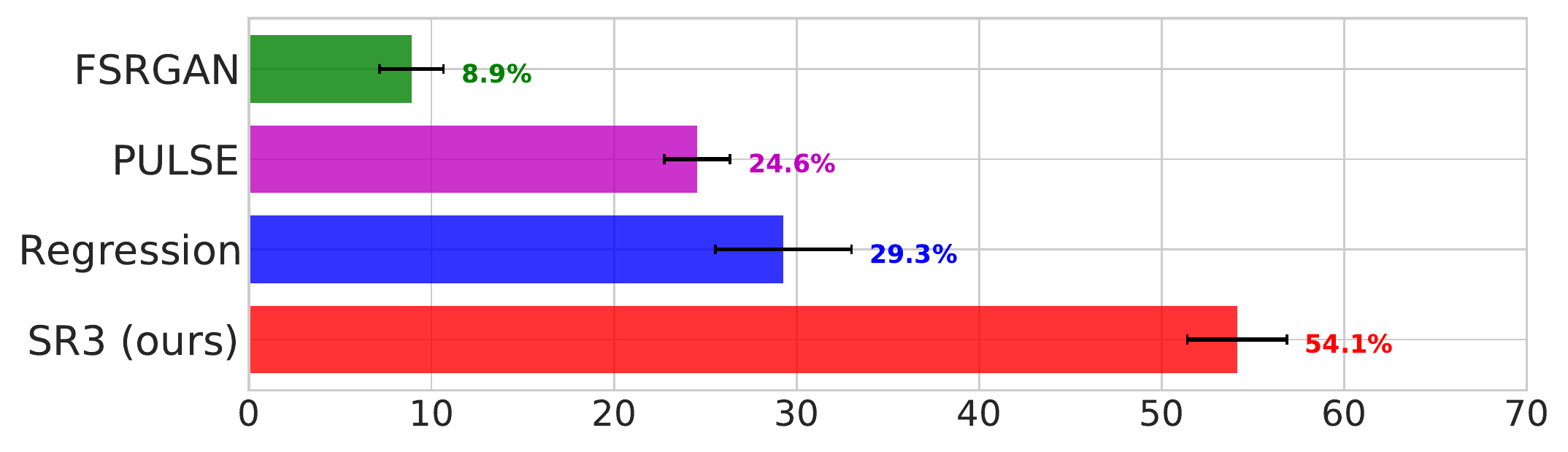}\\[.1cm]
{Fool rates ($3$ sec display w/o inputs, $16\!\times\!16 \to 128\!\times\!128$)}\\
\includegraphics[width=0.44\textwidth]{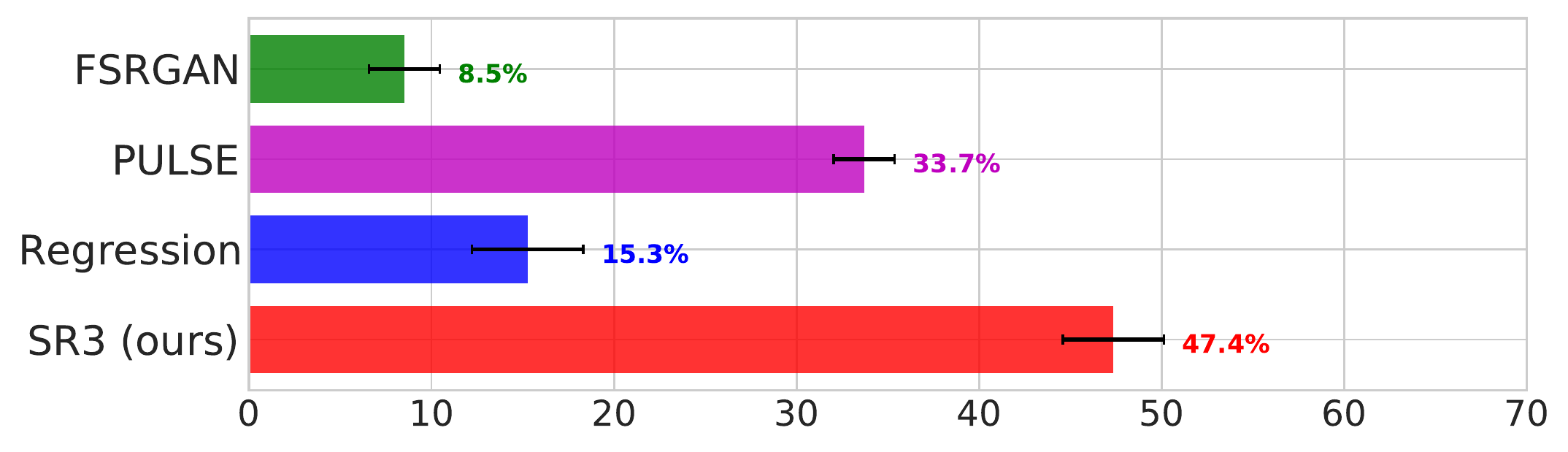}
\end{center}
\vspace*{-0.5cm}
\caption{Face super-resolution human fool rates (higher is better, photo-realistic samples yield a fool rate of 50\%).
Outputs of 4 models are compared against ground truth. (top) Subjects are shown low-resolution inputs.
(bottom) Inputs are not shown.}
\vspace{-.1cm}
\label{fig:face_foolrate}
\end{figure}

\subsubsection{Human Evaluation (2AFC)}

In this work, we are primarily interested in photo-realistic super-resolution with large magnification factors. Accordingly, we resort to direct human evaluation. 
While mean opinion score (MOS) is commonly used to measure image quality in this context, forced choice pairwise comparison has been found to be a more reliable 
method for such subjective quality assessments~\cite{mantiuk2012comparison}. Furthermore, standard MOS studies do not capture consistency between low-resolution inputs and high-resolution outputs.
We use a 2-alternative forced-choice
(2AFC) paradigm to measure how well humans can discriminate true
images from those generated from a model.
In Task-1 subjects were shown a low resolution input in between two high-resolution images, one being the real image (ground truth),
and the other generated from the model.
Subjects were asked \textit{``Which of the two images is a better high quality version of the low resolution image in the middle?"}
This task takes into account both image quality and consistency with the low resolution input. 
Task-2 is similar to Task-1, except that 
the low-resolution image was not shown, so subjects only had to select the image that was more photo-realistic.  
They were asked  \textit{``Which image would you guess is from a camera?"}
Subjects viewed images for 3 seconds before responding, in both tasks. The source code for human evaluation can be found here \footnote{\url{https://tinyurl.com/sr3-human-eval-code}}.

The subject {\it fool rate} is the fraction of trials on which a subject selects the model output over ground truth. 
Our fool rates for each model are based on 50 subjects, each of whom were shown 50 of the 100 images in the test set.
Figure \ref{fig:face_foolrate} shows the fool rates for  Task-1 (top), and for Task-2 (bottom).
In both experiments, the fool rate of \modelname is close to 50\%, indicating that \modelname produces images that are both photo-realistic and faithful to the low-resolution inputs.
We find similar fool rates over a wide range of viewing durations up to 12 seconds.

The fool rates for FSRGAN and PULSE in Task-1 are lower than the Regression baseline and \modelname.
We speculate that the PULSE optimization has failed to converge to high resolution images sufficiently close to the inputs. 
Indeed, when asked solely about image quality in Task-2 (Fig.\ \ref{fig:face_foolrate} (bottom)), the PULSE fool rate increases significantly.

The fool rate for the Regression baseline is lower in Task-2 (Fig.\ \ref{fig:face_foolrate} (bottom)) than Task-1. 
The regression model tends to generate images that are blurry, but nevertheless faithful to the low resolution input. 
We speculate that in Task-1, given the inputs, subjects are influenced by consistency, while in Task-2, ignoring consistency, they instead focus on image sharpness.
SR3 and Regression samples used for human evaluation are provided here \footnote{\url{https://tinyurl.com/sr3-outputs}}.
\begin{figure}[t]
\small
\begin{center}
{Fool rates ($3$ sec display w/ inputs, $64\!\times\!64 \to 256\!\times\!256$)} \\
\includegraphics[width=0.45\textwidth]{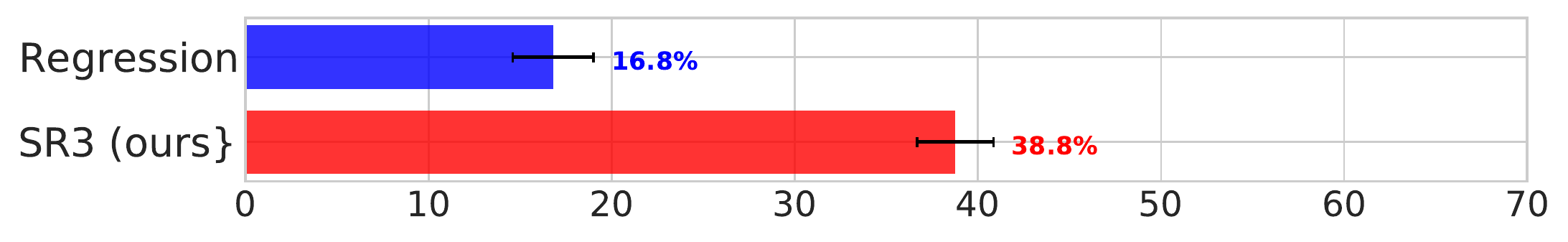}\\[.2cm]
{Fool rates ($3$ sec display w/o inputs, $64\!\times\!64 \to 256\!\times\!256$)}\\
\includegraphics[width=0.45\textwidth]{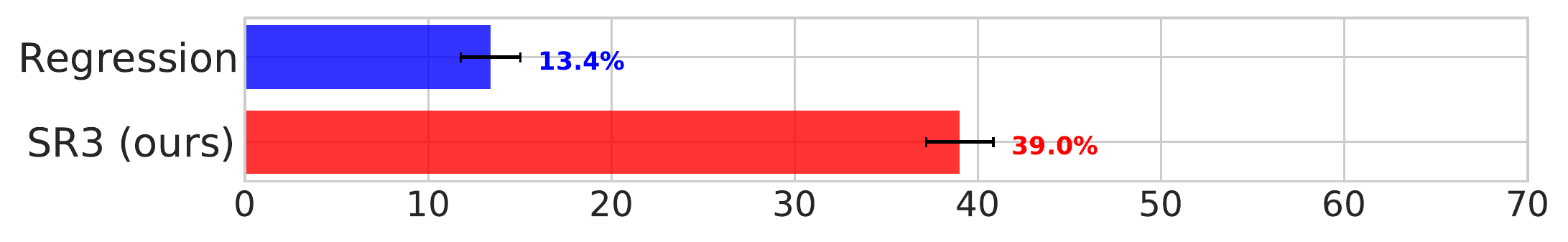} \\[.1cm]
\end{center}
\vspace*{-0.5cm}
\caption{ImageNet super-resolution fool rates (higher is better, photo-realistic samples yield a fool rate of 50\%).
SR3 and Regression outputs are compared against ground truth. (top) Subjects are shown low-resolution inputs.
(bottom) Inputs are not shown.}
\label{fig:imgenet_foolrate}
\end{figure}

We conduct similar human evaluation studies on
natural images comparing \modelname and the regression baseline on ImageNet. Figure \ref{fig:imgenet_foolrate} shows the results for Task-1 (top) and task-2 (bottom). In both tasks with natural images, SR3 achieves a human subject fool rate is close to 40\%. Like the face image experiments in Fig.\ \ref{fig:face_foolrate}, here again we find that the Regression baseline yields a lower fool rate in Task-2, where the low resolution image is not shown. Again we speculate that this is a result of a somewhat simpler task (looking at 2 rather than 3 images), and the fact that subjects can focus solely on image artifacts, such as blurriness, without having to worry about consistency between model output and the low resolution input.



To further appreciate the experimental results it is
useful to visually compare outputs of different models 
on the same inputs, as in 
Figure \ref{fig:16x_128x_faces_with_baselines}.
FSRGAN exhibits distortion in face region and struggles with generating glasses properly (e.g., top row). It also fails to recover
texture details in the hair region (see bottom row).
PULSE often produces images that differ significantly from the 
input image, both in the shape of the face and the background, and sometimes in gender too (see bottom row) presumably
due to failure of the optimization to find a sufficiently good minima.
As noted above, our Regression baseline produces results consistent to the input, however they are 
typically quite blurry.  By comparison, the \modelname results are consistent with the input and contain more detailed image structure.

\subsection{Cascaded High-Resolution Image Synthesis}
\vspace*{-0.1cm}

We study {\em cascaded} image generation, where SR3 models at different scales are chained together with unconditional generative models, enabling high-resolution image synthesis. 
Cascaded generation allows one to train different models in parallel, and each model in the cascade solves a
simpler task, requiring fewer parameters and less computation for training. 
Inference with cascaded models is also more efficient, especially for iterative refinement models. 
With cascaded generation we found it effective to use 
more refinement steps at low-resolutions, and fewer  steps at higher resolutions. This was much more 
efficient than generating directly at high resolution
without sacrificing image quality.


\begin{figure}[t]
\setlength{\tabcolsep}{2pt}
\begin{center}
\begin{tabular}{cc}
\includegraphics[width=0.23\textwidth]{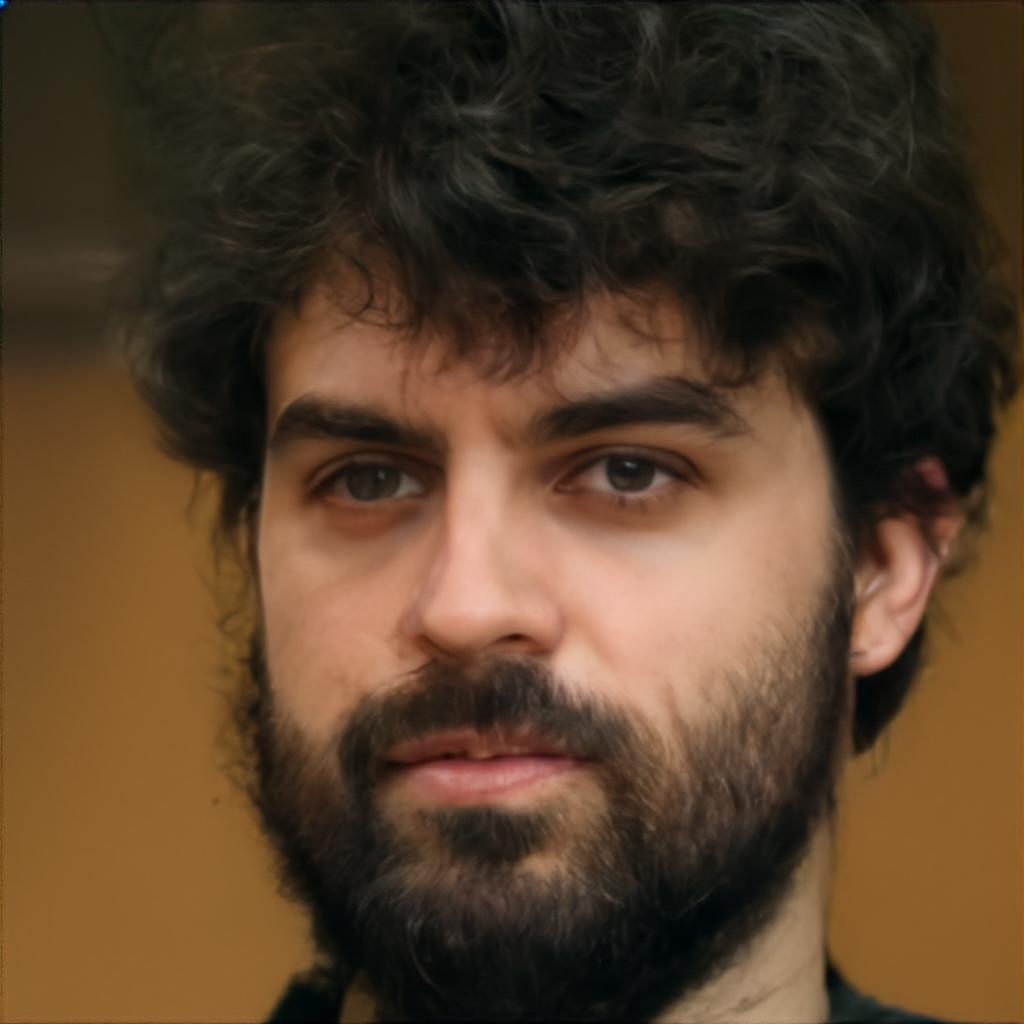} &
\includegraphics[width=0.23\textwidth]{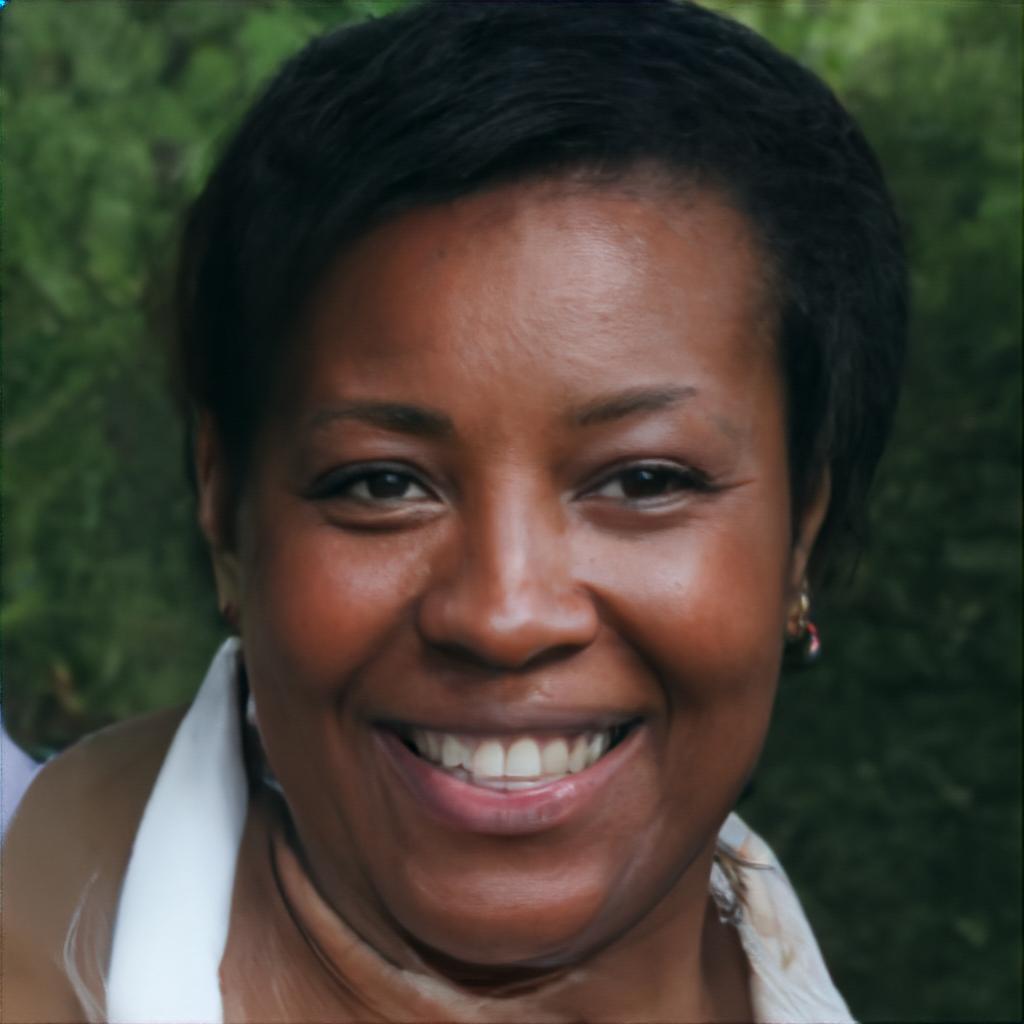} \\

\includegraphics[width=0.23\textwidth]{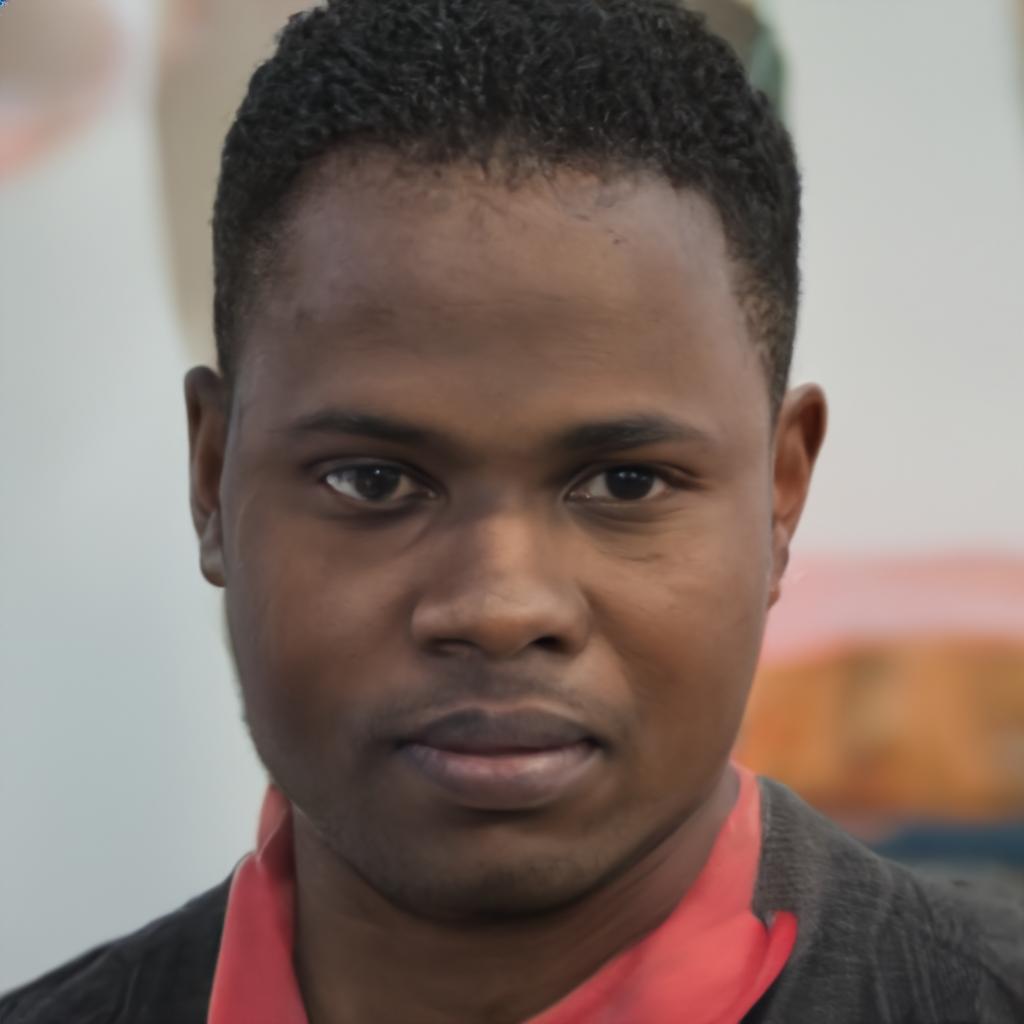} &
\includegraphics[width=0.23\textwidth]{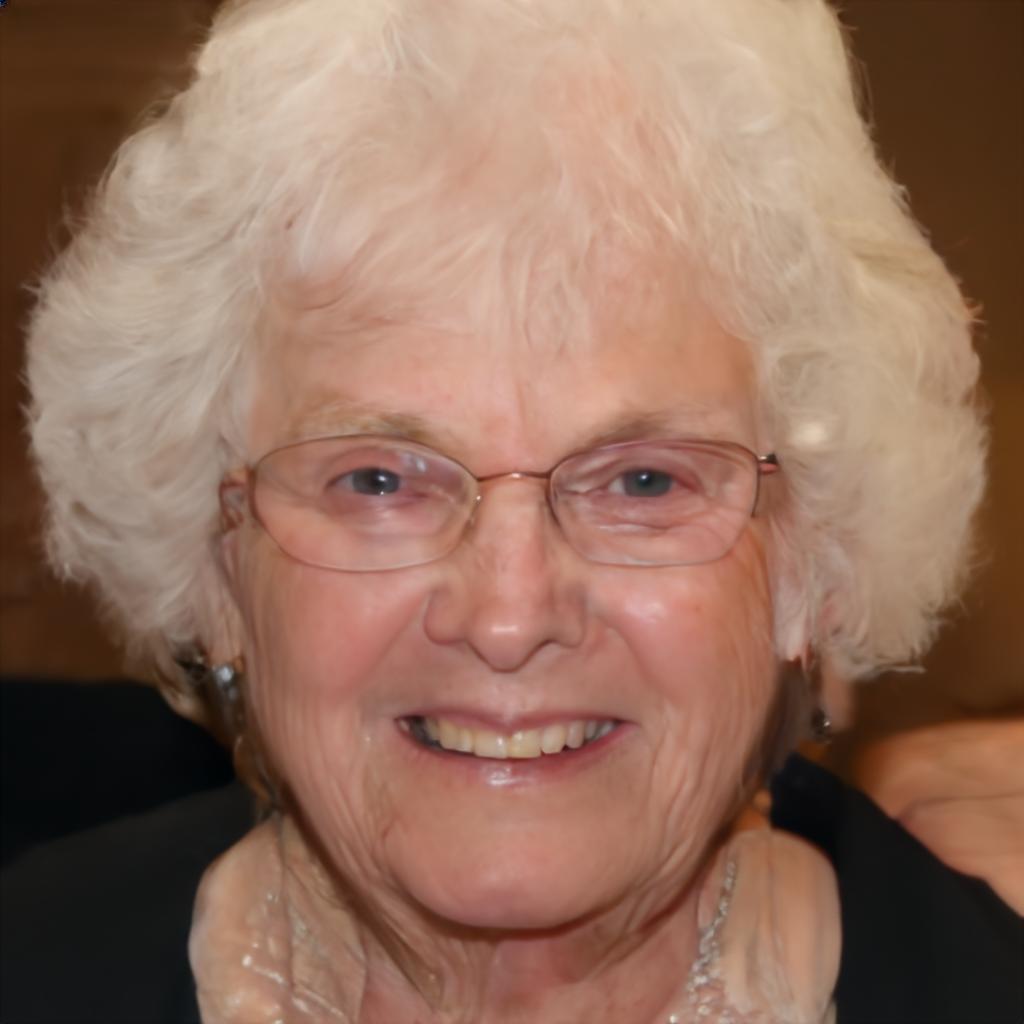} \\
\end{tabular}

\begin{tabular}{cccc}
\includegraphics[width=0.112\textwidth]{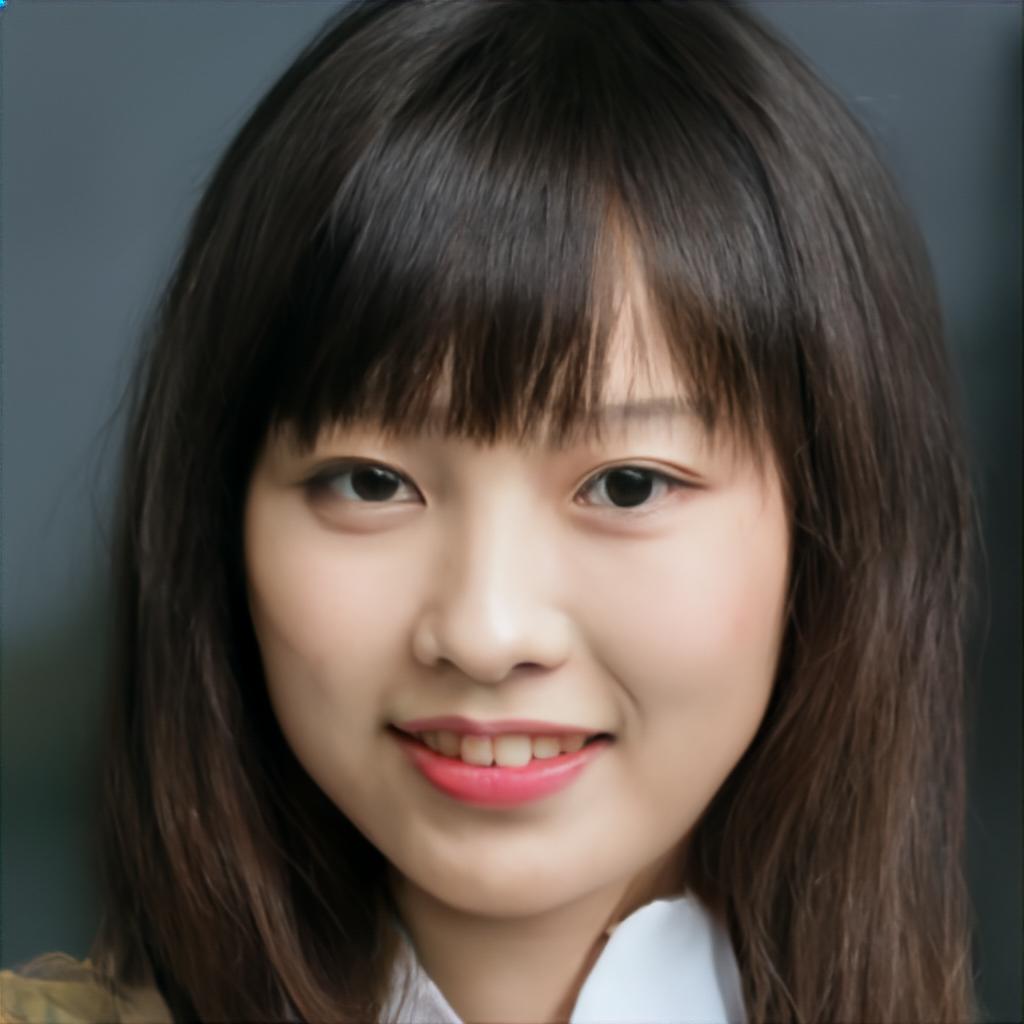} &
\includegraphics[width=0.112\textwidth]{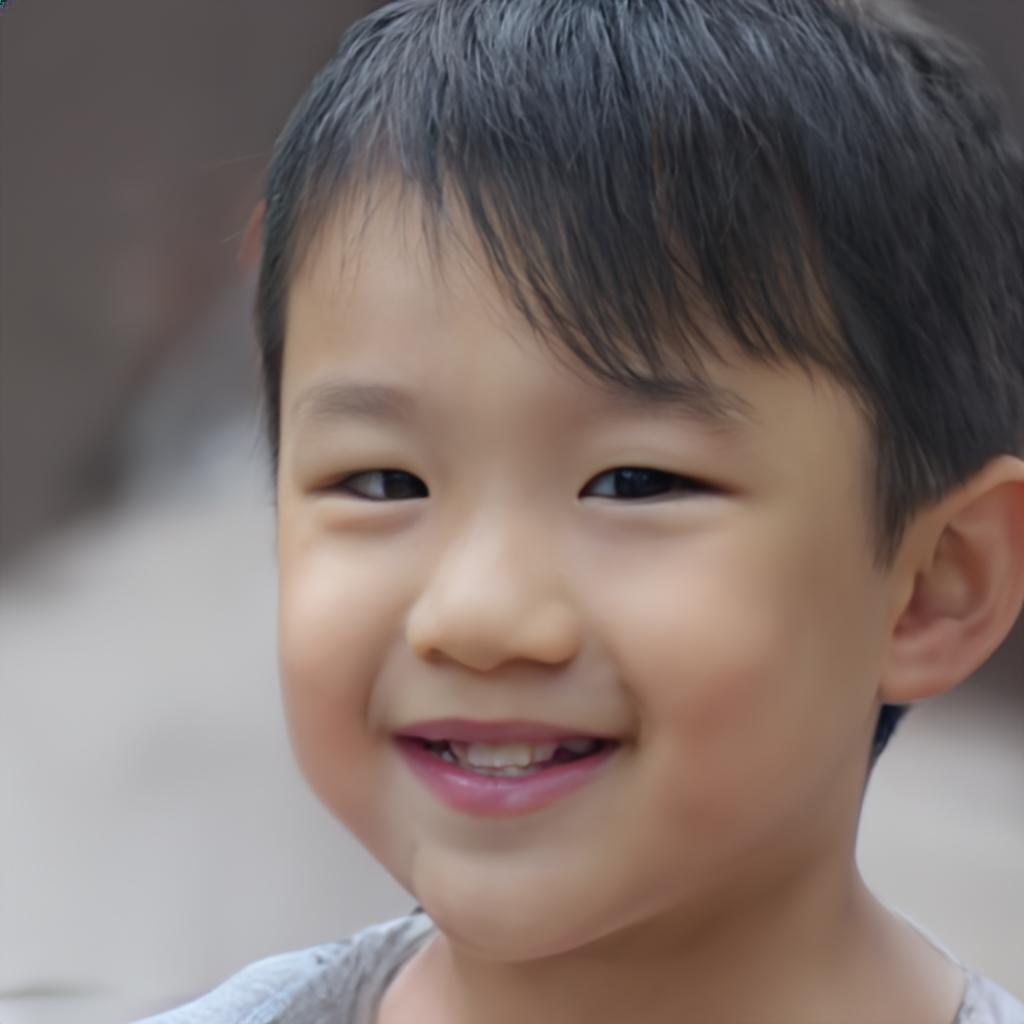} &
\includegraphics[width=0.112\textwidth]{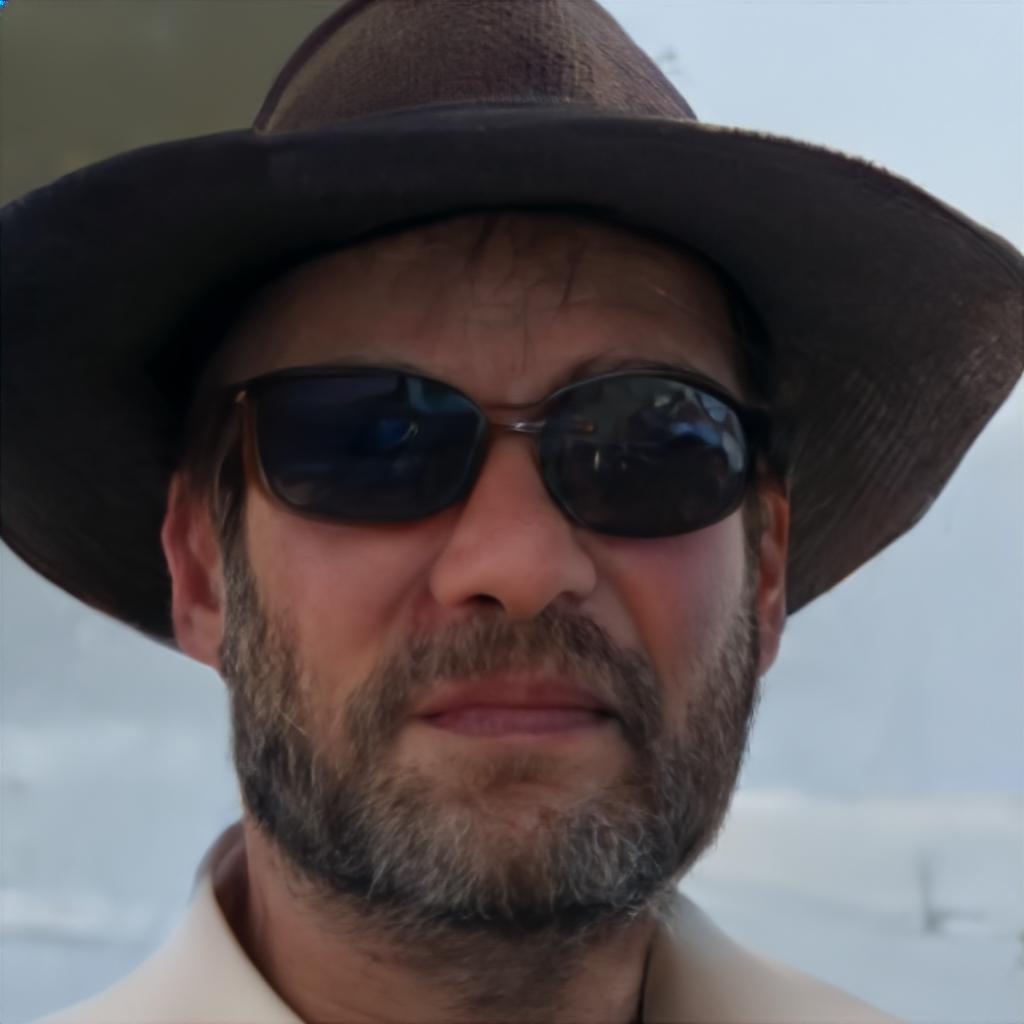} &
\includegraphics[width=0.112\textwidth]{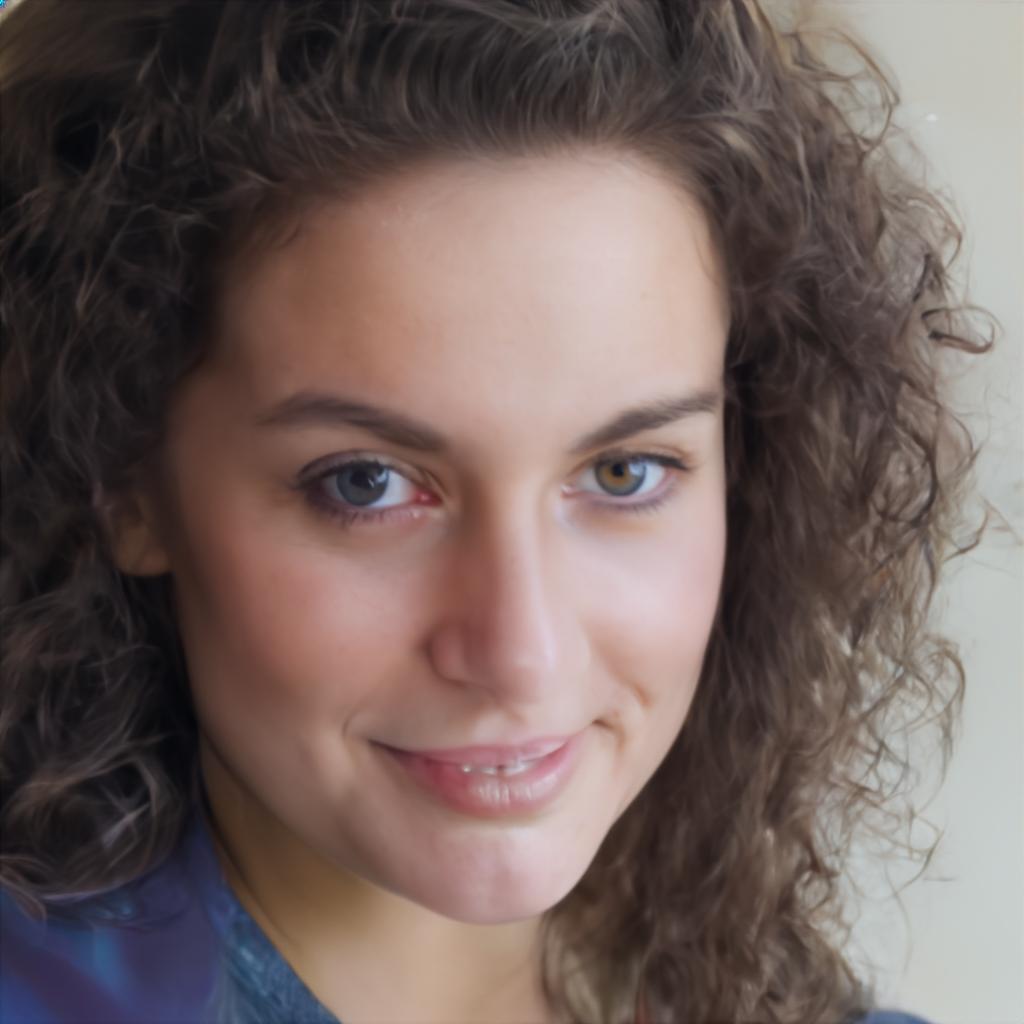} \\

\includegraphics[width=0.112\textwidth]{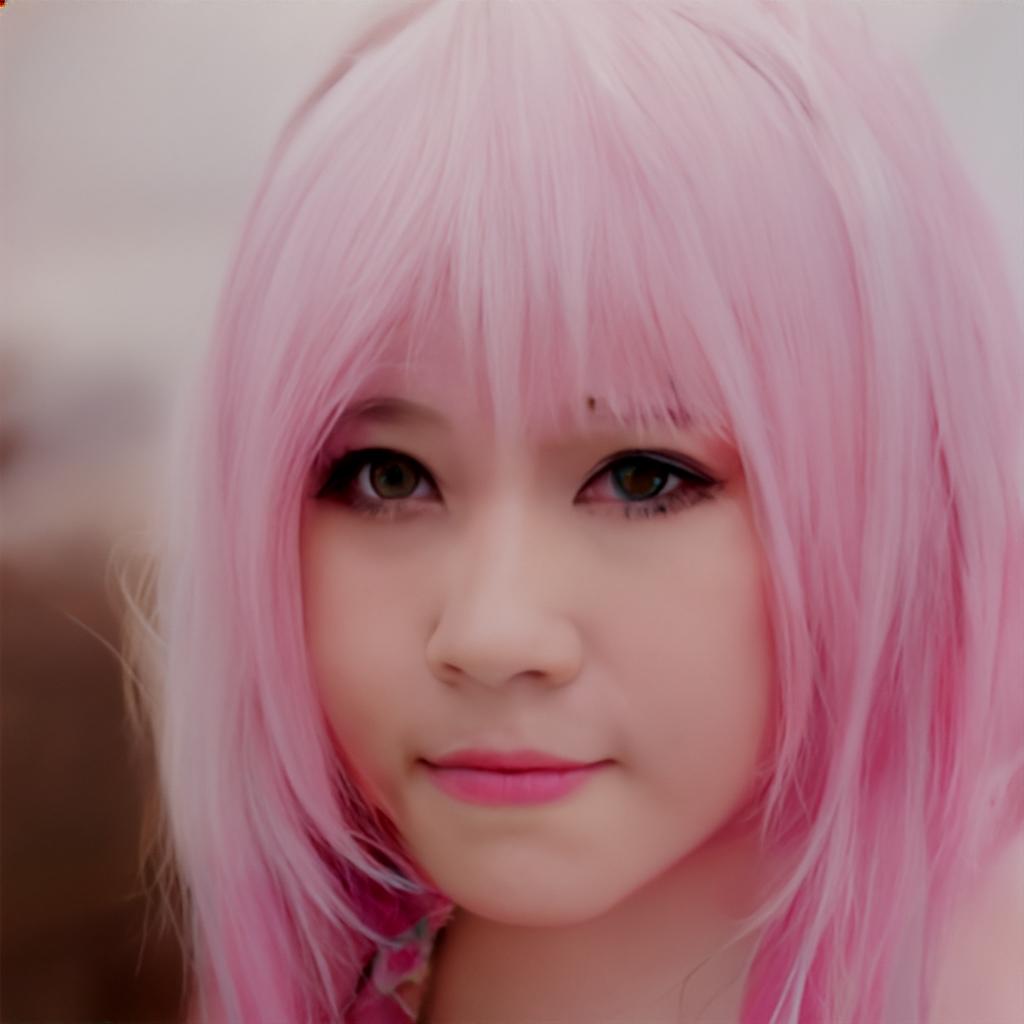} &
\includegraphics[width=0.112\textwidth]{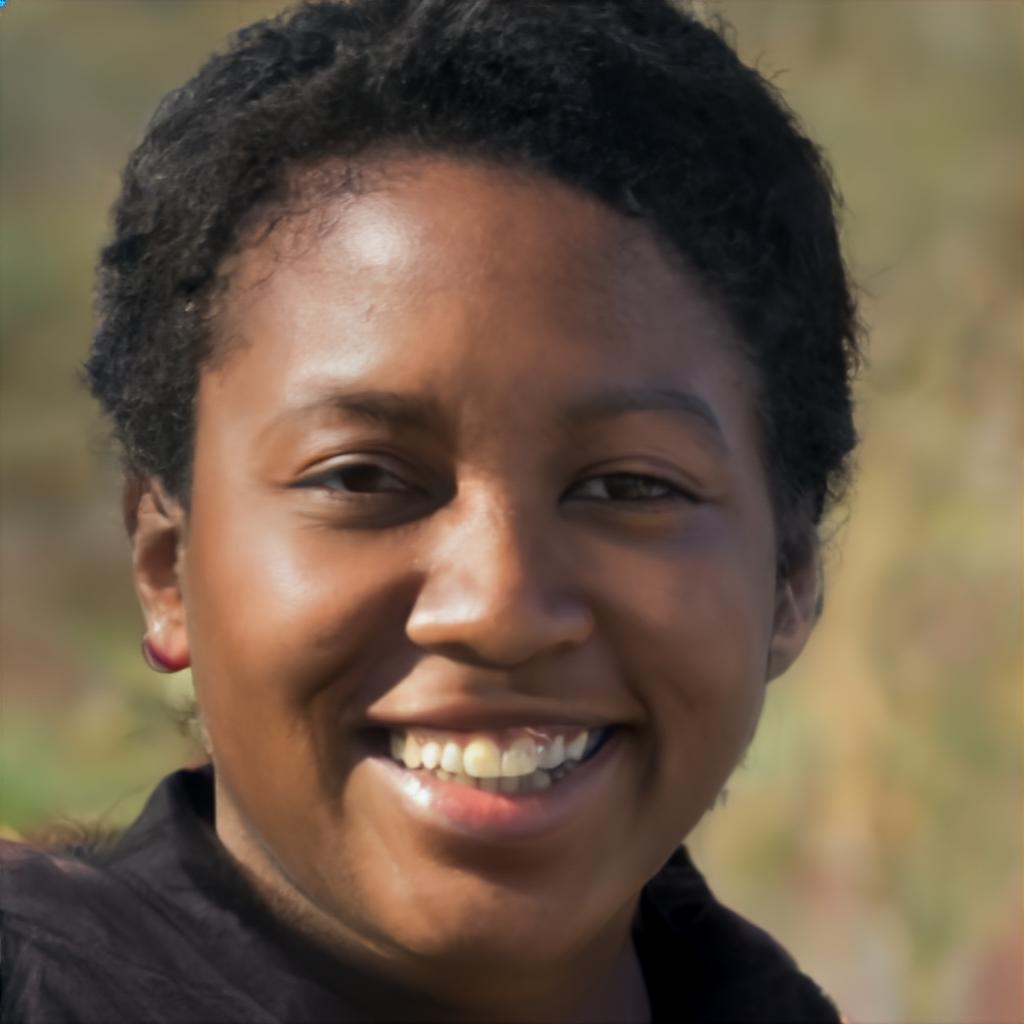} &
\includegraphics[width=0.112\textwidth]{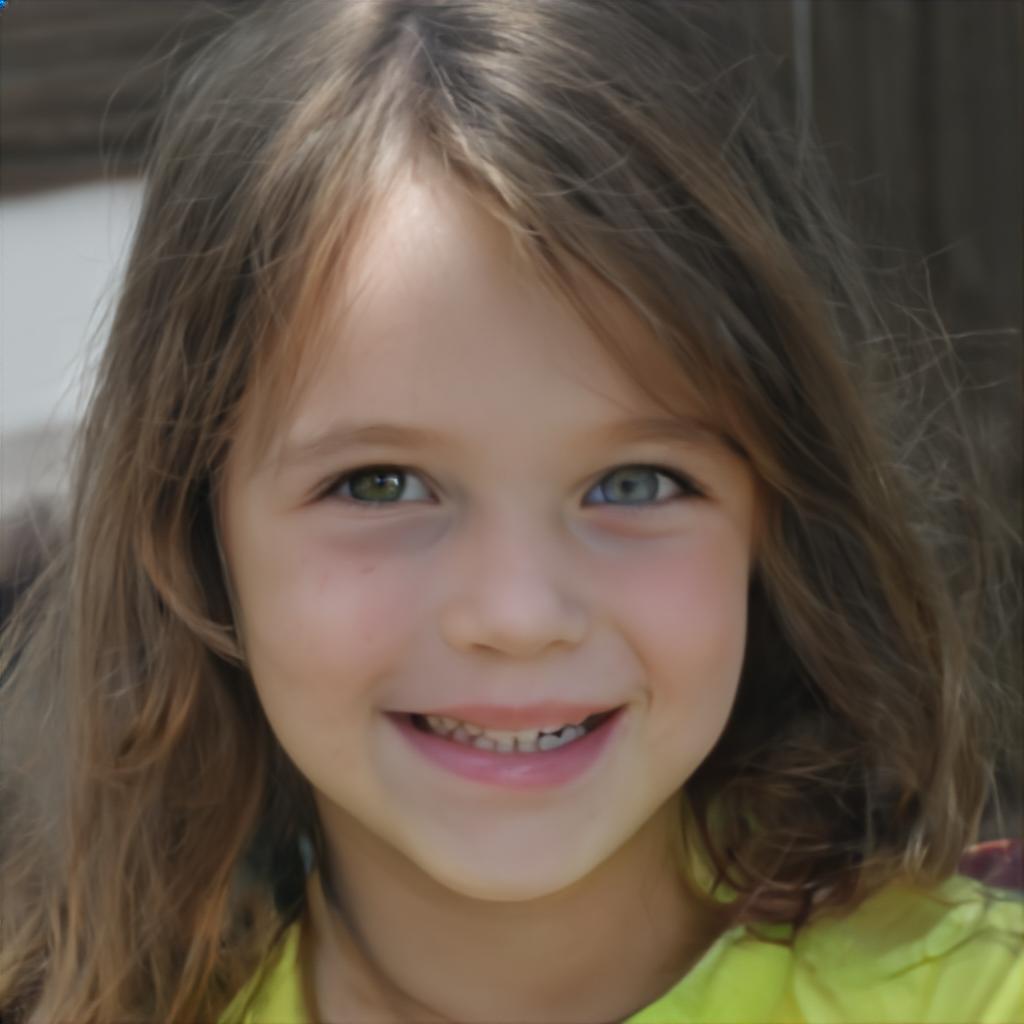} &
\includegraphics[width=0.112\textwidth]{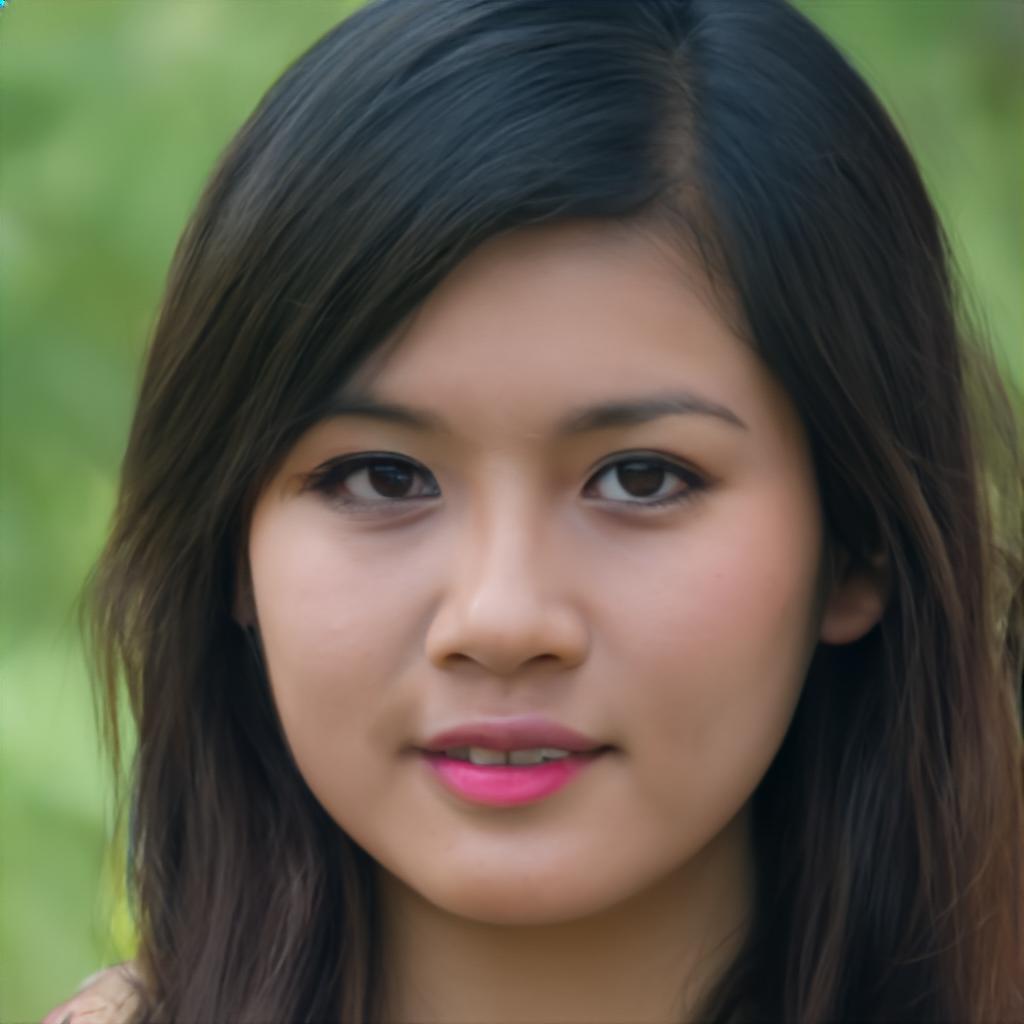} 
\end{tabular}
\end{center}
\vspace*{-0.5cm}
\caption{Synthetic 1024$\times$1024 face images. We first sample from an unconditional 64$\times$64 diffusion model, then pass the samples through two 4$\times$ \modelname models, \ie~ 64$\times$64 $\rightarrow$ 256$\times$256 $\rightarrow$ 1024$\times$1024. 
Additional samples in Appendix \ref{fig:1024x_cascade2}, \ref{fig:1024x_cascade3} and \ref{fig:1024x_cascade4}.
 }
\label{fig:1024x_cascade}
\vspace*{-0.35cm}
\end{figure}

\begin{figure}[t]
\setlength{\tabcolsep}{1.25pt}
\begin{center}
\begin{tabular}{cccc}
\includegraphics[width=0.115\textwidth]{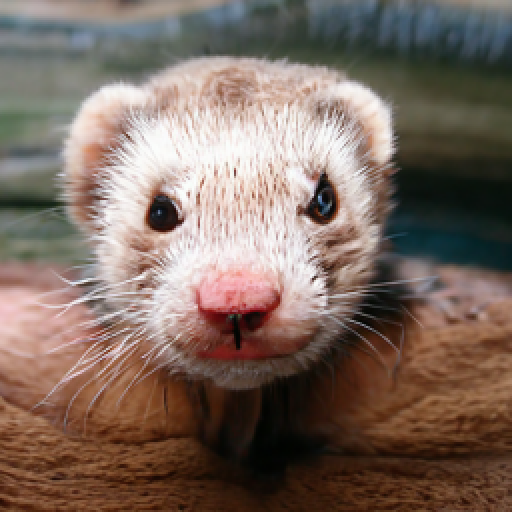} & 
\includegraphics[width=0.115\textwidth]{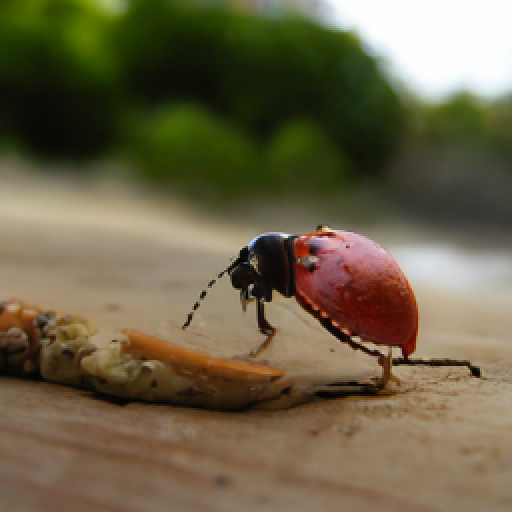} & 
\includegraphics[width=0.115\textwidth]{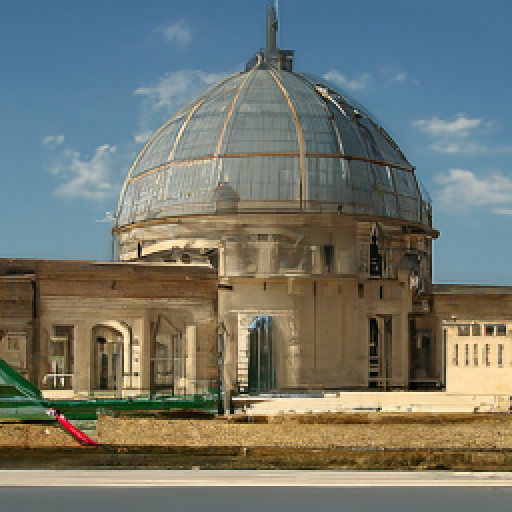} & 
\includegraphics[width=0.115\textwidth]{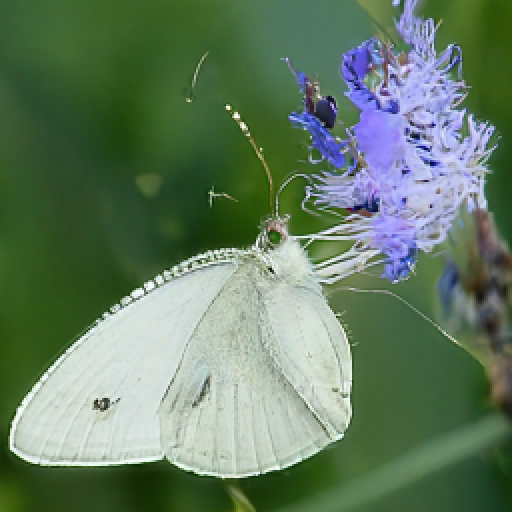} \\

\includegraphics[width=0.115\textwidth]{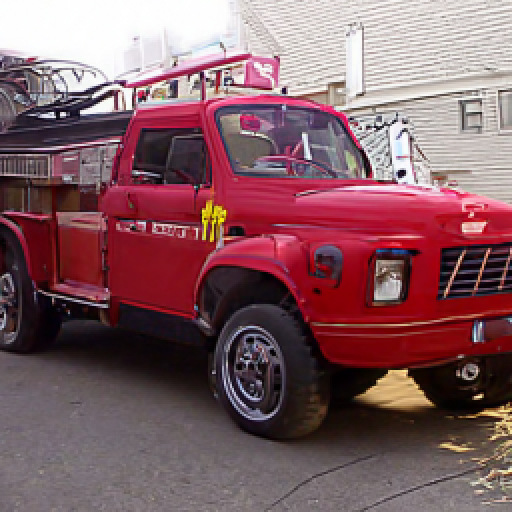} &
\includegraphics[width=0.115\textwidth]{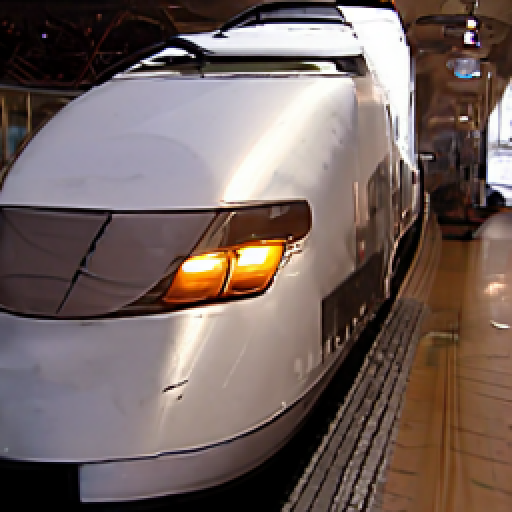} & 
\includegraphics[width=0.115\textwidth]{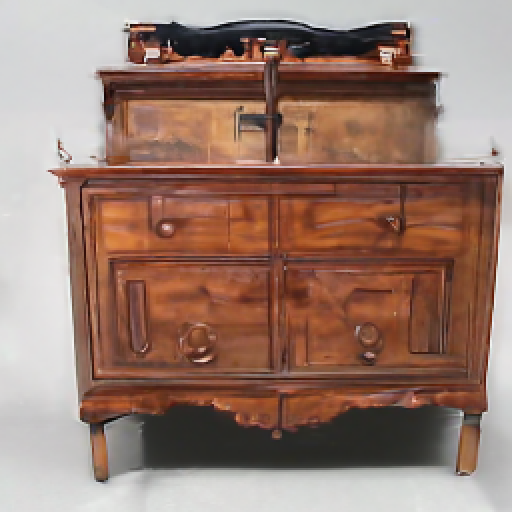} & 
\includegraphics[width=0.115\textwidth]{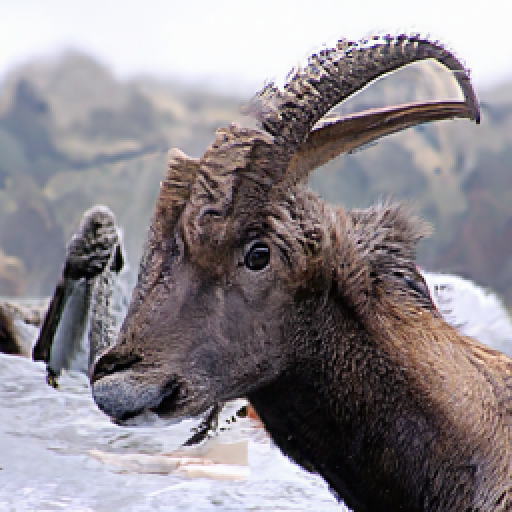} \\

\includegraphics[width=0.115\textwidth]{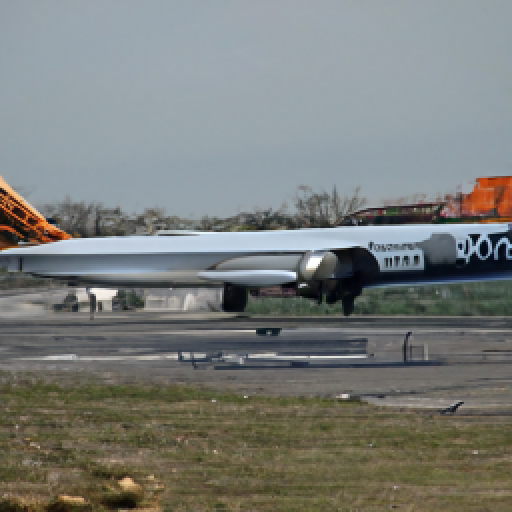} &
\includegraphics[width=0.115\textwidth]{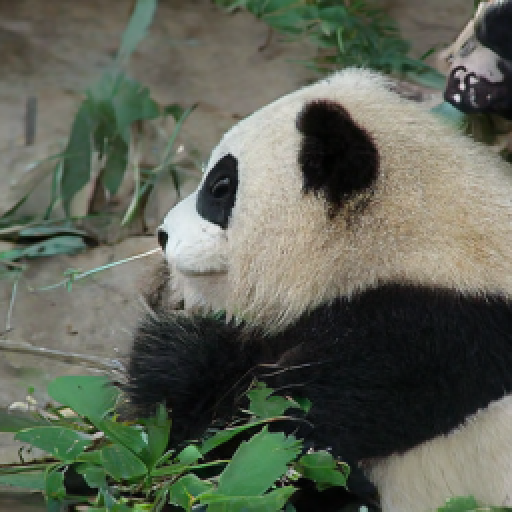} &
\includegraphics[width=0.115\textwidth]{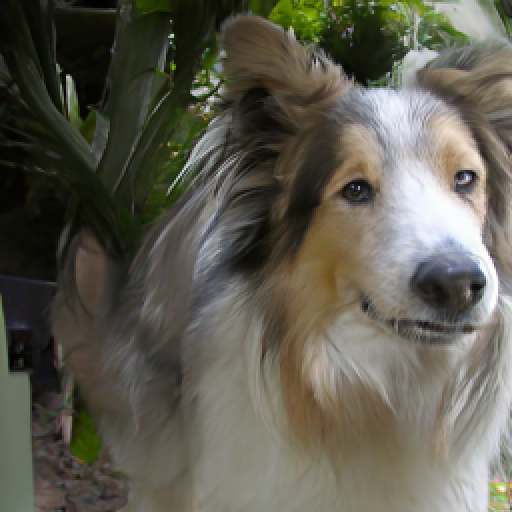} & 
\includegraphics[width=0.115\textwidth]{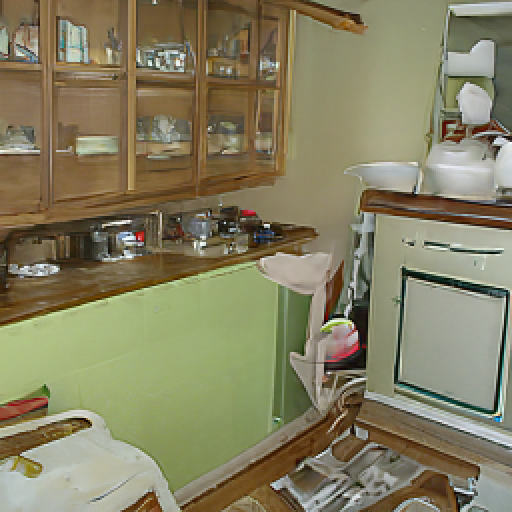} \\

\includegraphics[width=0.115\textwidth]{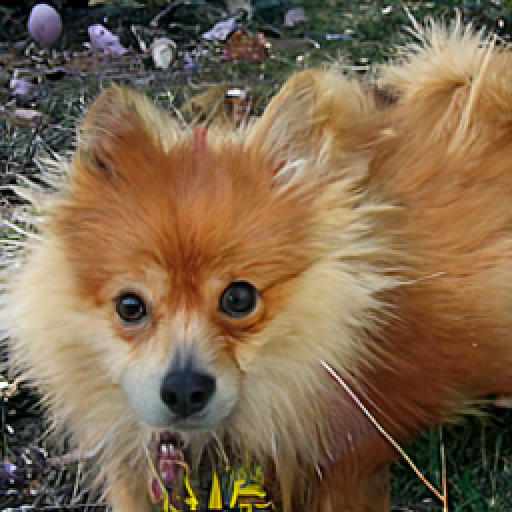} &
\includegraphics[width=0.115\textwidth]{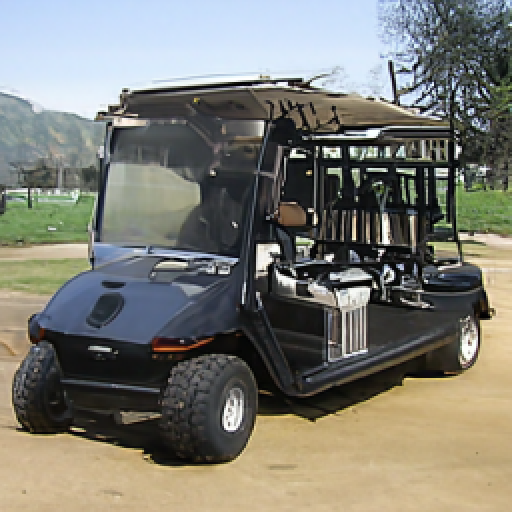} & 
\includegraphics[width=0.115\textwidth]{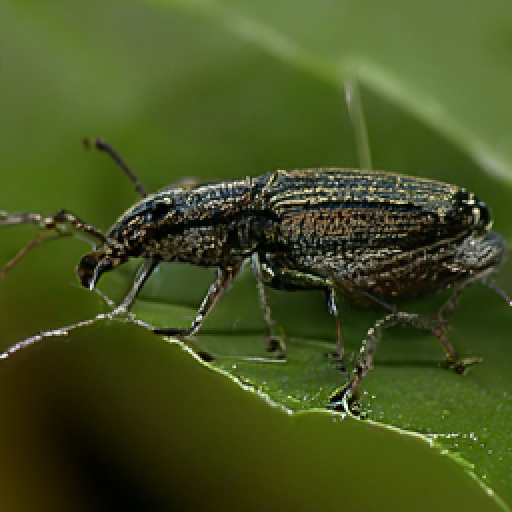} & 
\includegraphics[width=0.115\textwidth]{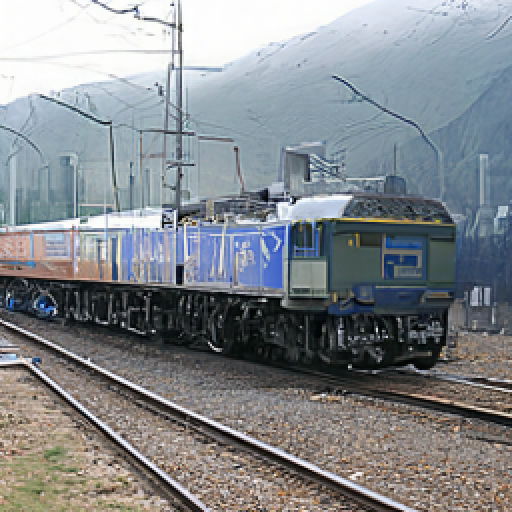} \\

\end{tabular}
\end{center}
\vspace*{-0.6cm}
\caption{Synthetic 256$\times$256 ImageNet images. We first draw a random label, then sample a 64$\times$64 image from a class-conditional diffusion model, and apply a 4$\times$ \modelname model to obtain 256$\times$256 images. Additional samples in Appendix \ref{fig:imagenet_256x_montage2} and \ref{fig:imagenet_256x_montage3}.}
\vspace*{-0.2cm}
\label{fig:imagenet_256x_montage}
\end{figure}

We train a DDPM~\cite{ho2020denoising} model for
unconditional $64 \!\times\! 64$ face generation.
Samples from this model are then fed to two 4$\times$
\modelname models, up-sampling to $256^2$ and then to $1024^2$ pixels.
Synthetic high-resolution face samples are shown in \Figref{fig:1024x_cascade}.
In addition, we train an Improved~DDPM~\cite{nichol2021improved} model on class-conditional $64 \!\times\! 64$ ImageNet, and we pass its generated samples to a 4$\times$ \modelname model yielding $256^2$ pixels. The 4$\times$ \modelname model is not conditioned on the class label. 
See \Figref{fig:imagenet_256x_montage} for representative samples.


\begin{table}[t]
    \vspace{0.1cm}
    \centering
    \begin{small}
    \begin{tabular}{lc}
    \toprule
    \bfseries Model &  \bfseries FID-50k \\
    \midrule
    \bfseries Prior Work \\
    \quad VQ-VAE-2 \cite{razavi2019generating} & 38.1 \\
    \quad BigGAN   (Truncation 1.0) \cite{brock2018large} & 7.4 \\
    \quad BigGAN   (Truncation 1.5) \cite{brock2018large} & 11.8 \\
    
    \bfseries Our Work \\
    \quad \modelname (Two Stage) & 11.3 \\
    \bottomrule
    \end{tabular}
    \end{small}
    \vspace{-.1cm}
    \caption{FID scores for class-conditional 256$\times$256 ImageNet.}
    \vspace{-.4cm}
    \label{tab:fid_imagenet_uncond}
\end{table}

Table \ref{tab:fid_imagenet_uncond} reports FID scores for the resulting class-conditional ImageNet samples. 
Our 2-stage model improves on VQ-VAE-2~\cite{razavi2019generating}, is comparable to deep BigGANs \cite{brock2018large} at truncation factor of 1.5 but underperforms them a truncation factor of $1.0$. 
Unlike BigGAN, our diffusion models do not provide a knob to control sample quality {\em vs.} sample diversity, and finding ways to do so is interesting avenue for future research.
Nichol and Dhariwal~\cite{nichol2021improved} concurrently trained cascaded generation models using super-resolution conditioned on class labels (our super-resolution is not conditioned on class labels),
and observed a similar trend in FID scores. 
The effectiveness of cascaded image generation indicates that \modelname models
are robust to the precise distribution of inputs (\ie~the specific form of anti-aliasing and downsampling). 

\textbf{Ablation Studies:} Table \ref{tab:fid_imagenet_ablation} shows  ablation studies on our $64\times64 \rightarrow 256\times256$ Imagenet SR3 model. In order to improve the robustness of the SR3 model, we experiment with use of data augmentation while training. Specifically, we trained the model with varying amounts of Gaussian Blurring noise added to the low resolution input image. No blurring is applied during inference. We find that this has a siginificant impact, improving the FID score roughly by 2 points. We also explore the choice of $L_p$ norm for the denoising objective (Equation \ref{eq:loss}). We find that $L_1$ norm gives slightly better FID scores than $L_2$.  

\begin{table}[t]
    \vspace{0.1cm}
    \centering
    \begin{small}
    \begin{tabular}{lc}
    \toprule
    \bfseries Model &  \bfseries FID-50k \\
    \midrule
    \bfseries Training with Augmentation \\
    \quad SR3  &  13.1 \\
    \quad SR3 (w/ Gaussian Blur)  & 11.3  \\
    \midrule
    \bfseries Objective $L_p$ Norm \\
    \quad SR3 ($L_2$)  &  11.8 \\
    \quad SR3 ($L_1$)  & 11.3  \\
    \bottomrule
    \end{tabular}
    \end{small}
    \vspace{-.1cm}
    \caption{Ablation study on SR3 model for class-conditional 256$\times$256 ImageNet.}
    \vspace{-.4cm}
    \label{tab:fid_imagenet_ablation}
\end{table}

\vspace{-.1cm}
\section{Discussion and Conclusion}
\vspace{-.1cm}

Bias is an important problem in all generative models. SR3 is no different, and suffers from bias issues. While in theory, our log-likelihood based objective is mode covering (\eg~unlike some GAN-based objectives), we believe it is likely our diffusion-based models drop modes. We observed some evidence of mode dropping, the model consistently generates nearly the same image output during sampling (when conditioned on the same input). We also observed the model to generate very continuous skin texture in face super-resolution, dropping moles, pimples and piercings found in the reference. SR3 should not be used for any real world super-resolution tasks, until these biases are thoroughly understood and mitigated.

In conclusion, SR3 is an approach to image super-resolution via iterative refinement. SR3 can be used in a cascaded fashion to generate high resolution super-resolution images, as well as unconditional samples when cascaded with a unconditional model. We demonstrate SR3 on face and natural image super-resolution at high resolution and high magnification ratios (\eg~64$\times$64$\rightarrow$256$\times$256 and 256$\times$256$\rightarrow$1024$\times$1024). SR3 achieves a human fool rate close to 50\%, suggesting photo-realistic outputs.

\vspace{-.1cm}
\section*{Acknowledgements}
\vspace{-.2cm}
We thank Jimmy Ba, Adji Bousso Dieng, Chelsea Finn, Geoffrey Hinton, Natacha Mainville, Shingai Manjengwa, and Ali Punjani for providing their face images on which we demonstrate the face SR3 results. We thank Ben Poole, Samy Bengio and the Google Brain team for research discussions and technical assistance. We also thank authors of \cite{menon2020pulse} for generously providing us with baseline super-resolution samples for human evaluation.




{\small
\bibliographystyle{ieee_fullname}
\bibliography{egbib}
}

\clearpage
\appendix

\onecolumn
\renewcommand\thefigure{\thesection.\arabic{figure}}
\renewcommand{\thetable}{A.\arabic{table}}
\counterwithin{figure}{section}
\counterwithin{table}{section}

\vspace*{0.5cm}
\begin{center}
    {\Large {\bf Appendix}}
\end{center}


\noindent
This appendix includes further details about the architecture of the models used for super-resolution.
It also formulates the training objective in terms of a variation bound and in terms of denoising score-matching.
We then provide additional experimental results to complement those in the main body of the paper.

\vspace*{0.25cm}
\section{Task Specific Architectural Details}

\begin{figure}[h]
\begin{center}
\begin{tikzpicture}[
scale=0.75, every node/.style={scale=0.75},
dot/.style = {circle, fill, minimum size=#1,
              inner sep=0pt, outer sep=0pt},
dot/.default = 2pt]

\tikzstyle{layer} = [rectangle, thick, minimum width=4cm, minimum height=0.5cm, align=center, draw=black]
\tikzstyle{arrow} = [thick,->]


\node[fill=green!20] (input_high_res1) at (-1.9,0) [layer,thick,minimum width=0.1cm,minimum height=5cm] {};
\node[label=below:{$\vx \cdot \vy_t$}, text width=0.0cm, minimum height=5cm] (temp) at (-1.775,0)  {};
\node[fill=blue!20] (input_low_res1) at (-1.65,0) [layer,thick,minimum width=0.1cm,minimum height=5cm] {};
\node[fill=red!15, label=below:{$128^2$, 128}] (first_res_d) at (-0.25,0) [layer,thick,minimum width=0.5cm,minimum height=5cm] {};
\draw[arrow] (input_low_res1.east) -- (first_res_d.west); 
\node[fill=red!15, label=below:{$64^2$, 256}] (second_res_d) at (1.2,0) [layer,thick,minimum width=0.75cm,minimum height=3cm] {};
\draw[arrow] (first_res_d.east) -- (second_res_d.west); 
\draw[arrow] (second_res_d.east) -- (1.9,0); 
\node[dot] at (2.1,0) {};
\node[dot] at (2.4,0) {};
\node[fill=red!15, label=below:{$8^2$, 1024}] (last_res_d) at (3.6,0) [layer,thick,minimum width=1.5cm,minimum height=1cm] {};
\draw[arrow] (2.6,0) -- (last_res_d.west); 
\node[fill=red!15, label=below:{$8^2$, 1024}] (first_res_u) at (5.6,0) [layer,thick,minimum width=1.5cm,minimum height=1cm] {};
\draw[arrow] (last_res_d.east) -- (first_res_u.west); 

\node[dot] at (6.8,0) {};
\node[dot] at (7.1,0) {};
\draw[arrow] (first_res_u.east) -- (6.6, 0);

\node[fill=red!15,label=below:{$64^2$, 256}] (second_res_u) at (8.0,0) [layer,thick,minimum width=0.75cm,minimum height=3cm] {};
\draw[arrow] (7.3,0) -- (second_res_u.west);
\node[fill=red!15,label=below:{$128^2$, 128}] (last_res_u) at (9.45,0) [layer,thick,minimum width=0.5cm,minimum height=5cm] {};
\draw[arrow] (second_res_u.east) -- (last_res_u.west);
\node[label=below:$\vy_{t-1}$,fill=blue!20] (output) at (10.85,0) [layer,thick,minimum width=0.15cm,minimum height=5cm] {};
\draw[arrow] (last_res_u.east) -- (output.west);

\draw[arrow] (first_res_d.north) to [out=30,in=150] (last_res_u.north);
\draw[arrow] (second_res_d.north) to [out=30,in=150] (second_res_u.north);

\end{tikzpicture}
\end{center}
\vspace*{-0.45cm}
\caption{Description of the U-Net architecture with skip connections. The low resolution input image $\vx$ is interpolated to the target high resolution, and concatenated with the noisy high resolution image $\vy_t$. We show the activation dimensions for the example task of 16$\times$16 $\rightarrow$ 128$\times$128 super resolution. }
\vspace*{-0.25cm}
\label{fig:architecture}
\end{figure}
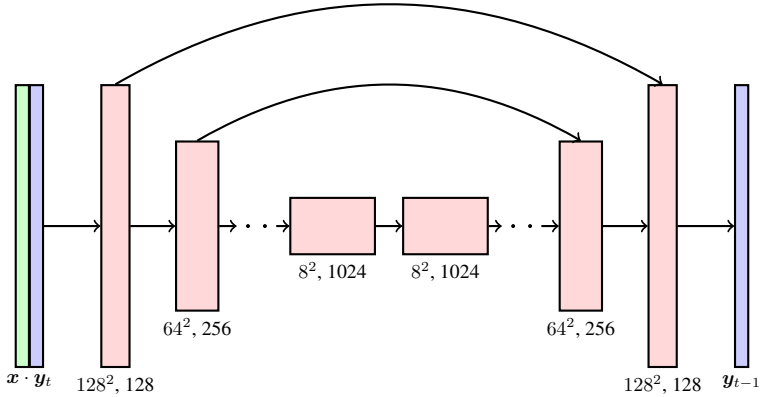

\noindent
Table \ref{tab:task_specific_arch} summarizes the primary architecture details for each super-resolution task. For a particular task, we use the same architecture for both SR3 and Regression models. Figure \ref{fig:architecture} describes our method of conditioning the diffusion model on the low resolution image. We first interpolate the low resolution image to the target high resolution, and then simply concatenate it with the input noisy high resolution image.

\label{sec:task-arch-details}
\begin{table}[h]
    \centering
    \begin{tabular}{c|c|c|c|c}
    \textbf{Task} & \textbf{Channel Dim} & \textbf{Depth Multipliers} & \textbf{\# ResNet Blocks} & \textbf{\# Parameters} \\
    \hline
    16 $\times$ 16 $\rightarrow$ 128 $\times$ 128 & 128 & \{1, 2, 4, 8, 8\} & 3 & 550M\\
    64 $\times$ 64 $\rightarrow$ 256 $\times$ 256 & 128 & \{1, 2, 4, 4, 8, 8\} & 3 & 625M\\
    64 $\times$ 64 $\rightarrow$ 512 $\times$ 512 & 64 & \{1, 2, 4, 8, 8, 16, 16\} & 3 & 625M\\
    256 $\times$ 256 $\rightarrow$ 1024 $\times$ 1024 & 16 & \{1, 2, 4, 8, 16, 32, 32, 32\} & 2 & 150M\\
    \end{tabular}
    \caption{Task specific architecture hyper-parameters for the U-Net model. Channel Dim is the dimension of the first U-Net layer, while the depth multipliers are the multipliers for subsequent resolutions. }
    \label{tab:task_specific_arch}
\end{table}

\vspace*{0.5cm}
\section{Justification of the Training Objective}
\label{sec:justification}

\subsection{A Variational Bound Perspective}

Following Ho \textit{et al.}~\cite{ho2020denoising}, we justify the choice of the training objective in \eqref{eq:loss} for the probabilistic model outlined in \eqref{eq:reverse_process} from a variational lower bound perspective.
If the forward diffusion process is viewed as a fixed approximate posterior to the inference process, one can derive the following variational lower bound on the marginal log-likelihood:
\begin{equation}
\begin{aligned}
    \mathbb{E}_{(\vx,\vy_0)}\log p_\theta(\vy_0 | \vx) \geq 
    \mathbb{E}_{\vx,\vy_0}\mathbb{E}_{q(\vy_{1:T}|\vy_0)}\bigg[ \log p(\vy_T) + \sum_{t \geq 1} \log \frac{p_{\theta} (\vy_{t-1} | \vy_t, \vx)}{q(\vy_t|\vy_{t-1})}  \bigg]~. \label{eq:vlb}
\end{aligned}
\end{equation}

Given the particular parameterization of the inference process outlined above, one can show~\cite{ho2020denoising} that the negative  variational lower bound
can be expressed as the following simplified loss, up to a constant weighting of each term for each time step:
\begin{align}
    \mathbb{E}_{\vx,\vy_0,\veps} \!\sum_{t=1}^T \frac{1}{T} \bigg\lVert \bm{\epsilon} - \bm{\epsilon}_{\theta}(\vx, \sqrt{\gamma_t} \vy_0 + \sqrt{1 - \gamma_t}\bm{\epsilon}, \gamma_t) \bigg\rVert^{2}_{2}
\end{align}
where ${\veps} \sim \stdnormal$. Note that this objective function corresponds to $L_2$ norm in \eqref{eq:loss}, and a characterization of $p(\gamma)$ in terms of 
a uniform distribution over $\{\gamma_1, \ldots, \gamma_T\}$.

\subsection{A Denoising Score-Matching Perspective}

Our approach is also linked to denoising score matching~\cite{hyvarinen2005estimation,vincent2011connection,raphan2011least,saremi-arxiv-2018} for training unnormalized energy functions for density estimation. These methods learn a parametric score function to approximate the gradient of the empirical data log-density. To make sure that the gradient of the data log-density is well-defined,
one often replaces each data point with a Gaussian distribution with a small variance. Song and Ermon~\cite{song-arxiv-2020} advocate for the use of a Multi-scale Guassian mixture as the target density, where each data point is perturbed with different amounts of Gaussian noise, so that Langevin dynamics starting from pure noise can still yield reasonable samples.

One can view our approach as a variant of denoising score matching in which the target density is given by a mixture of
$q(\vty | \vy_0, \gamma) = \mathcal{N}(\vty \,|\, \sqrt{\gamma}\vy_0, 1-\gamma)$ for different values of $\vy_0$ and $\gamma$.
Accordingly, the gradient of data log-density is given by
\begin{equation}
    \frac{\mathrm{d} \log q(\vty \mid \vy_0, \gamma)}{\mathrm{d} \vty} ~=~ -\frac{\vty - \sqrt{\gamma}\vy_0}{\sqrt{1-\gamma}} ~=~ -\veps~,
\end{equation}
which is used as the regression target of our model.

\vspace*{0.25cm}
\section{Additional Experimental Results}

The following figures show more examples of SR3 on faces, natural images, and samples from unconditional
generative models, architectural details, more details about the formulation of the
loss. 
We first show more examples of SR3, augmenting the results shown in Figures \ref{fig:64x_512x_faces}, \ref{fig:64x_256x_natural_images}, \ref{fig:16x_128x_faces_with_baselines}. We then show cascaded generation of $1024\times1024$ face images, followed by more unconditional samples for $1024\times1024$ faces, and  $256\times256$ class conditional ImageNet samples.

\vspace*{0.15cm}
\noindent
\newpage
\begin{center}
{\large \bf Face Super-Resolution \ 64$\times$64  $\rightarrow$ 512$\times$512}\\
\end{center}

\begin{figure}[H]
\vspace*{-0.4cm}
\setlength{\tabcolsep}{2pt}
\begin{center}
\begin{tabular}{cccc}
{\small Bicubic} & {\small Regression} & {\small \modelname (ours)} & {\small Reference} \\
{\includegraphics[width=0.19\textwidth]{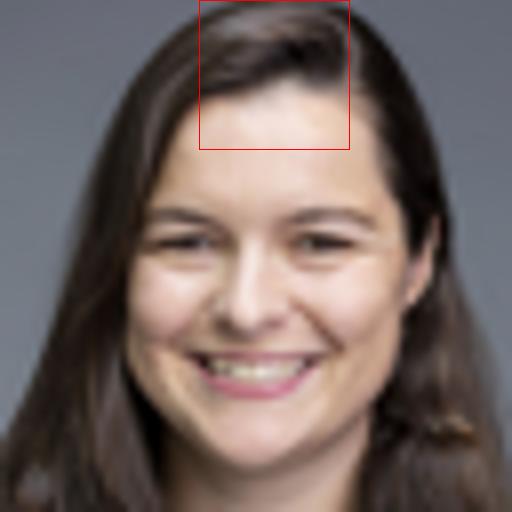}}&
{\includegraphics[width=0.19\textwidth]{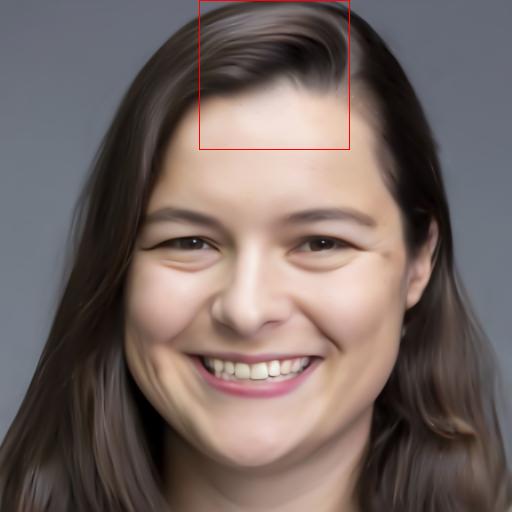}}&
{\includegraphics[width=0.19\textwidth]{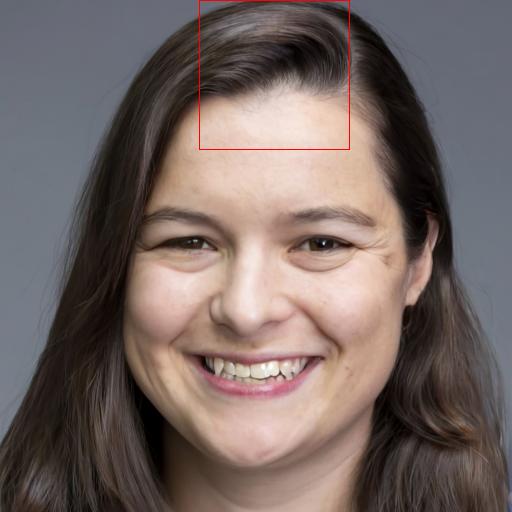}} &
{\includegraphics[width=0.19\textwidth]{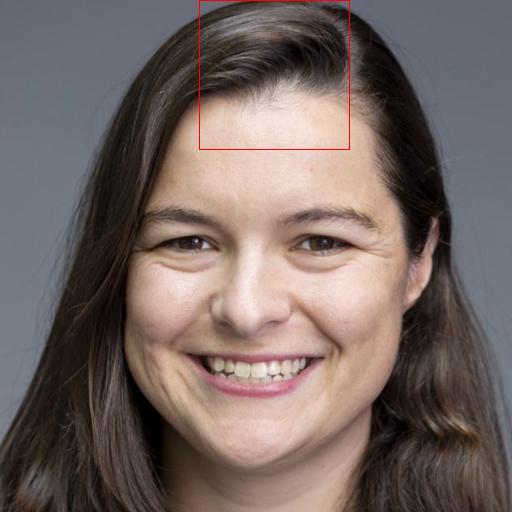}} \\

\medskip

{\includegraphics[width=0.19\textwidth]{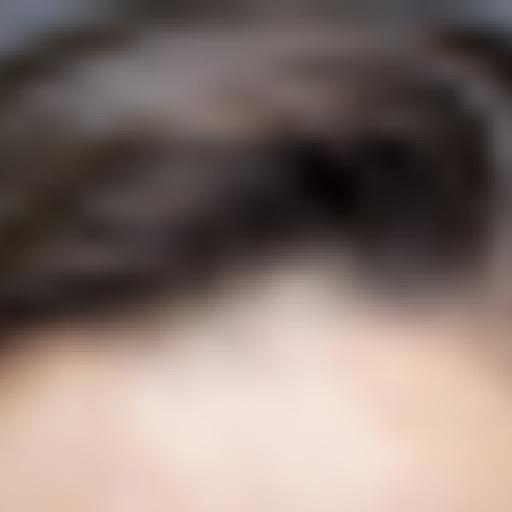}}&
{\includegraphics[width=0.19\textwidth]{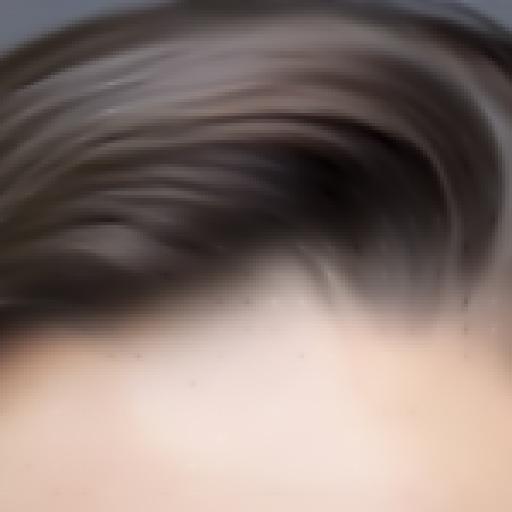}}&
{\includegraphics[width=0.19\textwidth]{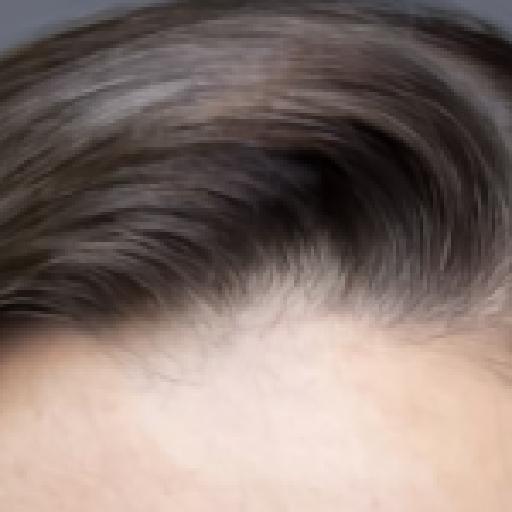}} &
{\includegraphics[width=0.19\textwidth]{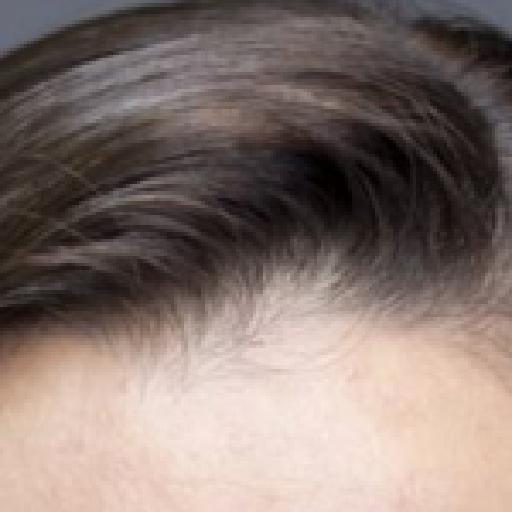}} \\

{\includegraphics[width=0.19\textwidth]{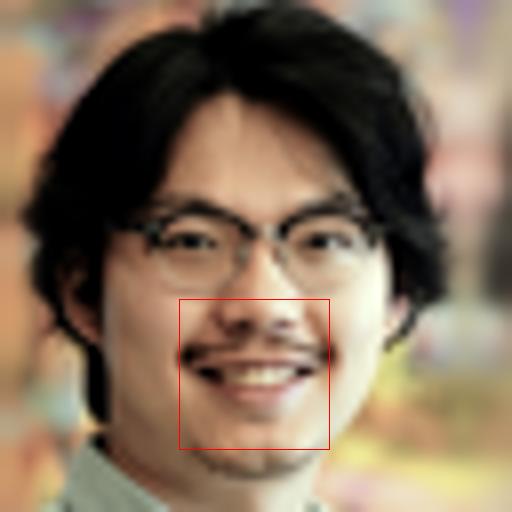}}&
{\includegraphics[width=0.19\textwidth]{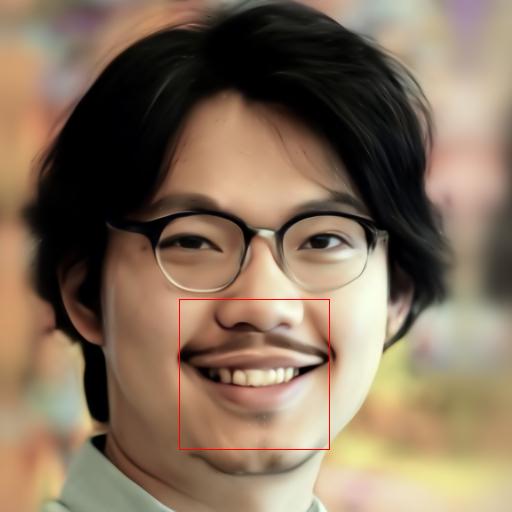}}&
{\includegraphics[width=0.19\textwidth]{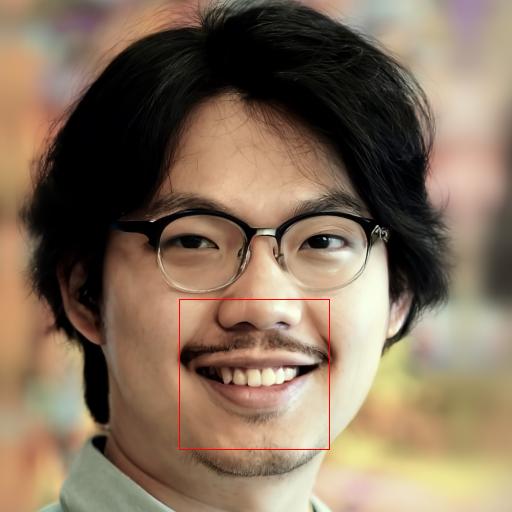}} &
{\includegraphics[width=0.19\textwidth]{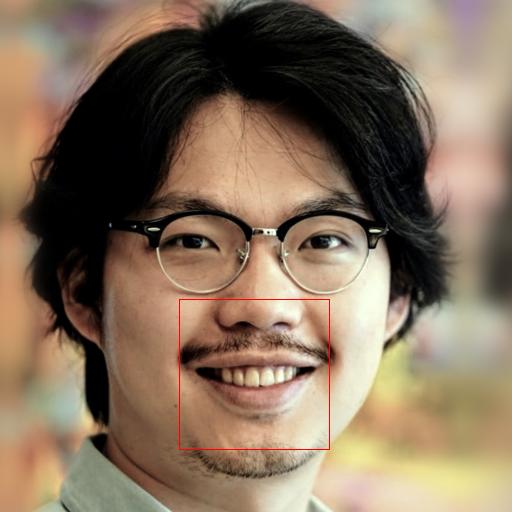}} \\

\medskip

{\includegraphics[width=0.19\textwidth]{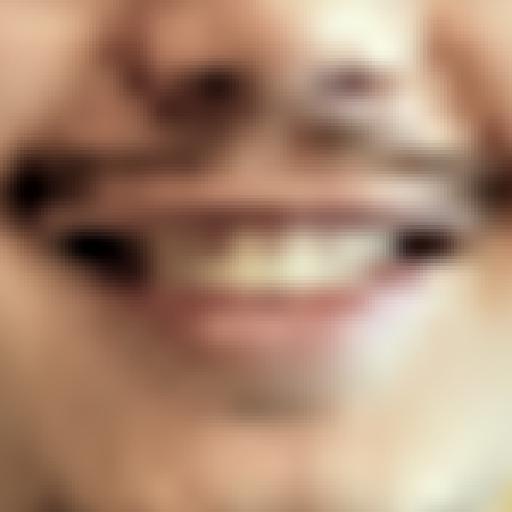}}&
{\includegraphics[width=0.19\textwidth]{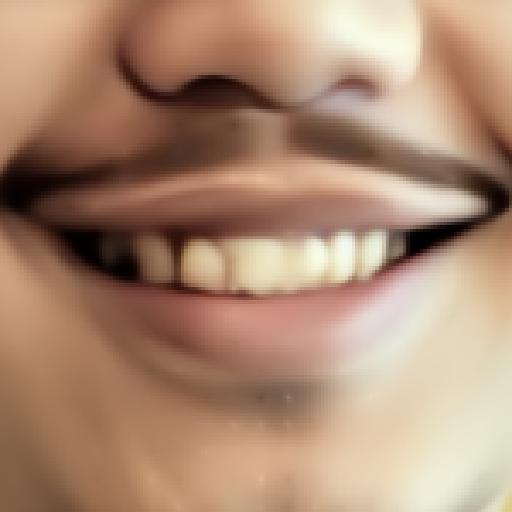}}&
{\includegraphics[width=0.19\textwidth]{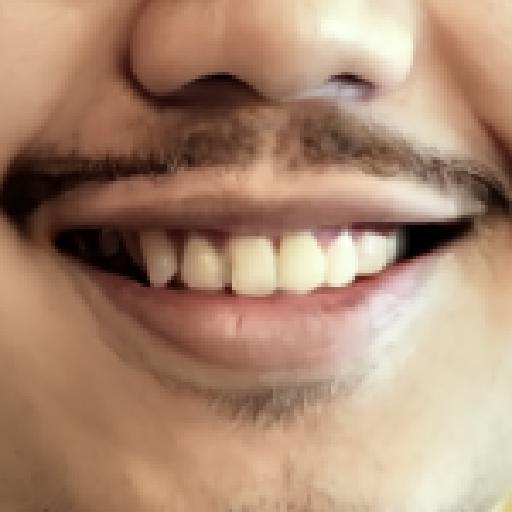}} &
{\includegraphics[width=0.19\textwidth]{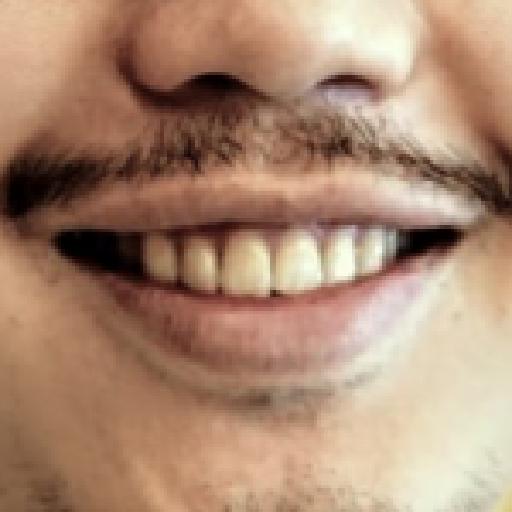}} \\

{\includegraphics[width=0.19\textwidth]{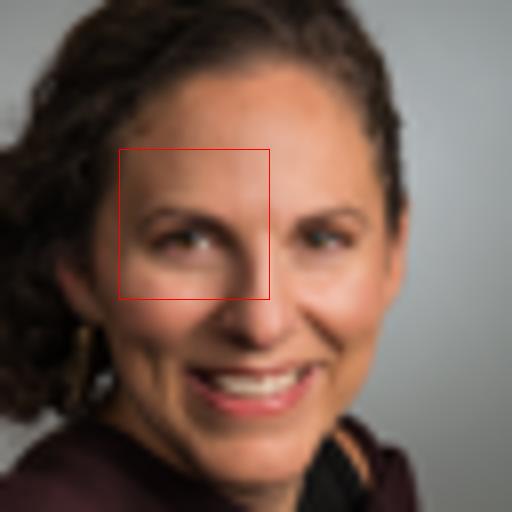}}&
{\includegraphics[width=0.19\textwidth]{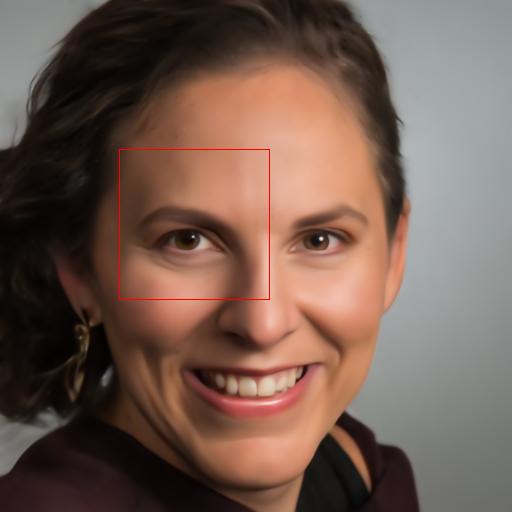}}&
{\includegraphics[width=0.19\textwidth]{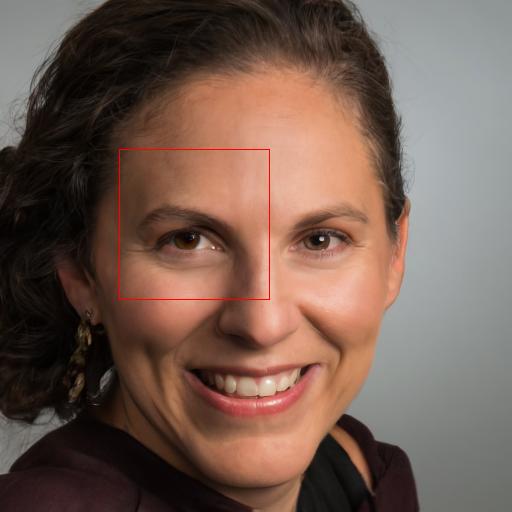}} &
{\includegraphics[width=0.19\textwidth]{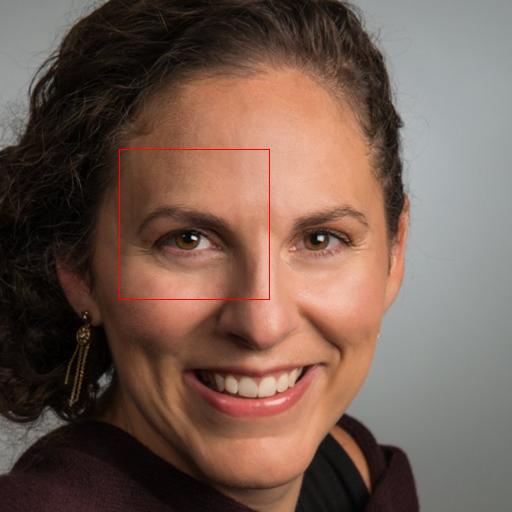}} \\

{\includegraphics[width=0.19\textwidth]{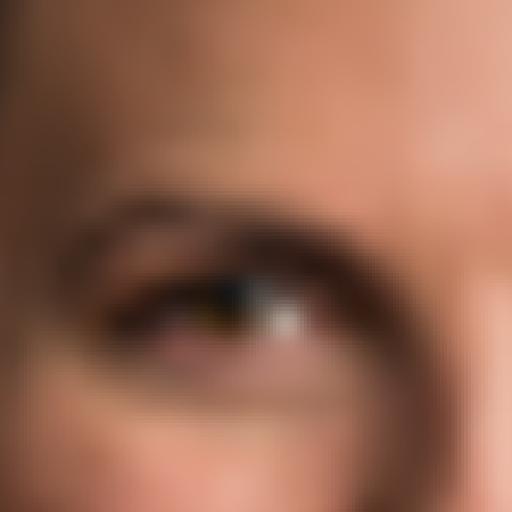}}&
{\includegraphics[width=0.19\textwidth]{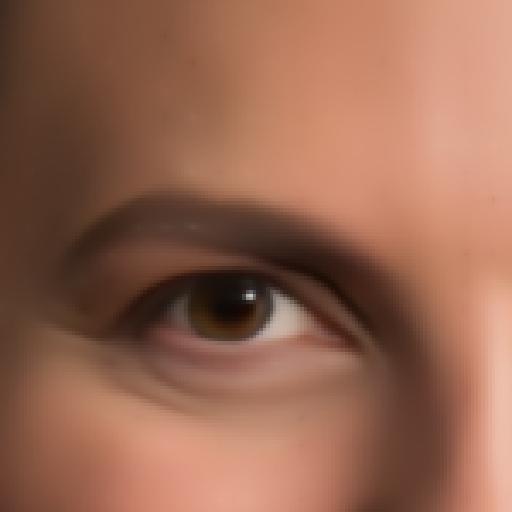}}&
{\includegraphics[width=0.19\textwidth]{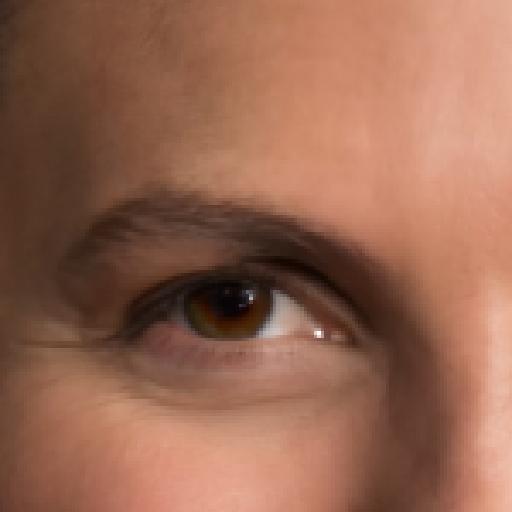}} &
{\includegraphics[width=0.19\textwidth]{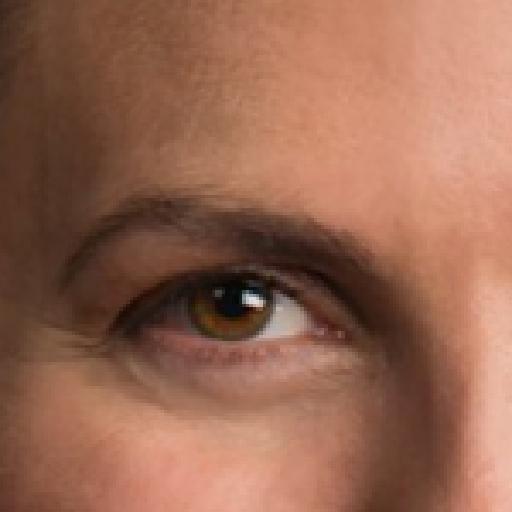}} \\

\end{tabular}
\end{center}
\vspace*{-0.35cm}
\caption{Additional results of a \modelname model (64$\times$64 $\rightarrow$ 512$\times$512), trained on FFHQ, and applied to images outside of the training set. We crop and align these faces to be consistent with FFHQ using the script provided \href{https://gist.github.com/lzhbrian/bde87ab23b499dd02ba4f588258f57d5}{here}.  \vspace*{-.8cm}
}
\label{fig:64x_512x_faces2_arxiv}
\end{figure}

\begin{figure}[H]
\vspace*{-0.4cm}
\setlength{\tabcolsep}{2pt}
\begin{center}
\begin{tabular}{cccc}
{\small Bicubic} & {\small Regression} & {\small \modelname (ours)} & {\small Reference} \\

{\includegraphics[width=0.19\textwidth]{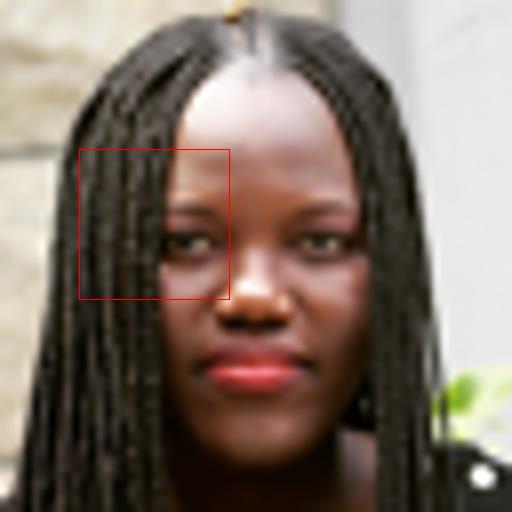}}&
{\includegraphics[width=0.19\textwidth]{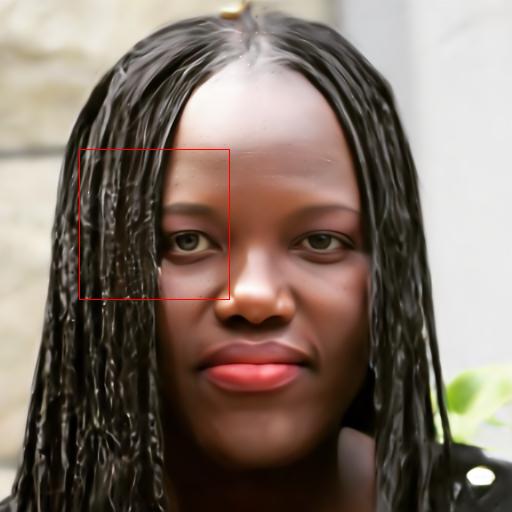}}&
{\includegraphics[width=0.19\textwidth]{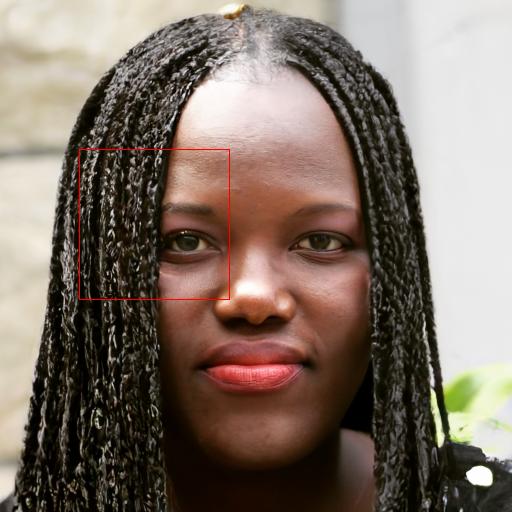}} &
{\includegraphics[width=0.19\textwidth]{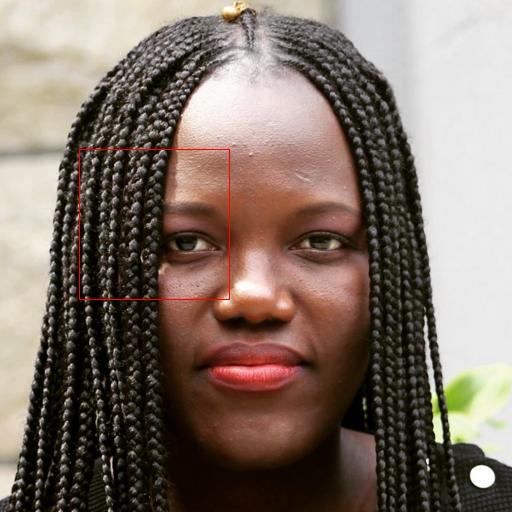}} \\

\medskip

{\includegraphics[width=0.19\textwidth]{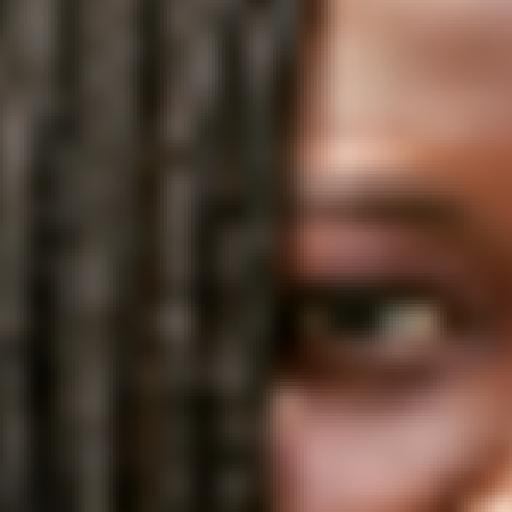}}&
{\includegraphics[width=0.19\textwidth]{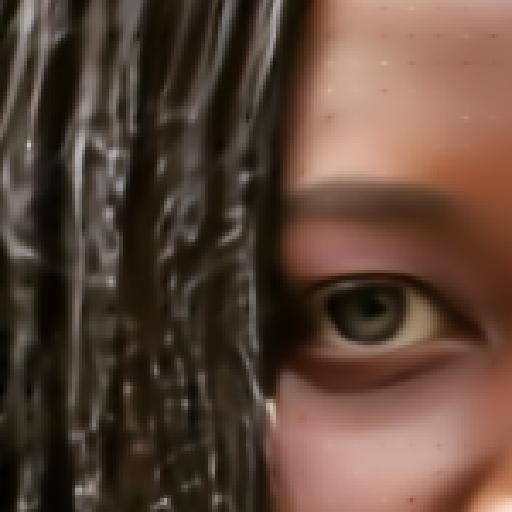}}&
{\includegraphics[width=0.19\textwidth]{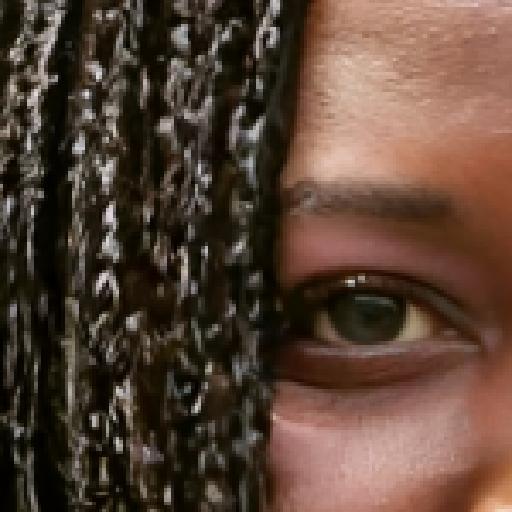}} &
{\includegraphics[width=0.19\textwidth]{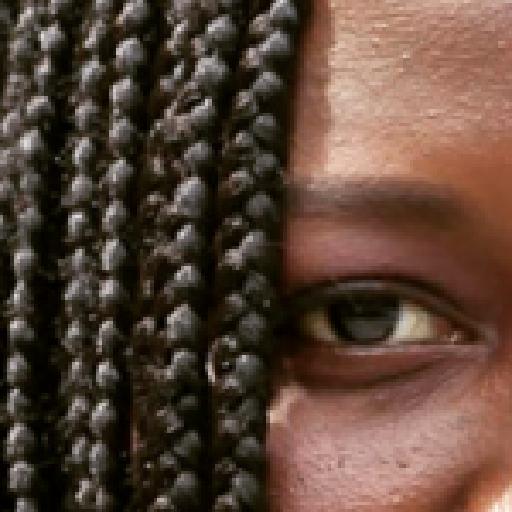}} \\

{\includegraphics[width=0.19\textwidth]{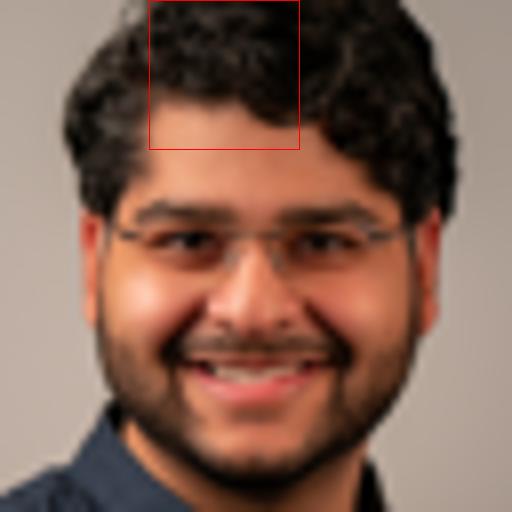}}&
{\includegraphics[width=0.19\textwidth]{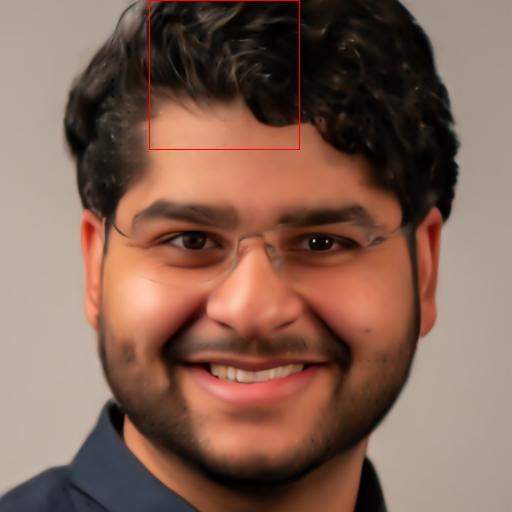}}&
{\includegraphics[width=0.19\textwidth]{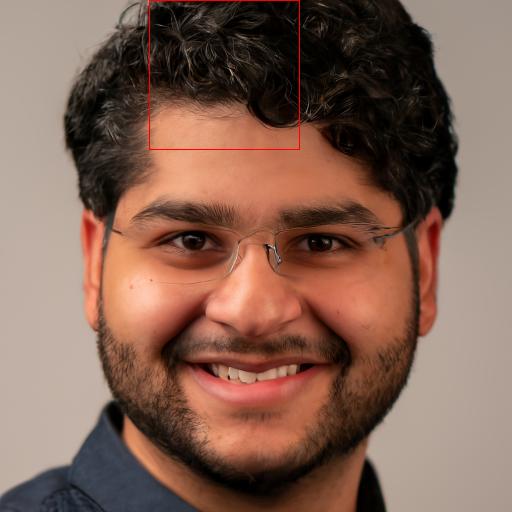}} &
{\includegraphics[width=0.19\textwidth]{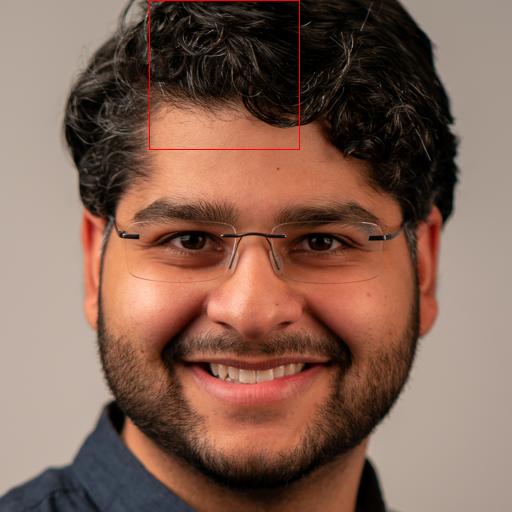}} \\

\medskip

{\includegraphics[width=0.19\textwidth]{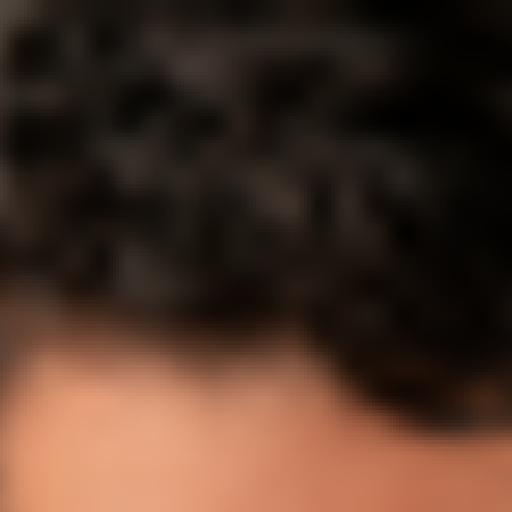}}&
{\includegraphics[width=0.19\textwidth]{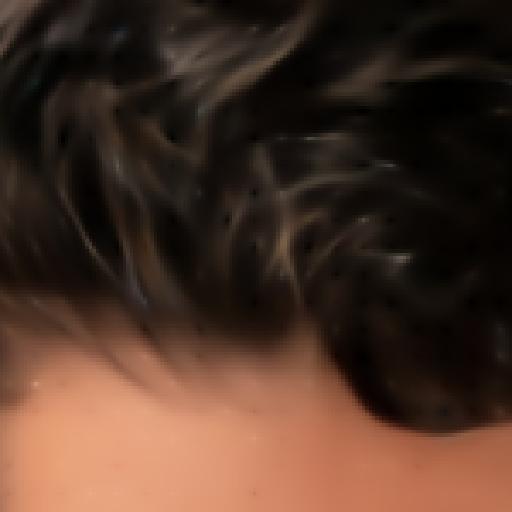}}&
{\includegraphics[width=0.19\textwidth]{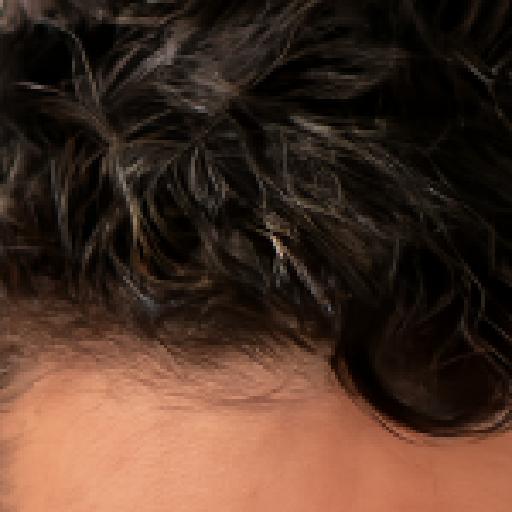}} &
{\includegraphics[width=0.19\textwidth]{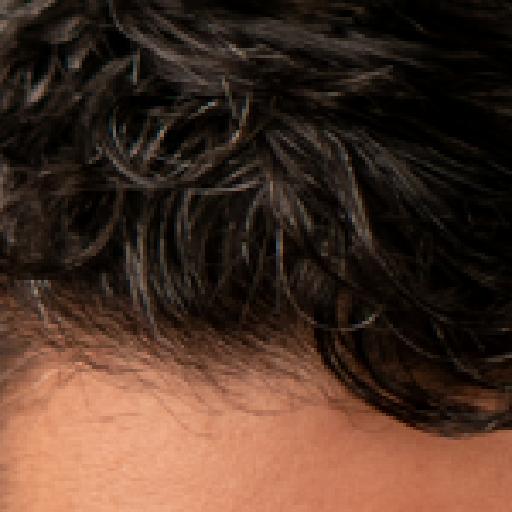}} \\

{\includegraphics[width=0.19\textwidth]{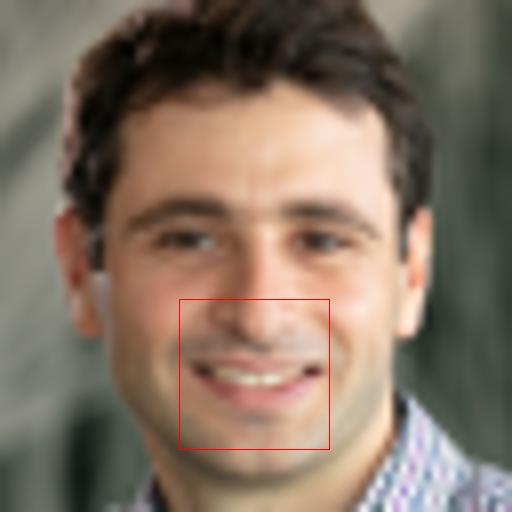}}&
{\includegraphics[width=0.19\textwidth]{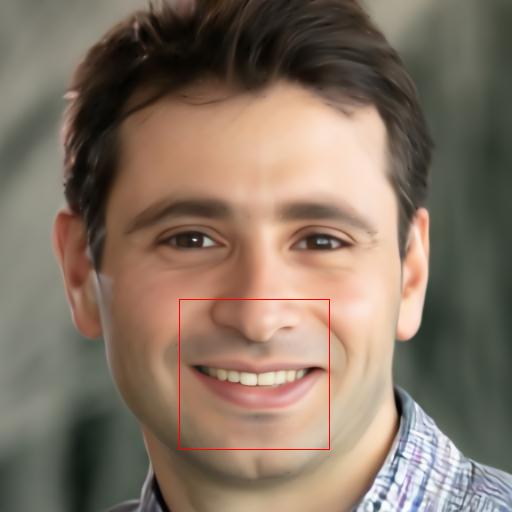}}&
{\includegraphics[width=0.19\textwidth]{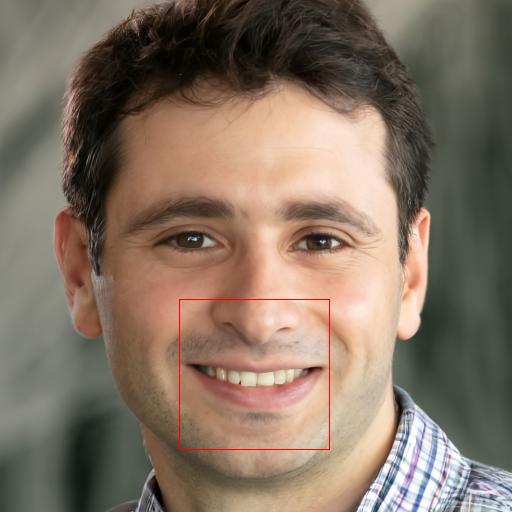}} &
{\includegraphics[width=0.19\textwidth]{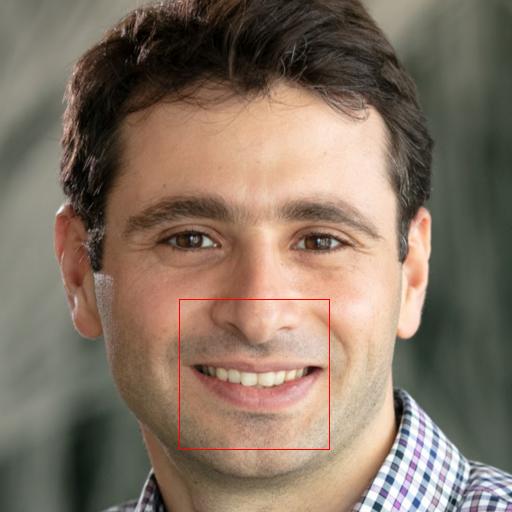}} \\

\medskip

{\includegraphics[width=0.19\textwidth]{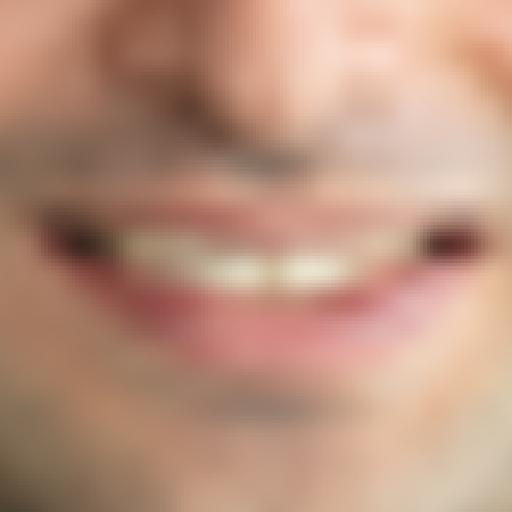}}&
{\includegraphics[width=0.19\textwidth]{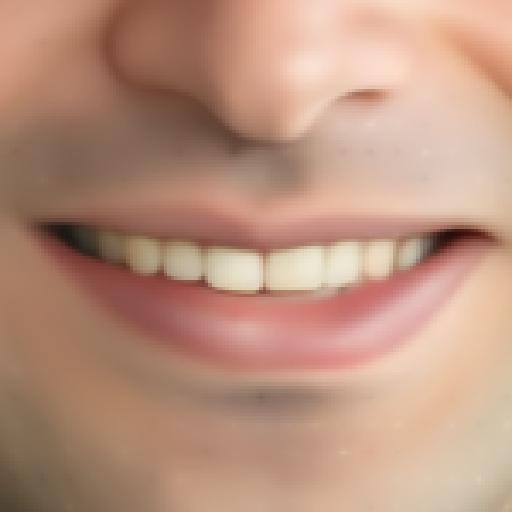}}&
{\includegraphics[width=0.19\textwidth]{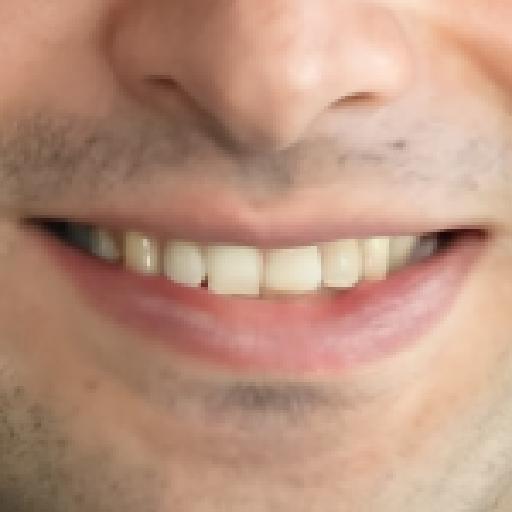}} &
{\includegraphics[width=0.19\textwidth]{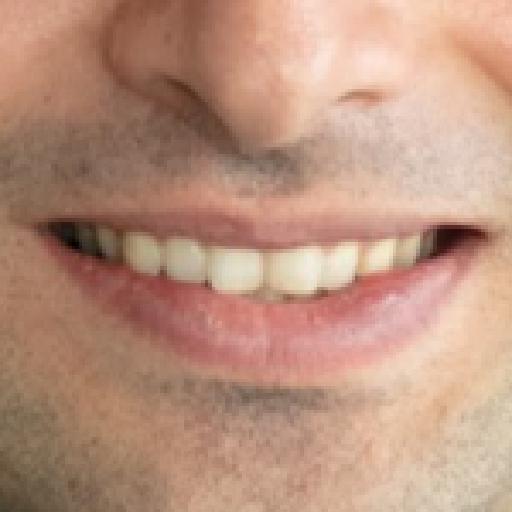}} \\

\end{tabular}
\end{center}
\vspace*{-0.35cm}
\caption{Additional results of a \modelname model (64$\times$64 $\rightarrow$ 512$\times$512), trained on FFHQ, and applied to images outside of the training set. We crop and align these faces to be consistent with FFHQ using the script provided \href{https://gist.github.com/lzhbrian/bde87ab23b499dd02ba4f588258f57d5}{here}.  \vspace*{-.45cm}
}
\label{fig:64x_512x_faces3_arxiv}
\end{figure}

\vfill

\newpage

\begin{center}
{\large \bf Natural Image Super-Resolution \ 64$\times$64  $\rightarrow$ 256$\times$256}\\
\end{center}

\begin{figure}[H]
\vspace*{-0.4cm}
\setlength{\tabcolsep}{2pt}
\begin{center}
\begin{tabular}{cccc}
{\small Bicubic} & 
{\small Regression} & {\small SR3 (ours)} & {\small Reference} \\
{\includegraphics[width=0.19\textwidth]{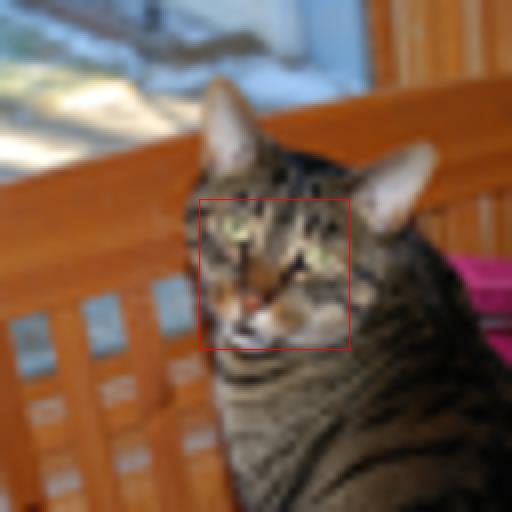}} &
{\includegraphics[width=0.19\textwidth]{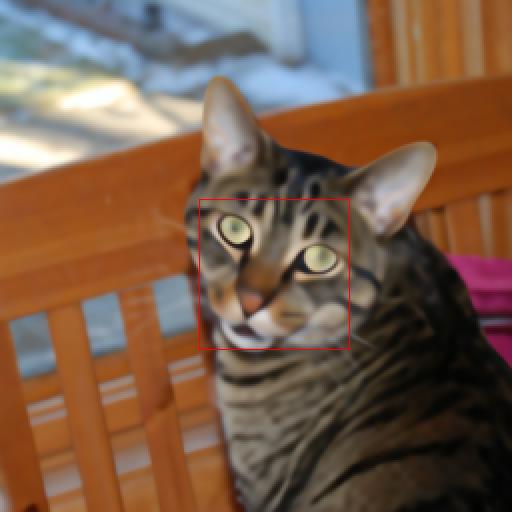}} &
{\includegraphics[width=0.19\textwidth]{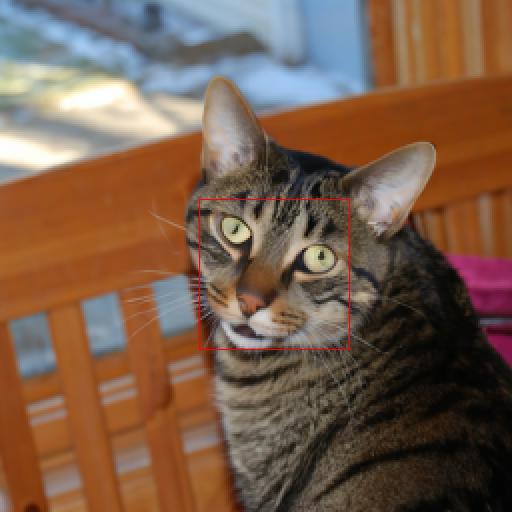}} &
{\includegraphics[width=0.19\textwidth]{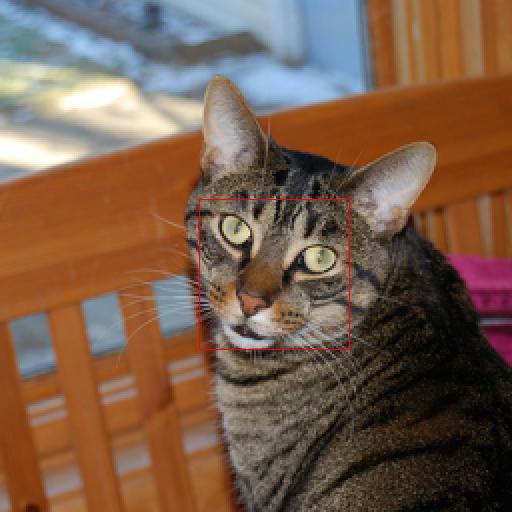}} \\

\medskip

{\includegraphics[width=0.19\textwidth]{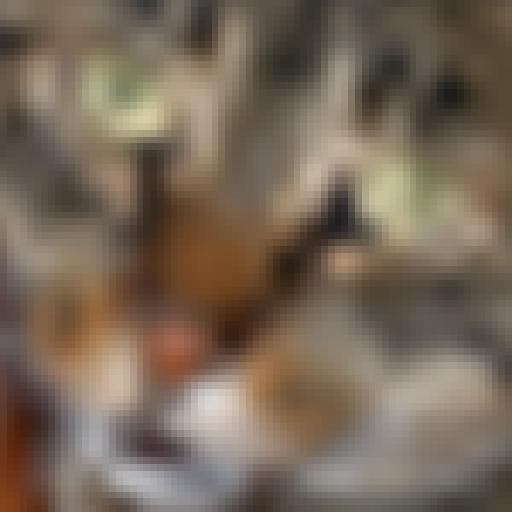}} &
{\includegraphics[width=0.19\textwidth]{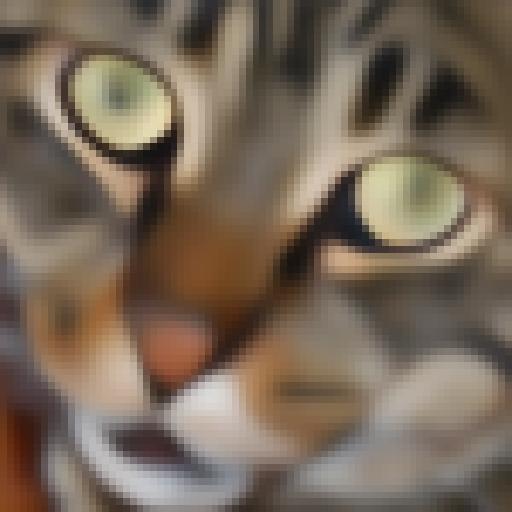}} &
{\includegraphics[width=0.19\textwidth]{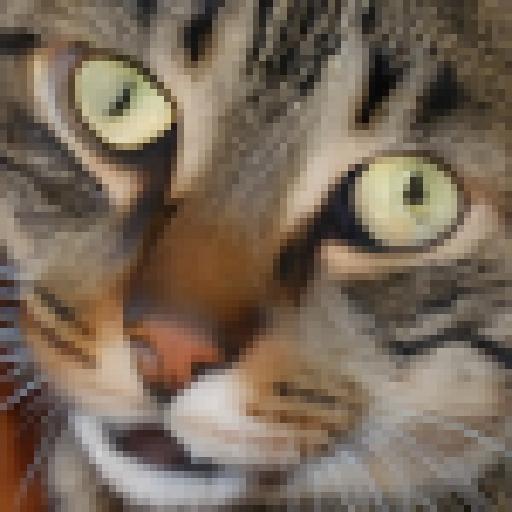}} &
{\includegraphics[width=0.19\textwidth]{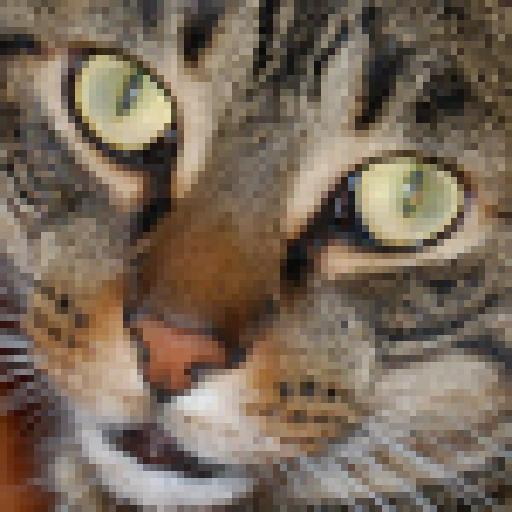}} \\

{\includegraphics[width=0.19\textwidth]{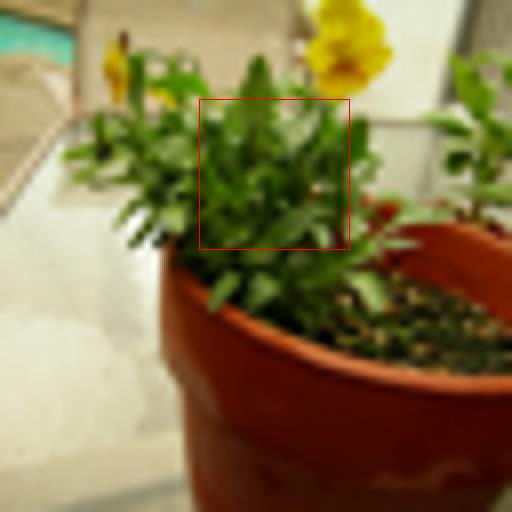}} &
{\includegraphics[width=0.19\textwidth]{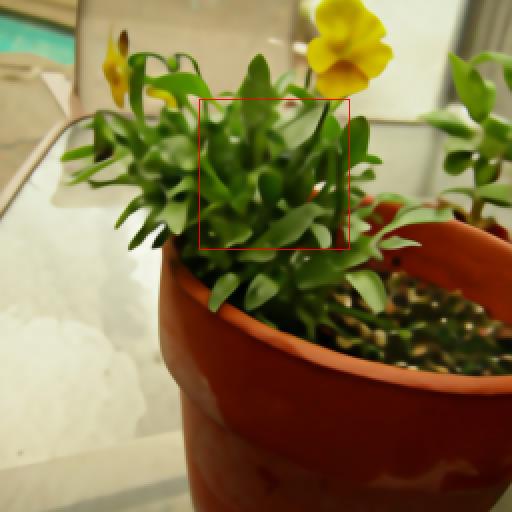}} &
{\includegraphics[width=0.19\textwidth]{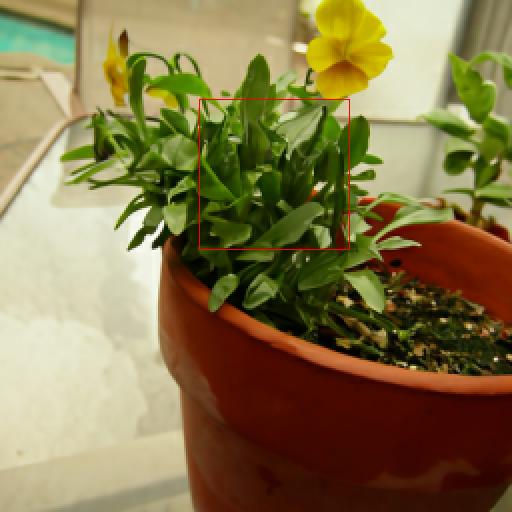}} &
{\includegraphics[width=0.19\textwidth]{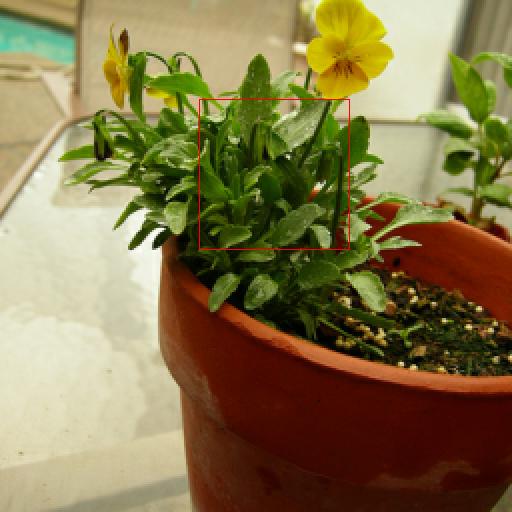}} \\

\medskip

{\includegraphics[width=0.19\textwidth]{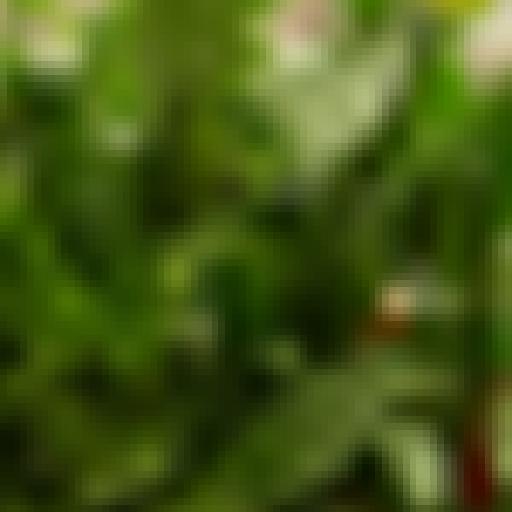}} &
{\includegraphics[width=0.19\textwidth]{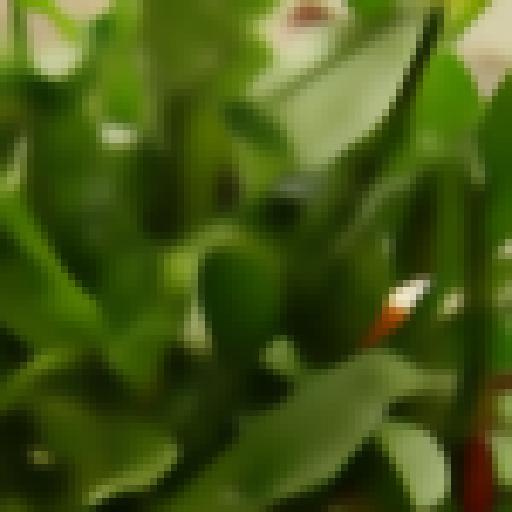}} &
{\includegraphics[width=0.19\textwidth]{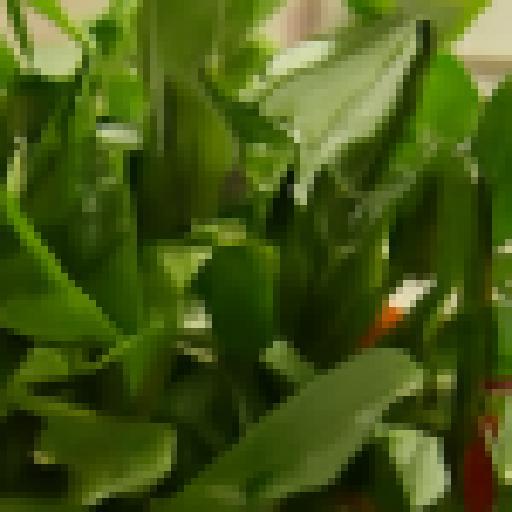}} &
{\includegraphics[width=0.19\textwidth]{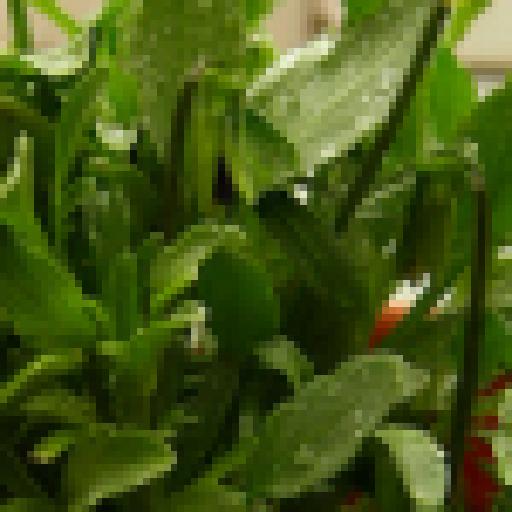}} \\

{\includegraphics[width=0.19\textwidth]{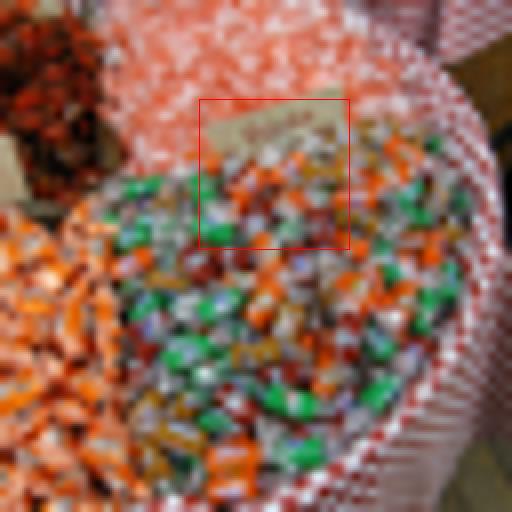}} &
{\includegraphics[width=0.19\textwidth]{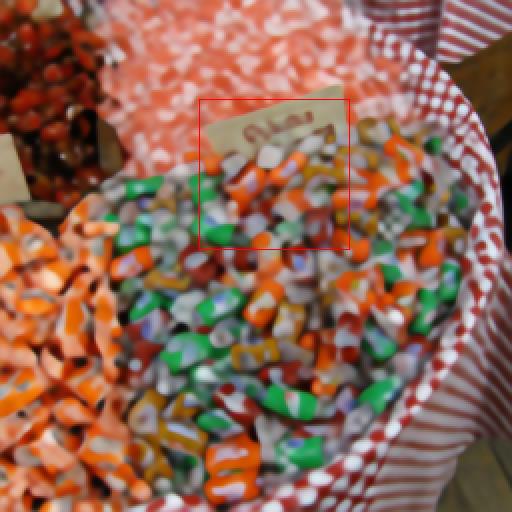}} &
{\includegraphics[width=0.19\textwidth]{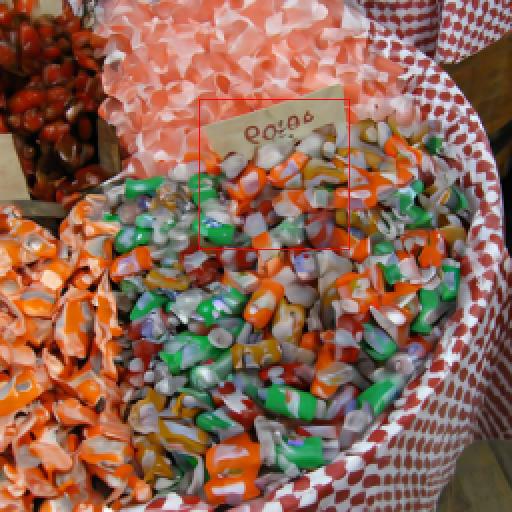}} &
{\includegraphics[width=0.19\textwidth]{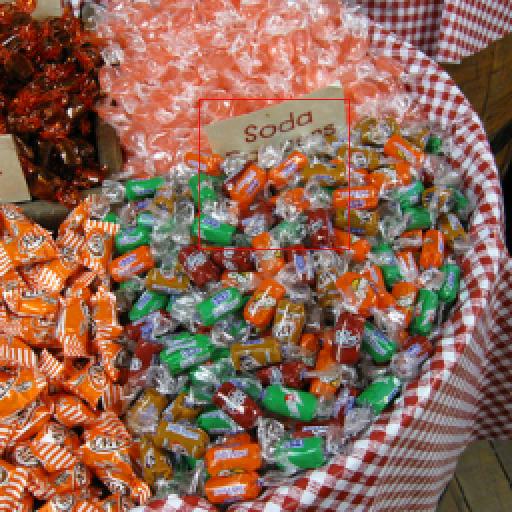}} \\

{\includegraphics[width=0.19\textwidth]{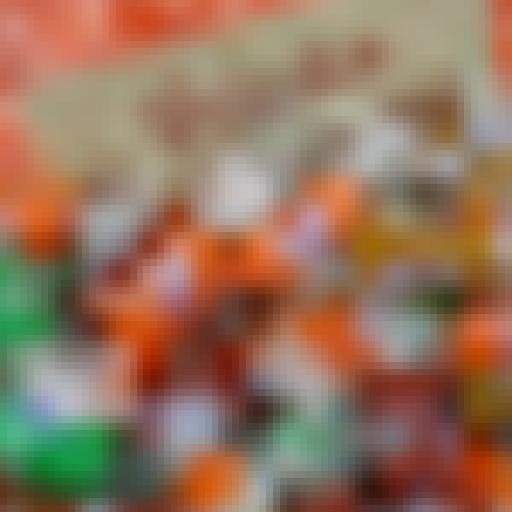}} &
{\includegraphics[width=0.19\textwidth]{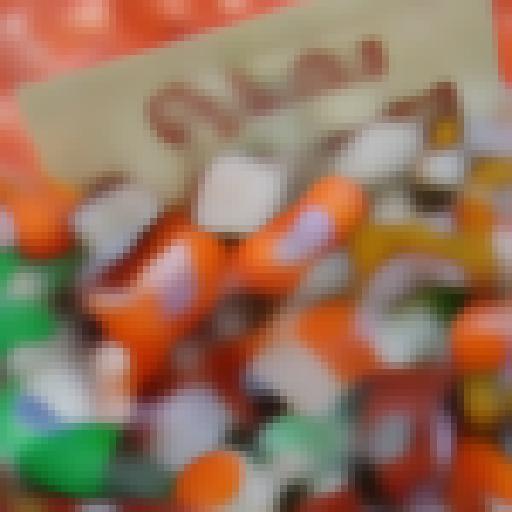}} &
{\includegraphics[width=0.19\textwidth]{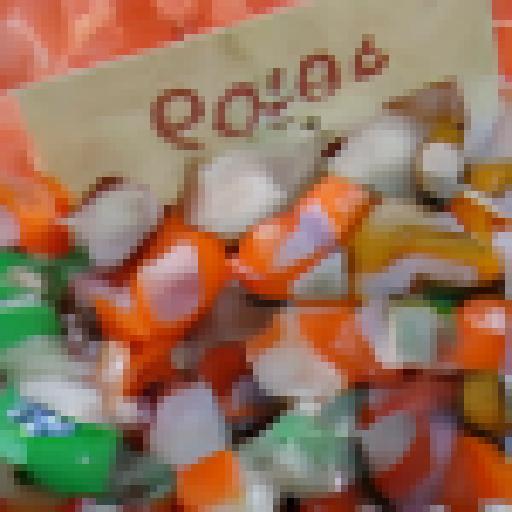}} &
{\includegraphics[width=0.19\textwidth]{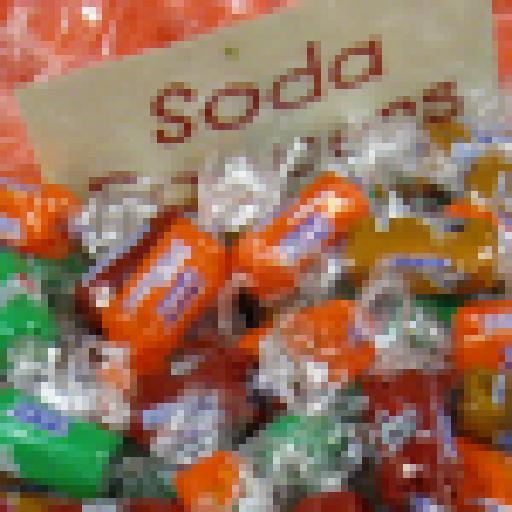}} \\
\end{tabular}
\end{center}
\vspace*{-0.4cm}
\caption{Additional results of a \modelname model (64$\times$64 $\rightarrow$ 256$\times$256), trained on ImageNet and evaluated on ImageNet test images.
\vspace*{-0.4cm}}
\label{fig:64x_256x_natural_images2}
\end{figure}

\begin{figure}[H]
\vspace*{-0.4cm}
\setlength{\tabcolsep}{2pt}
\begin{center}
\begin{tabular}{cccc}
{\small Bicubic} & 
{\small Regression} & {\small SR3 (ours)} & {\small Reference} \\
{\includegraphics[width=0.19\textwidth]{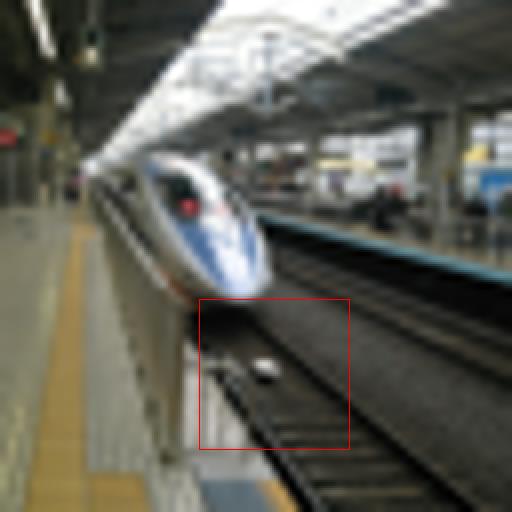}} &
{\includegraphics[width=0.19\textwidth]{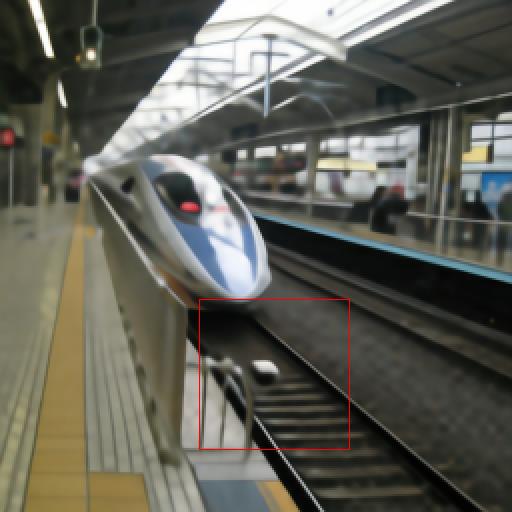}} &
{\includegraphics[width=0.19\textwidth]{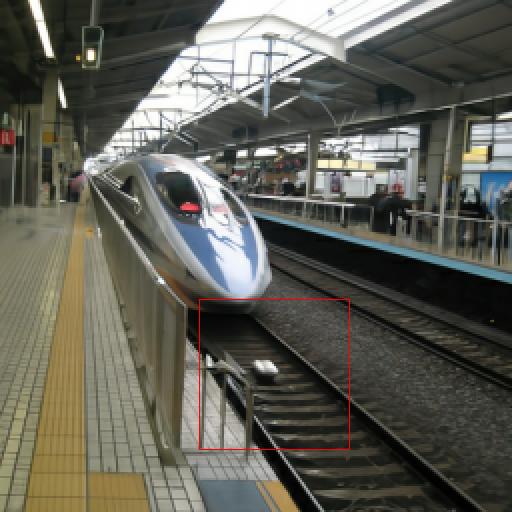}} &
{\includegraphics[width=0.19\textwidth]{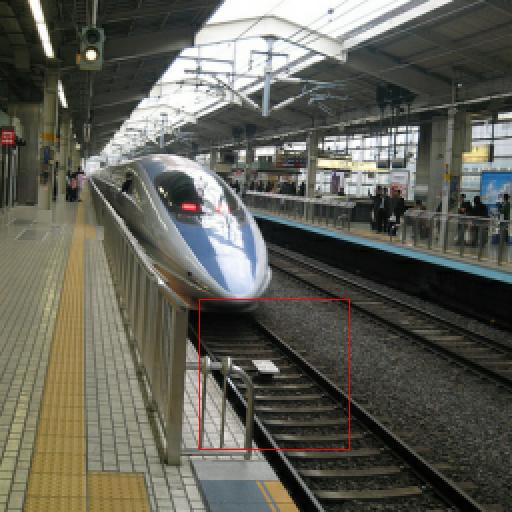}} \\

\medskip

{\includegraphics[width=0.19\textwidth]{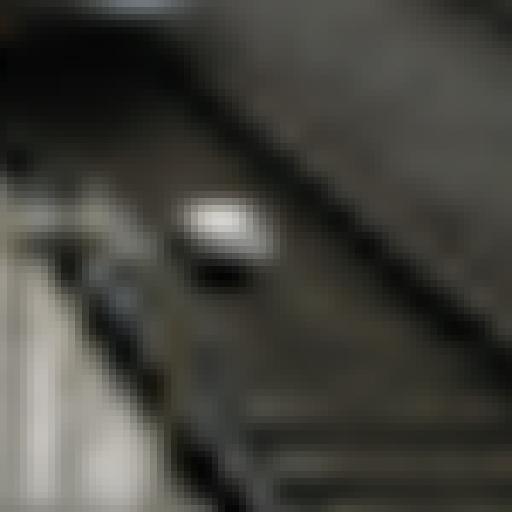}} &
{\includegraphics[width=0.19\textwidth]{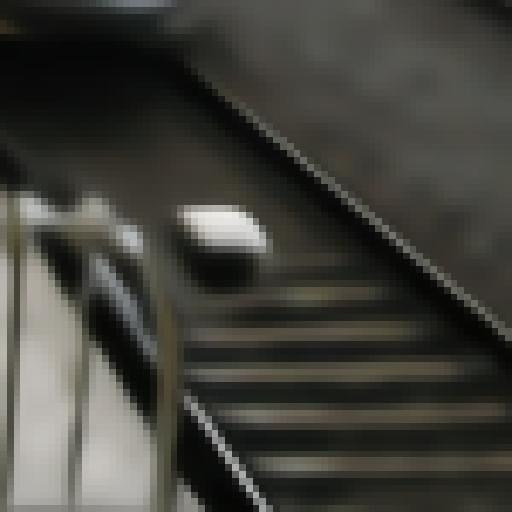}} &
{\includegraphics[width=0.19\textwidth]{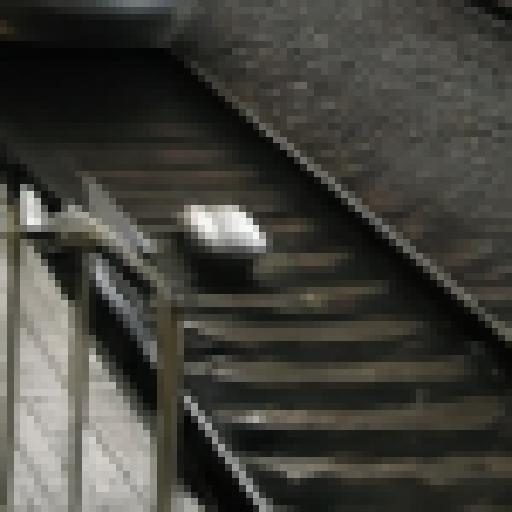}} &
{\includegraphics[width=0.19\textwidth]{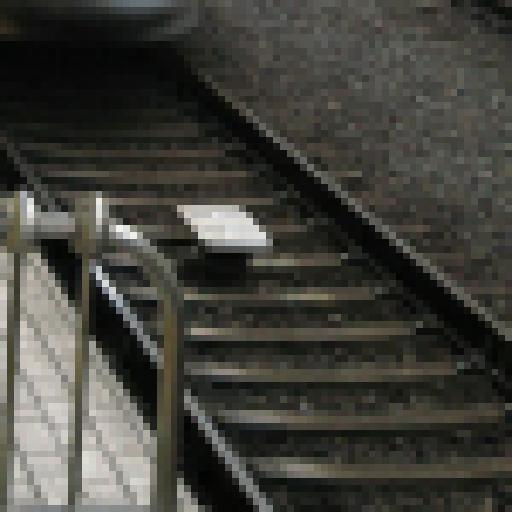}} \\

{\includegraphics[width=0.19\textwidth]{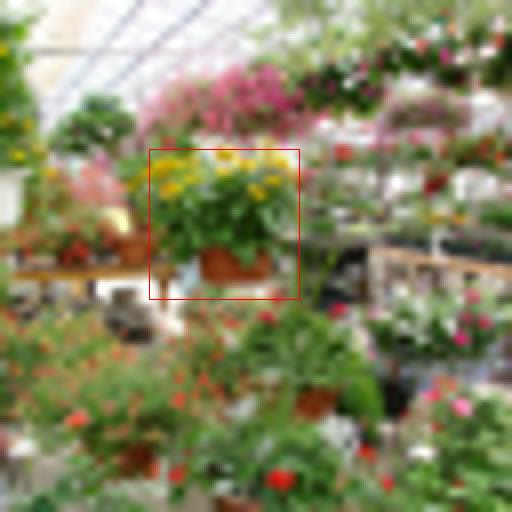}} &
{\includegraphics[width=0.19\textwidth]{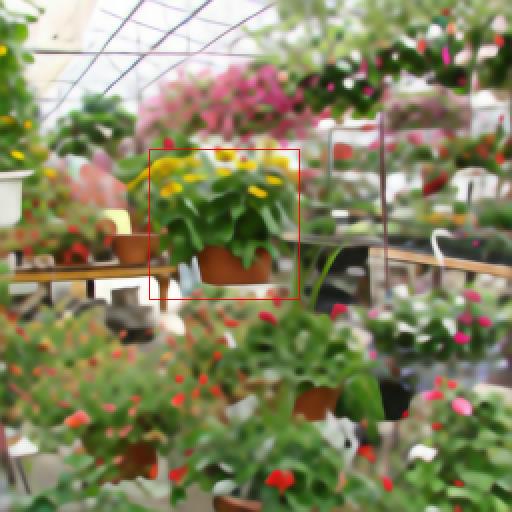}} &
{\includegraphics[width=0.19\textwidth]{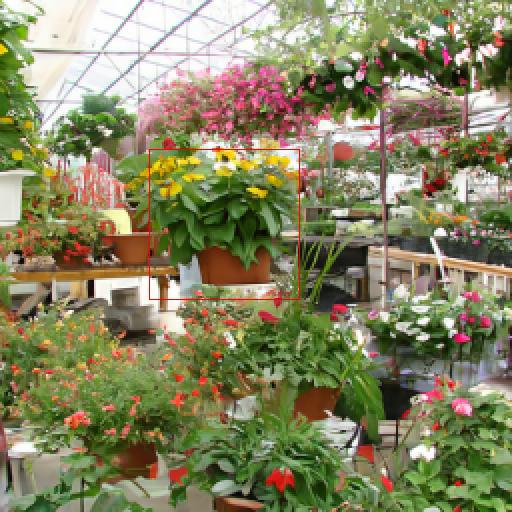}} &
{\includegraphics[width=0.19\textwidth]{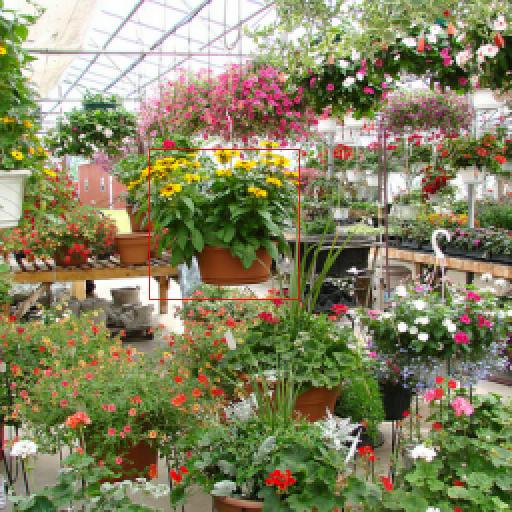}} \\

\medskip

{\includegraphics[width=0.19\textwidth]{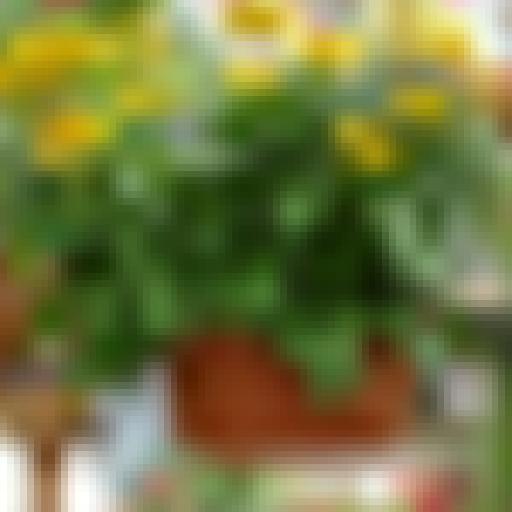}} &
{\includegraphics[width=0.19\textwidth]{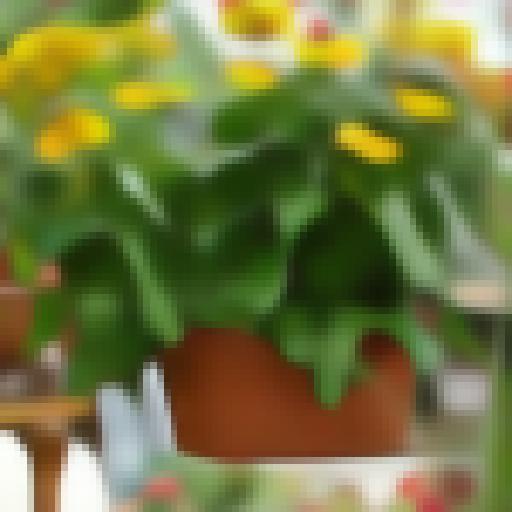}} &
{\includegraphics[width=0.19\textwidth]{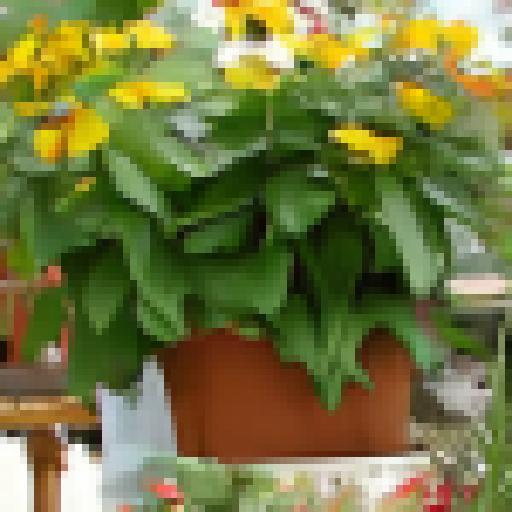}} &
{\includegraphics[width=0.19\textwidth]{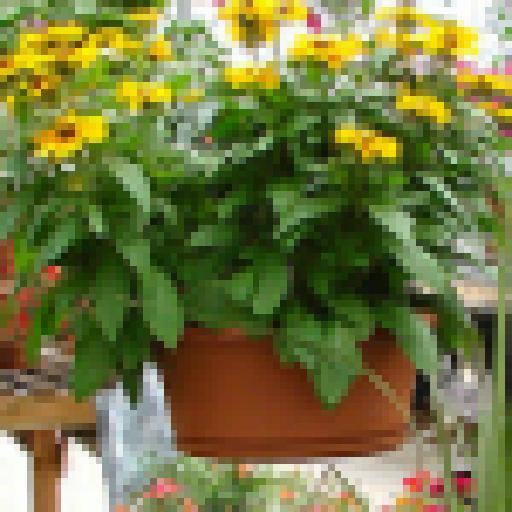}} \\

{\includegraphics[width=0.19\textwidth]{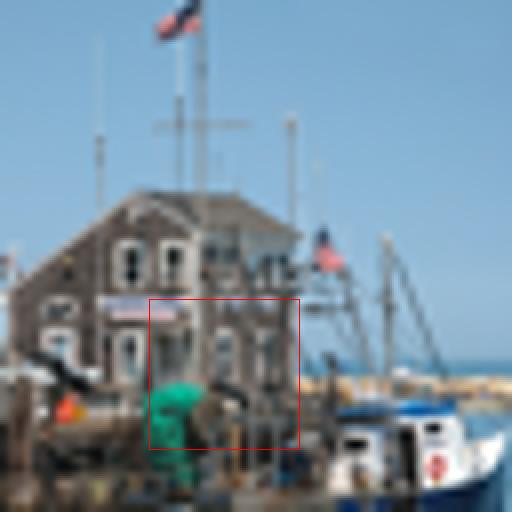}} &
{\includegraphics[width=0.19\textwidth]{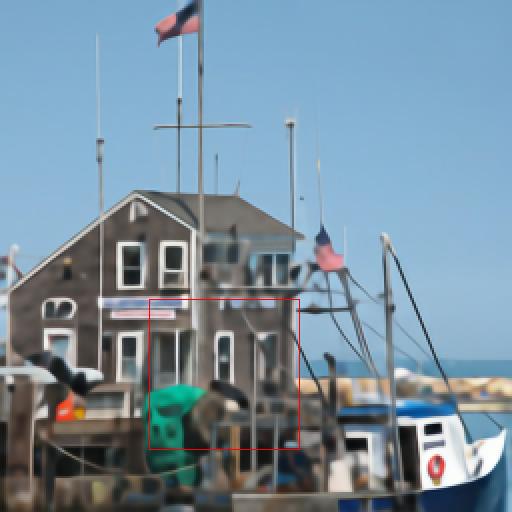}} &
{\includegraphics[width=0.19\textwidth]{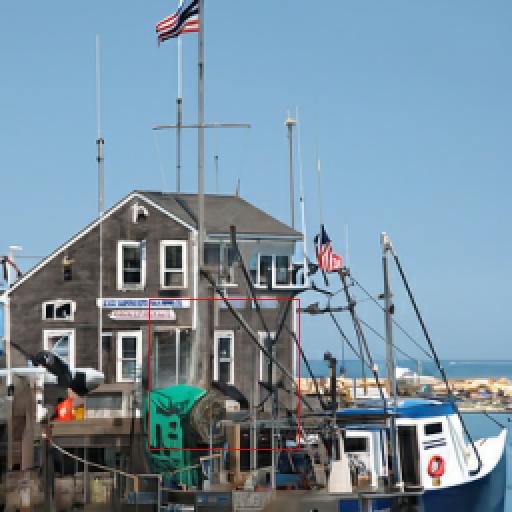}} &
{\includegraphics[width=0.19\textwidth]{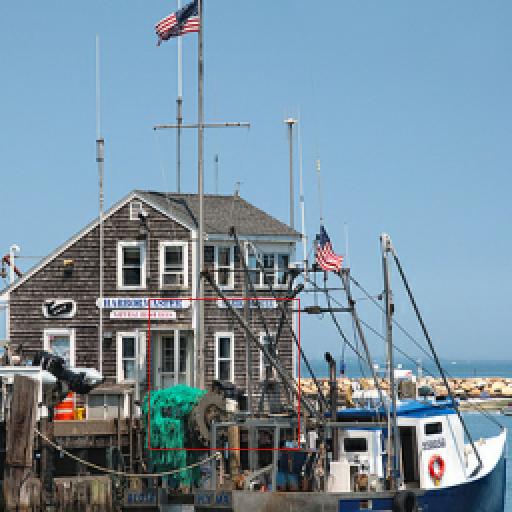}} \\

{\includegraphics[width=0.19\textwidth]{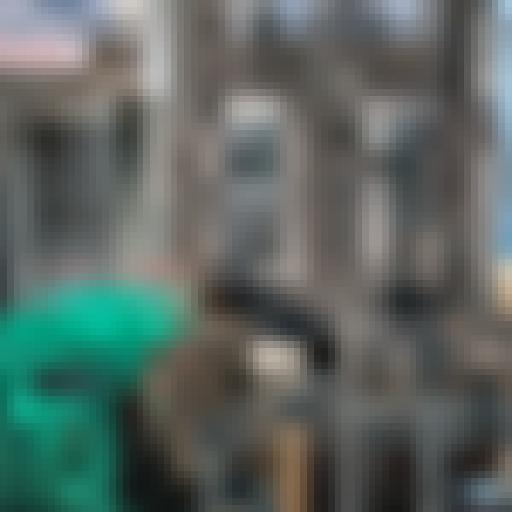}} &
{\includegraphics[width=0.19\textwidth]{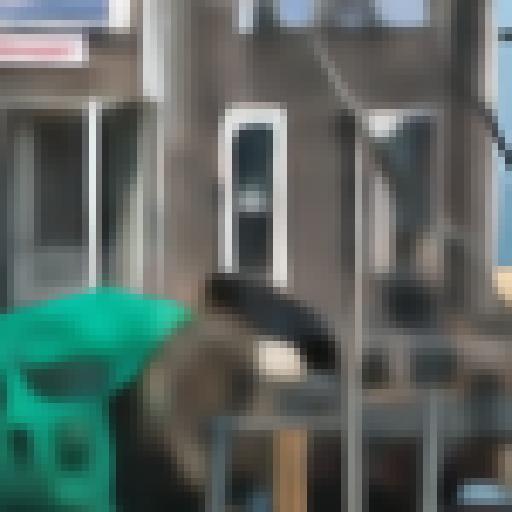}} &
{\includegraphics[width=0.19\textwidth]{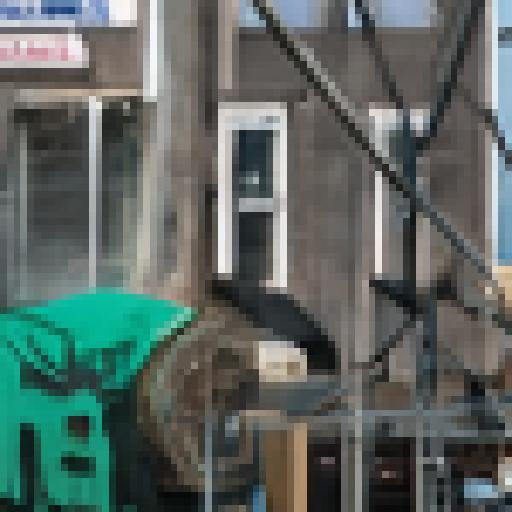}} &
{\includegraphics[width=0.19\textwidth]{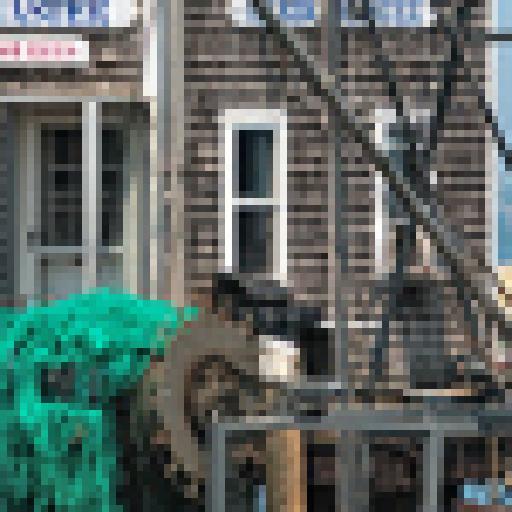}} \\
\end{tabular}
\end{center}
\vspace*{-0.4cm}
\caption{Additional results of a \modelname model (64$\times$64 $\rightarrow$ 256$\times$256), trained on ImageNet and evaluated on ImageNet test images.
\vspace*{-0.4cm}}
\label{fig:64x_256x_natural_images3}
\end{figure}

\vfill
\newpage

\begin{center}
{\large \bf Benchmark Comparison on  Test Faces \  16$\times$16  $\rightarrow$ 128$\times$128}
\end{center}

\begin{figure}[h]
\vspace*{-0.25cm}
\setlength{\tabcolsep}{2pt}
\begin{center}
\begin{tabular}{cccccc}
{\small Bicubic} & {\small FSRGAN \cite{chen2018fsrnet}}  & {\small PULSE \cite{menon2020pulse}} &  {\small Regression} & {\small \modelname} \\
{\includegraphics[width=0.14\textwidth]{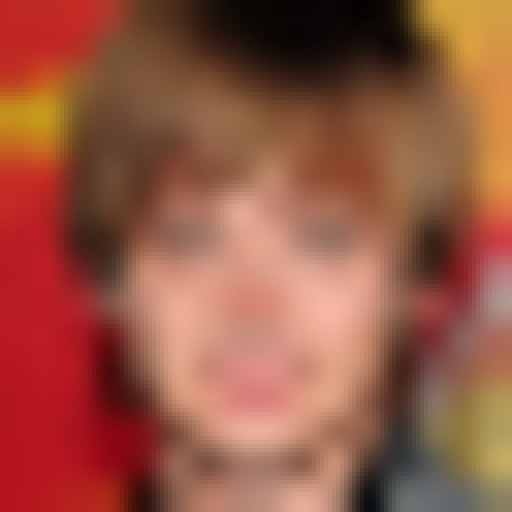}} &
{\includegraphics[width=0.14\textwidth]{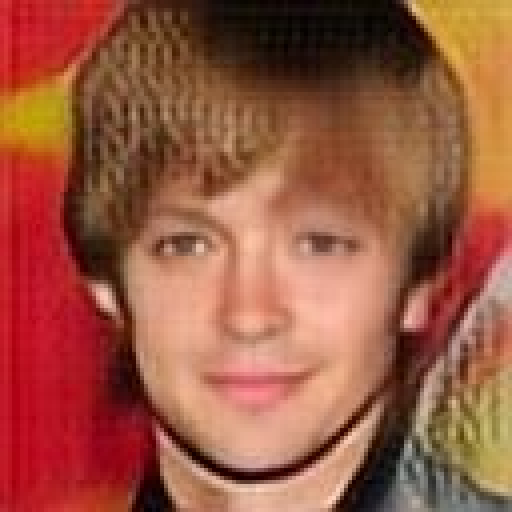}} &
{\includegraphics[width=0.14\textwidth]{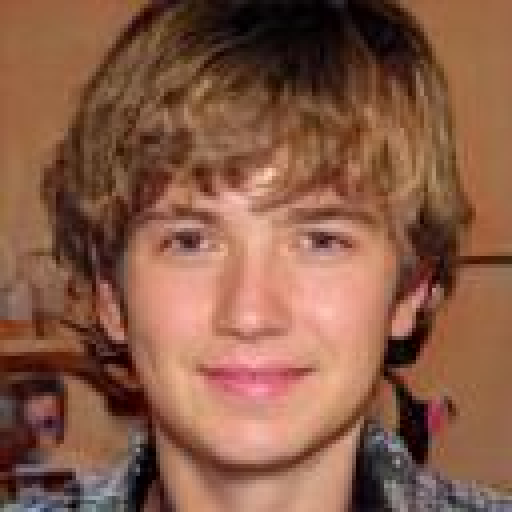}} &
{\includegraphics[width=0.14\textwidth]{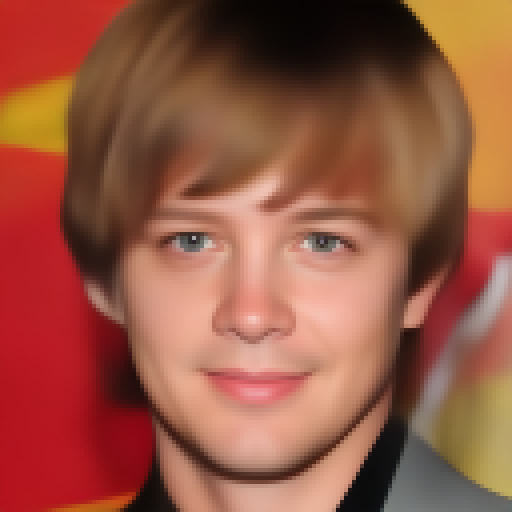}} &
{\includegraphics[width=0.14\textwidth]{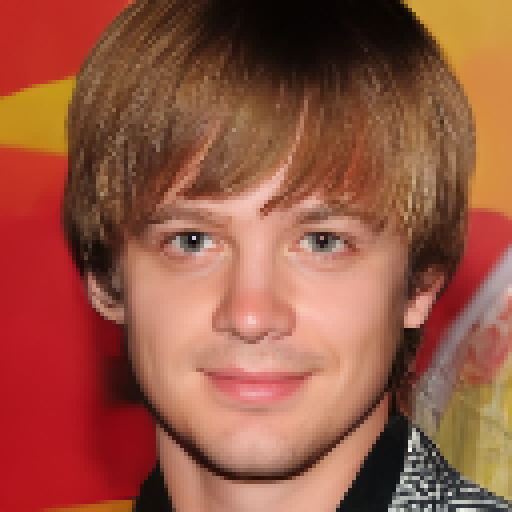}} \\
{\includegraphics[width=0.14\textwidth]{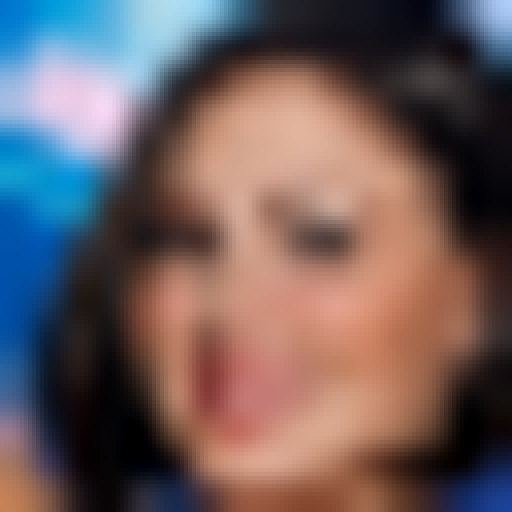}} &
{\includegraphics[width=0.14\textwidth]{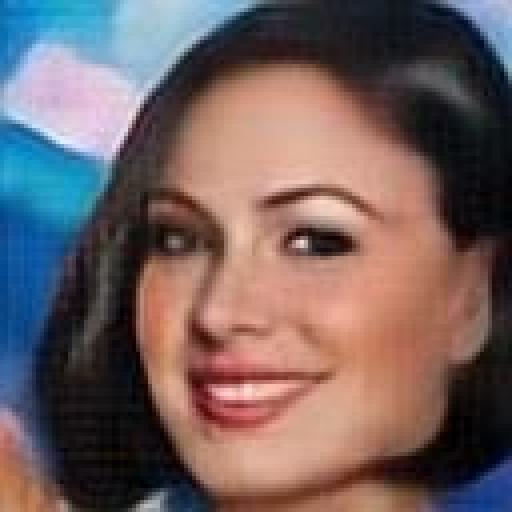}} &
{\includegraphics[width=0.14\textwidth]{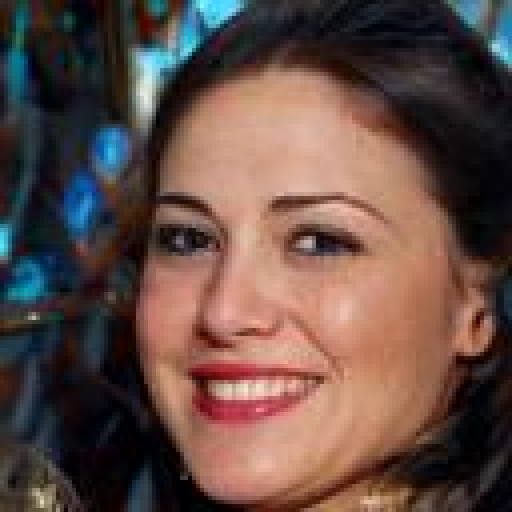}} &
{\includegraphics[width=0.14\textwidth]{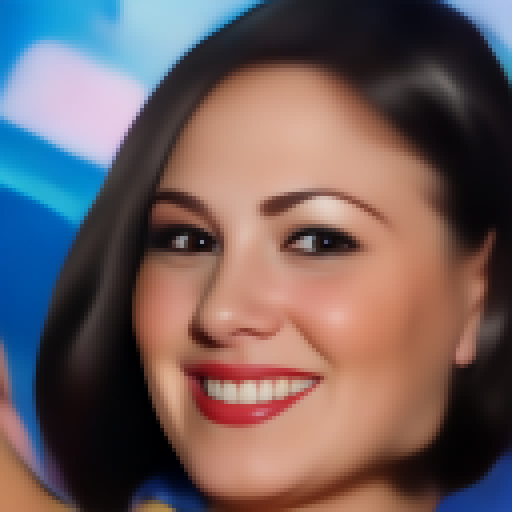}} &
{\includegraphics[width=0.14\textwidth]{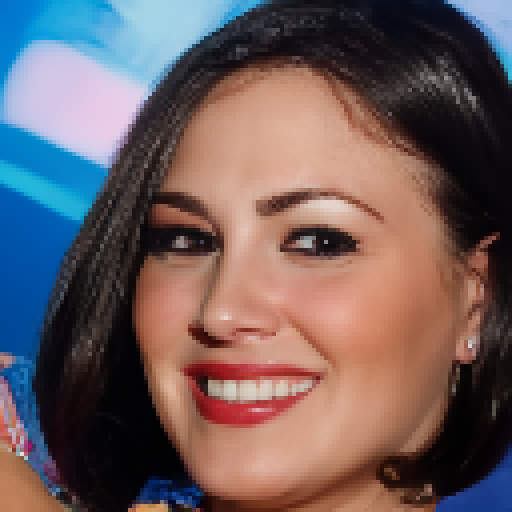}} \\
{\includegraphics[width=0.14\textwidth]{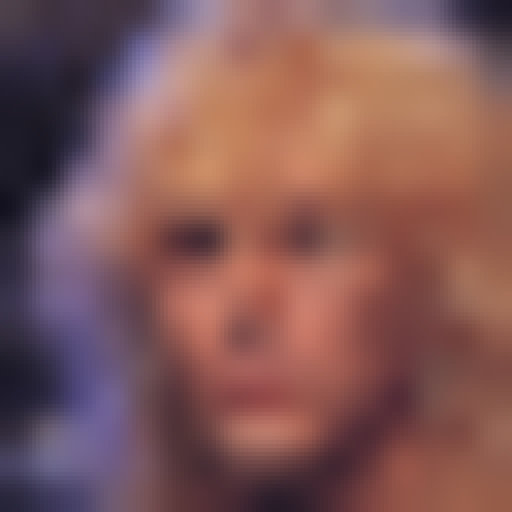}} &
{\includegraphics[width=0.14\textwidth]{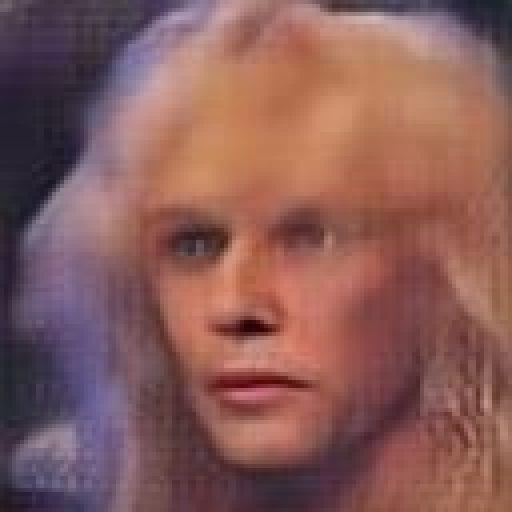}} &
{\includegraphics[width=0.14\textwidth]{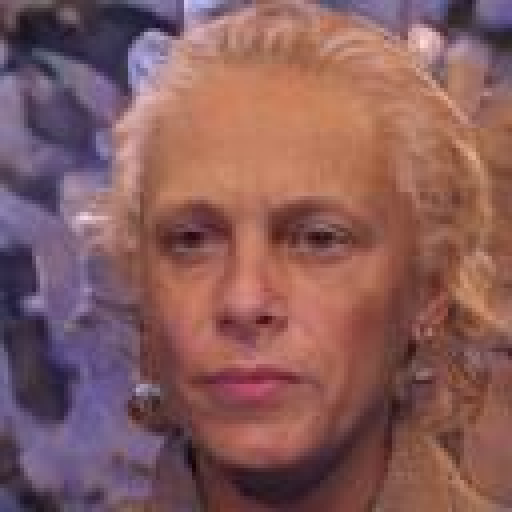}} &
{\includegraphics[width=0.14\textwidth]{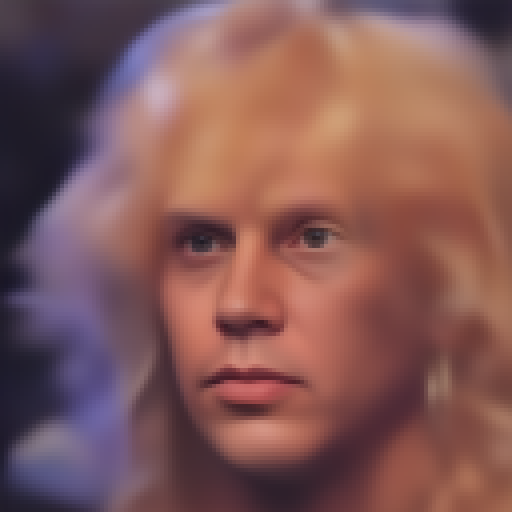}} &
{\includegraphics[width=0.14\textwidth]{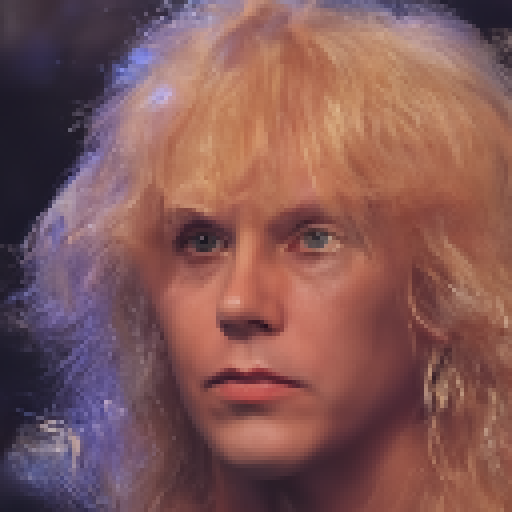}} \\
{\includegraphics[width=0.14\textwidth]{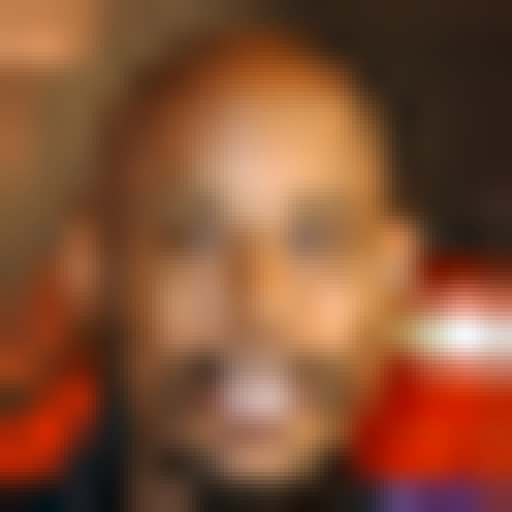}} &
{\includegraphics[width=0.14\textwidth]{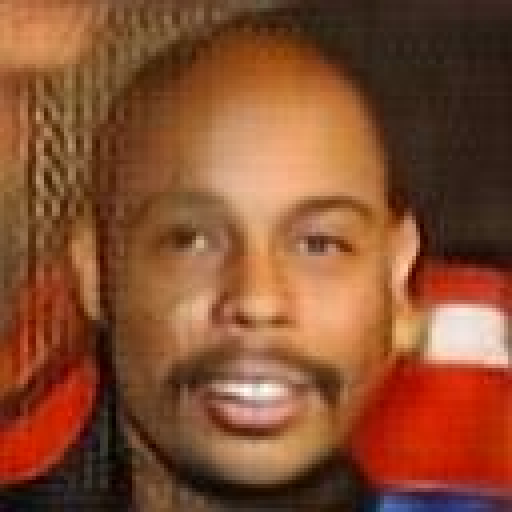}} &
{\includegraphics[width=0.14\textwidth]{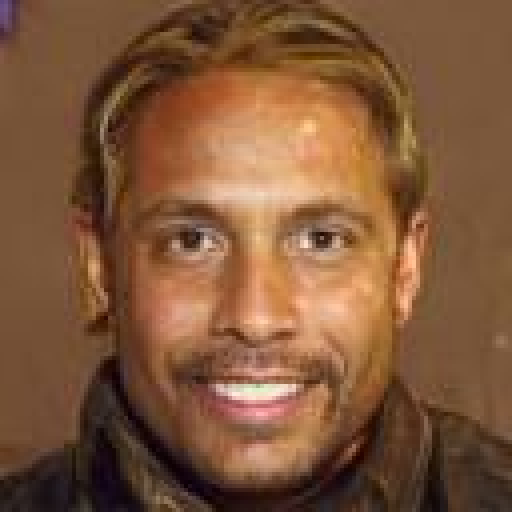}} &
{\includegraphics[width=0.14\textwidth]{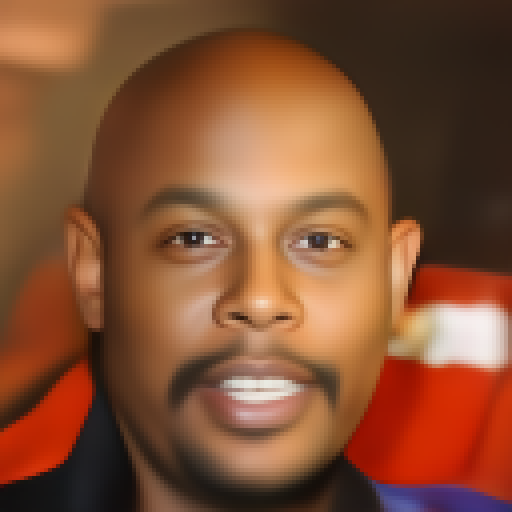}} &
{\includegraphics[width=0.14\textwidth]{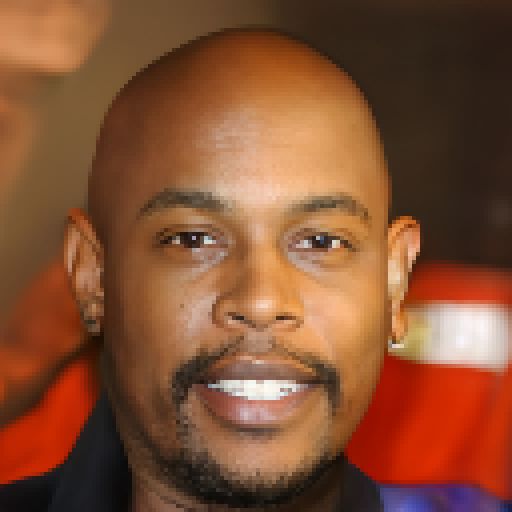}} \\
{\includegraphics[width=0.14\textwidth]{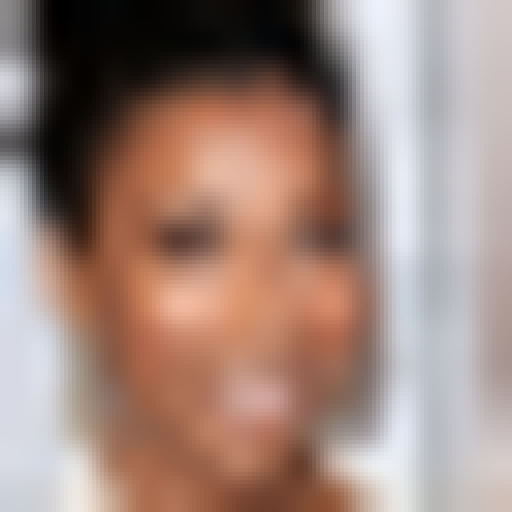}} &
{\includegraphics[width=0.14\textwidth]{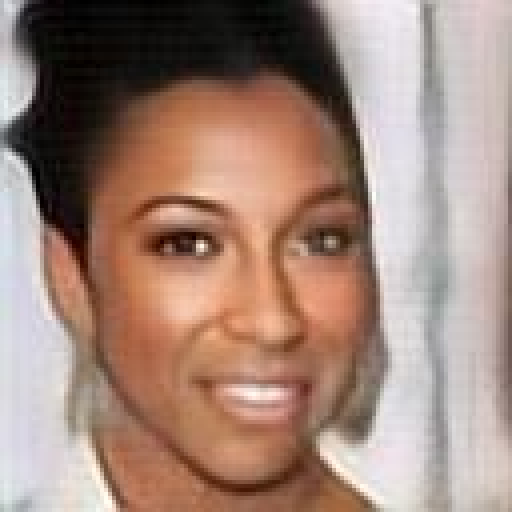}} &
{\includegraphics[width=0.14\textwidth]{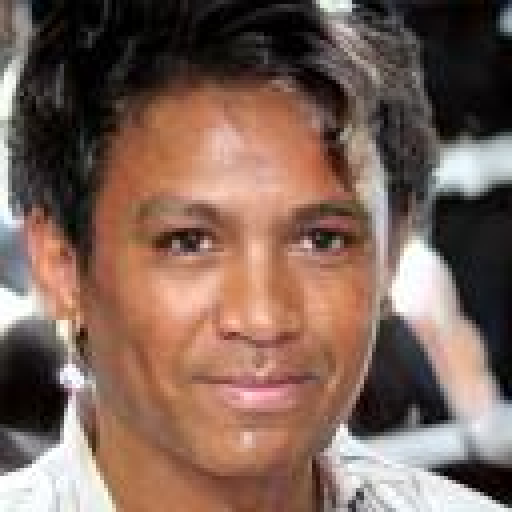}} &
{\includegraphics[width=0.14\textwidth]{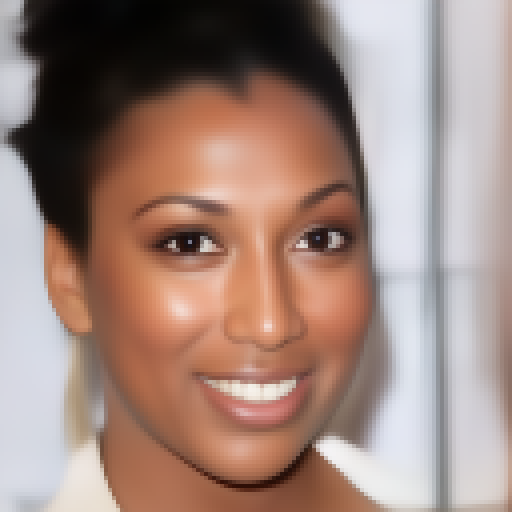}} &
{\includegraphics[width=0.14\textwidth]{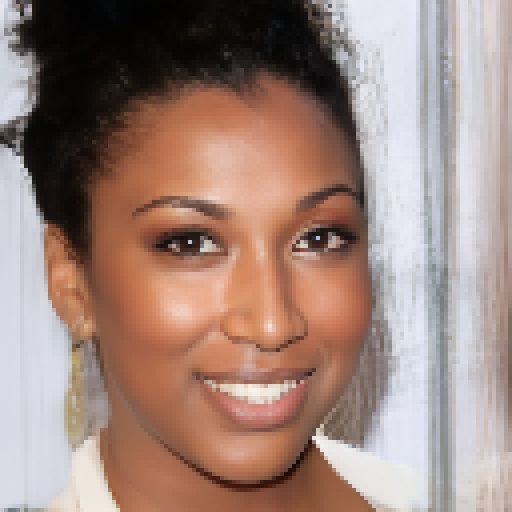}} \\
{\includegraphics[width=0.14\textwidth]{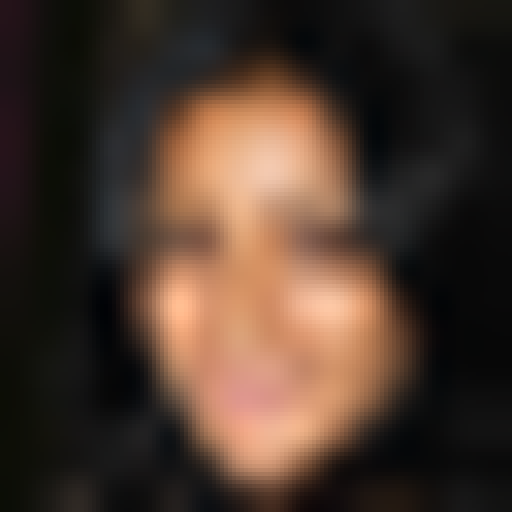}} &
{\includegraphics[width=0.14\textwidth]{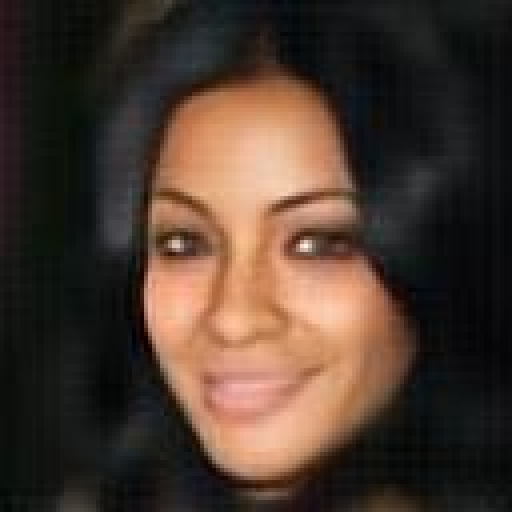}} &
{\includegraphics[width=0.14\textwidth]{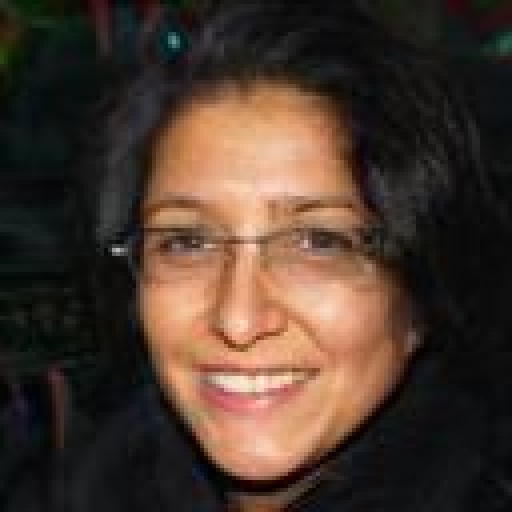}} &
{\includegraphics[width=0.14\textwidth]{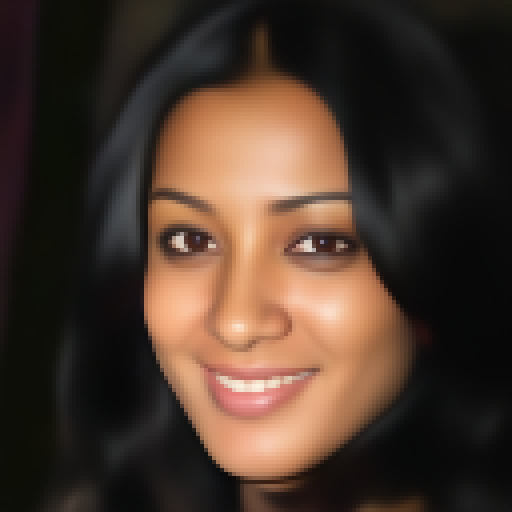}} &
{\includegraphics[width=0.14\textwidth]{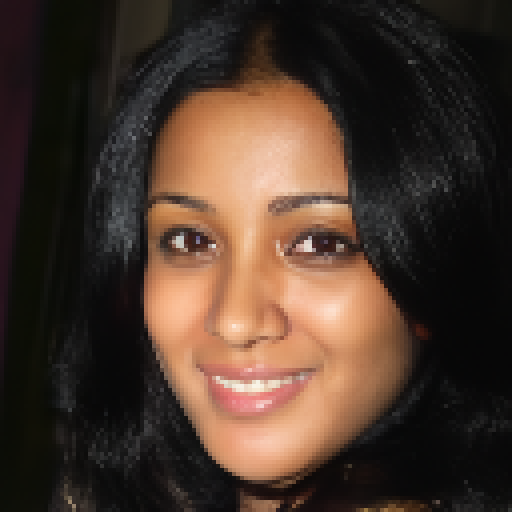}} \\
{\includegraphics[width=0.14\textwidth]{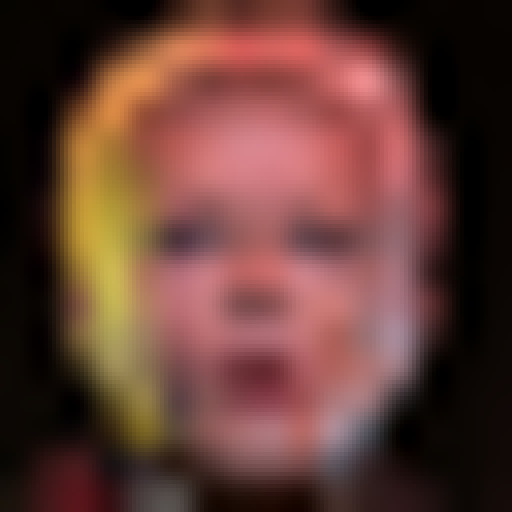}} &
{\includegraphics[width=0.14\textwidth]{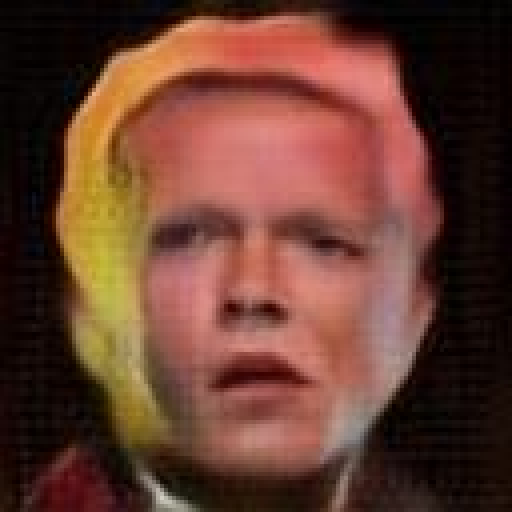}} &
{\includegraphics[width=0.14\textwidth]{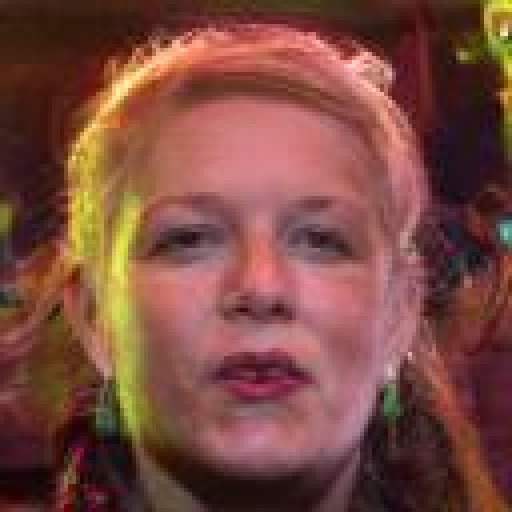}} &
{\includegraphics[width=0.14\textwidth]{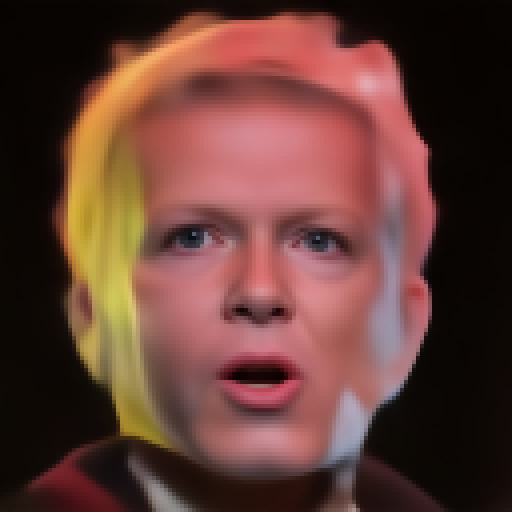}} &
{\includegraphics[width=0.14\textwidth]{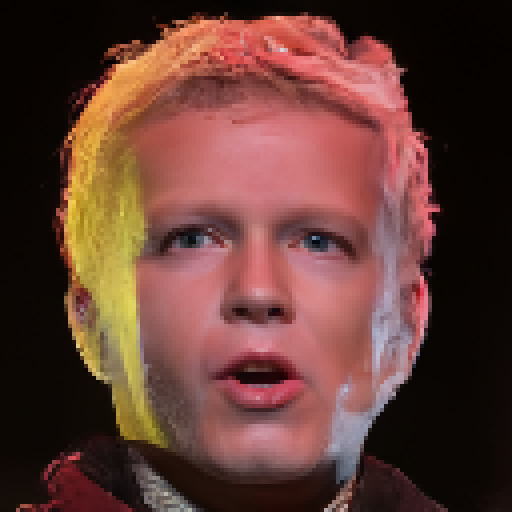}} \\
{\includegraphics[width=0.14\textwidth]{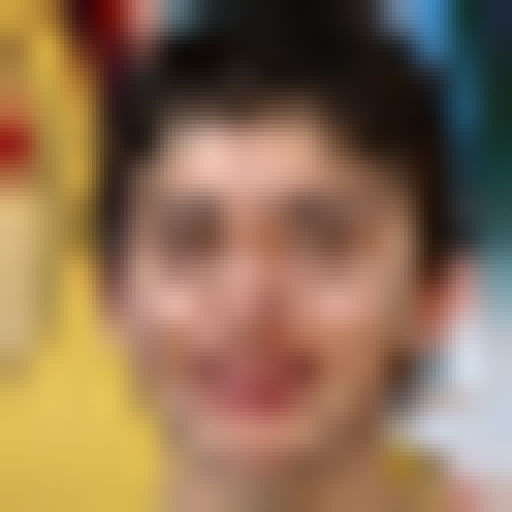}} &
{\includegraphics[width=0.14\textwidth]{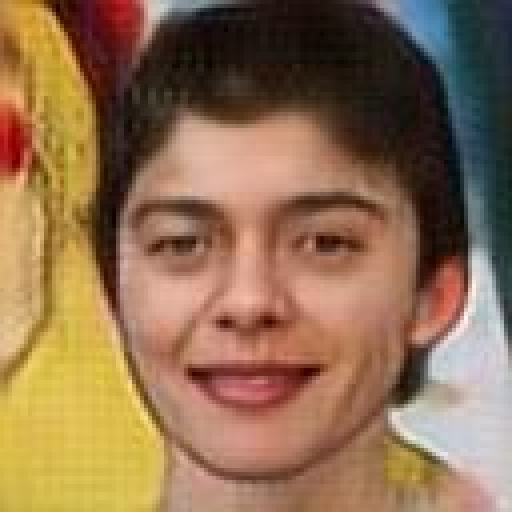}} &
{\includegraphics[width=0.14\textwidth]{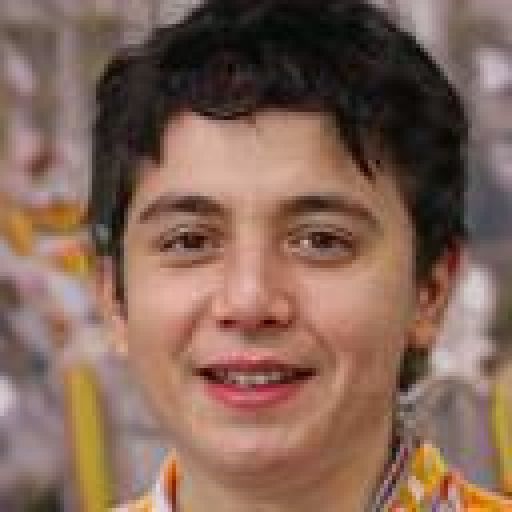}} &
{\includegraphics[width=0.14\textwidth]{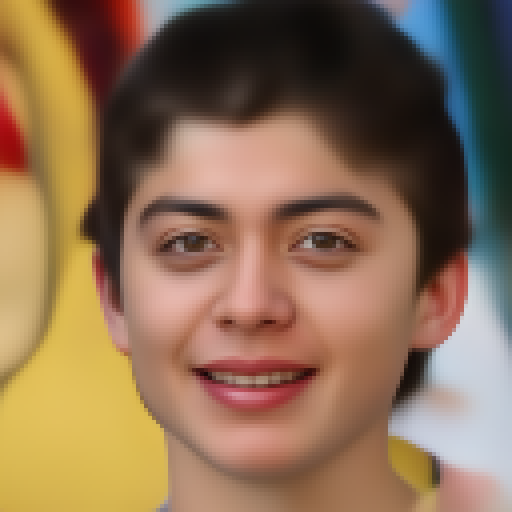}} &
{\includegraphics[width=0.14\textwidth]{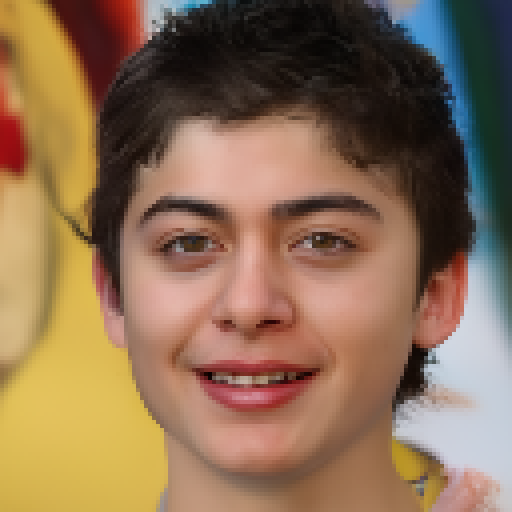}}
\end{tabular}
\end{center}
\vspace*{-0.25cm}
\caption{Additional results showing the comparison between different methods on the 
16$\times$16 $\rightarrow$ 128$\times$128
face super-resolution task.
\vspace*{-0.3cm}}
\label{fig:16x_128x_faces_with_baselines2}
\end{figure}

\vfill

\newpage

\newpage
\begin{center}
{\large \bf Cascaded Face Generation 1024$\times$1024}
\vspace*{-0.1cm}
\end{center}

\begin{figure}[H]
    \vspace*{-0.35cm}
    \centering
    \begin{tabular}{ccc}
    \small
    \scriptsize $(64\!\times\!64)$ & \scriptsize $(256\!\times\!256)$ & \scriptsize $(1024\!\times\!1024)$\\
    \raisebox{.27\textwidth}{\includegraphics[width=.09\textwidth]{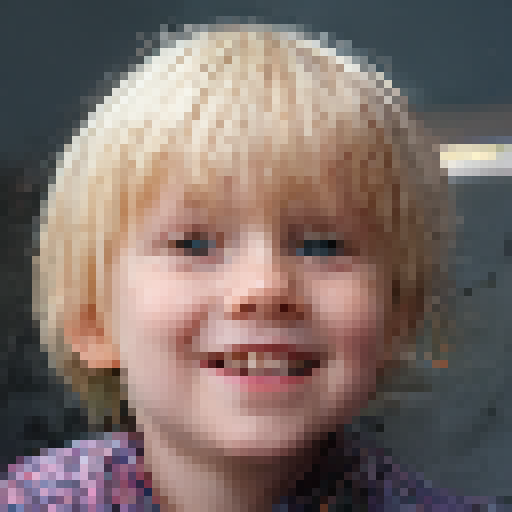}} &
    \raisebox{.18\textwidth}{\includegraphics[width=.18\textwidth]{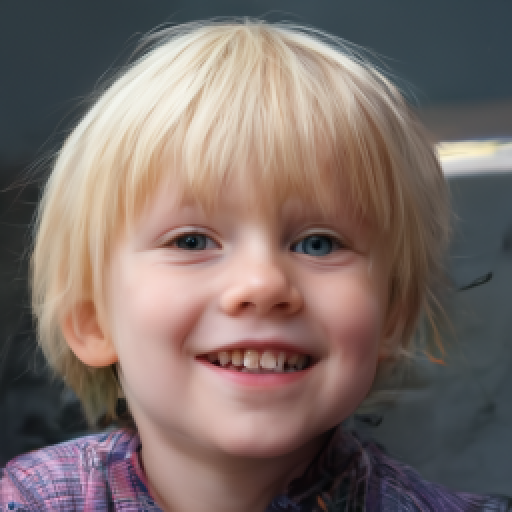}} &
    \includegraphics[width=.36\textwidth]{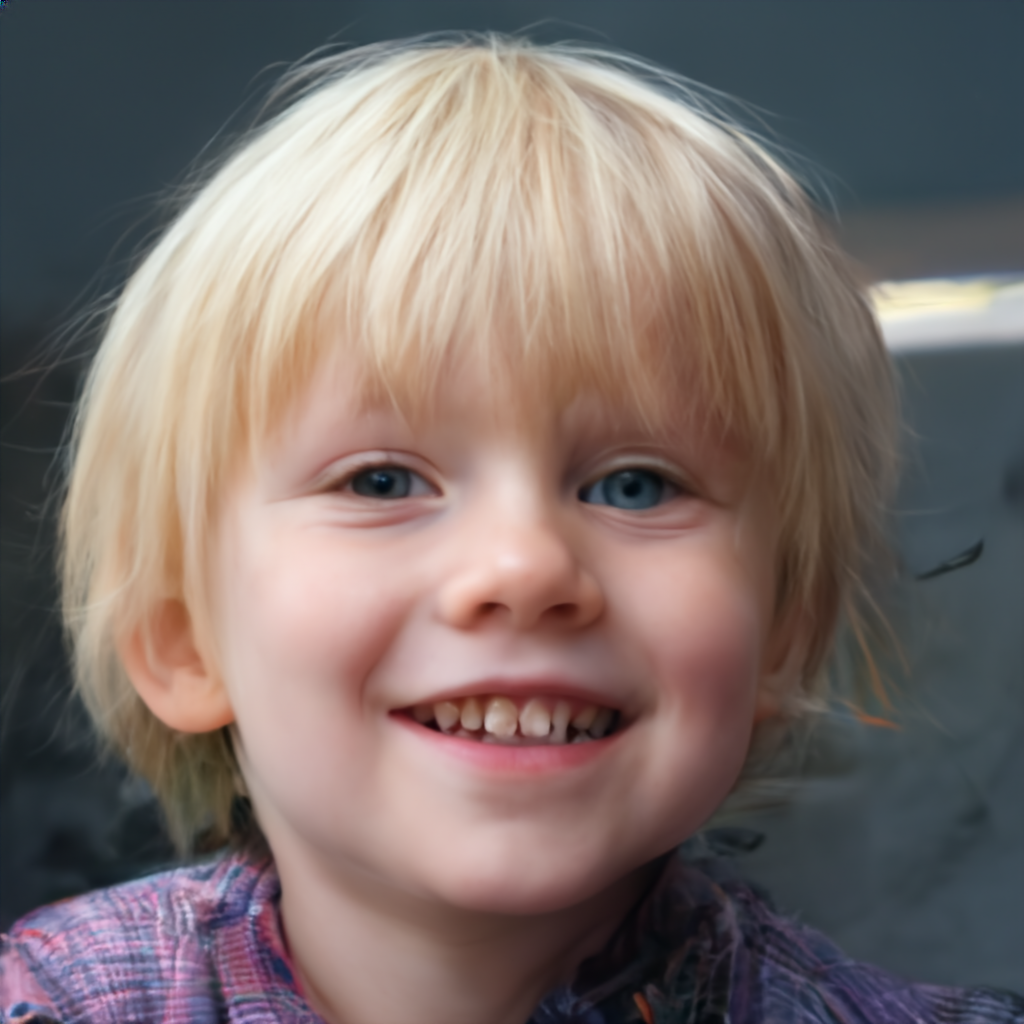}
    \end{tabular}
    
    \medskip
    \begin{tabular}{ccc}
    \small
    \raisebox{.27\textwidth}{\includegraphics[width=.09\textwidth]{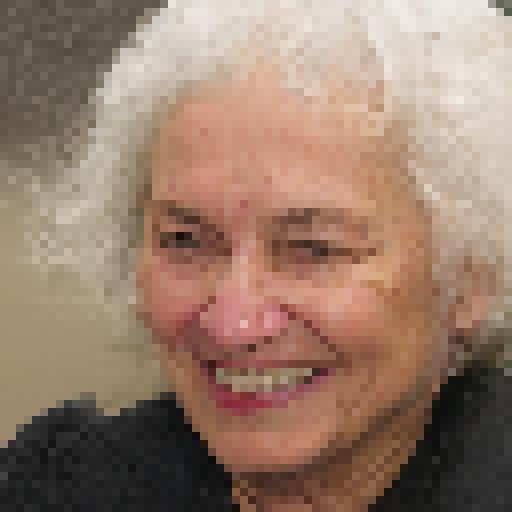}} &
    \raisebox{.18\textwidth}{\includegraphics[width=.18\textwidth]{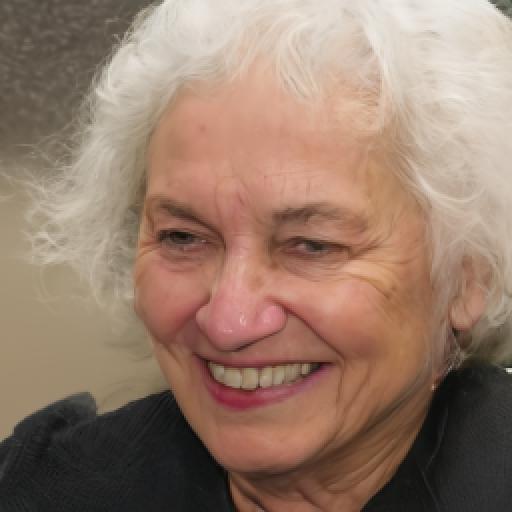}} &
    \includegraphics[width=.36\textwidth]{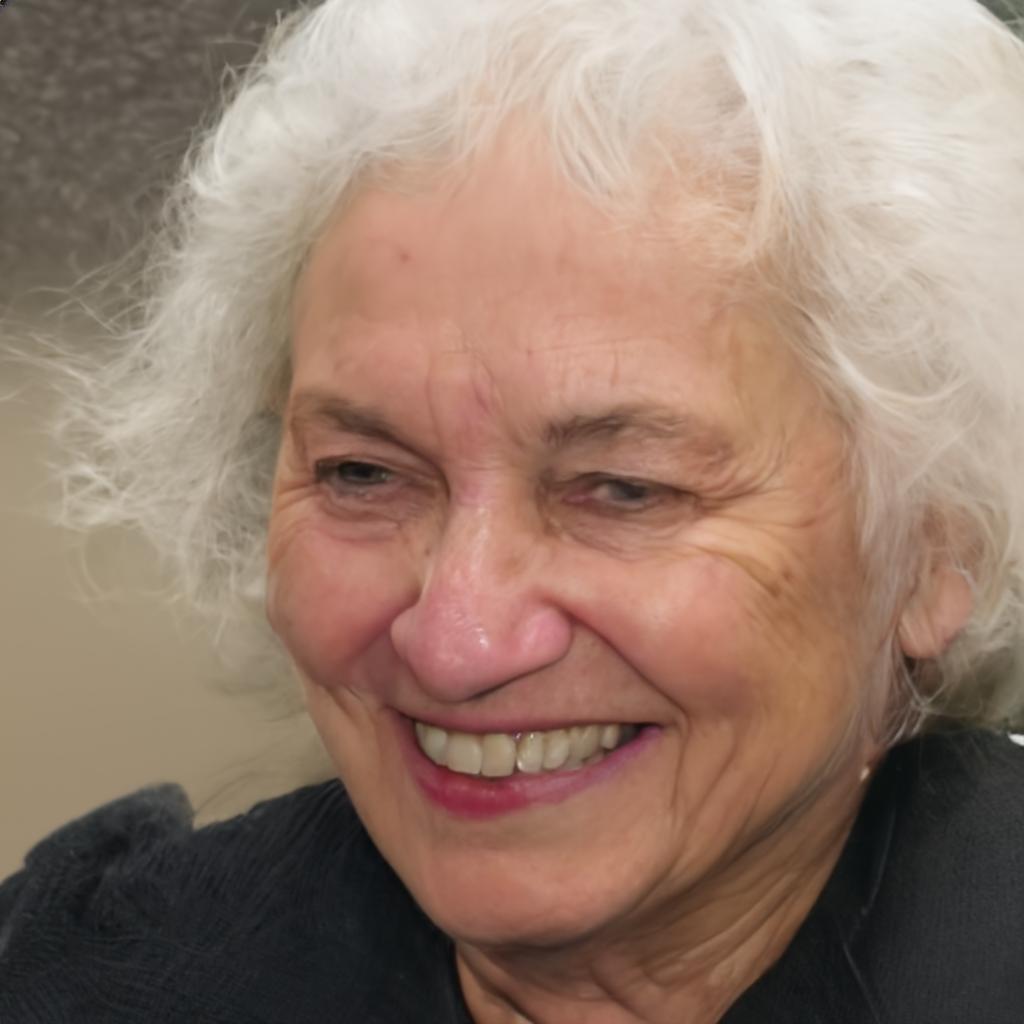}
    \end{tabular}
    
    \medskip
     \begin{tabular}{ccc}
    \small
    \raisebox{.27\textwidth}{\includegraphics[width=.09\textwidth]{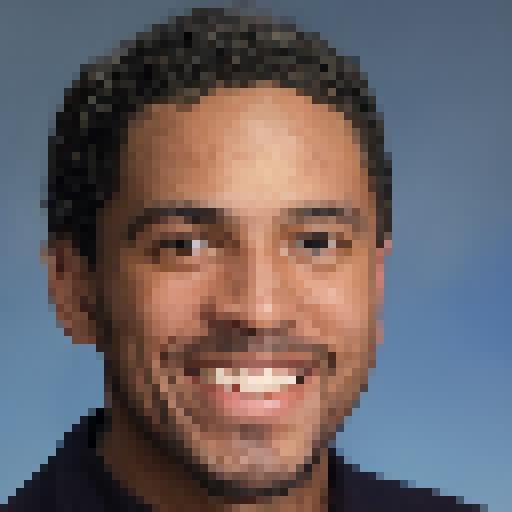}} &
    \raisebox{.18\textwidth}{\includegraphics[width=.18\textwidth]{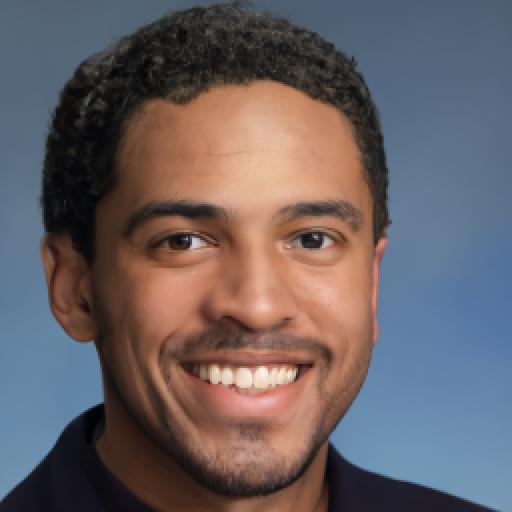}} &
    \includegraphics[width=.36\textwidth]{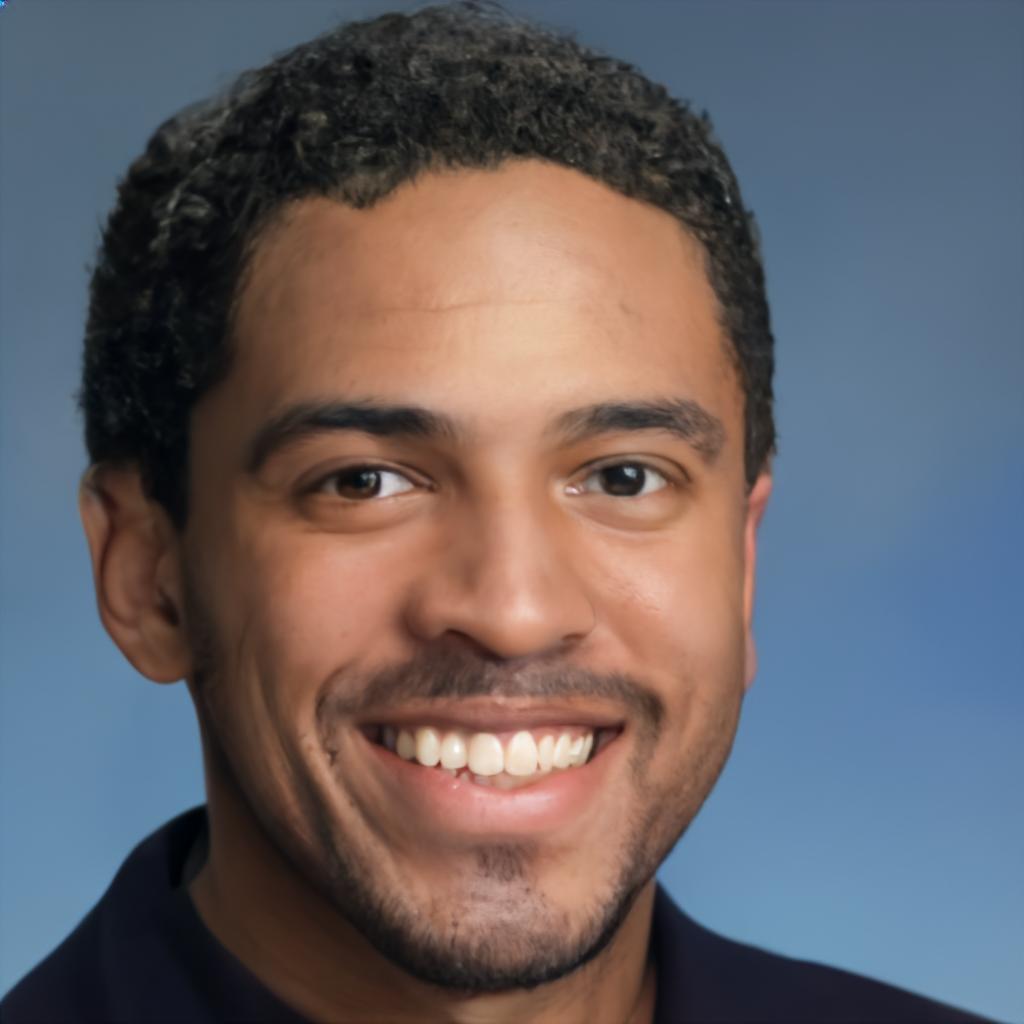}
    \end{tabular}
    \caption{Cascaded generation on faces using an unconditional model chained with two \modelname models.}
    \label{fig:my_label}
\end{figure}



\vfill


\newpage
\begin{center}
{\large \bf Unconditional Face Samples \  1024$\times$1024}
\end{center}

\begin{figure}[H]
\vspace*{-0.2cm}
\setlength{\tabcolsep}{2pt}
\begin{center}
\begin{tabular}{cccccc}
\multicolumn{3}{c}{\includegraphics[width=0.515\textwidth]{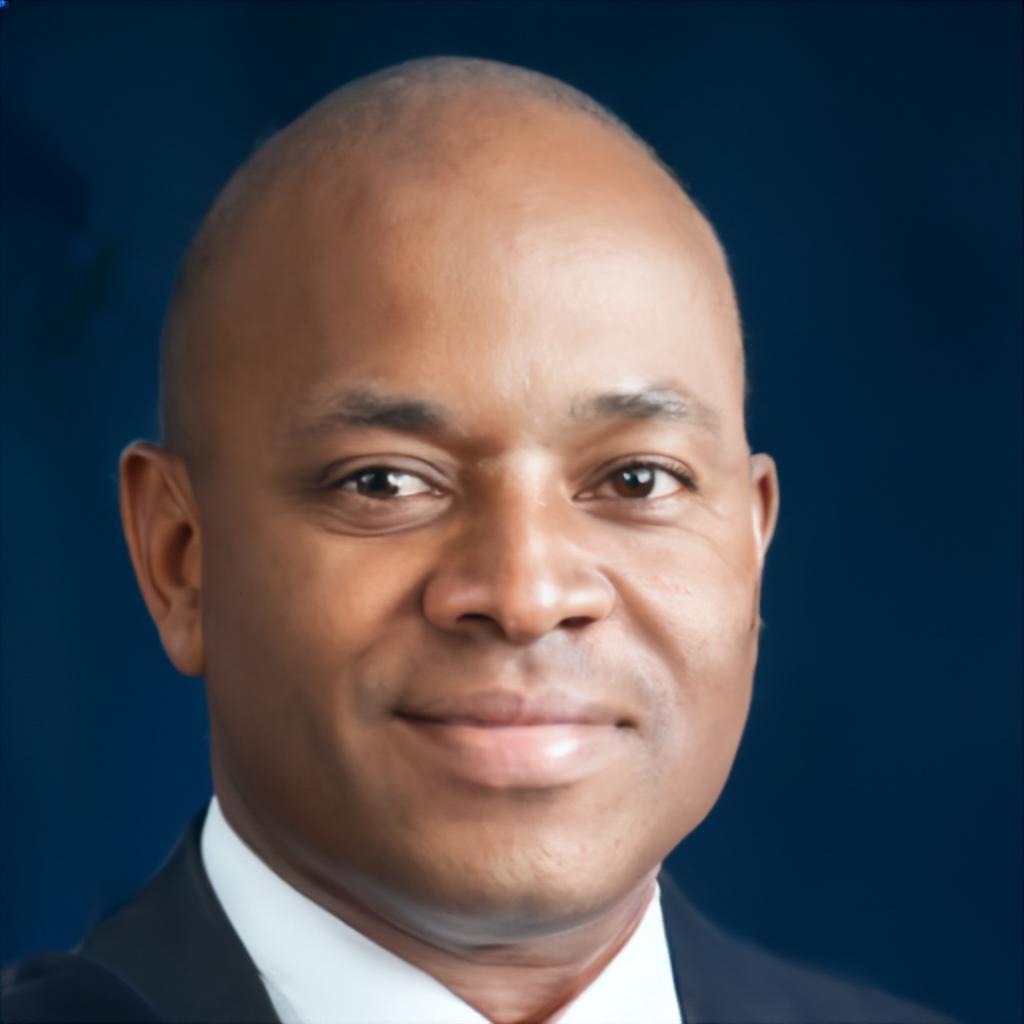}} &
\multicolumn{3}{c}{\includegraphics[width=0.515\textwidth]{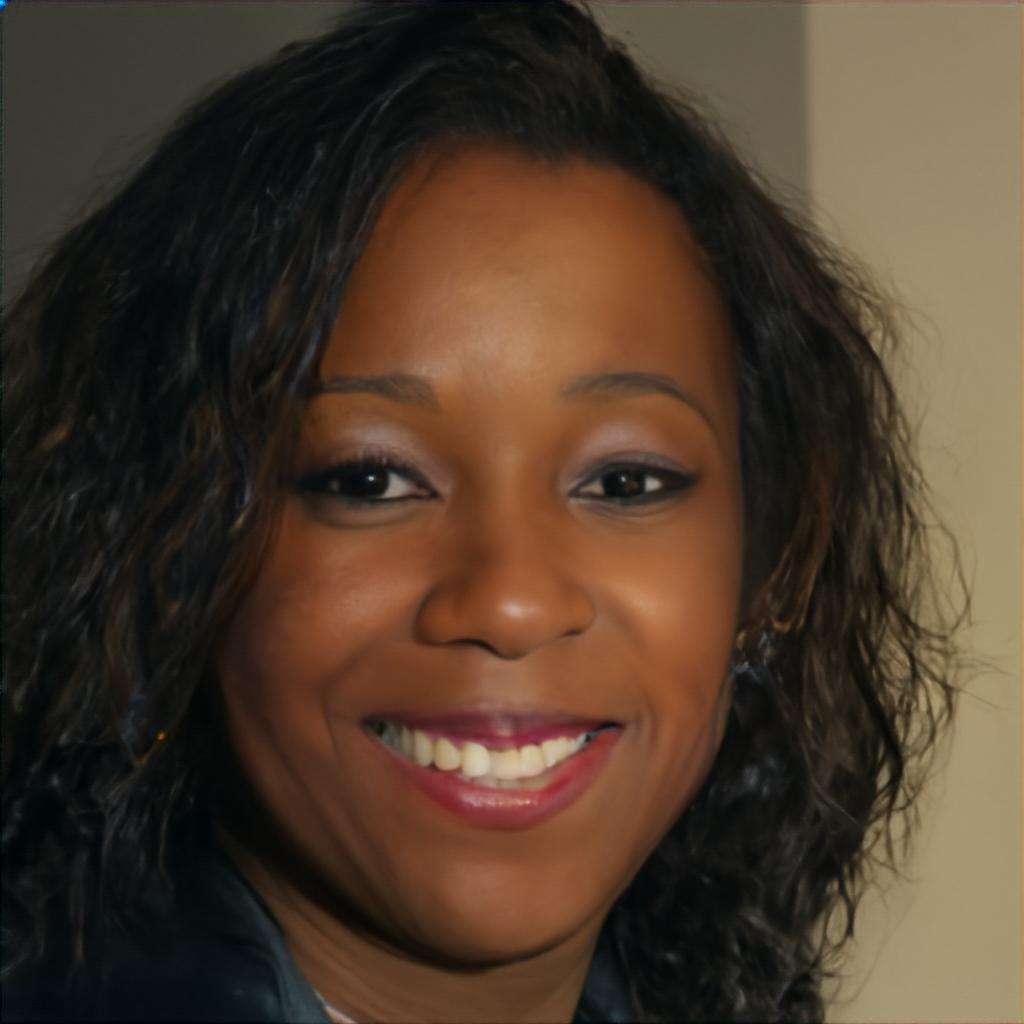}}\\

\multicolumn{2}{c}{\includegraphics[width=0.34\textwidth]{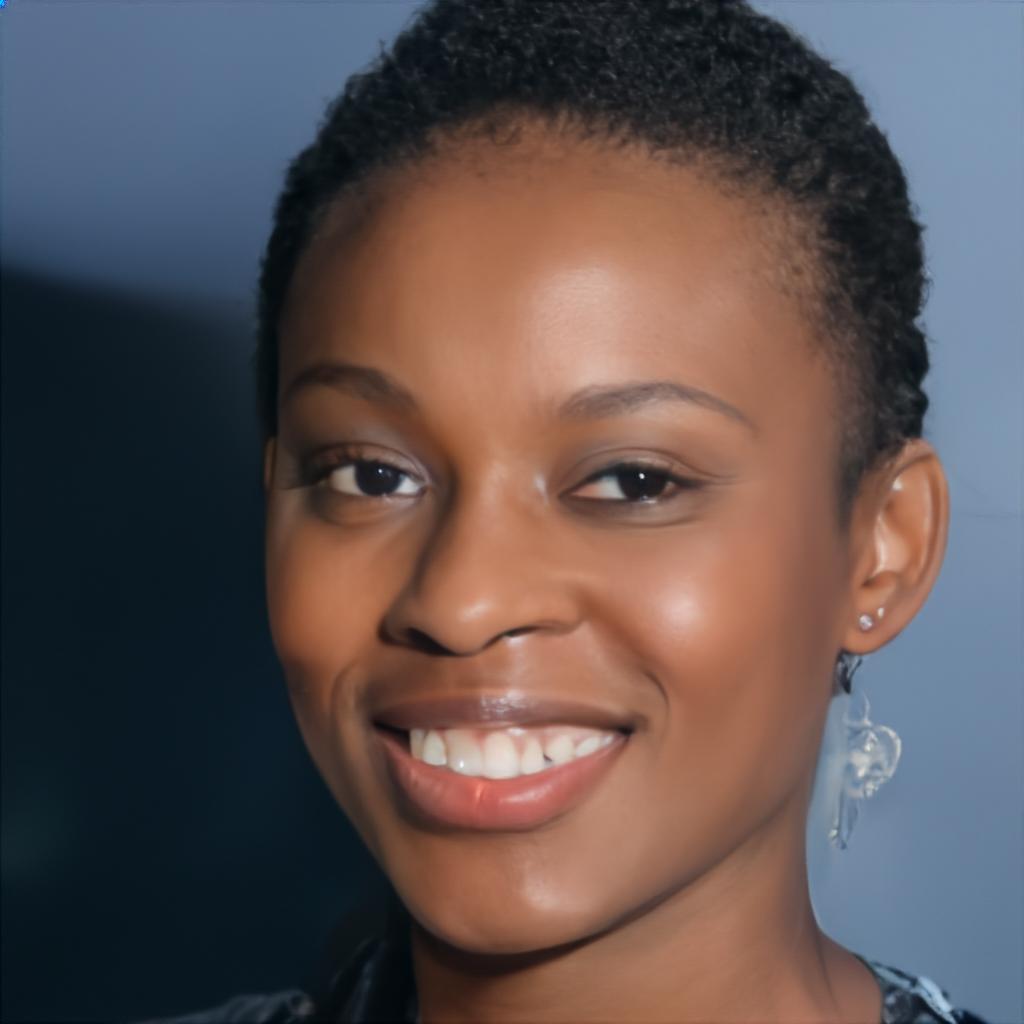}} &
\multicolumn{2}{c}{\includegraphics[width=0.34\textwidth]{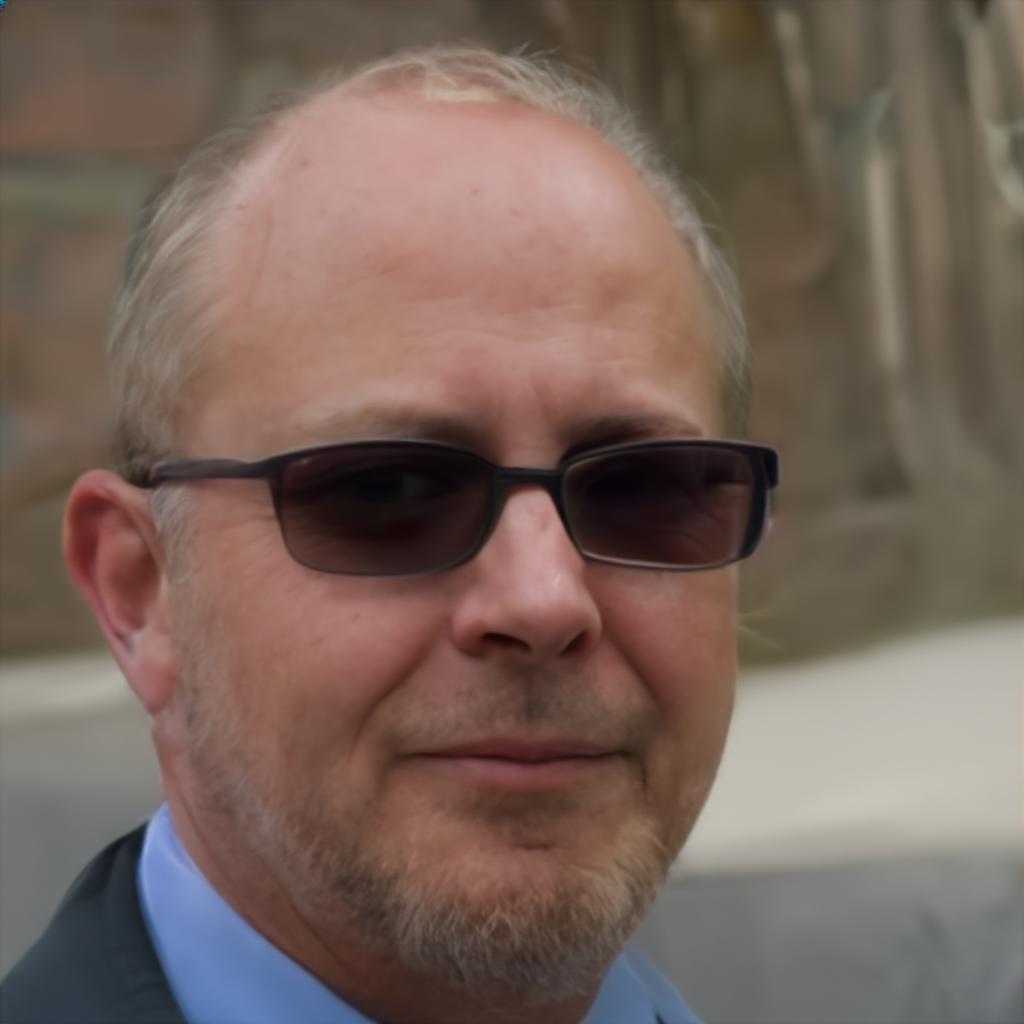}} &
\multicolumn{2}{c}{\includegraphics[width=0.34\textwidth]{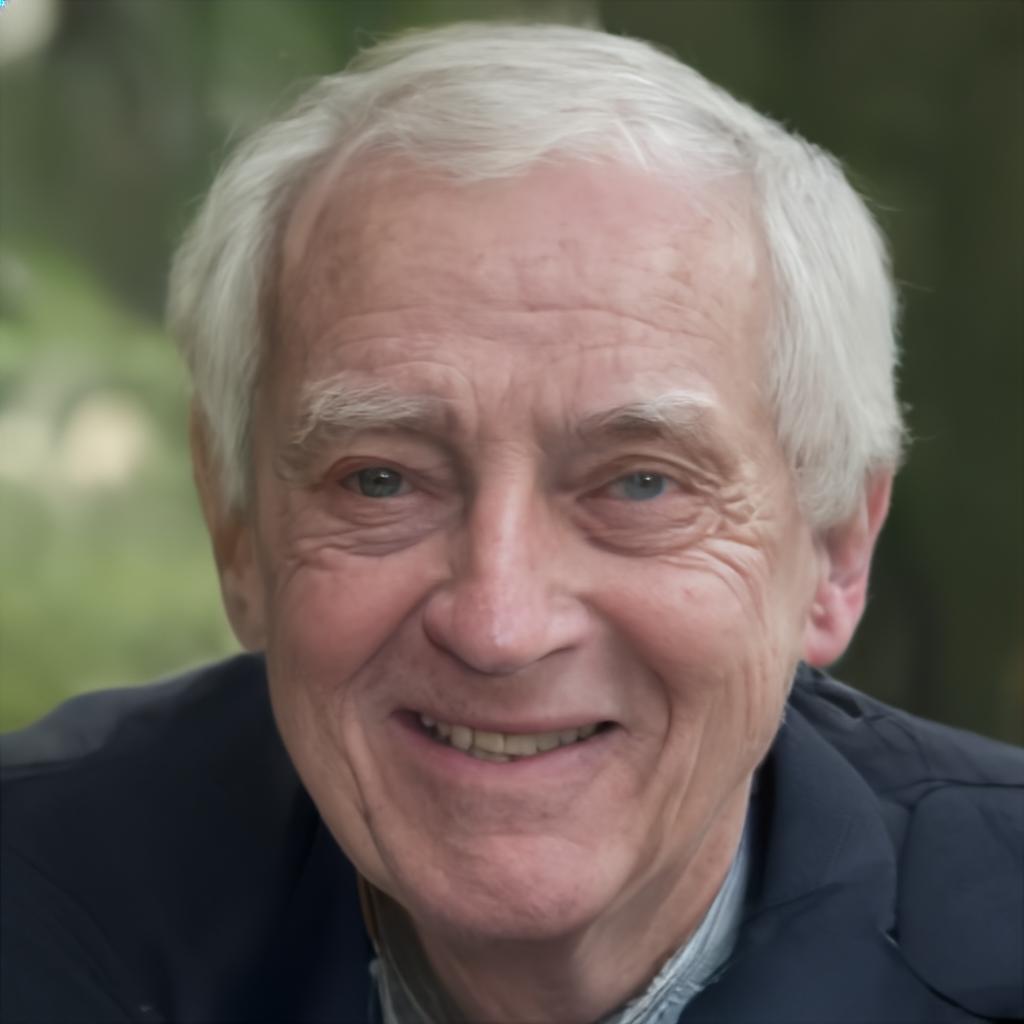}} \\

\multicolumn{2}{c}{\includegraphics[width=0.34\textwidth]{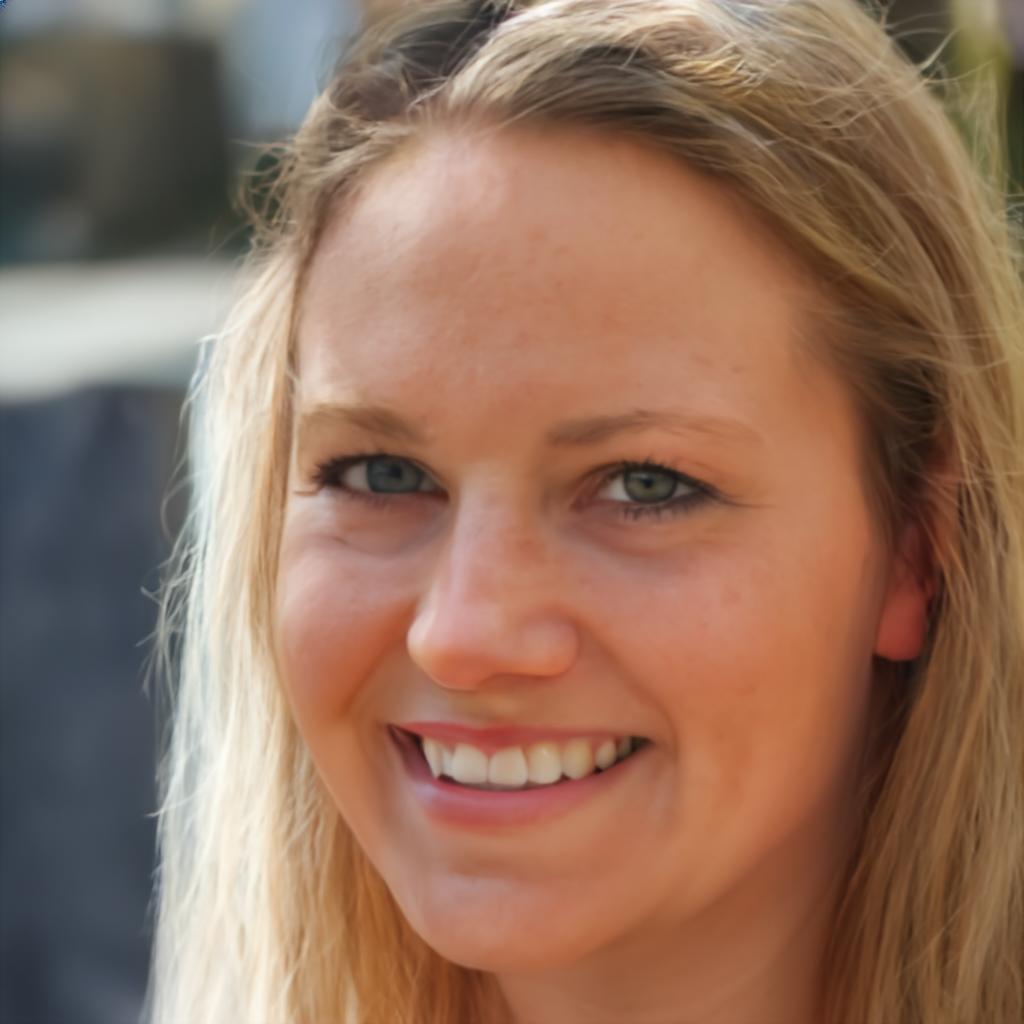}} &
\multicolumn{2}{c}{\includegraphics[width=0.34\textwidth]{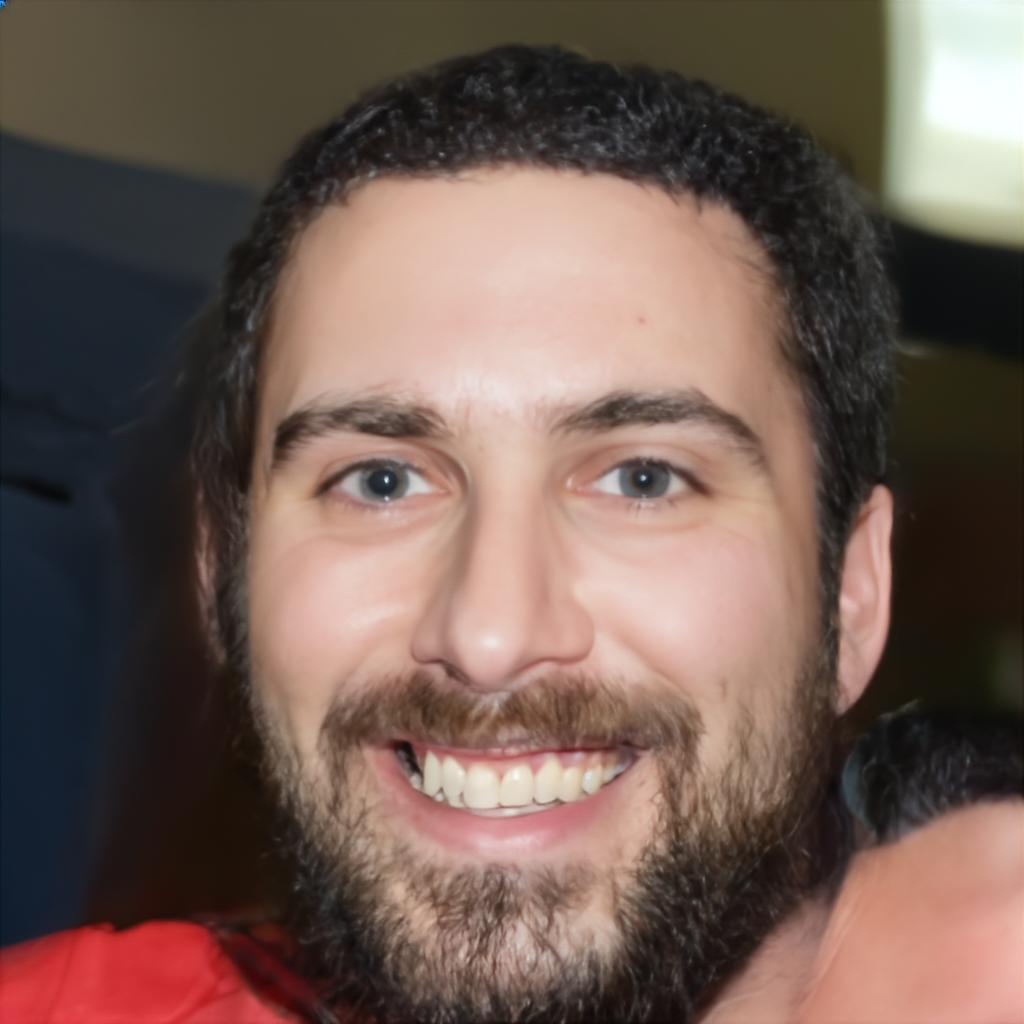}} &
\multicolumn{2}{c}{\includegraphics[width=0.34\textwidth]{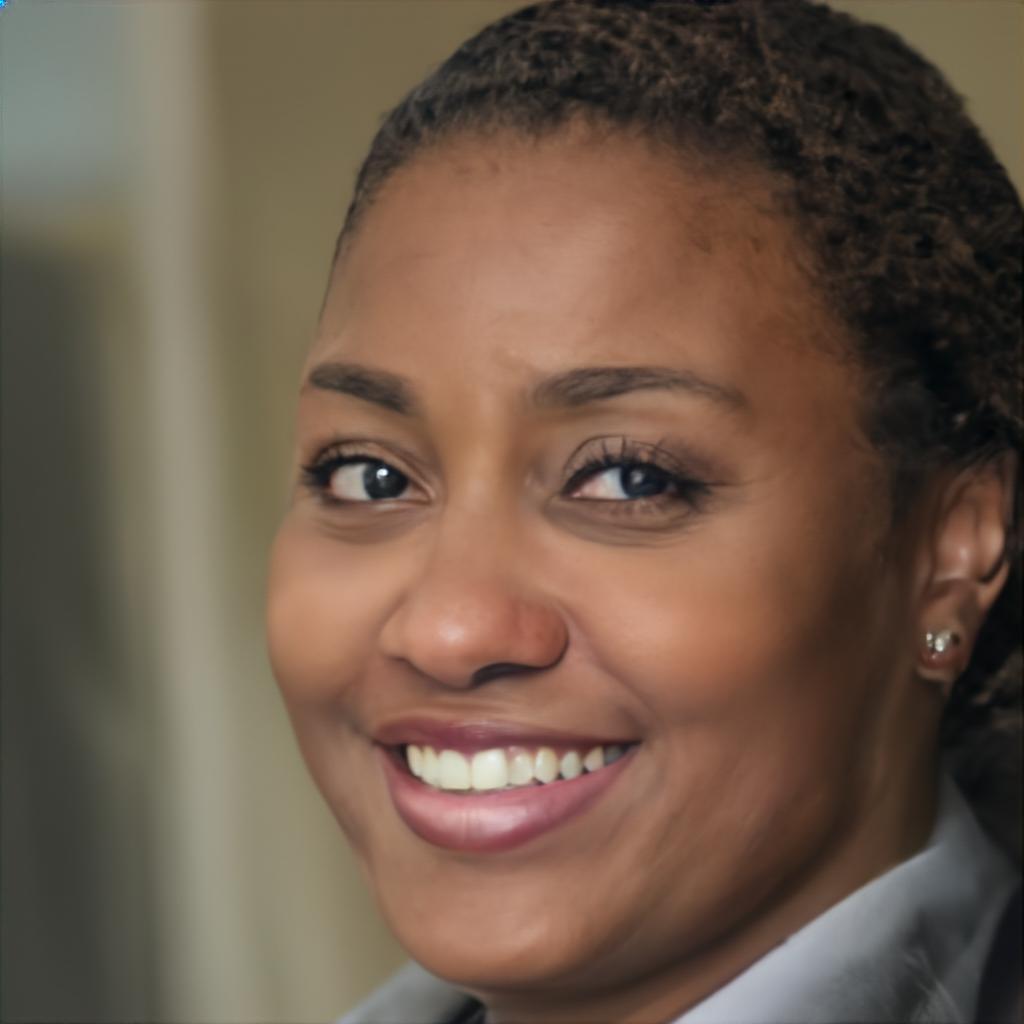}} \\
\end{tabular}
\end{center}
\vspace*{-0.4cm}
\caption{Additional Synthetic 1024$\times$1024 faces images. We first sample from an unconditional 64$\times$64 diffusion model, then pass the samples through two 4$\times$ \modelname models, \ie~ 64$\times$64 $\rightarrow$ 256$\times$256 $\rightarrow$ 1024$\times$1024.
\vspace*{-.4cm}
}
\label{fig:1024x_cascade2}
\vspace*{-0.35cm}
\end{figure}

\begin{figure}[H]
\setlength{\tabcolsep}{2pt}
\begin{center}
\begin{tabular}{cccccc}
\multicolumn{3}{c}{\includegraphics[width=0.515\textwidth]{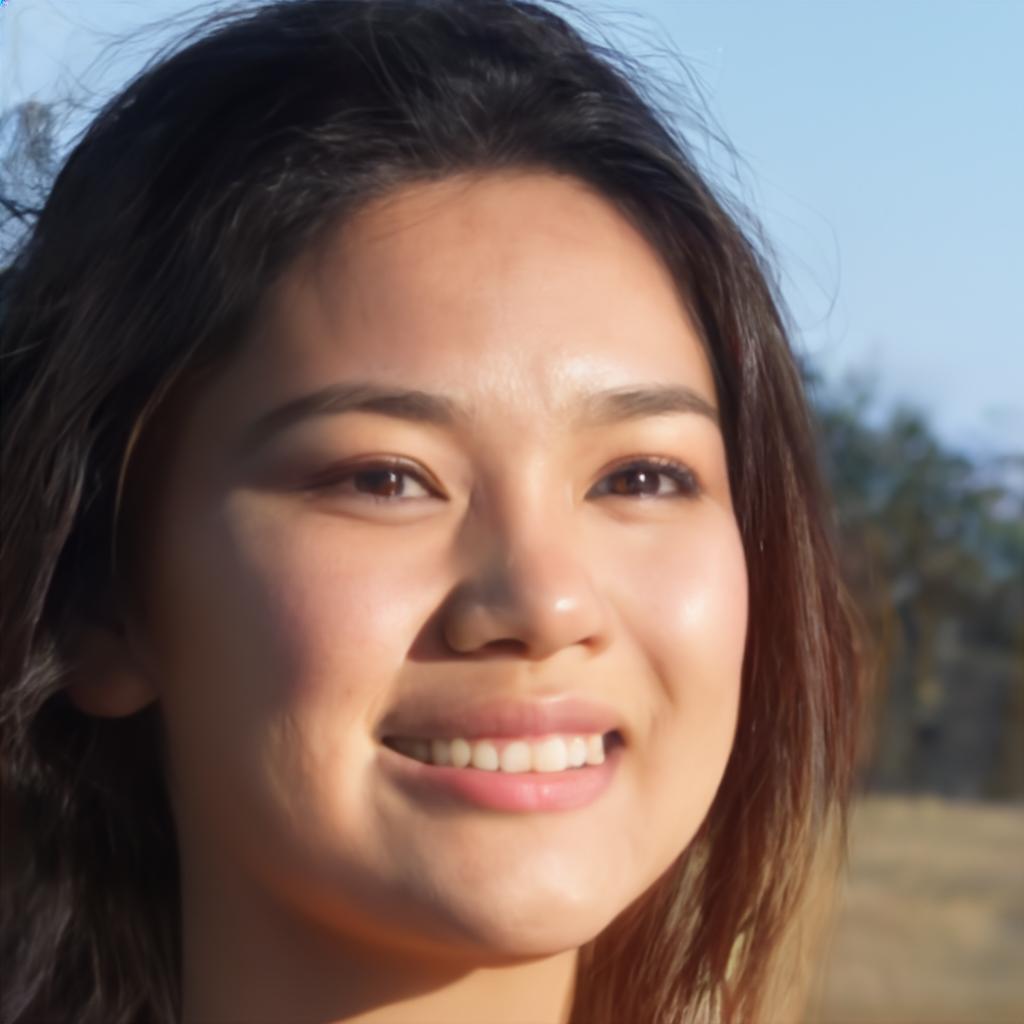}} &
\multicolumn{3}{c}{\includegraphics[width=0.515\textwidth]{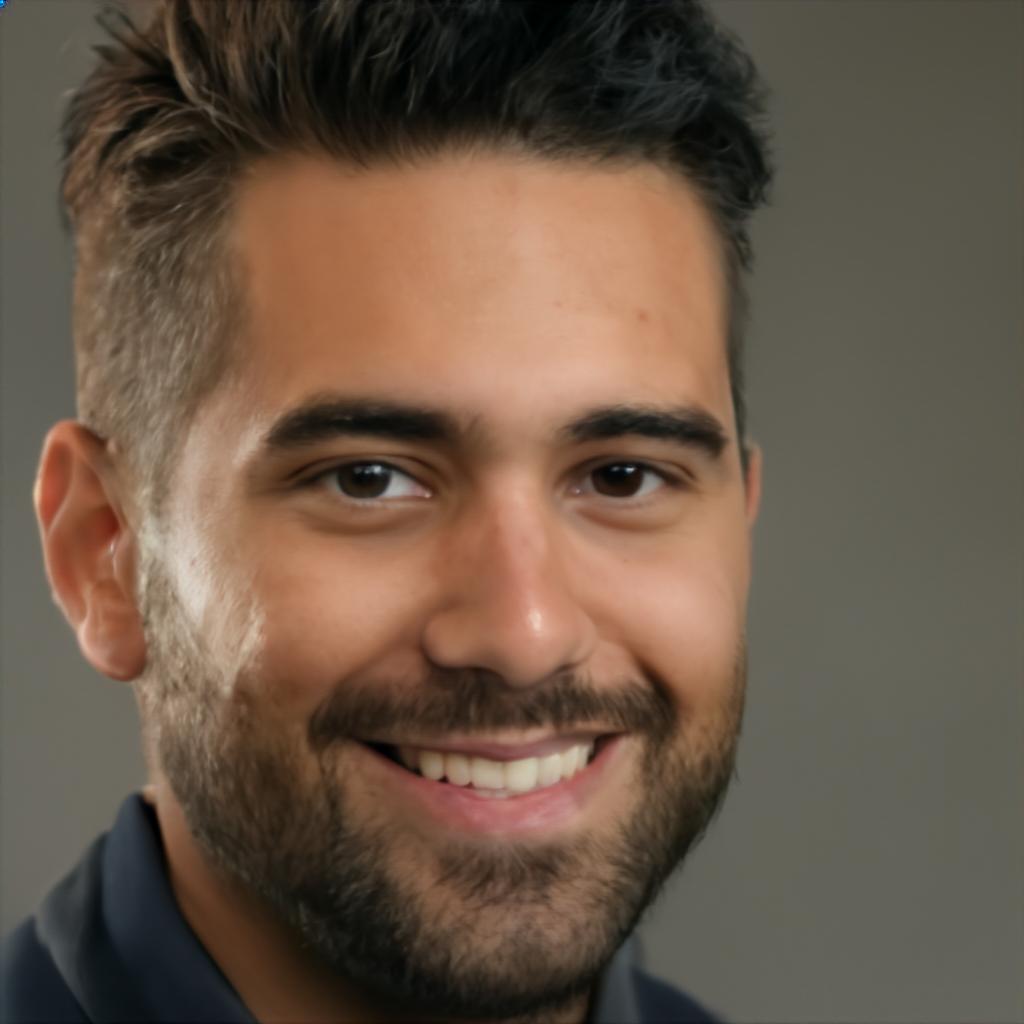}}\\

\multicolumn{2}{c}{\includegraphics[width=0.34\textwidth]{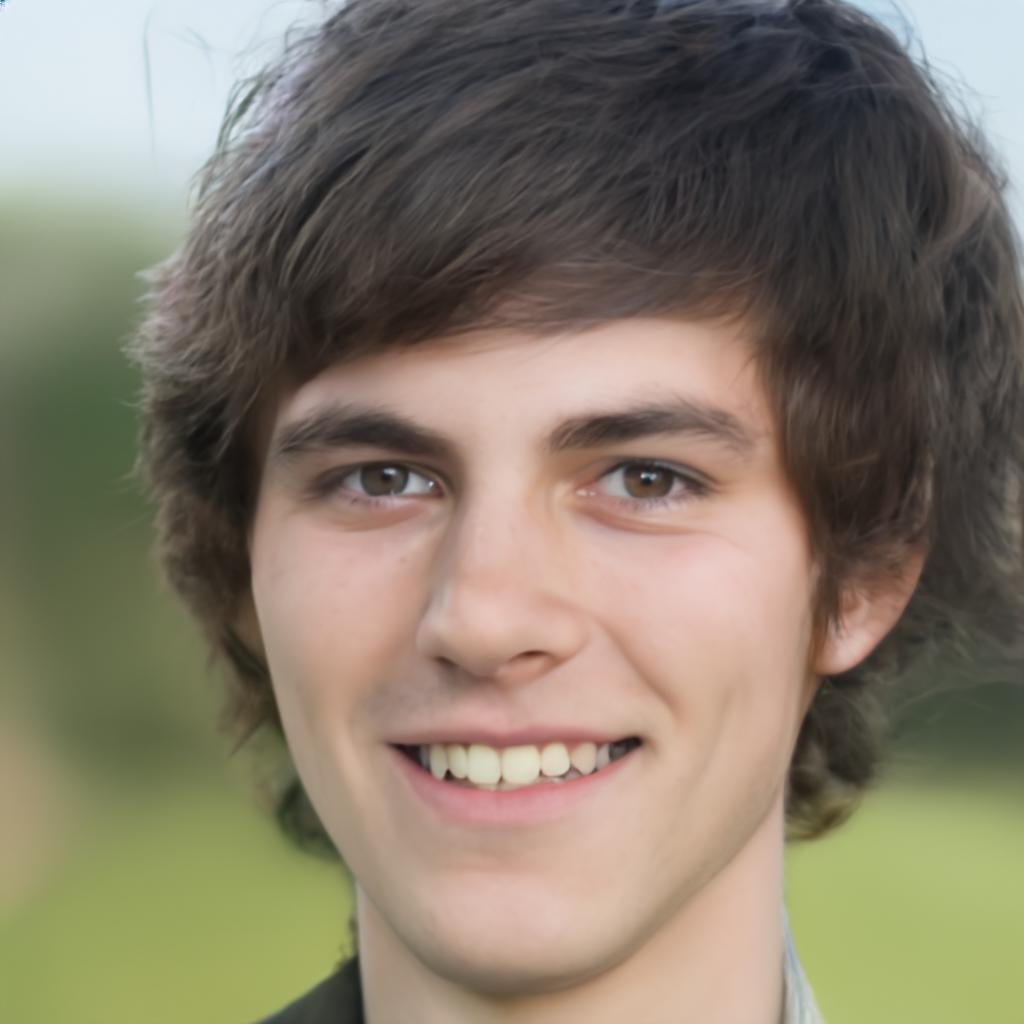}} &
\multicolumn{2}{c}{\includegraphics[width=0.34\textwidth]{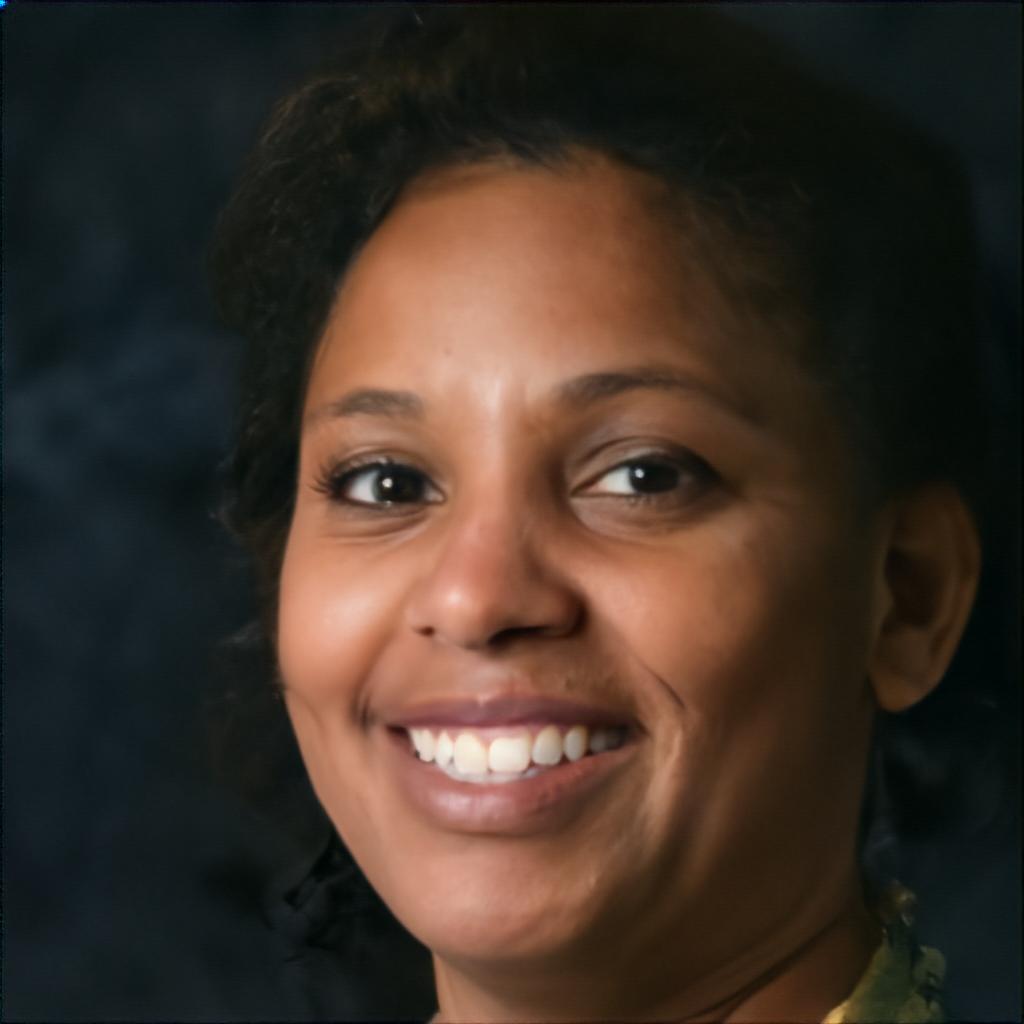}} &
\multicolumn{2}{c}{\includegraphics[width=0.34\textwidth]{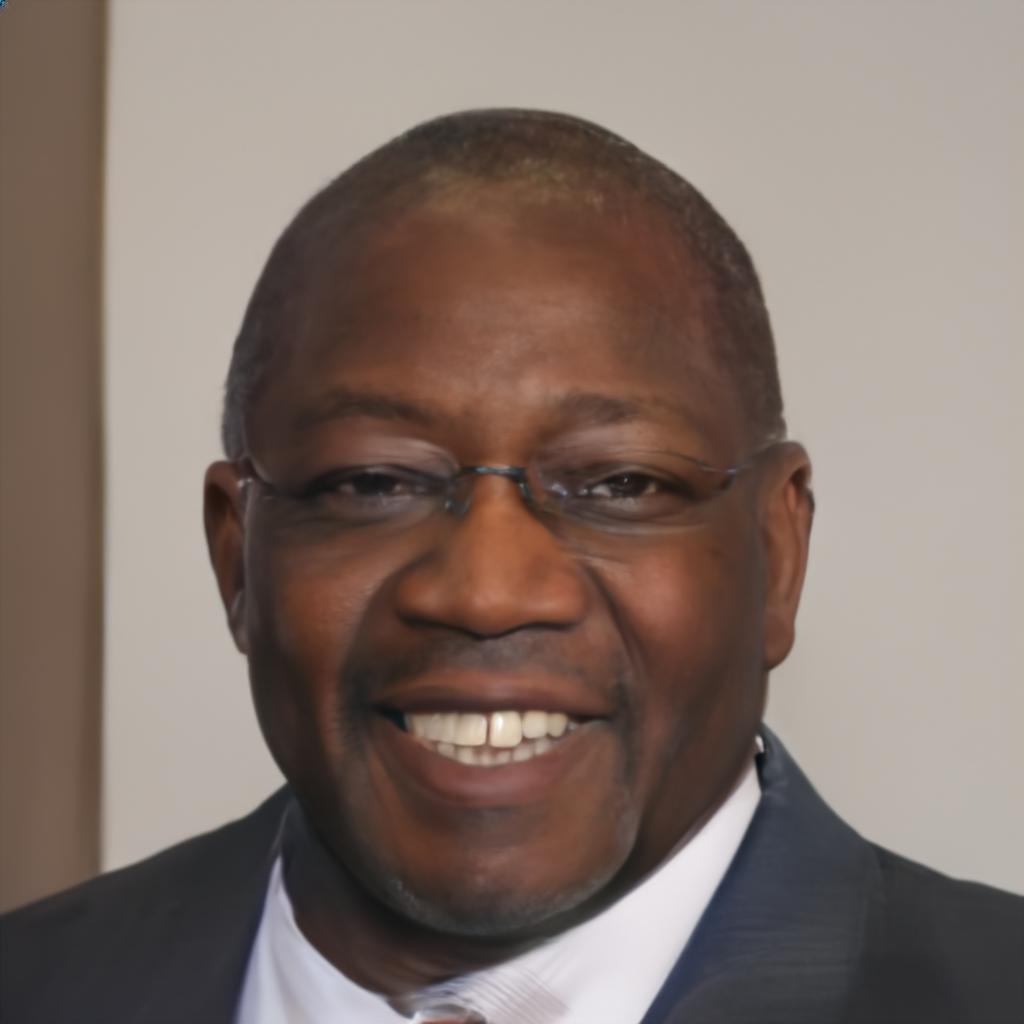}} \\

\multicolumn{2}{c}{\includegraphics[width=0.34\textwidth]{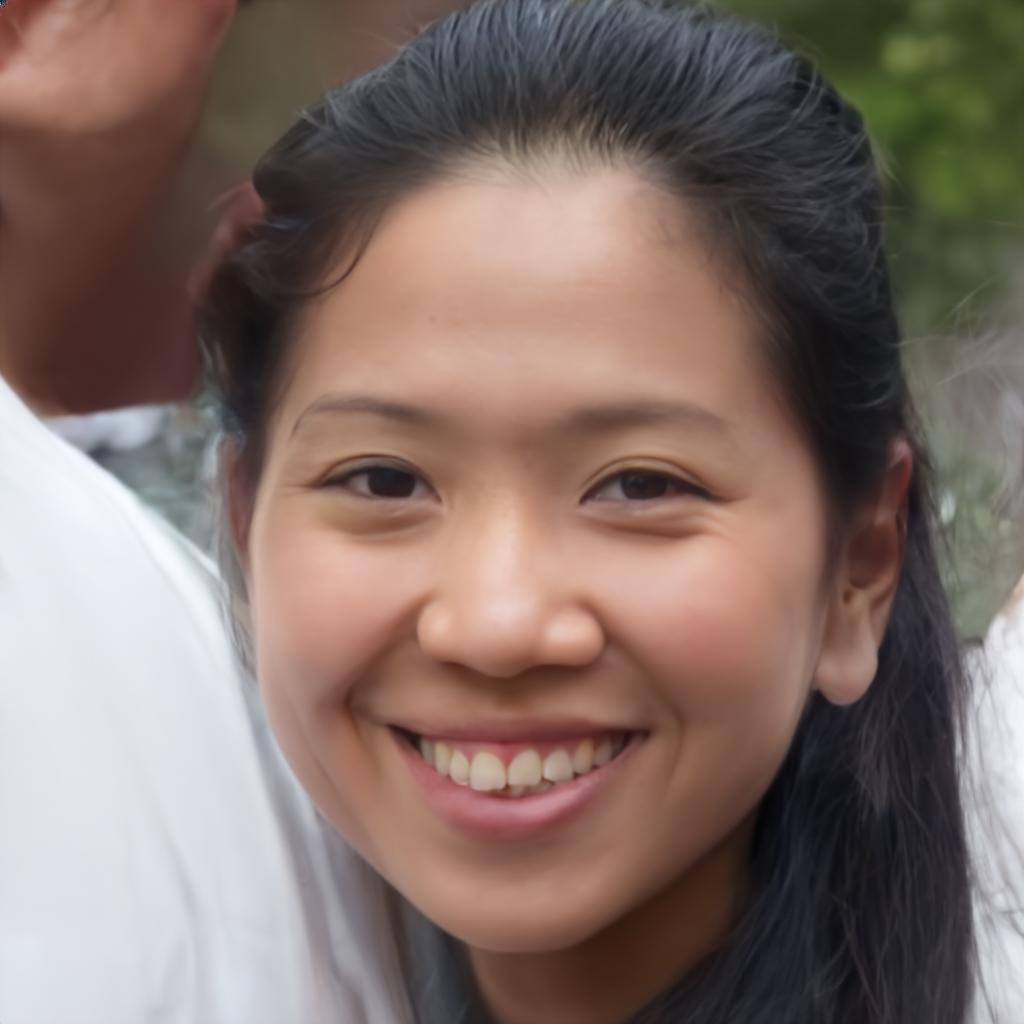}} &
\multicolumn{2}{c}{\includegraphics[width=0.34\textwidth]{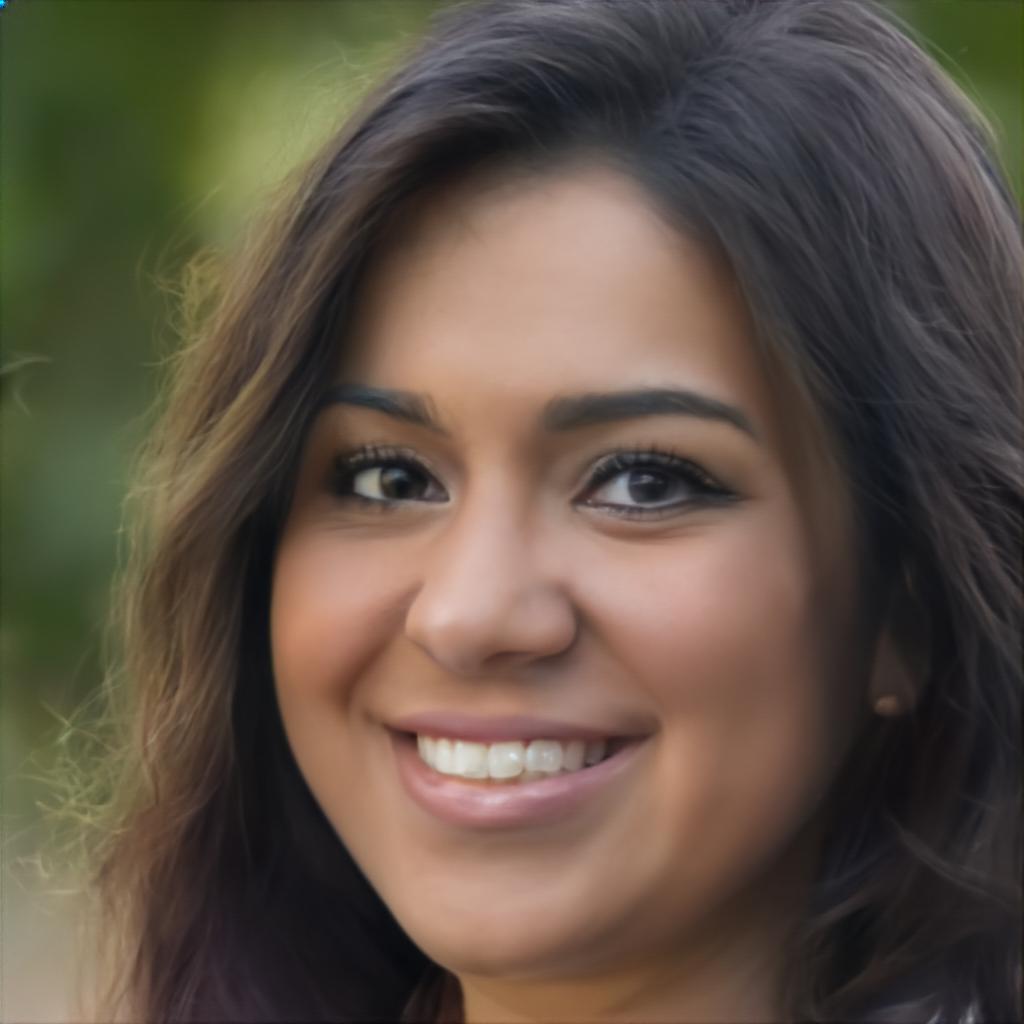}} &
\multicolumn{2}{c}{\includegraphics[width=0.34\textwidth]{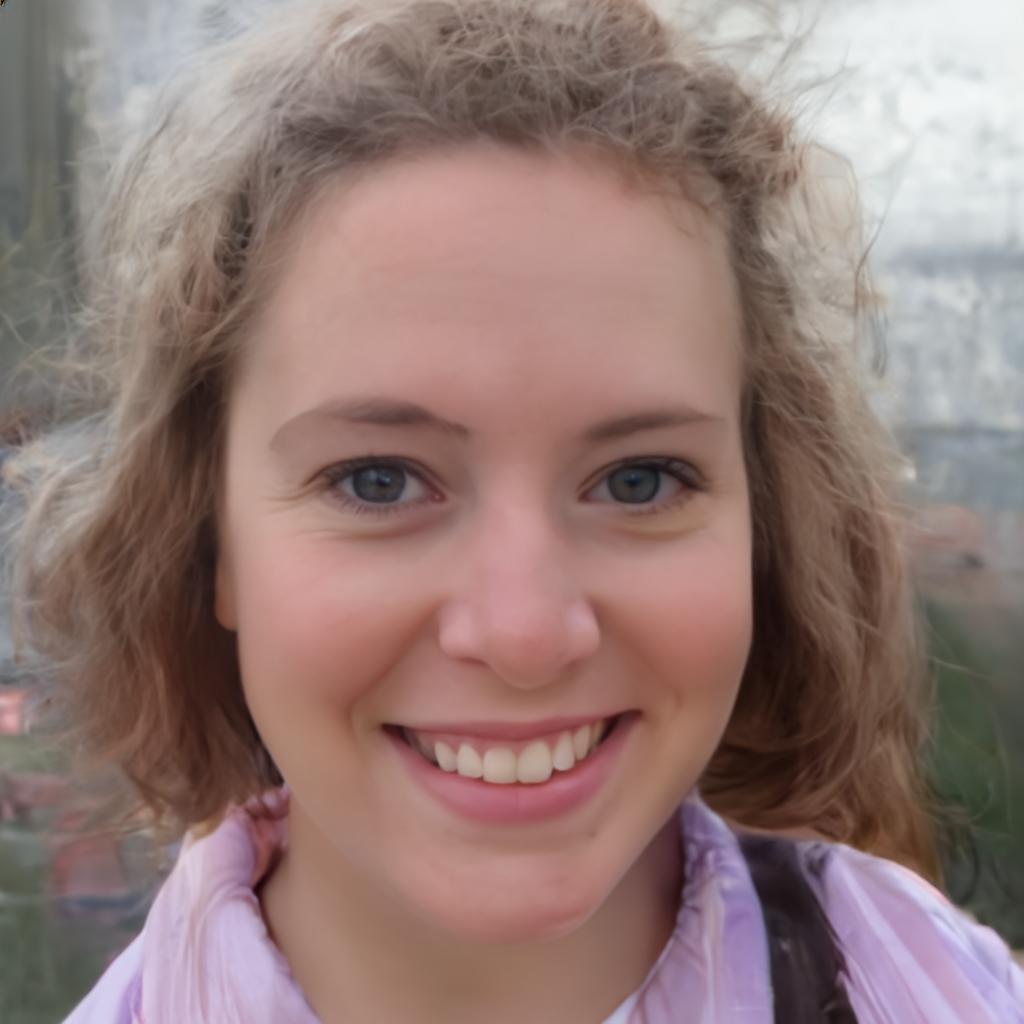}} \\
\end{tabular}
\end{center}
\vspace*{-0.4cm}
\caption{Additional Synthetic 1024$\times$1024 faces images. We first sample from an unconditional 64$\times$64 diffusion model, then pass the samples through two 4$\times$ \modelname models, \ie~ 64$\times$64 $\rightarrow$ 256$\times$256 $\rightarrow$ 1024$\times$1024.
\vspace*{-.4cm}
}
\label{fig:1024x_cascade3}
\vspace*{-0.35cm}
\end{figure}

\begin{figure}[H]
\setlength{\tabcolsep}{2pt}
\begin{center}
\begin{tabular}{cccccc}

\includegraphics[width=0.15\textwidth]{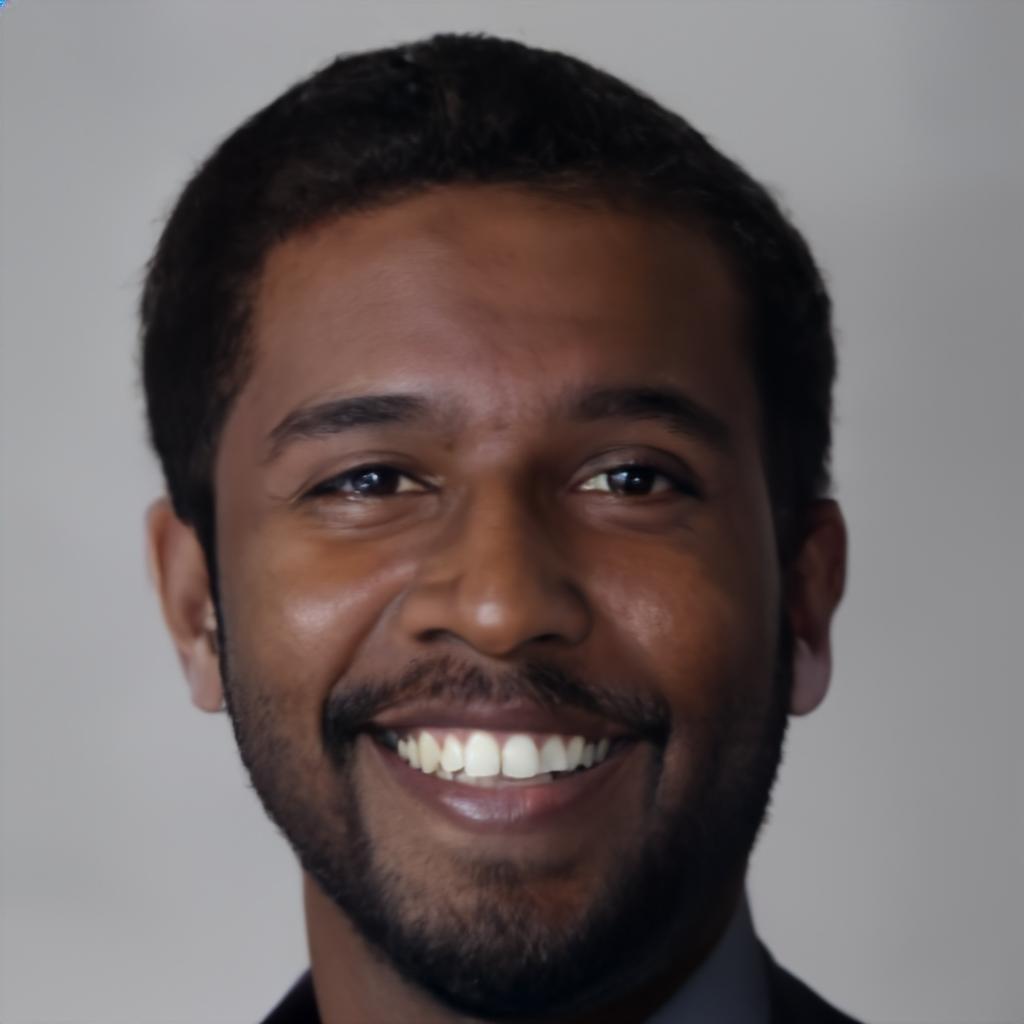} &
\includegraphics[width=0.15\textwidth]{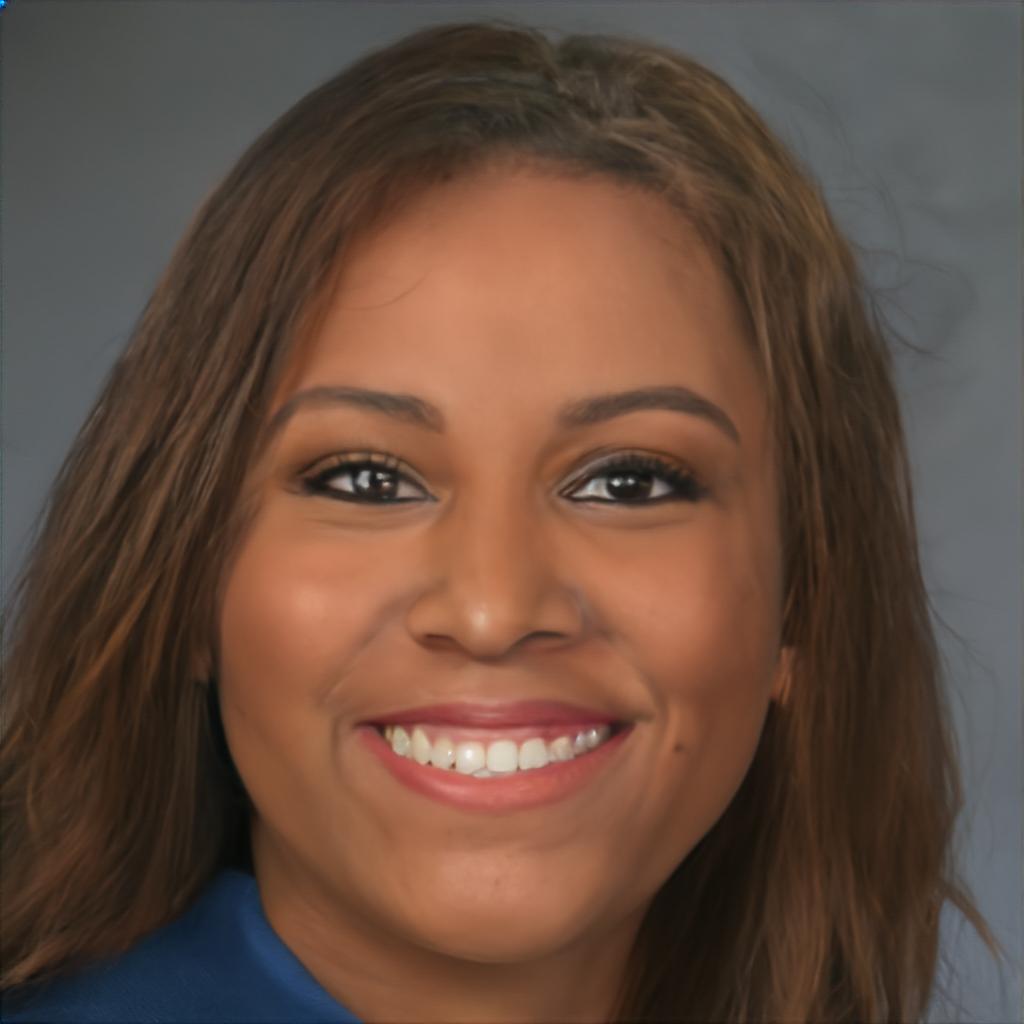} &
\includegraphics[width=0.15\textwidth]{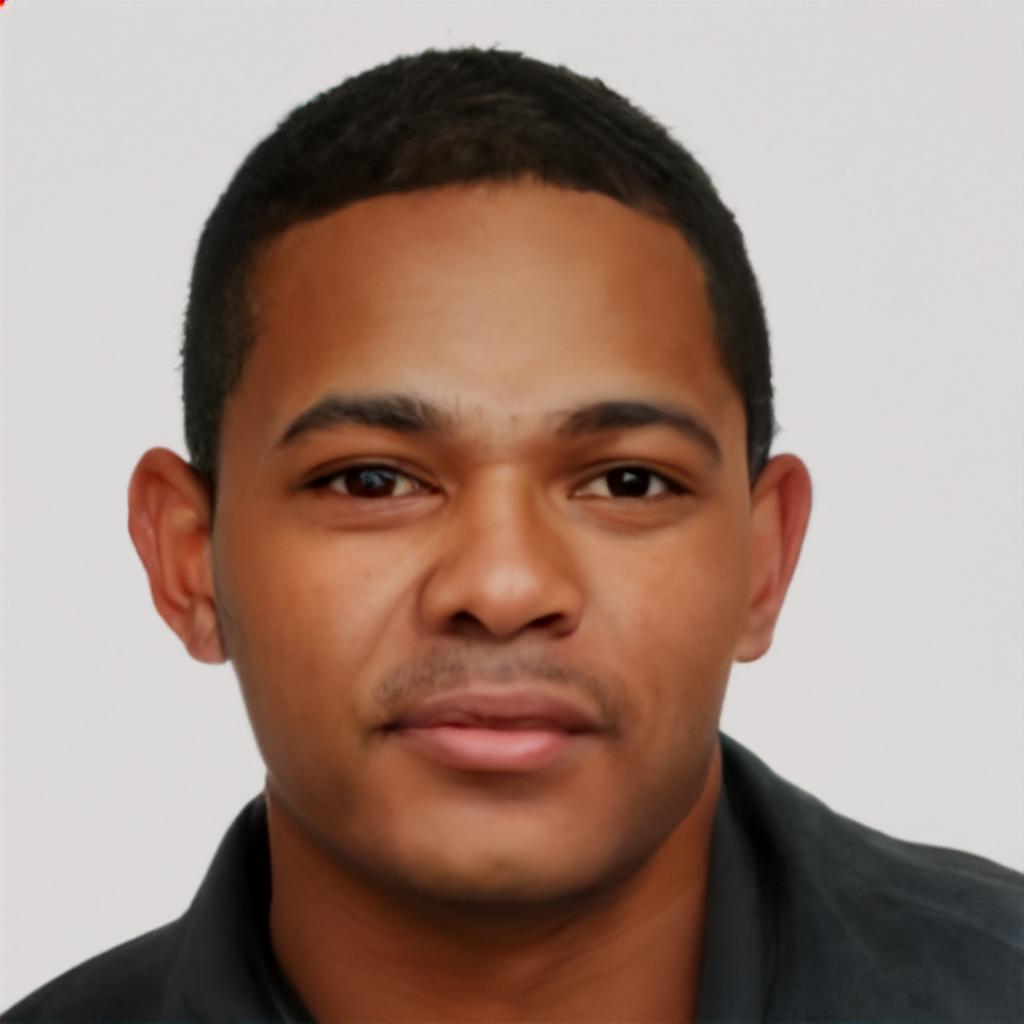} &
\includegraphics[width=0.15\textwidth]{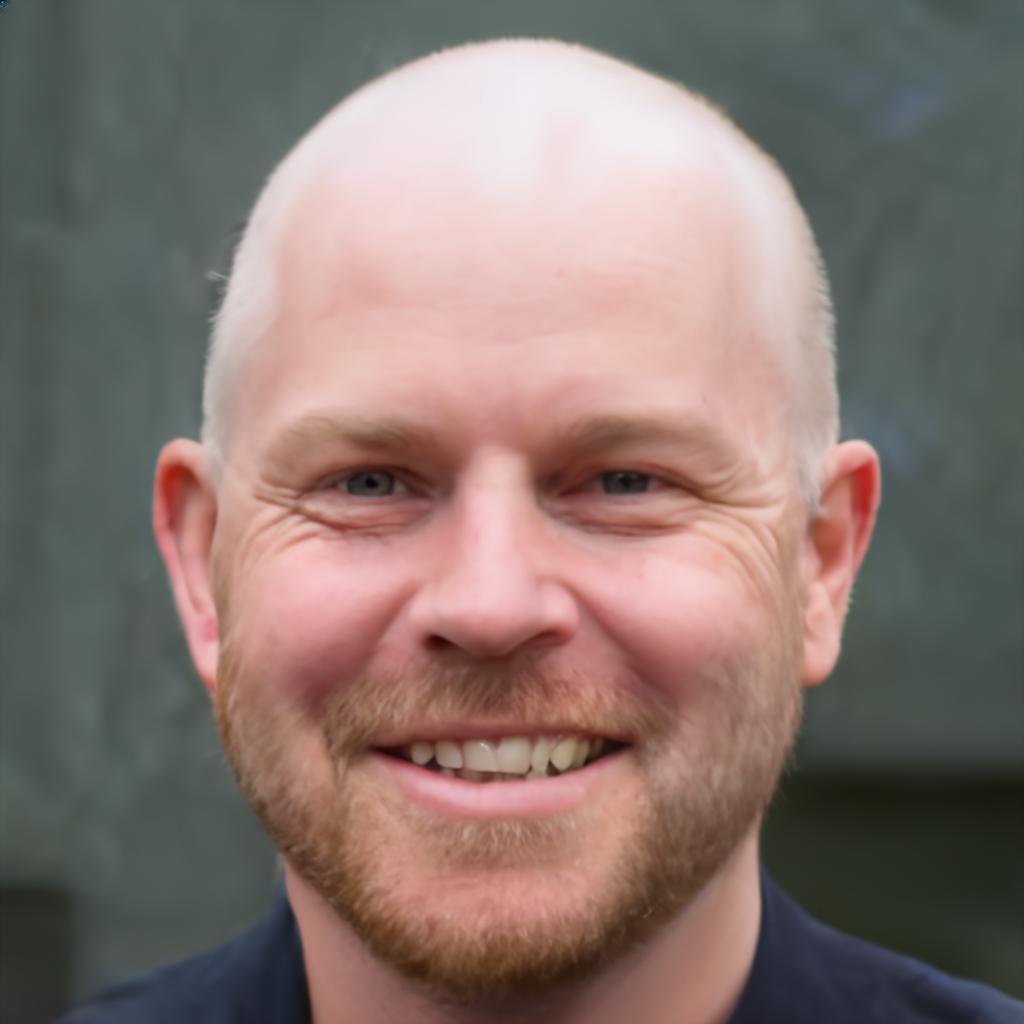} &
\includegraphics[width=0.15\textwidth]{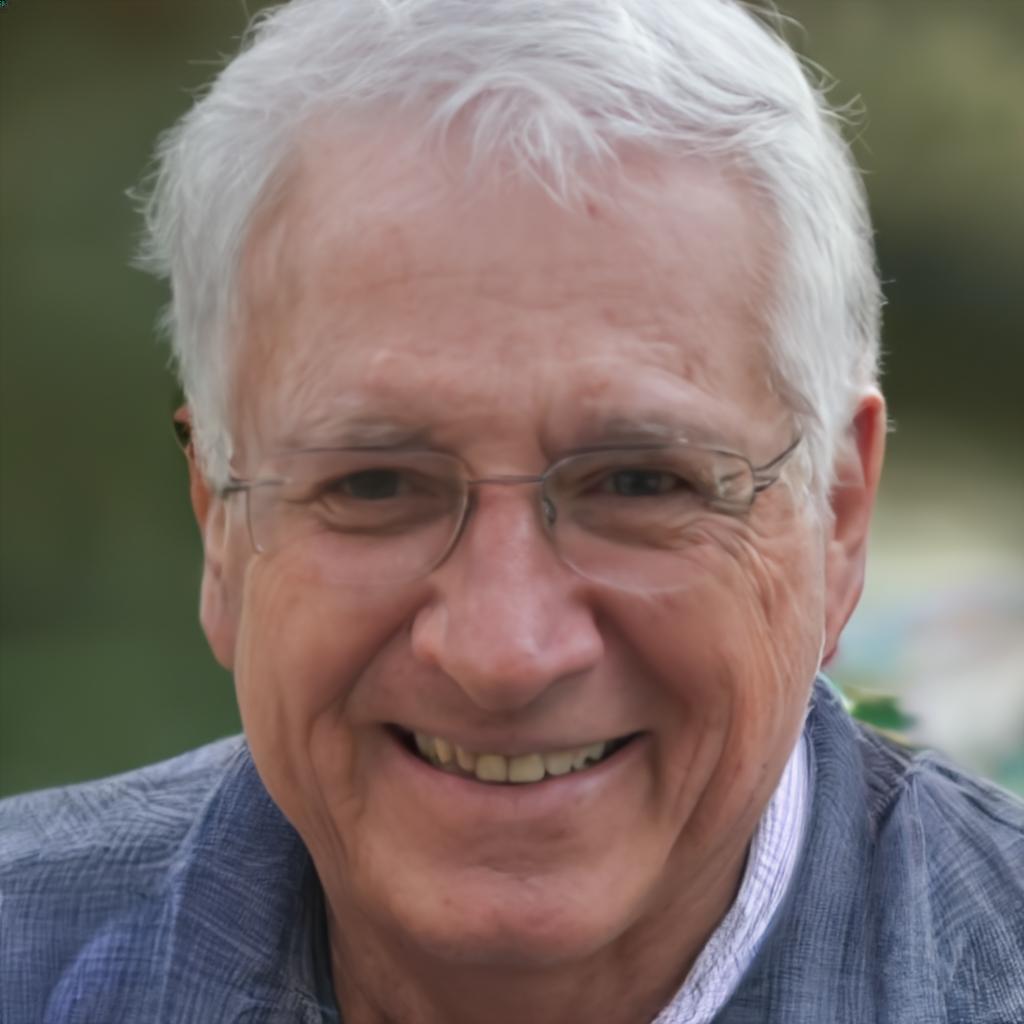} &
\includegraphics[width=0.15\textwidth]{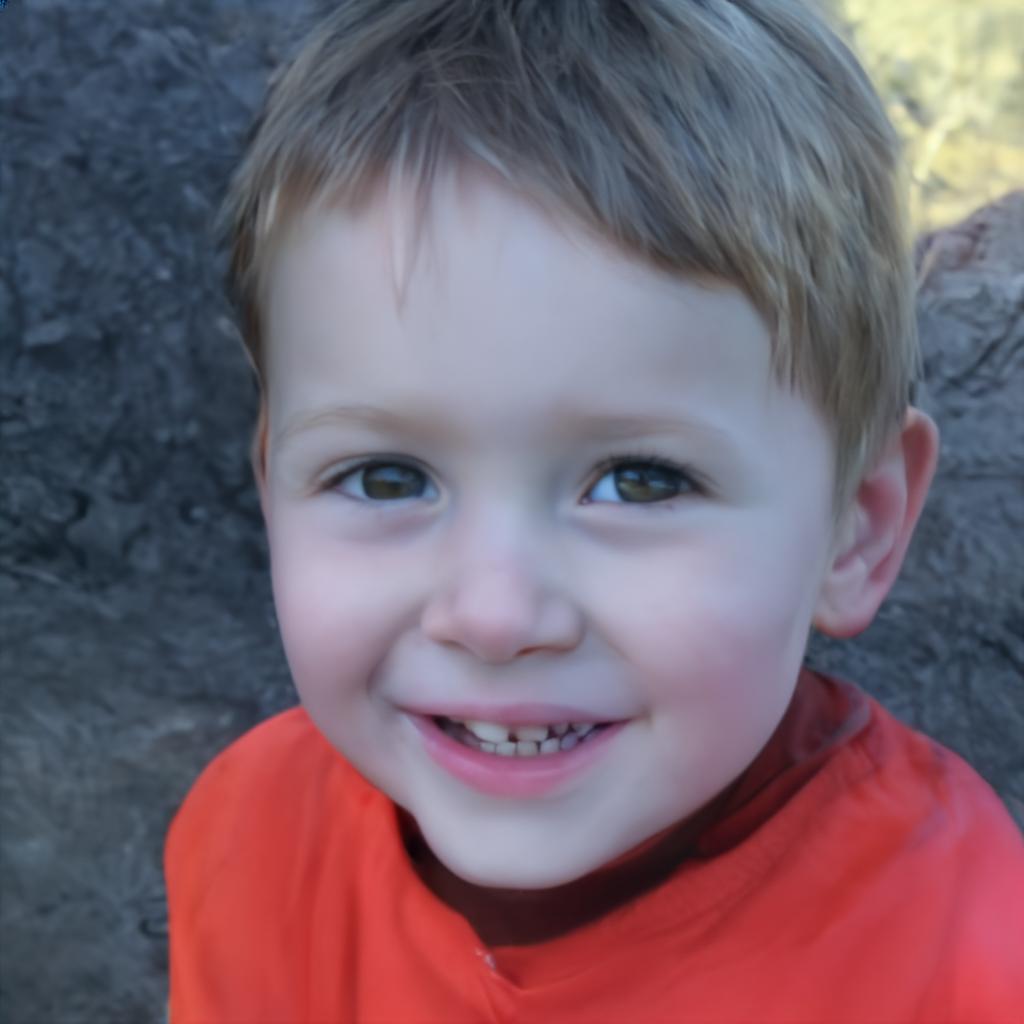} \\

\includegraphics[width=0.15\textwidth]{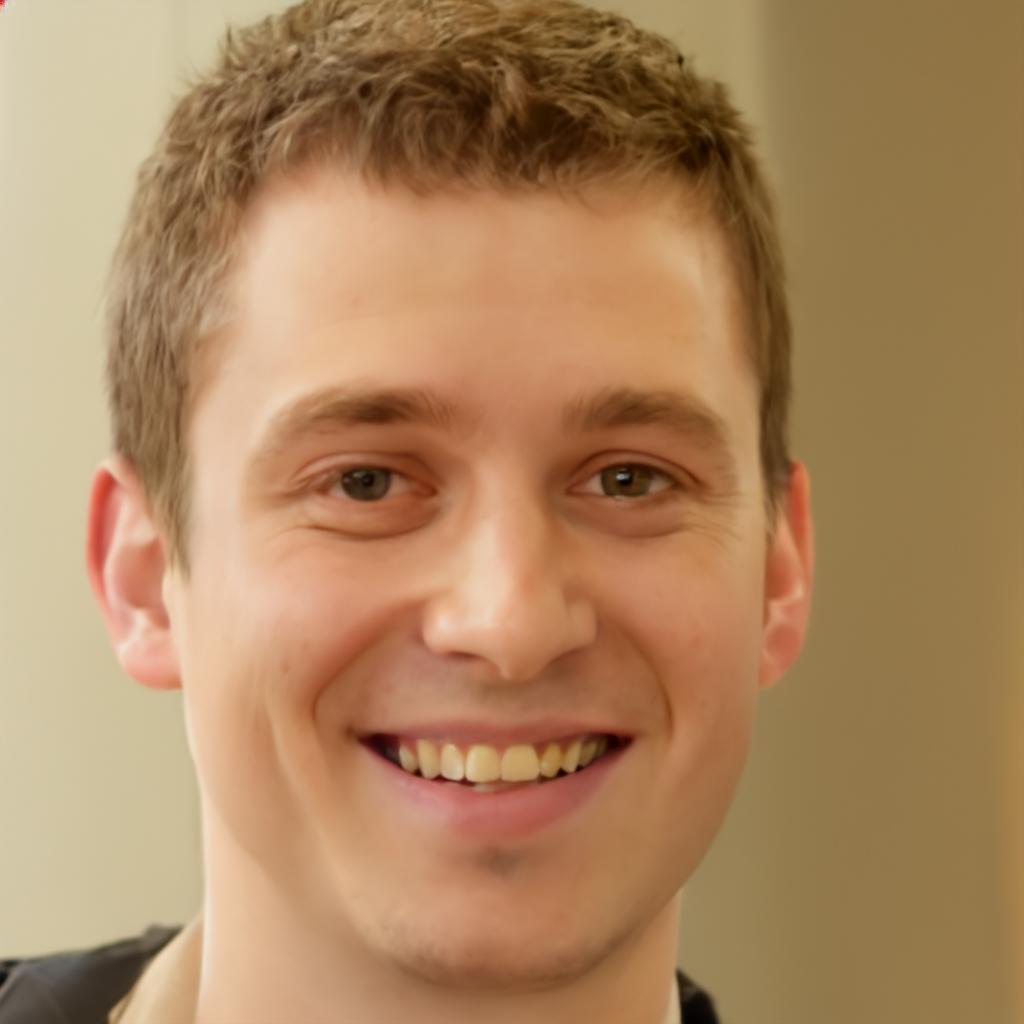} &
\includegraphics[width=0.15\textwidth]{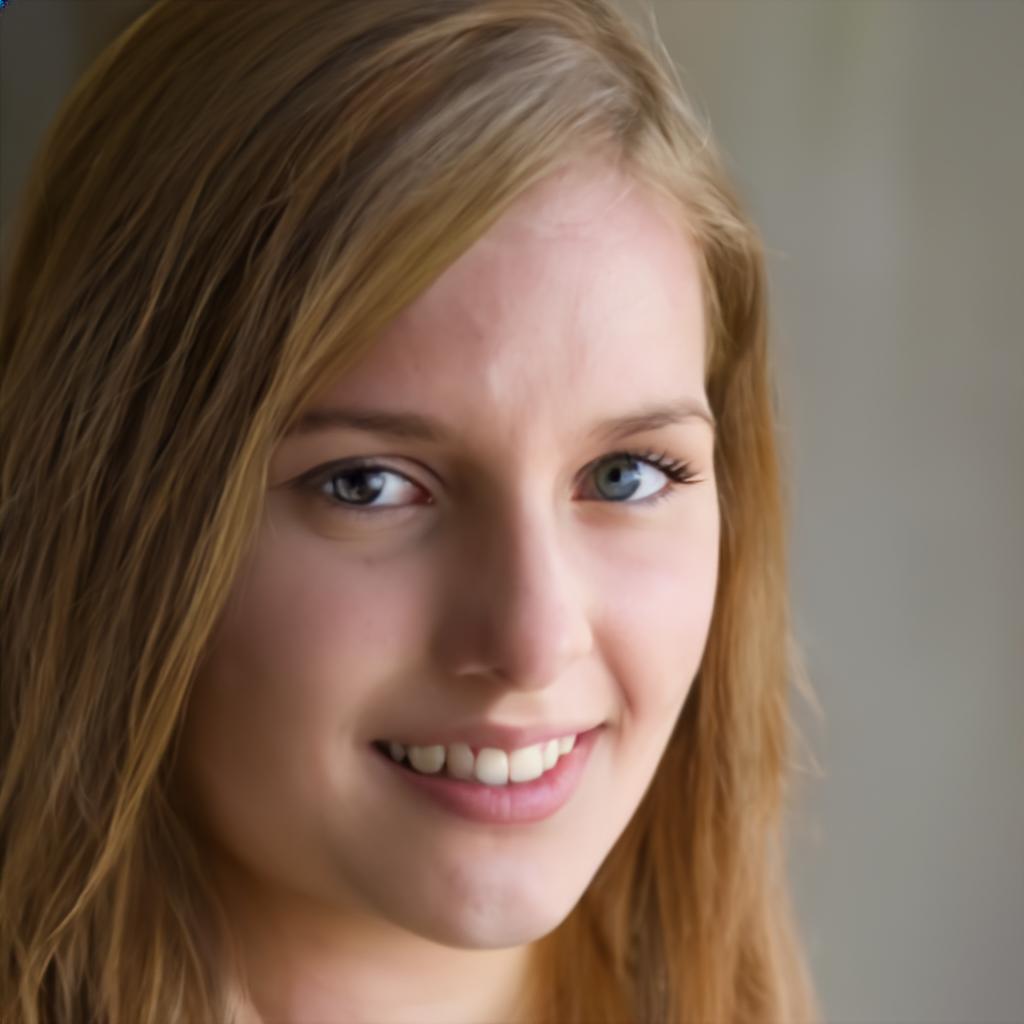} &
\includegraphics[width=0.15\textwidth]{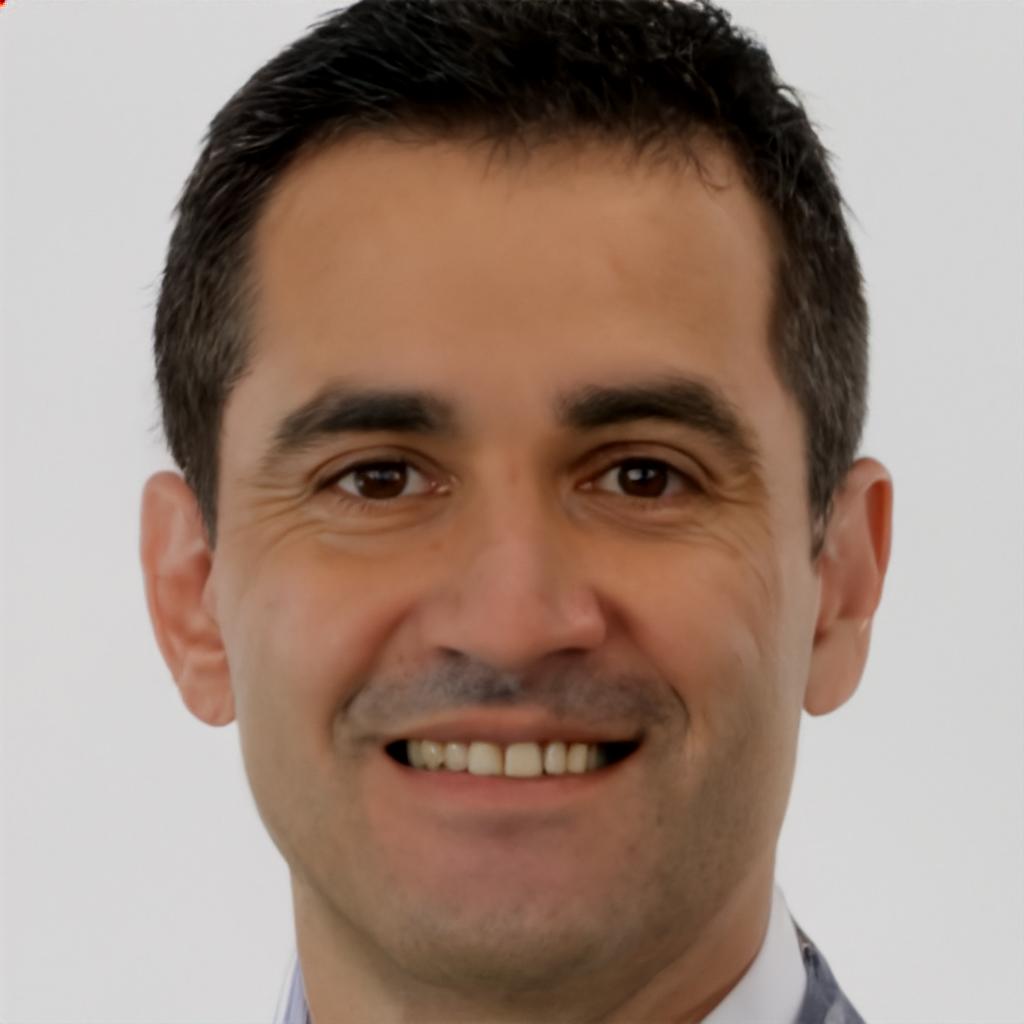} &
\includegraphics[width=0.15\textwidth]{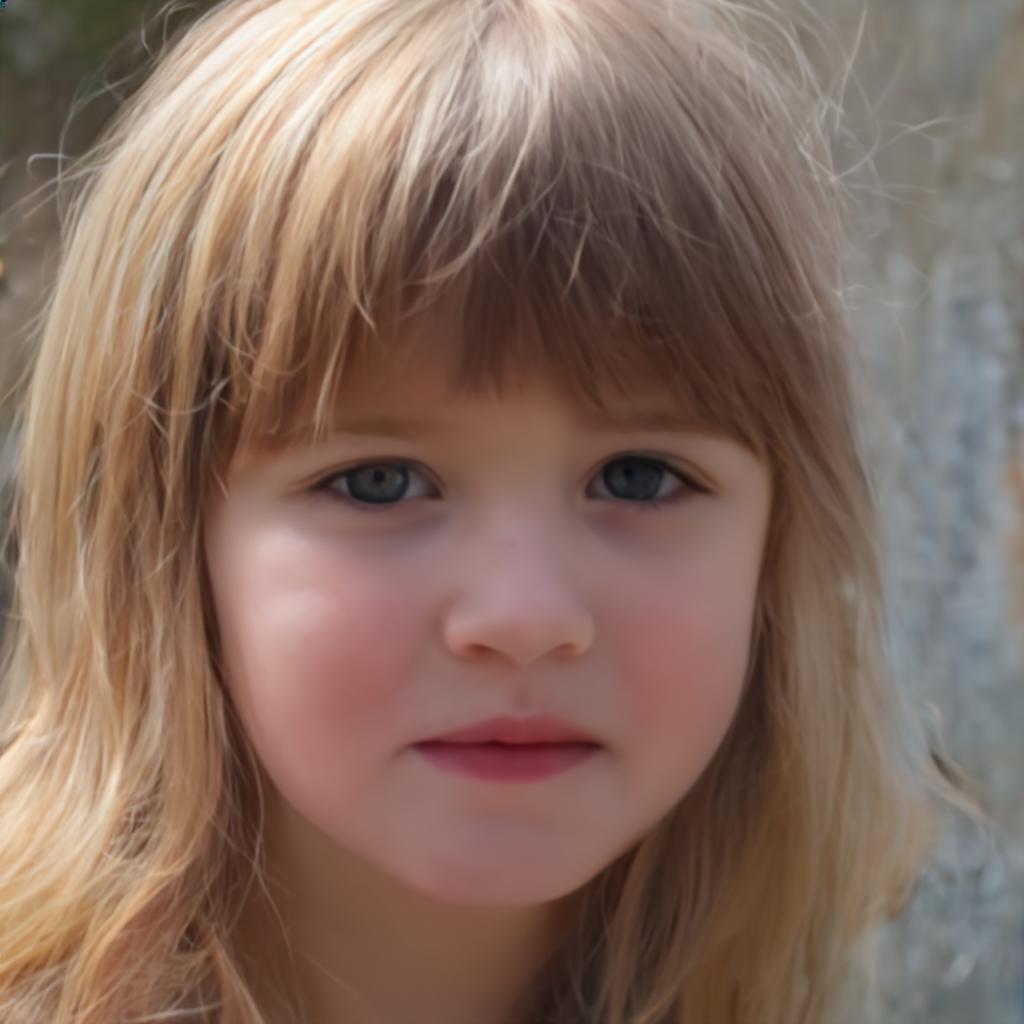} &
\includegraphics[width=0.15\textwidth]{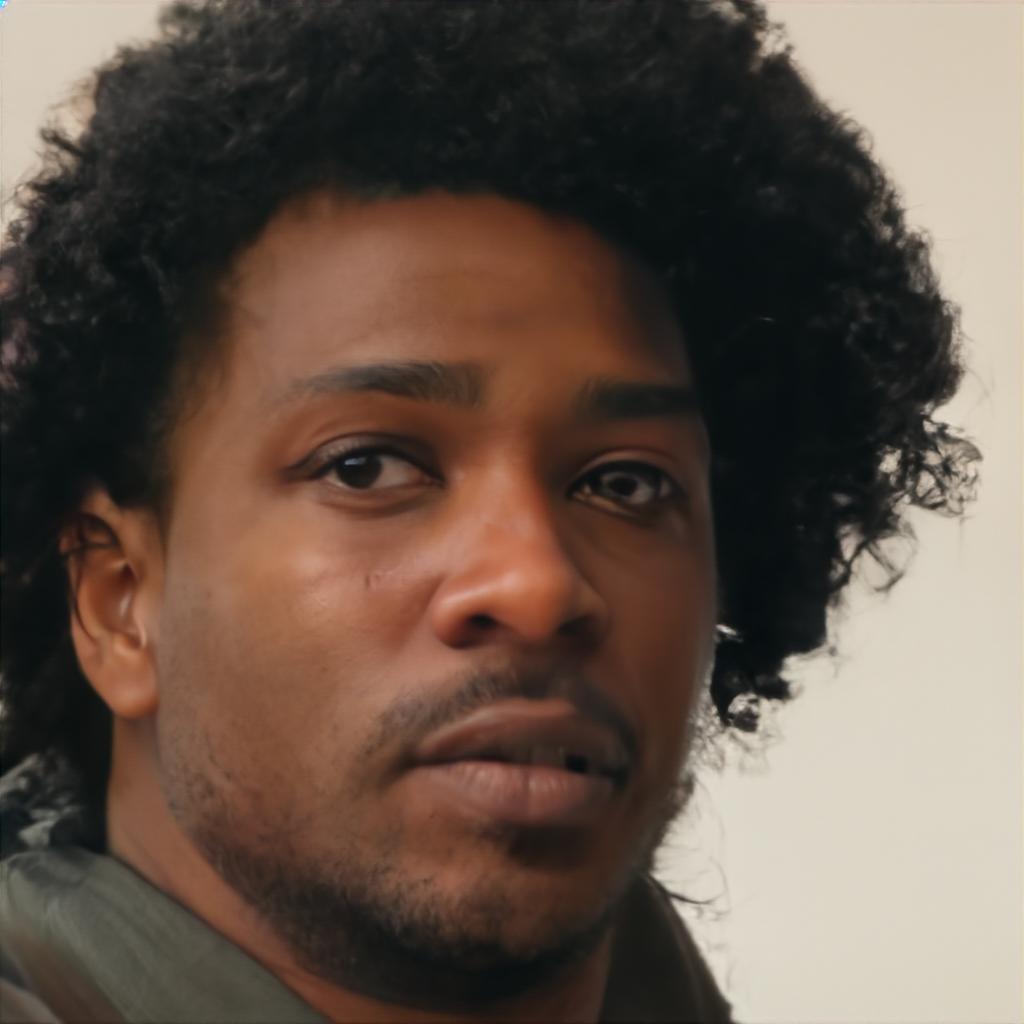} &
\includegraphics[width=0.15\textwidth]{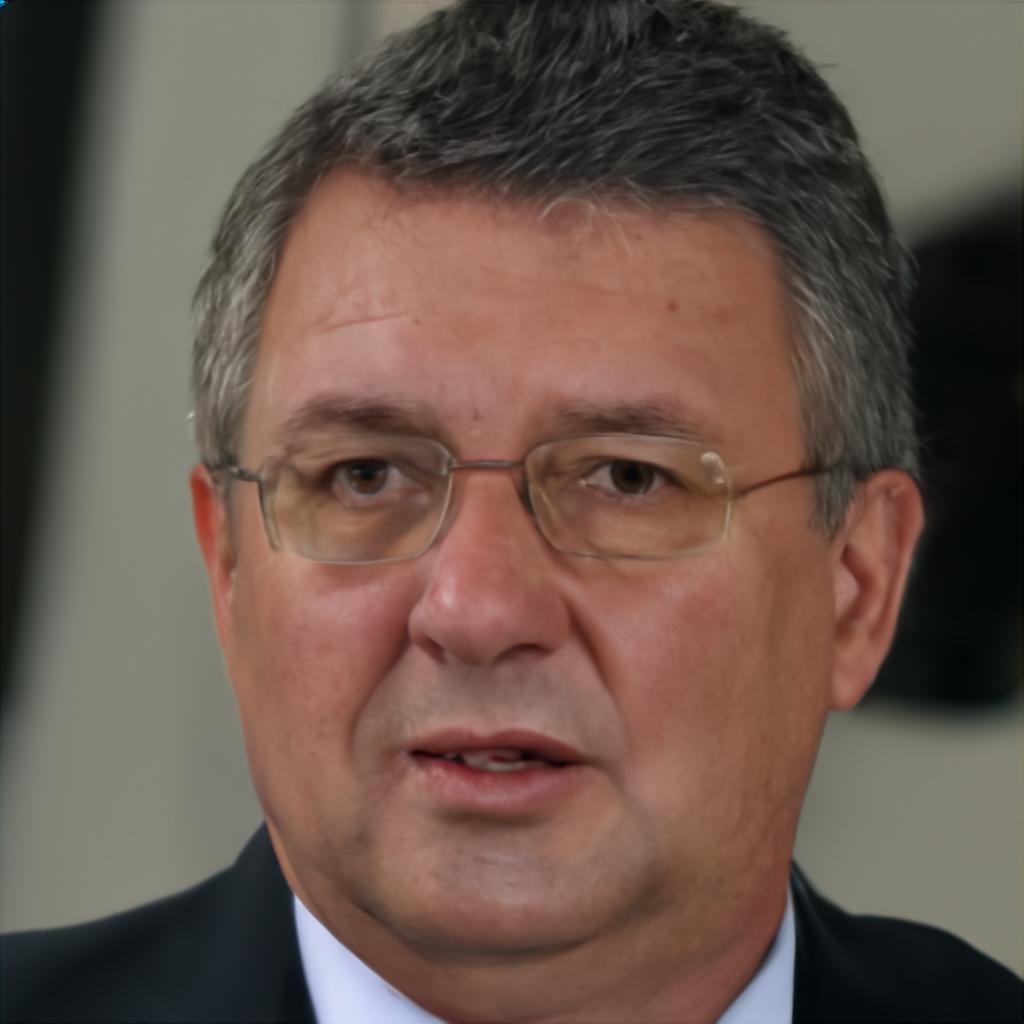} \\

\includegraphics[width=0.15\textwidth]{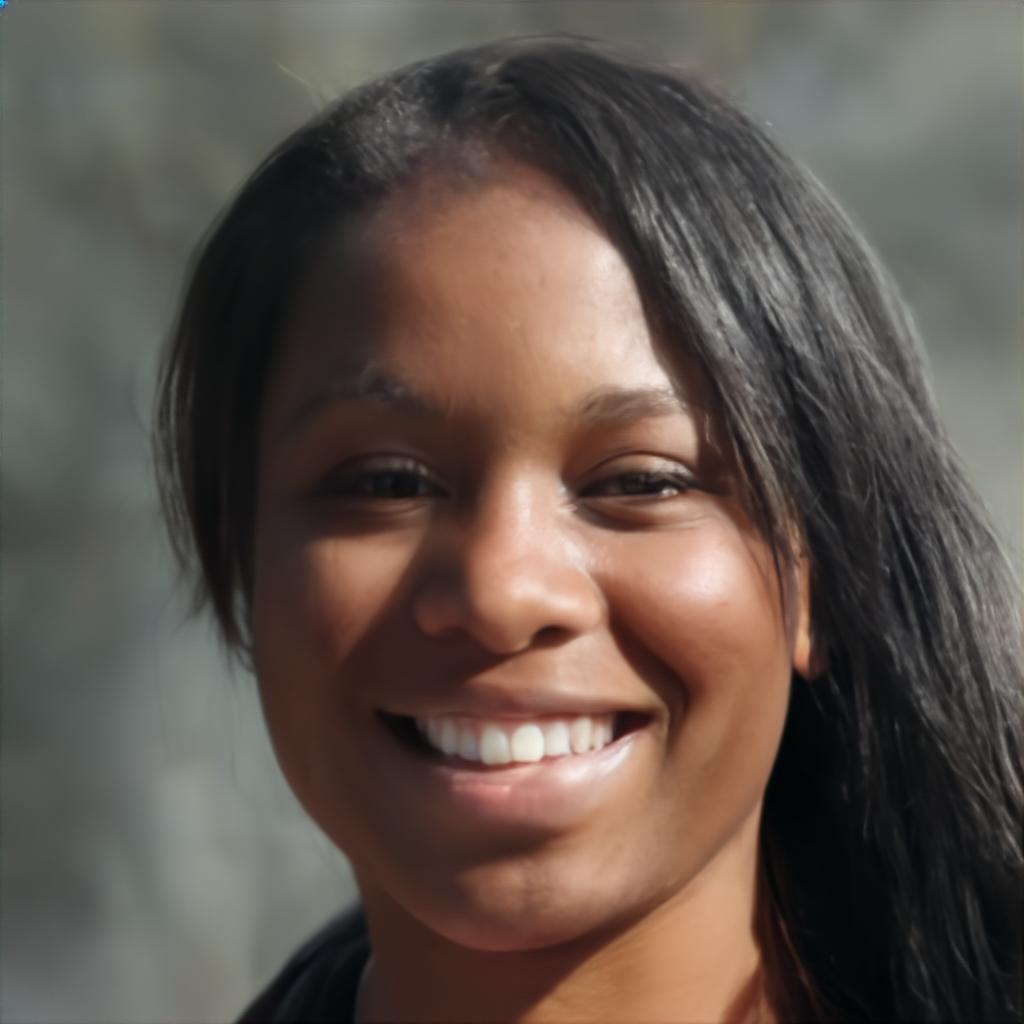} &
\includegraphics[width=0.15\textwidth]{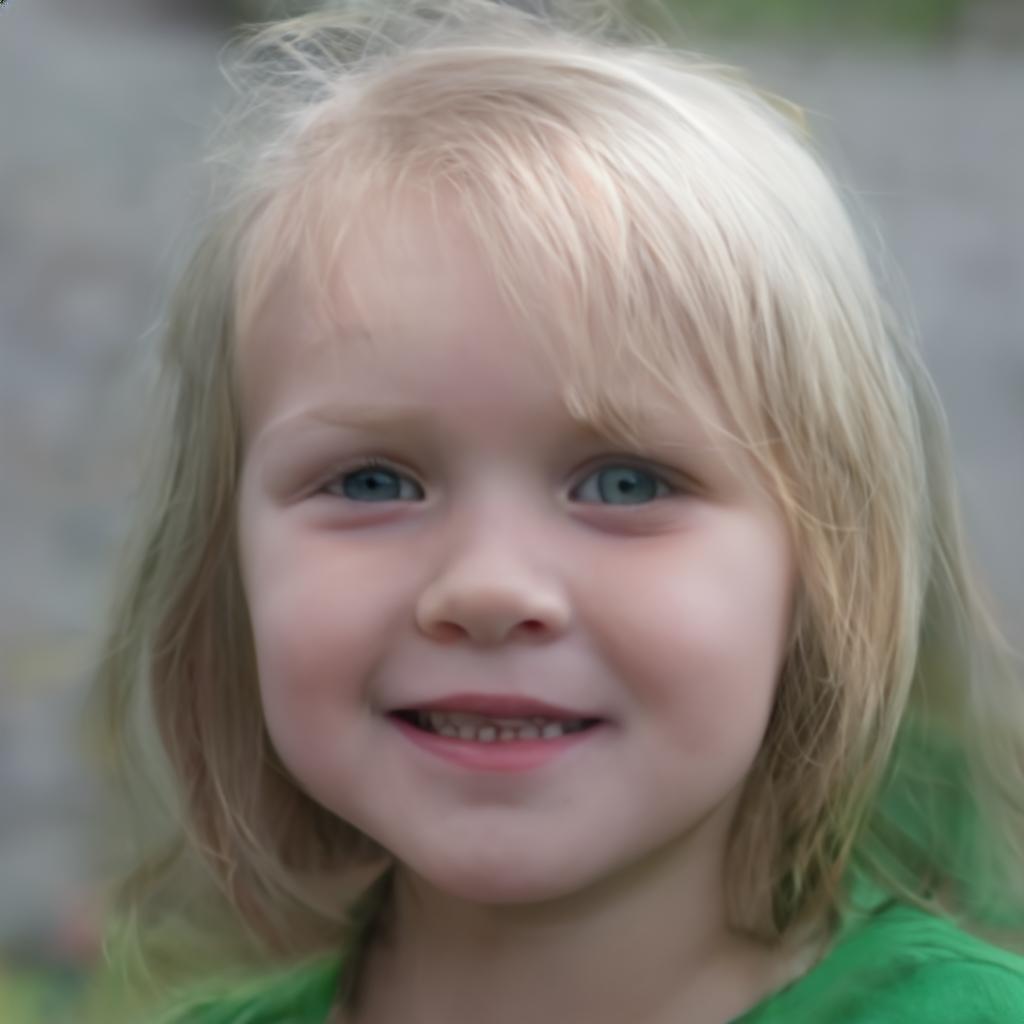} &
\includegraphics[width=0.15\textwidth]{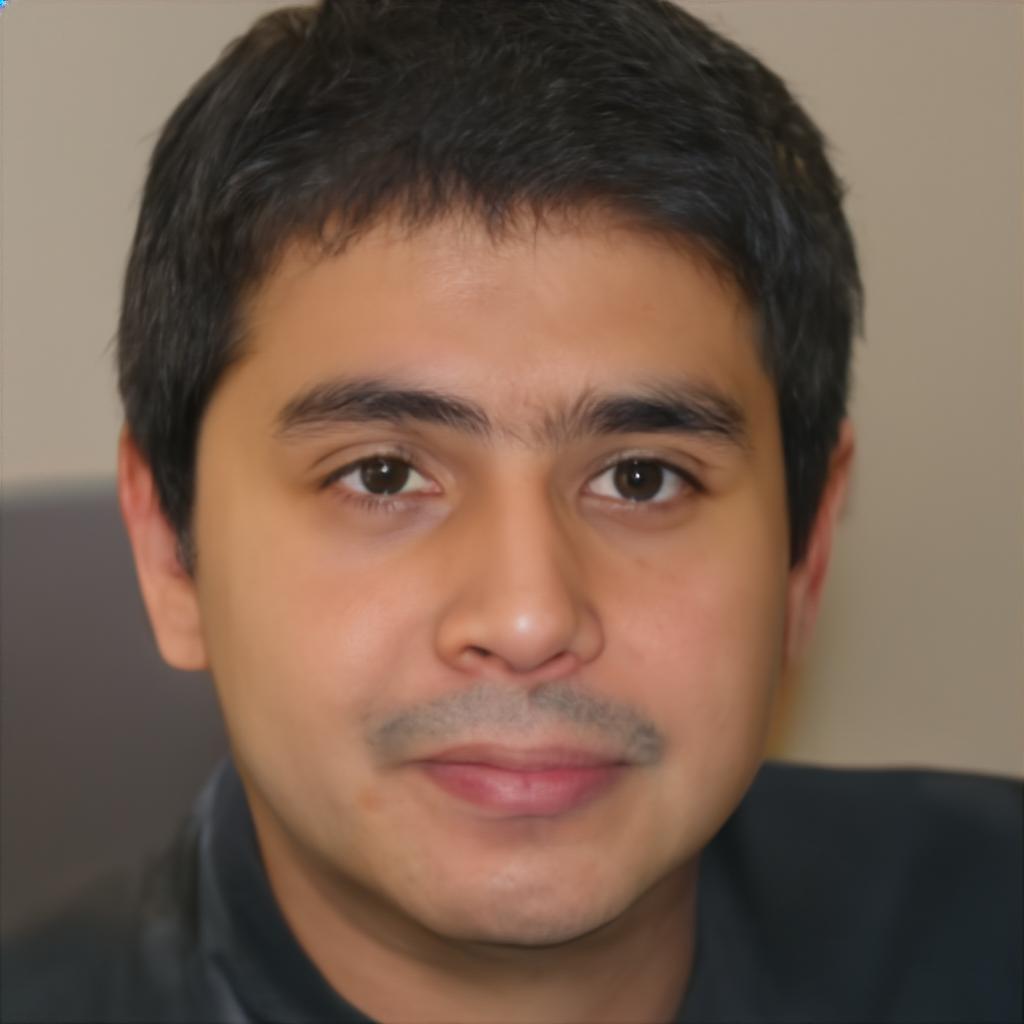} &
\includegraphics[width=0.15\textwidth]{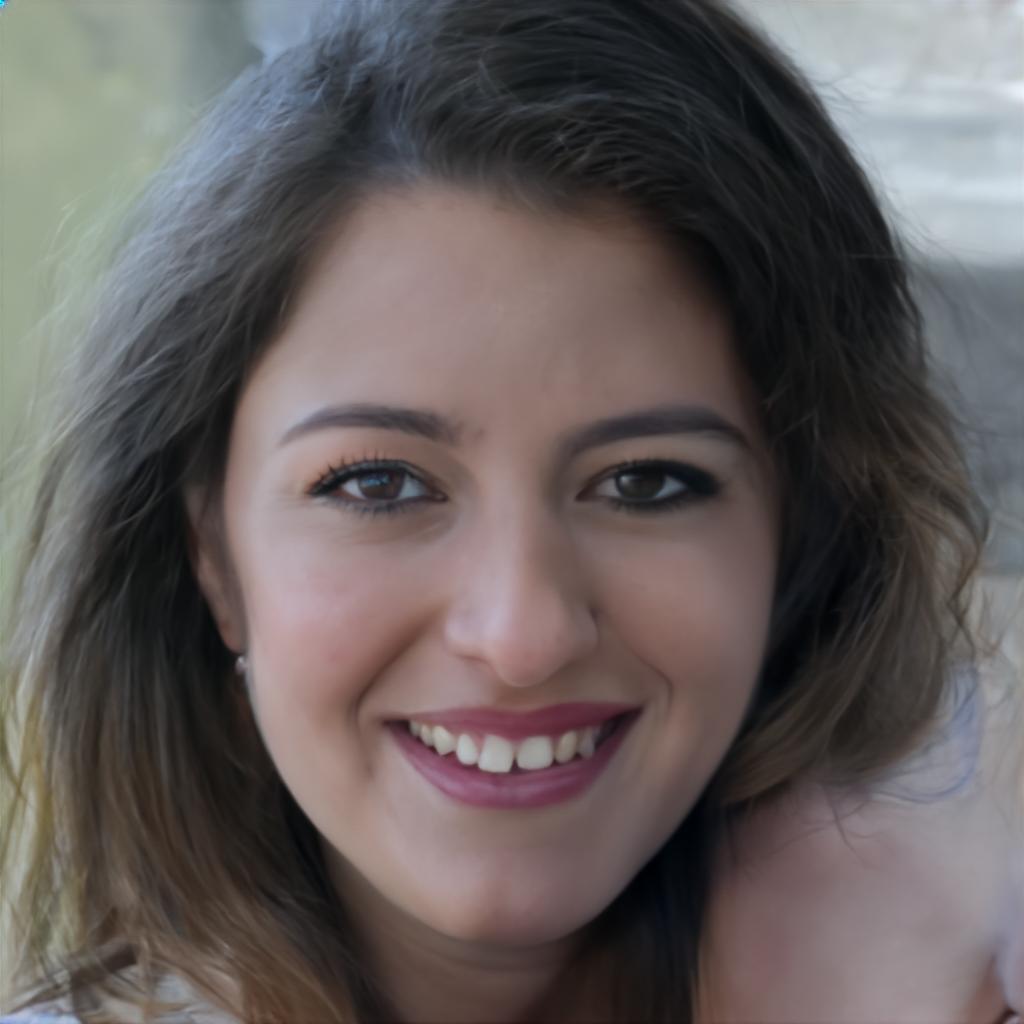} &
\includegraphics[width=0.15\textwidth]{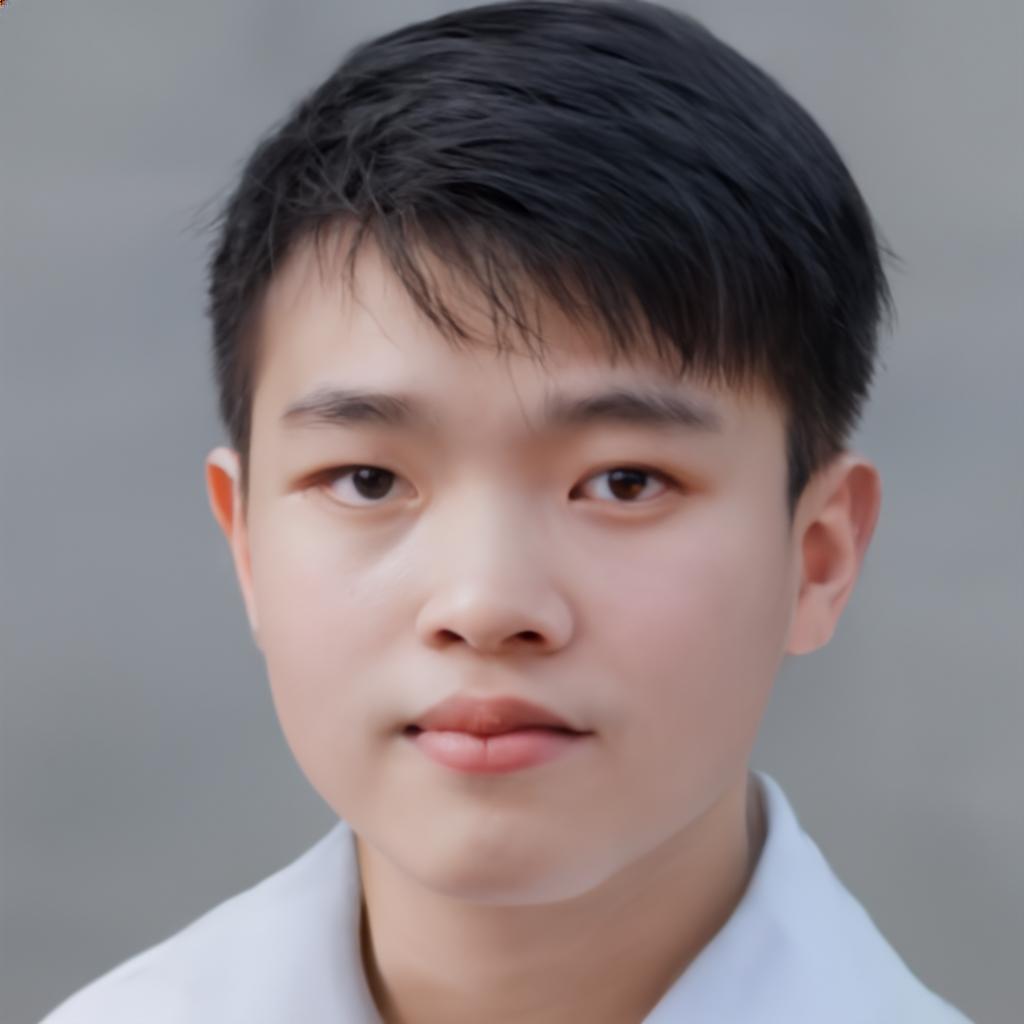} &
\includegraphics[width=0.15\textwidth]{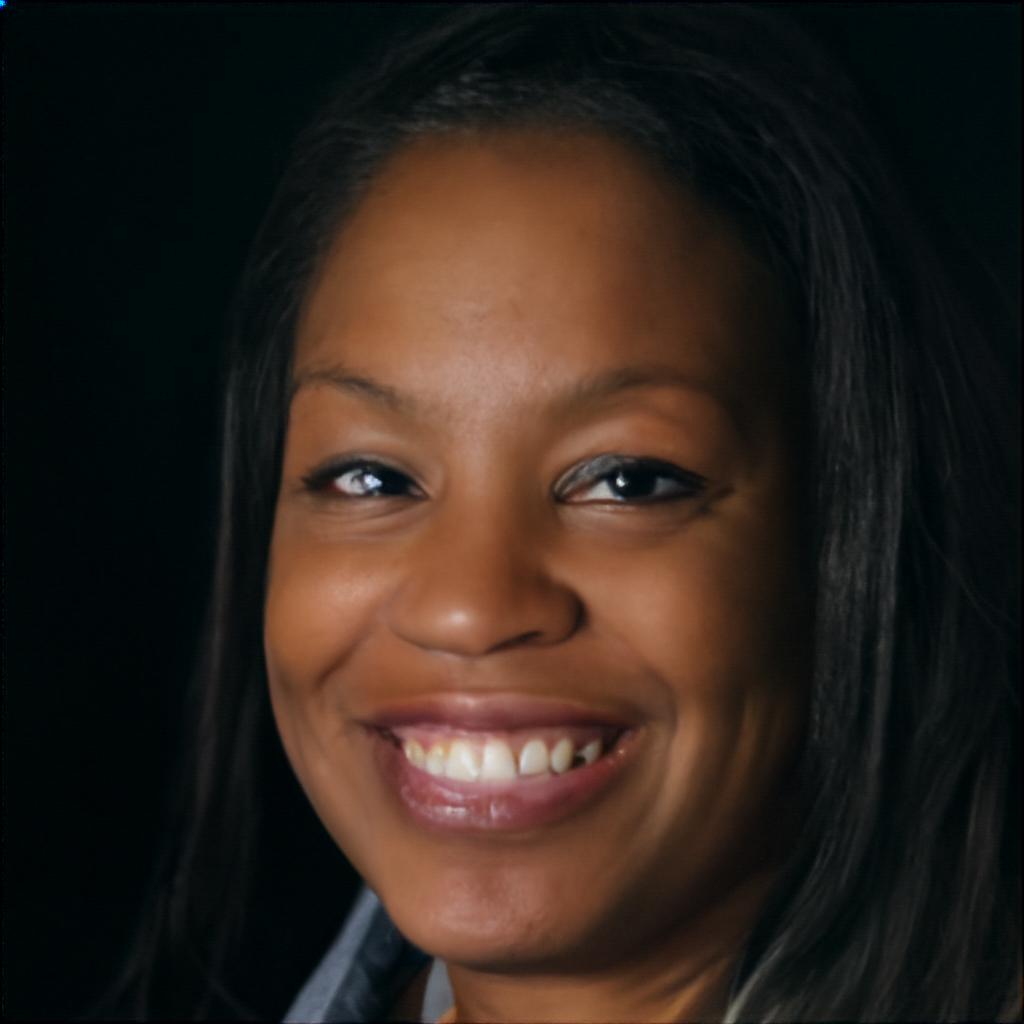} \\

\includegraphics[width=0.15\textwidth]{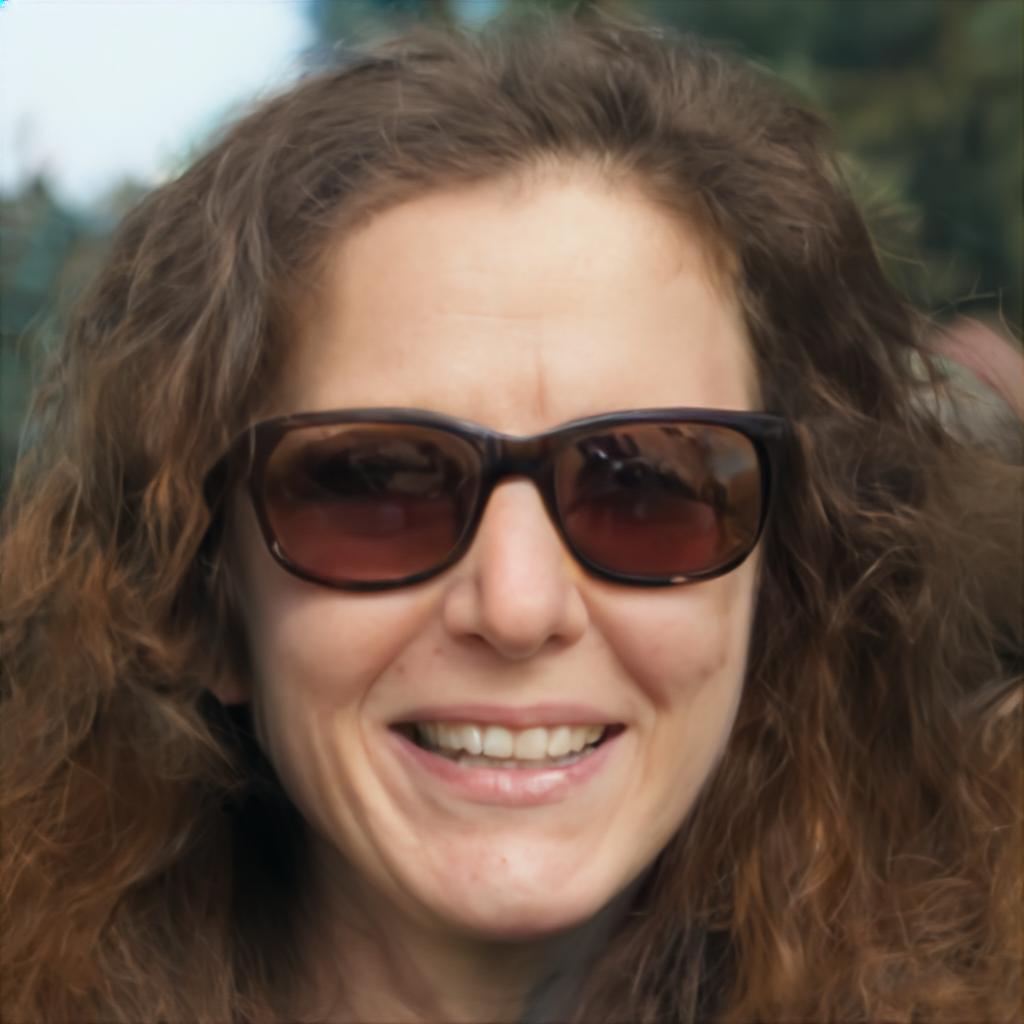} &
\includegraphics[width=0.15\textwidth]{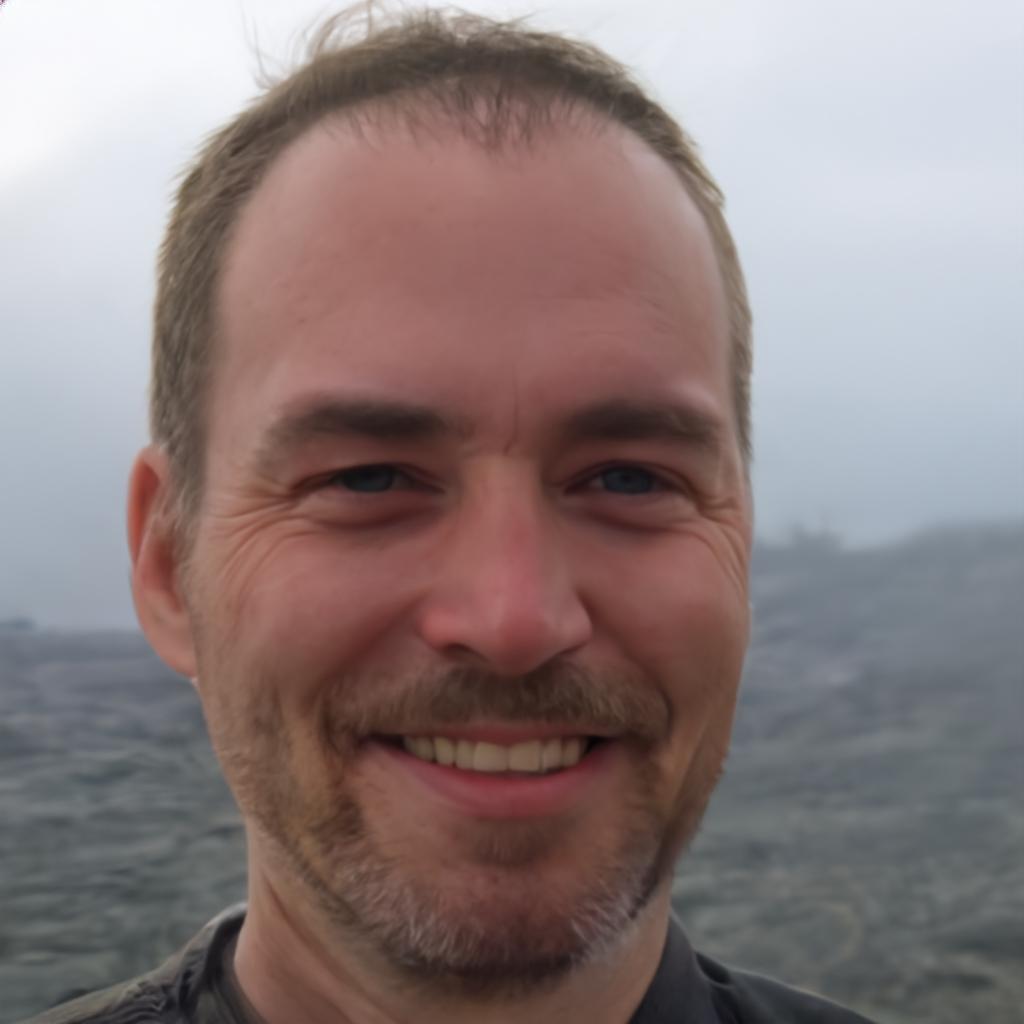} &
\includegraphics[width=0.15\textwidth]{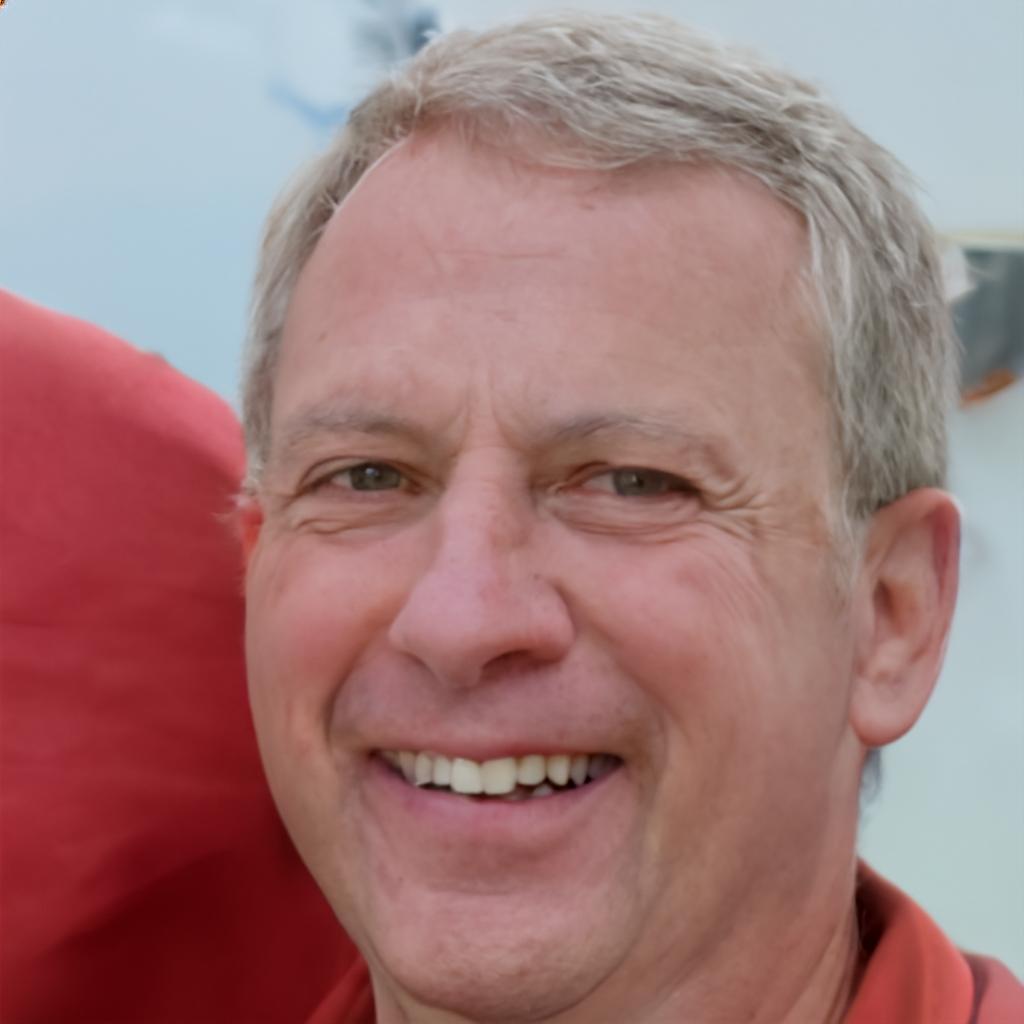} &
\includegraphics[width=0.15\textwidth]{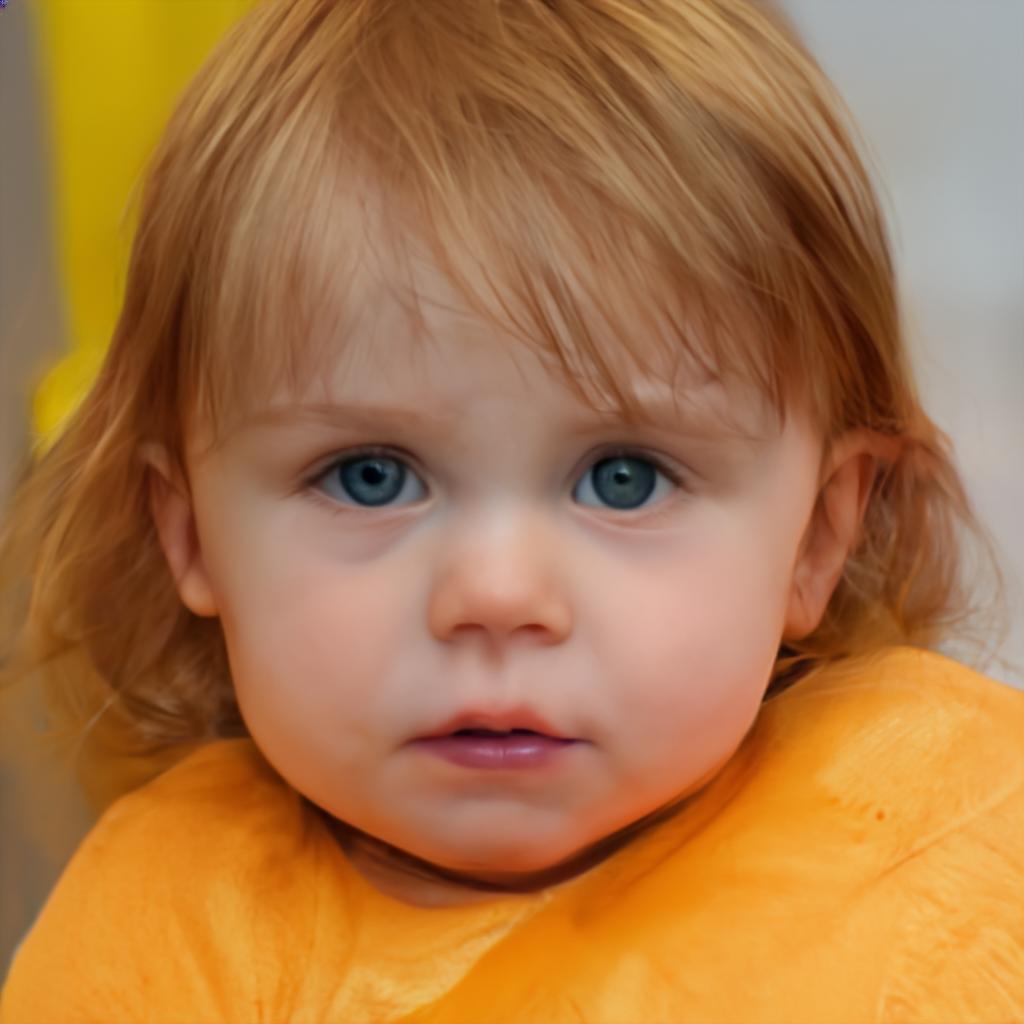} &
\includegraphics[width=0.15\textwidth]{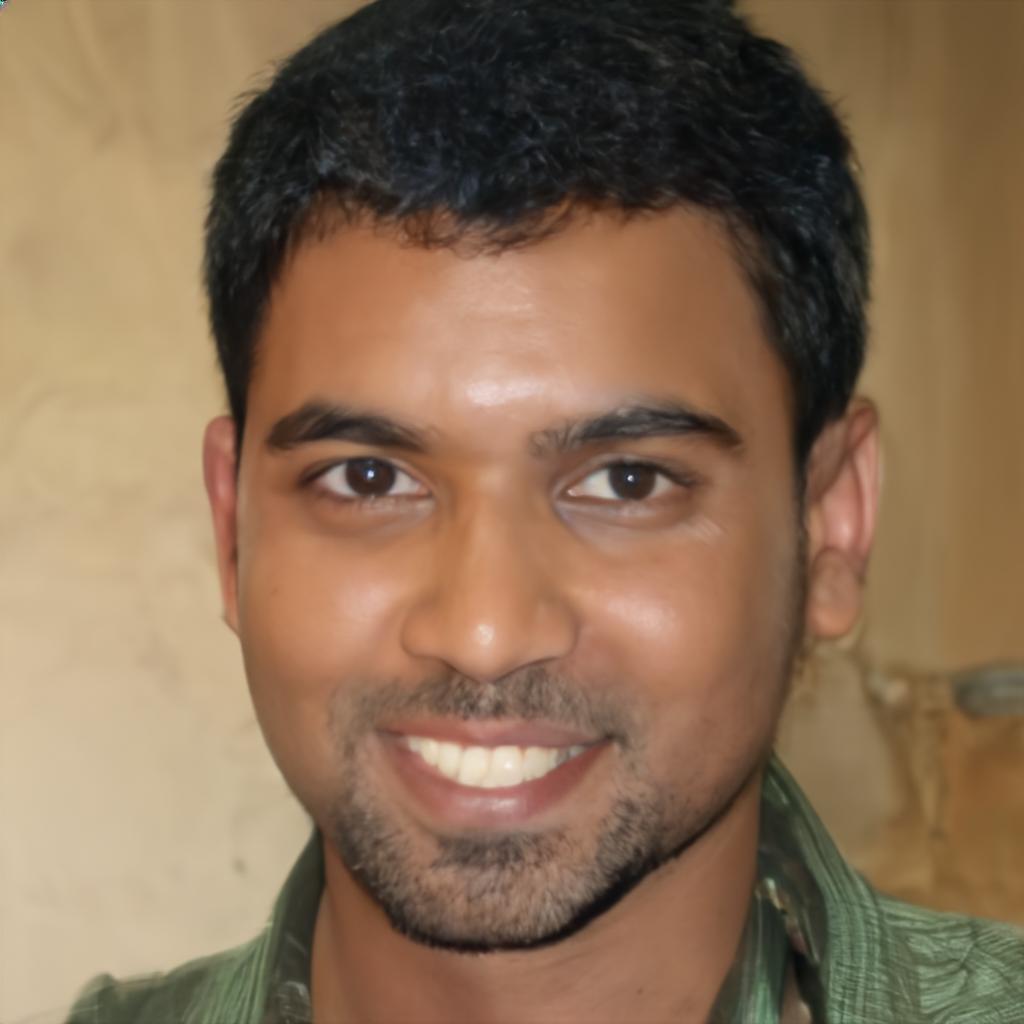} &
\includegraphics[width=0.15\textwidth]{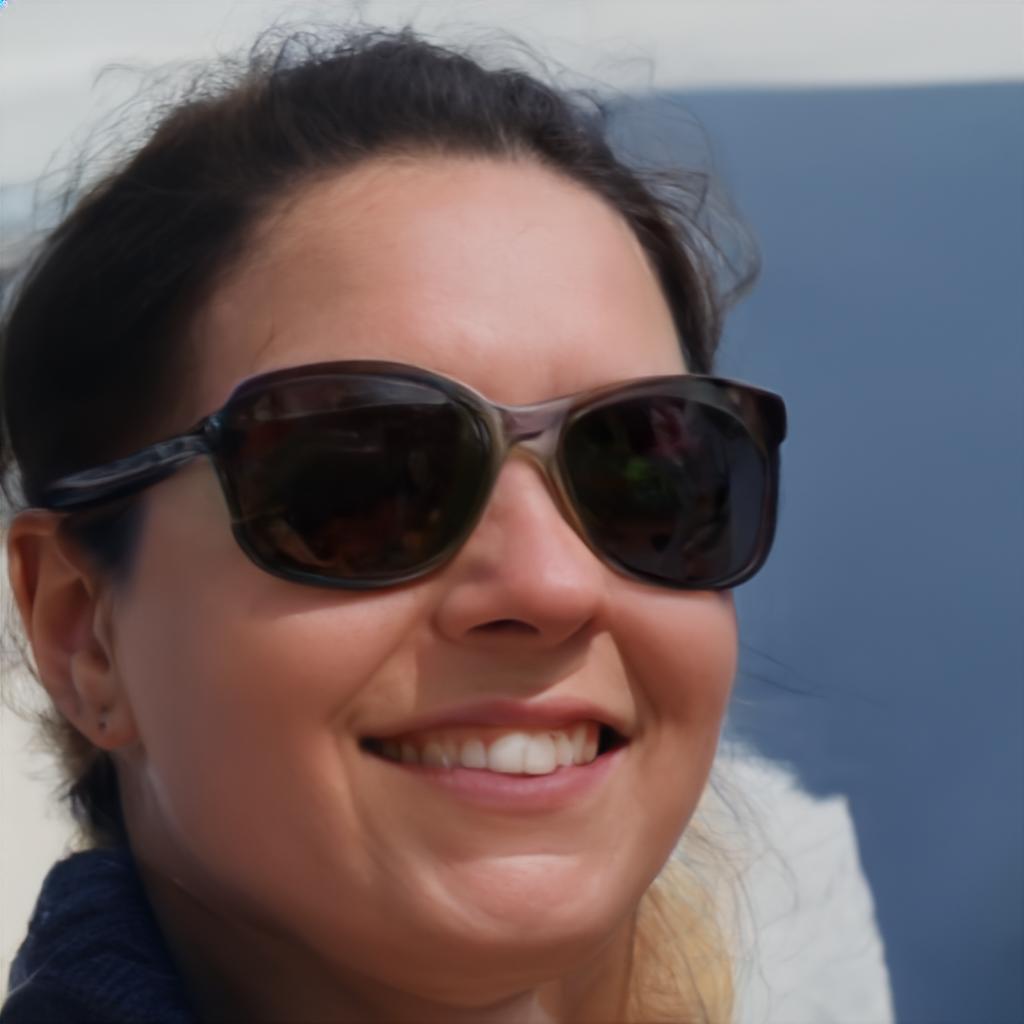} \\

\includegraphics[width=0.15\textwidth]{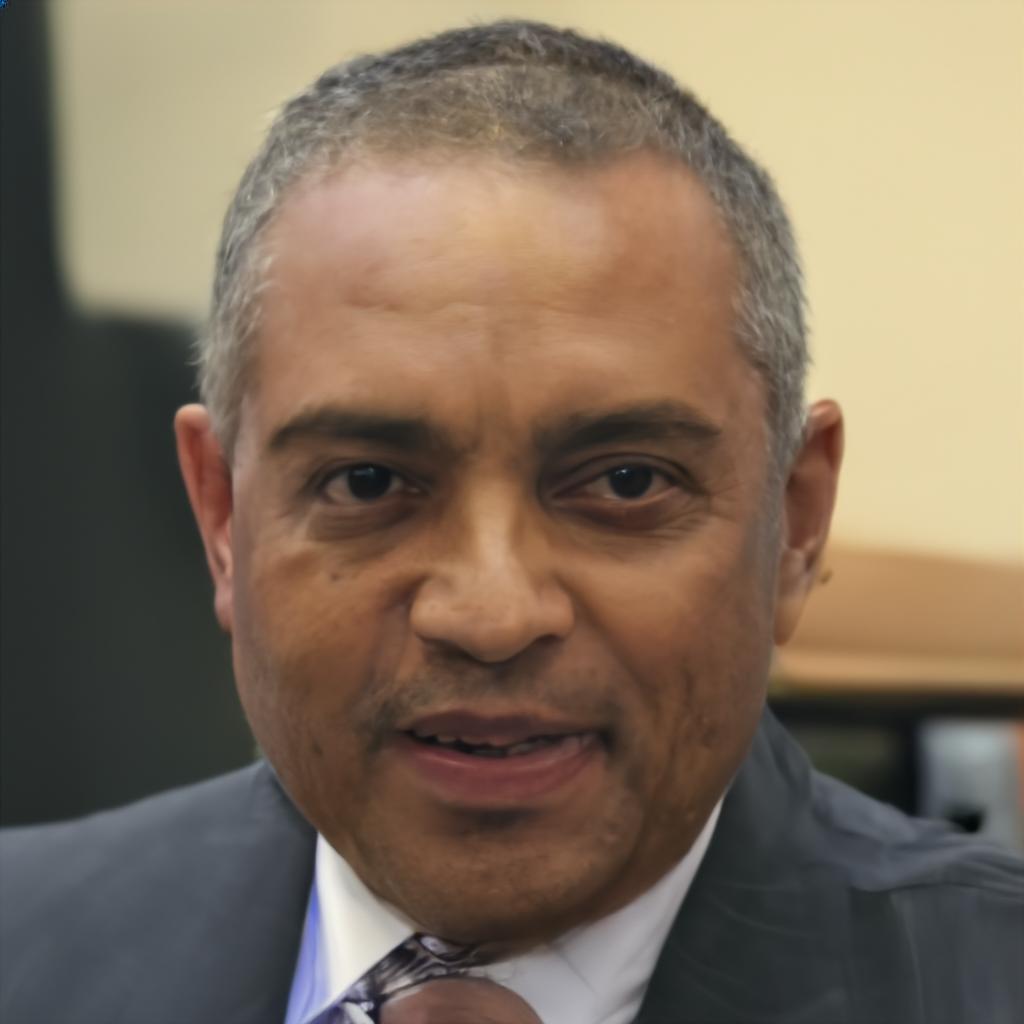} &
\includegraphics[width=0.15\textwidth]{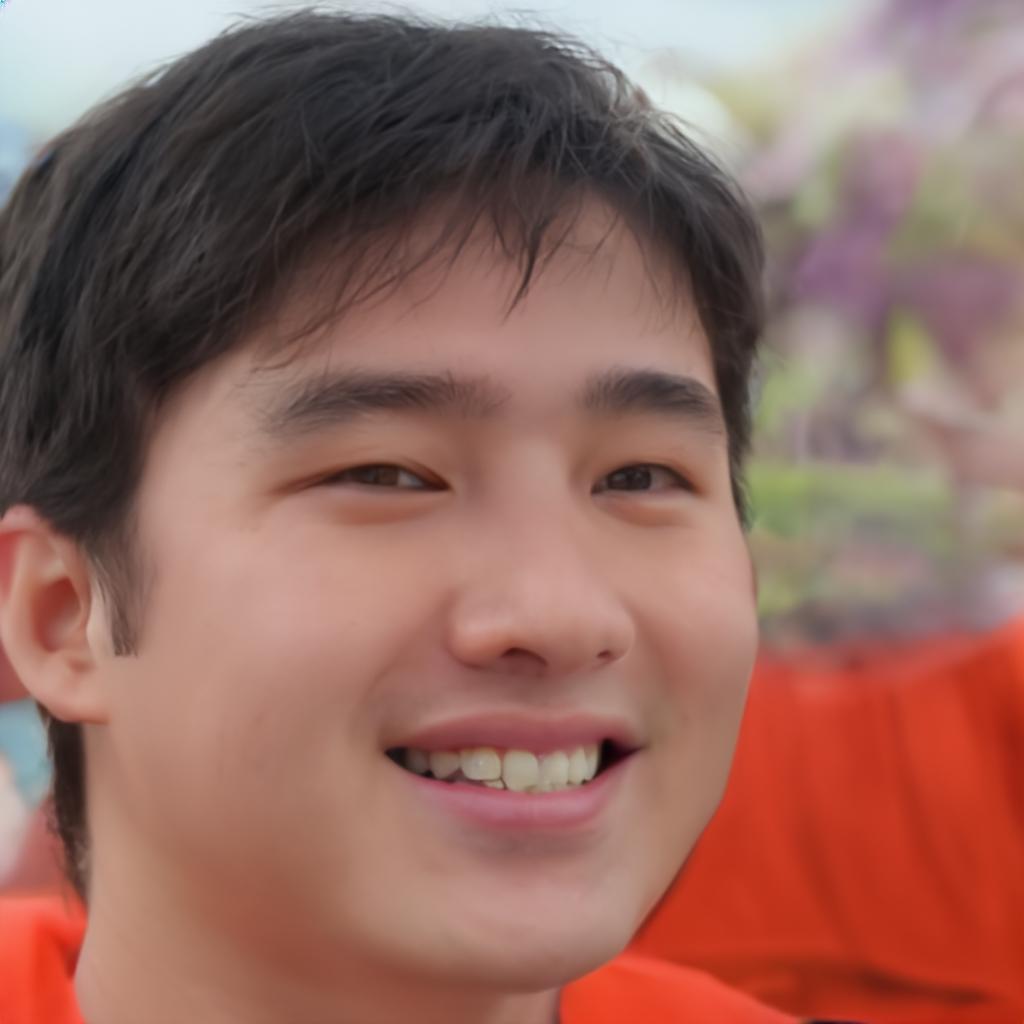} &
\includegraphics[width=0.15\textwidth]{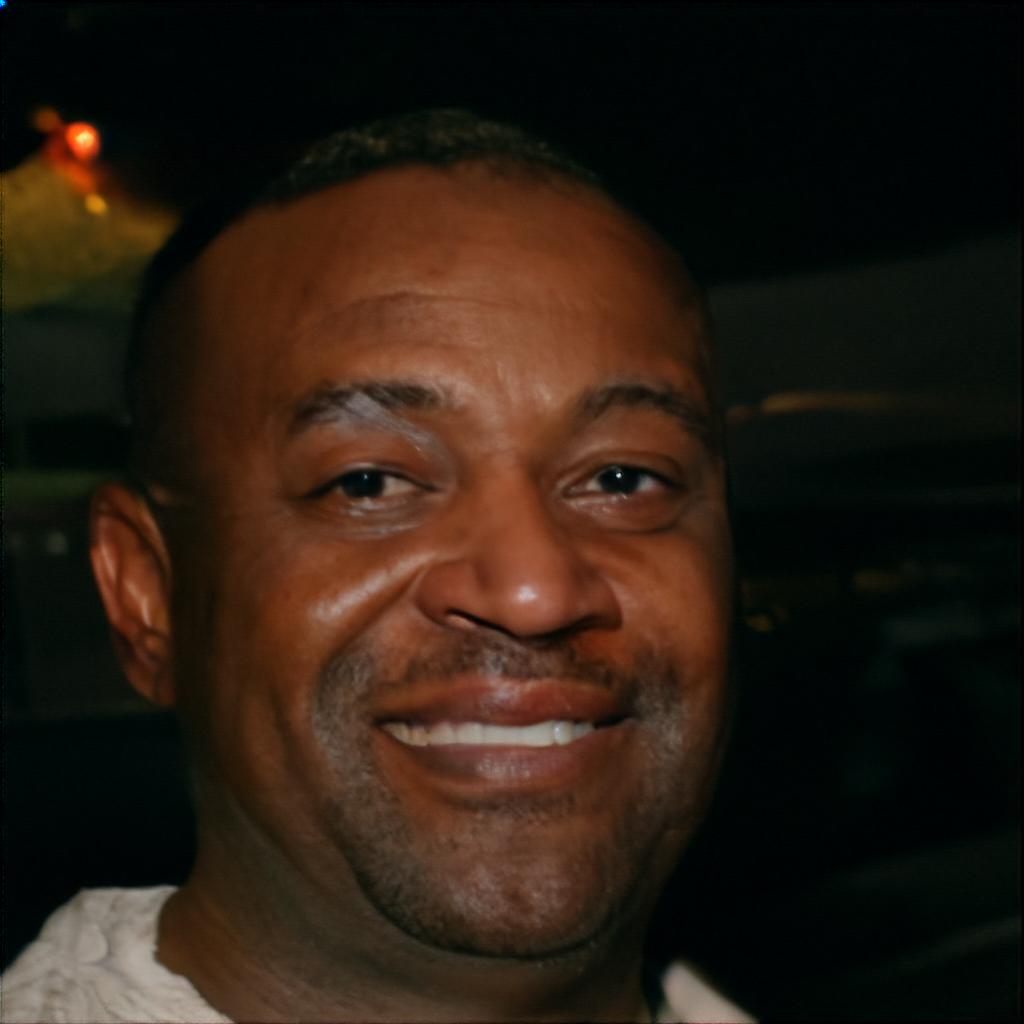} &
\includegraphics[width=0.15\textwidth]{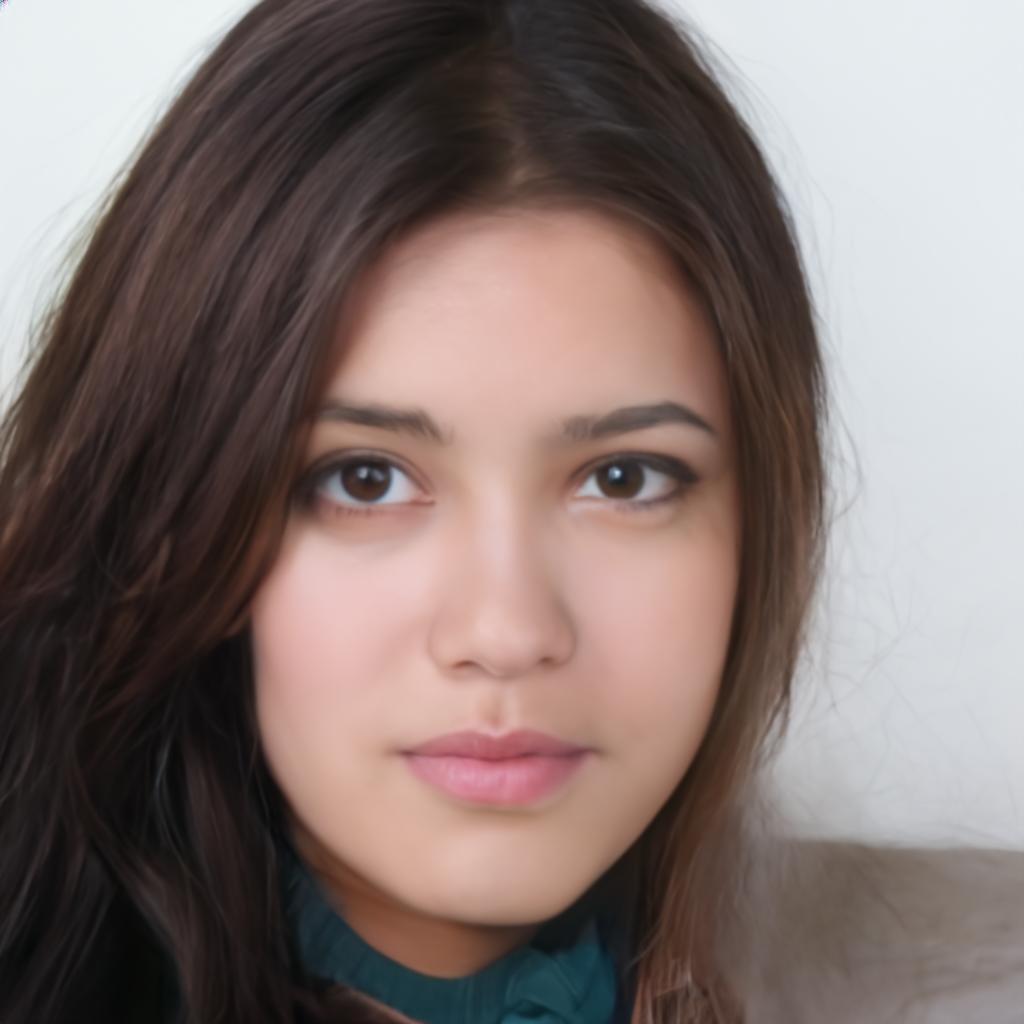} &
\includegraphics[width=0.15\textwidth]{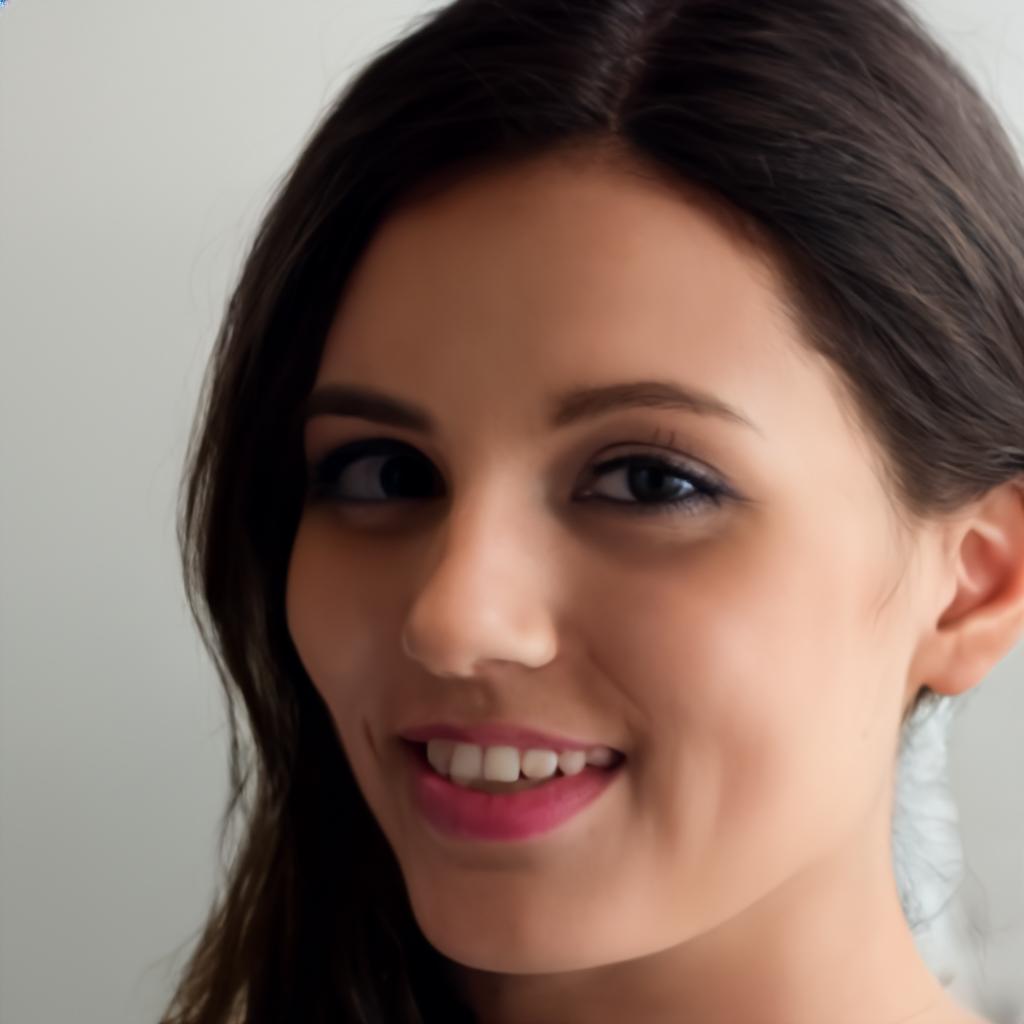} &
\includegraphics[width=0.15\textwidth]{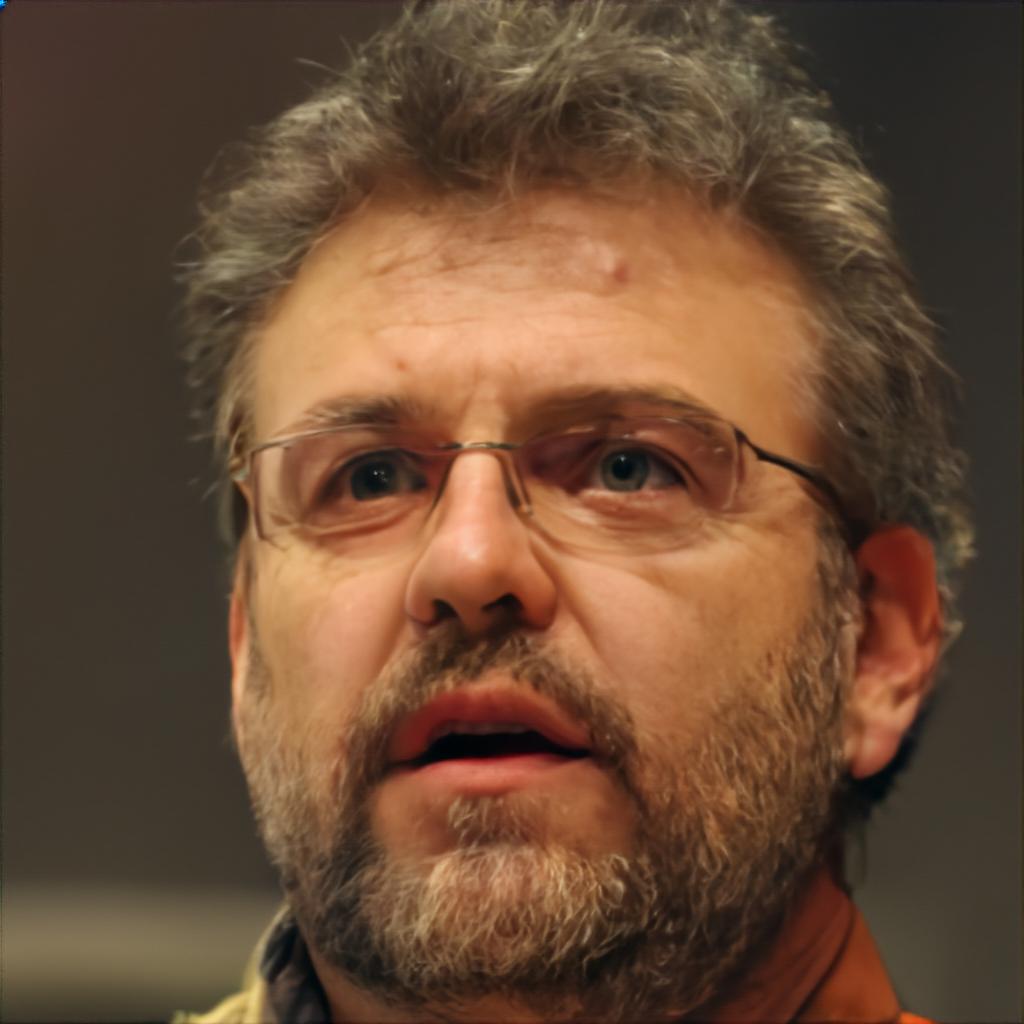} \\

\includegraphics[width=0.15\textwidth]{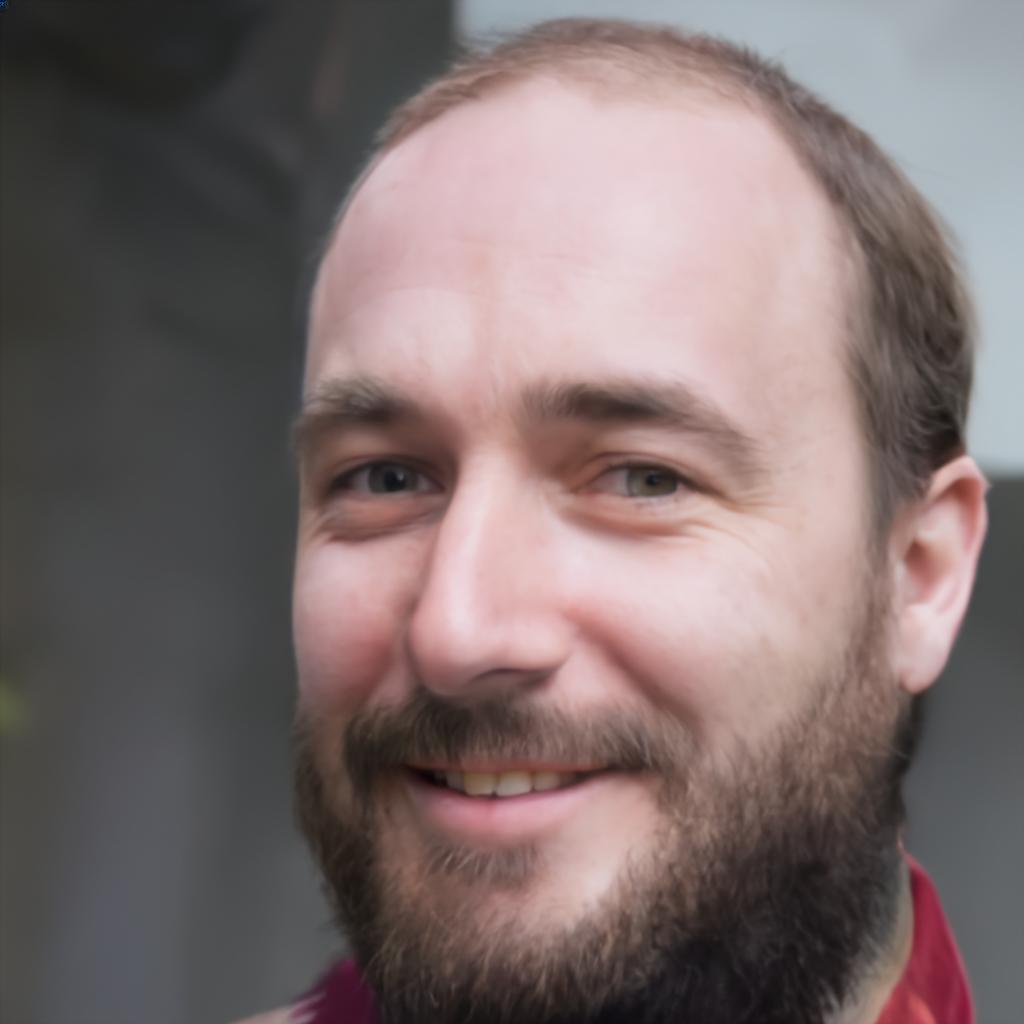} &
\includegraphics[width=0.15\textwidth]{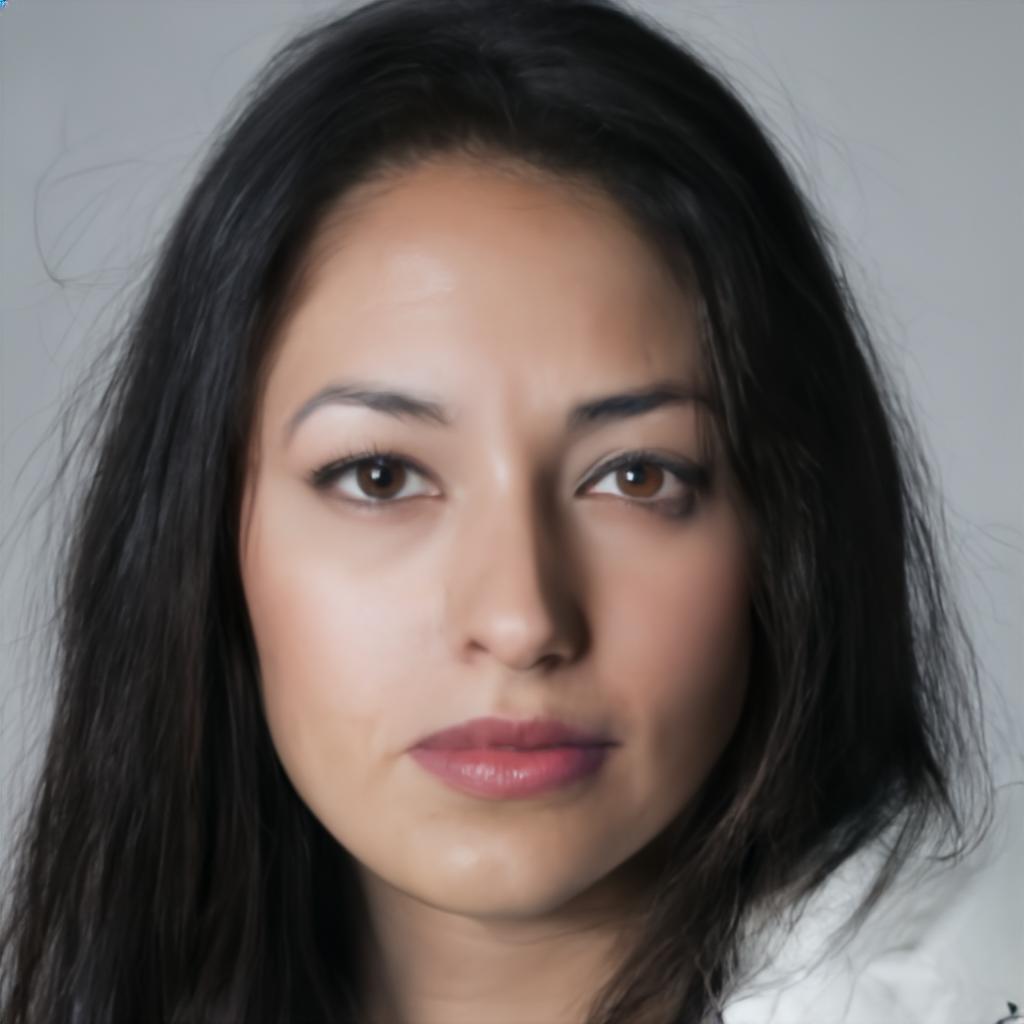} &
\includegraphics[width=0.15\textwidth]{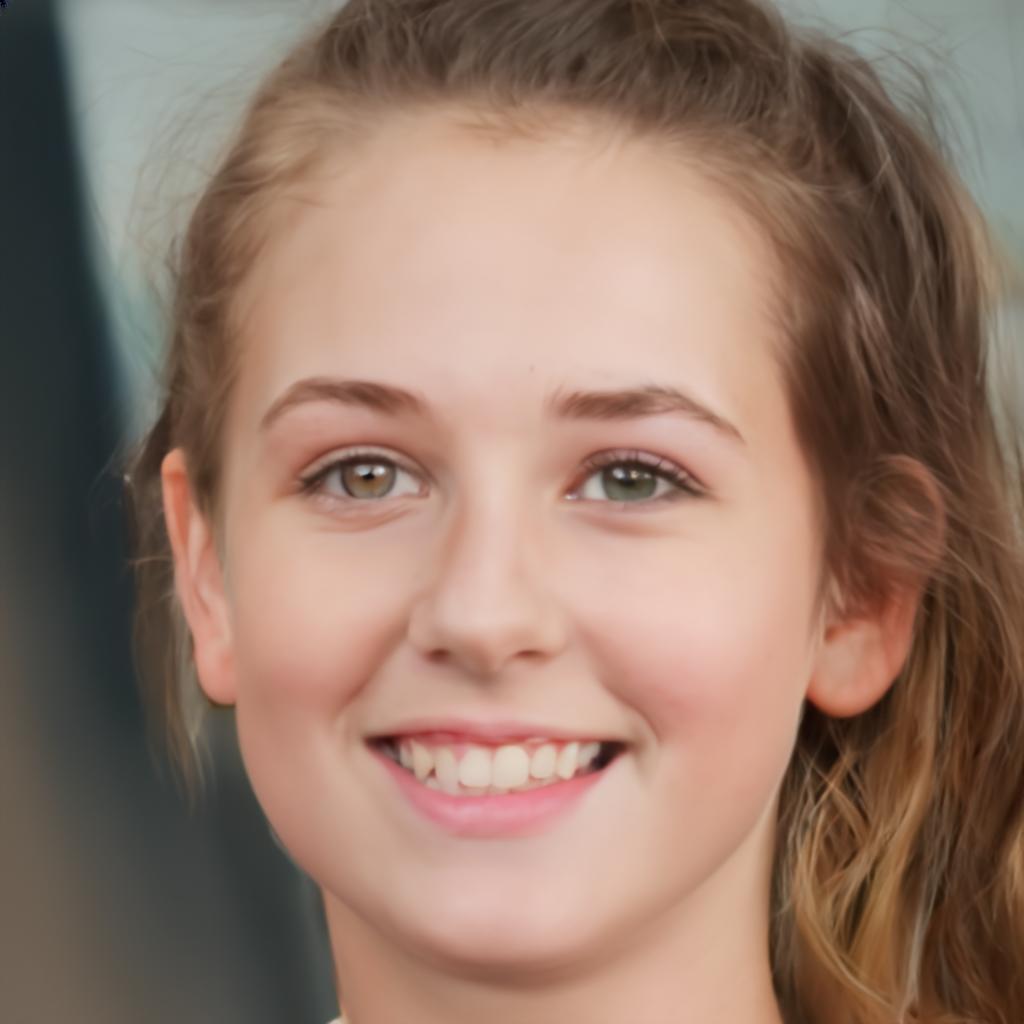} &
\includegraphics[width=0.15\textwidth]{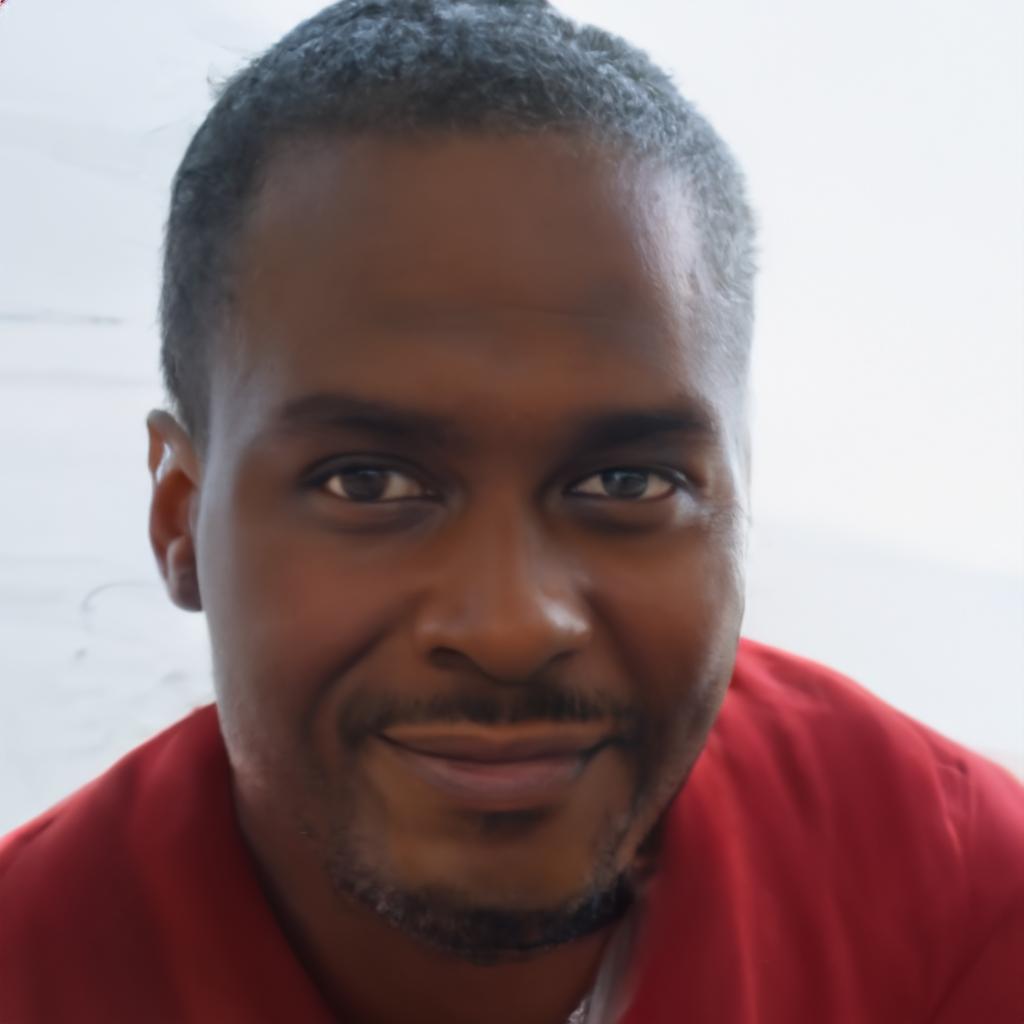} &
\includegraphics[width=0.15\textwidth]{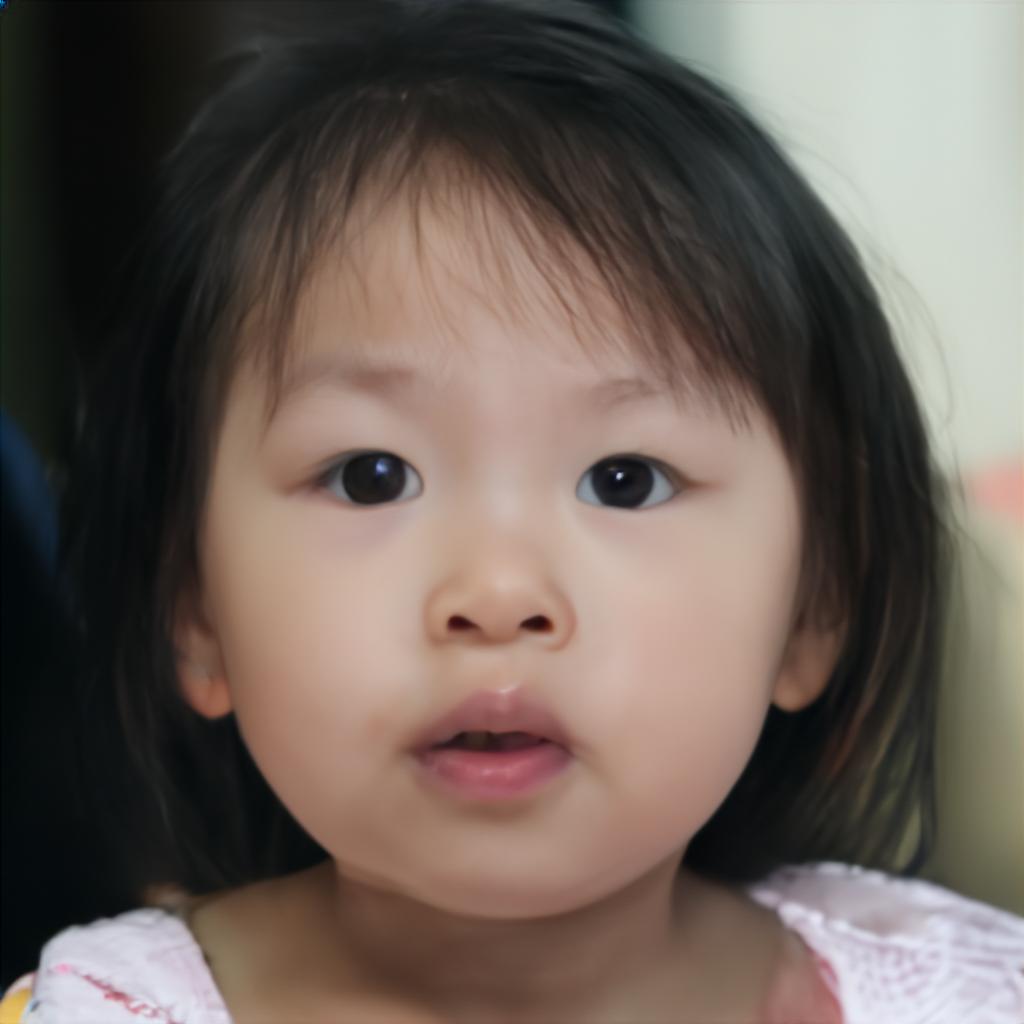} &
\includegraphics[width=0.15\textwidth]{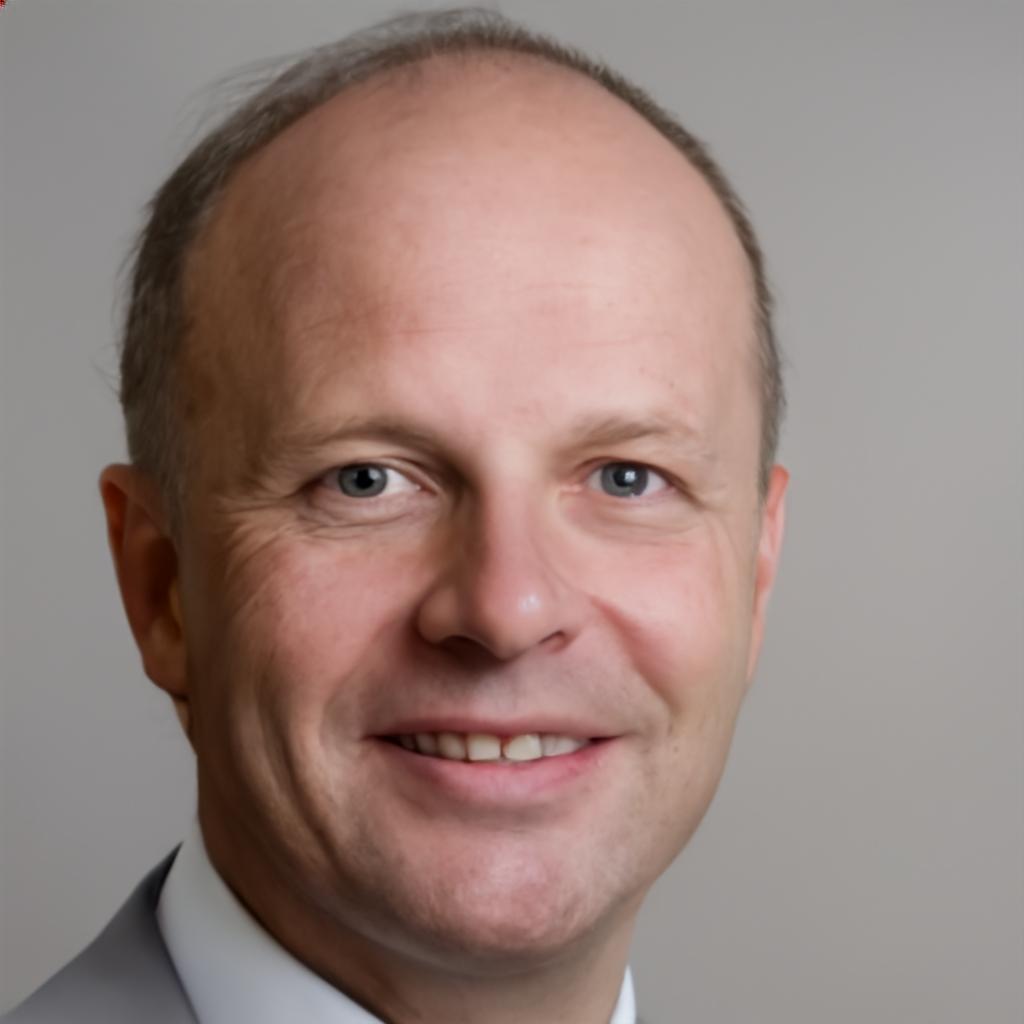} \\

\includegraphics[width=0.15\textwidth]{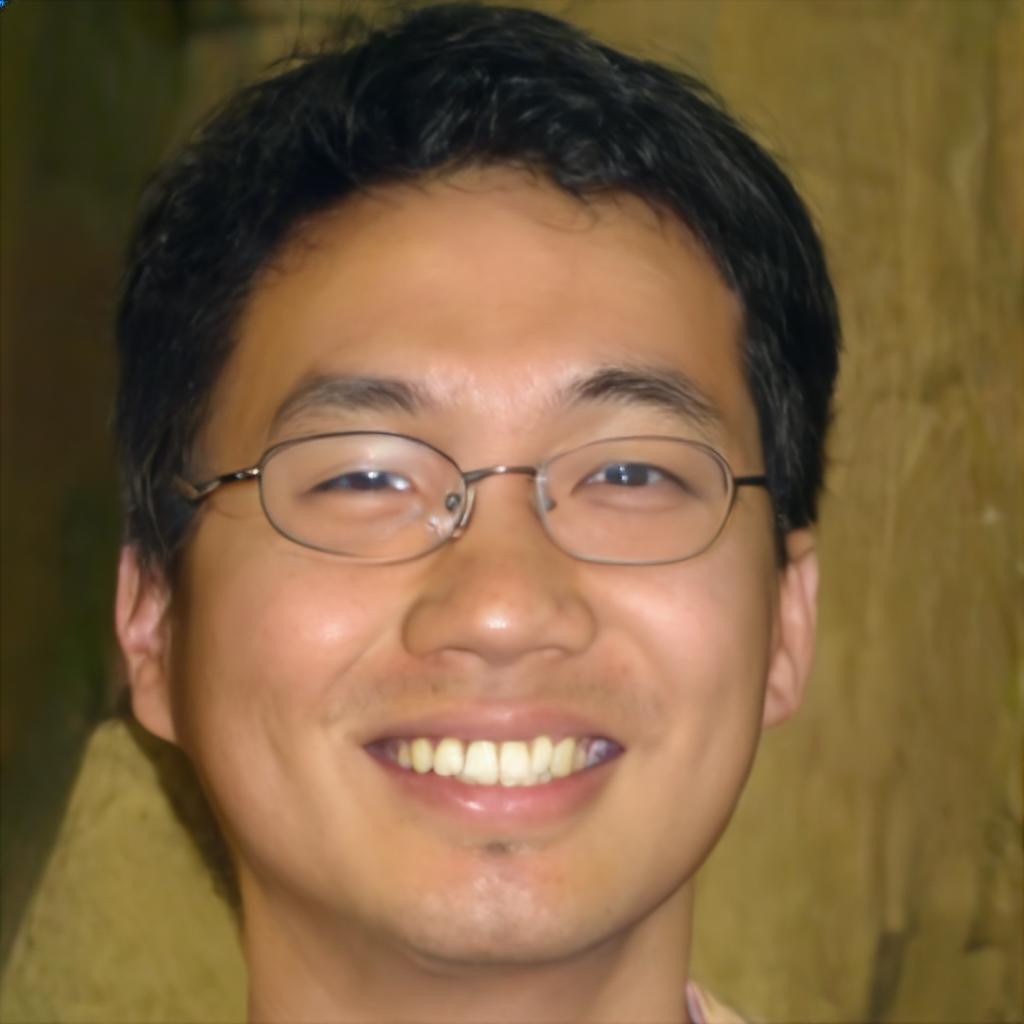} &
\includegraphics[width=0.15\textwidth]{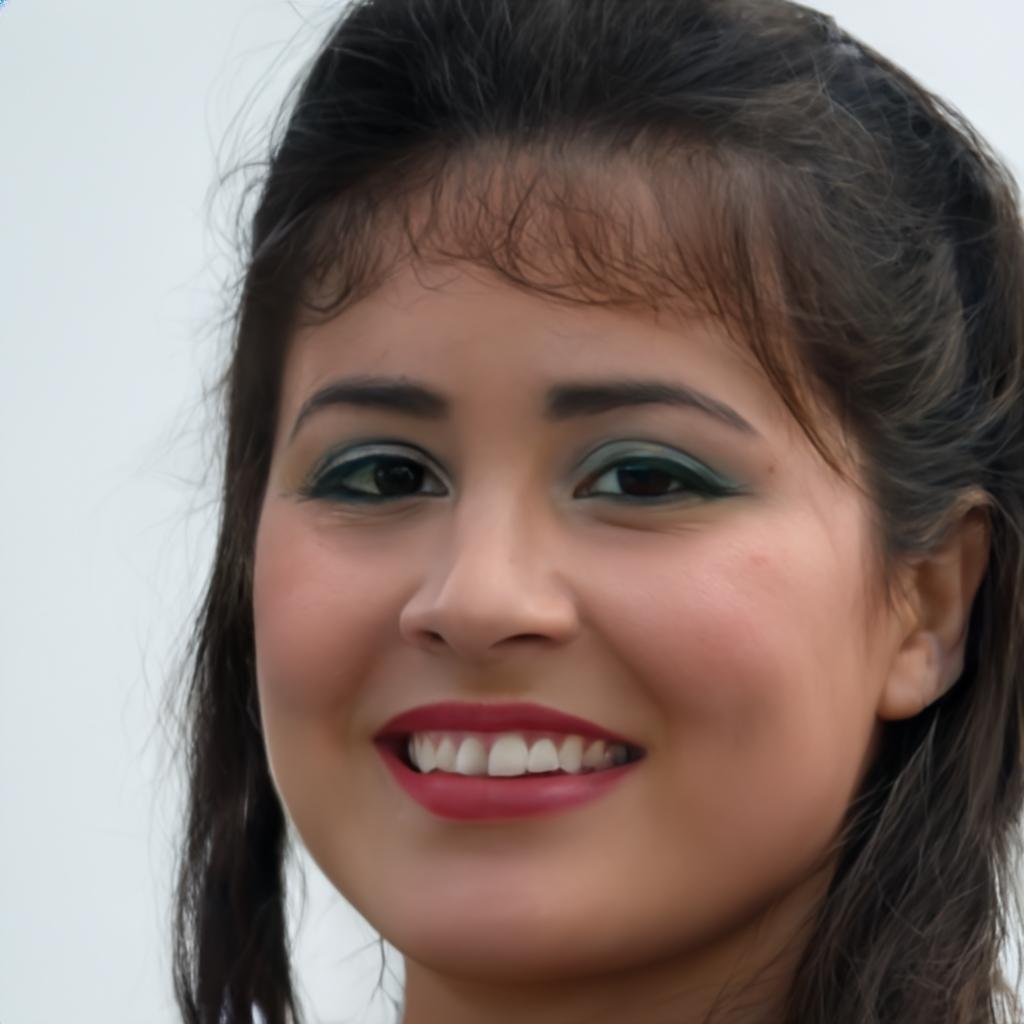} &
\includegraphics[width=0.15\textwidth]{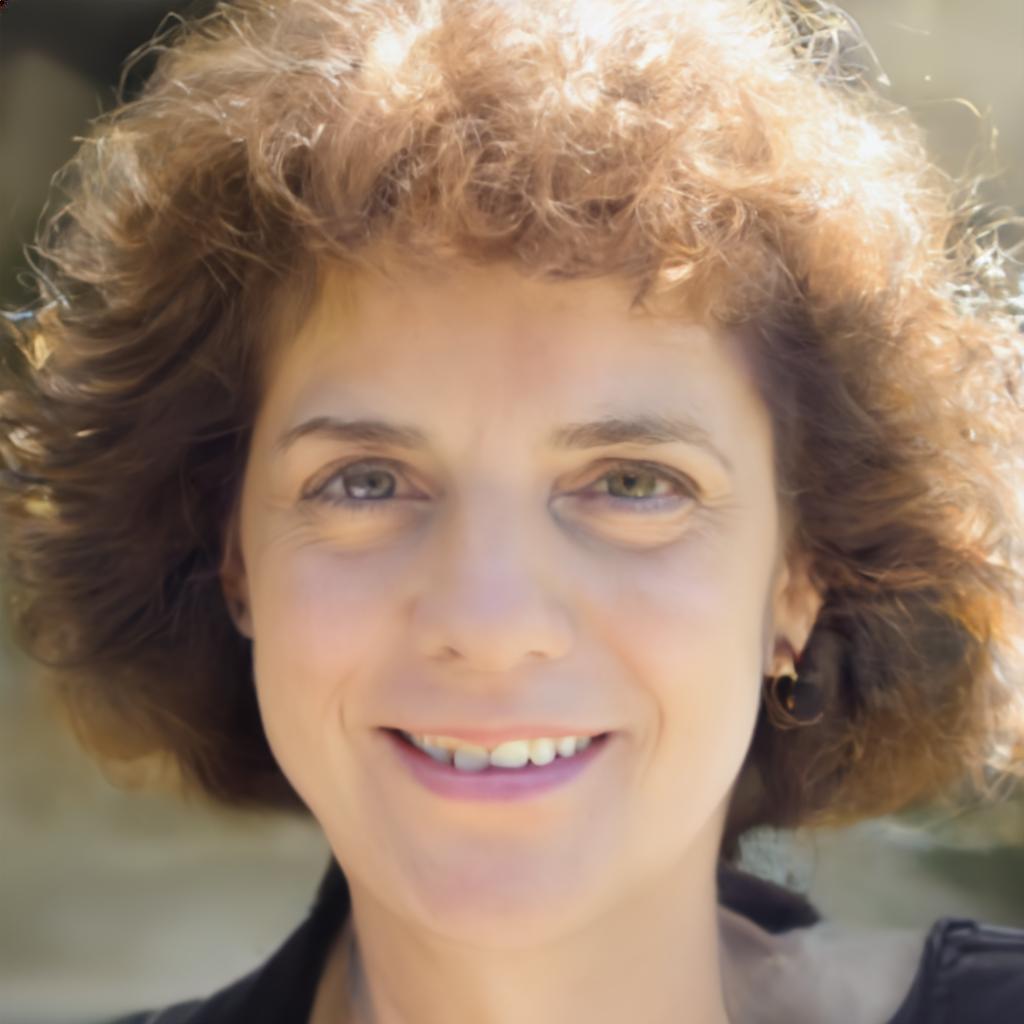} &
\includegraphics[width=0.15\textwidth]{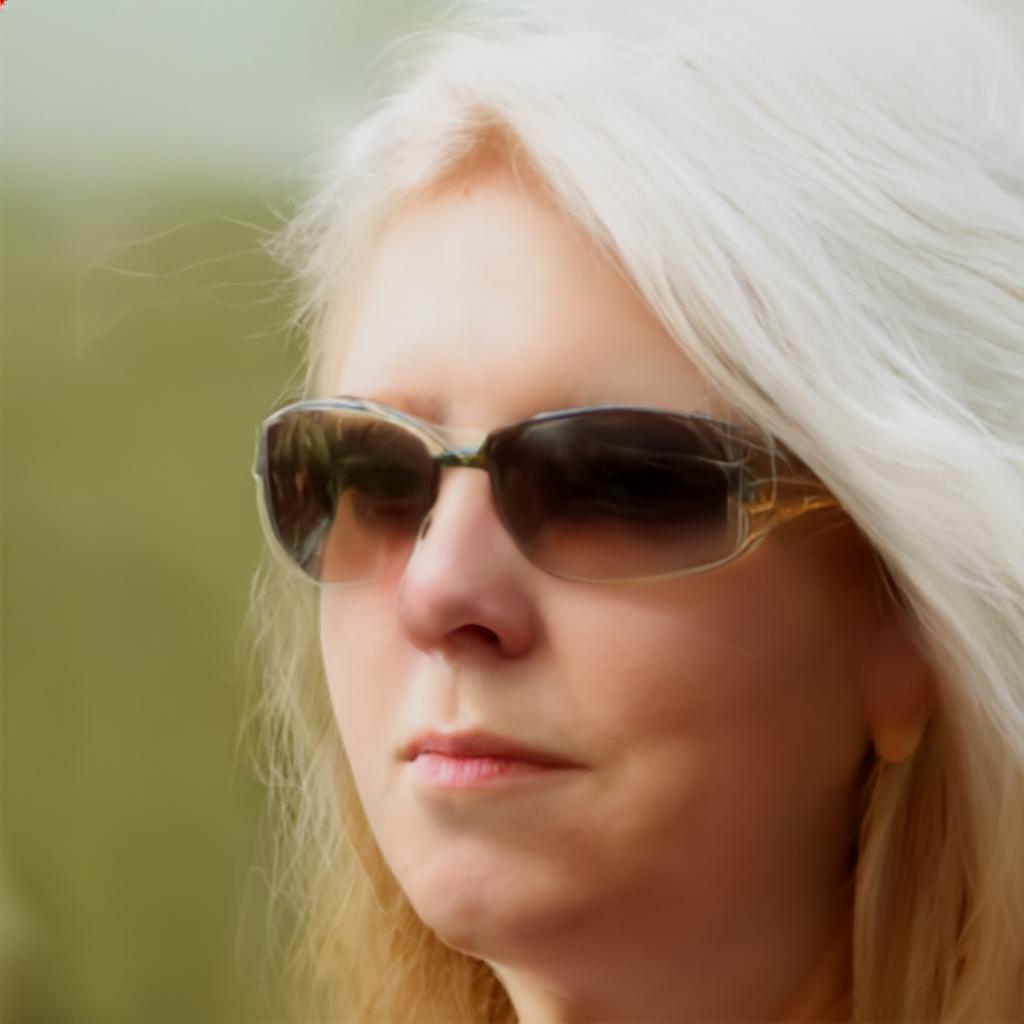} &
\includegraphics[width=0.15\textwidth]{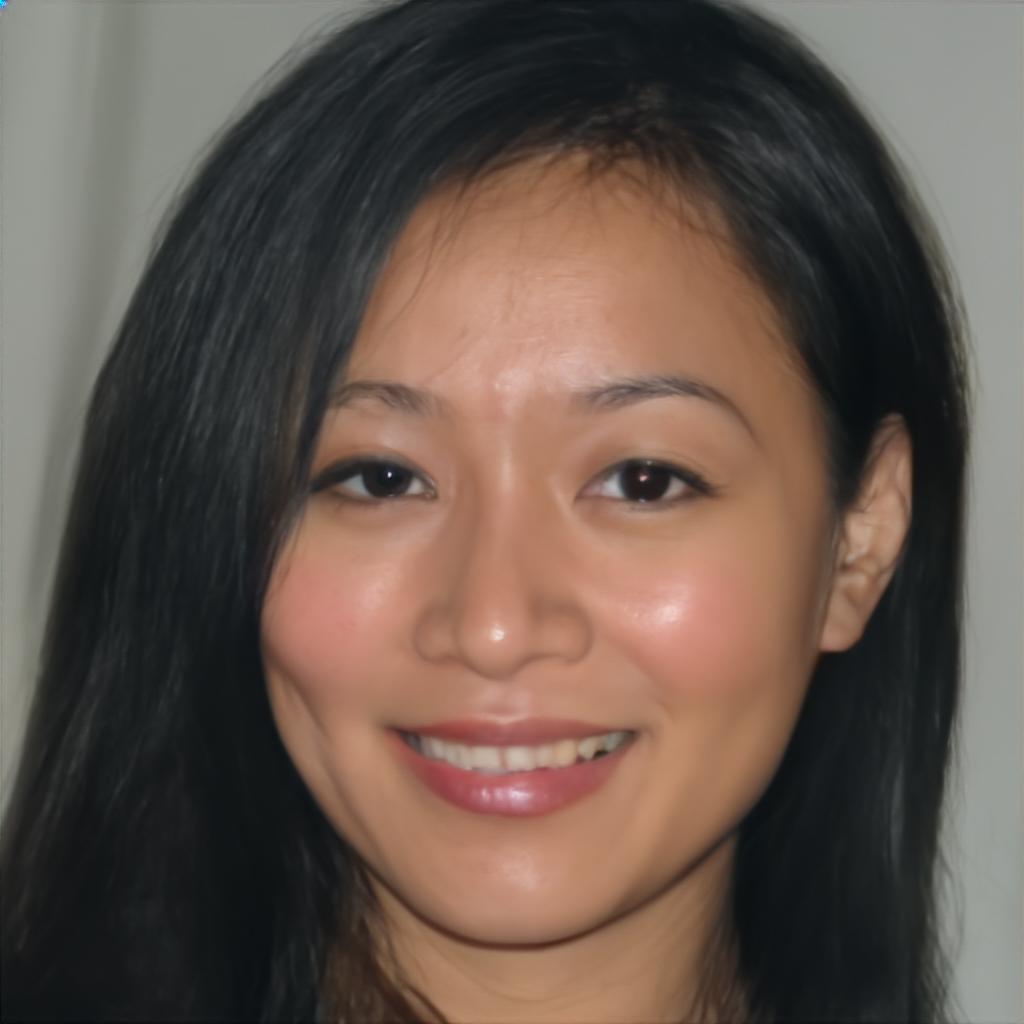} &
\includegraphics[width=0.15\textwidth]{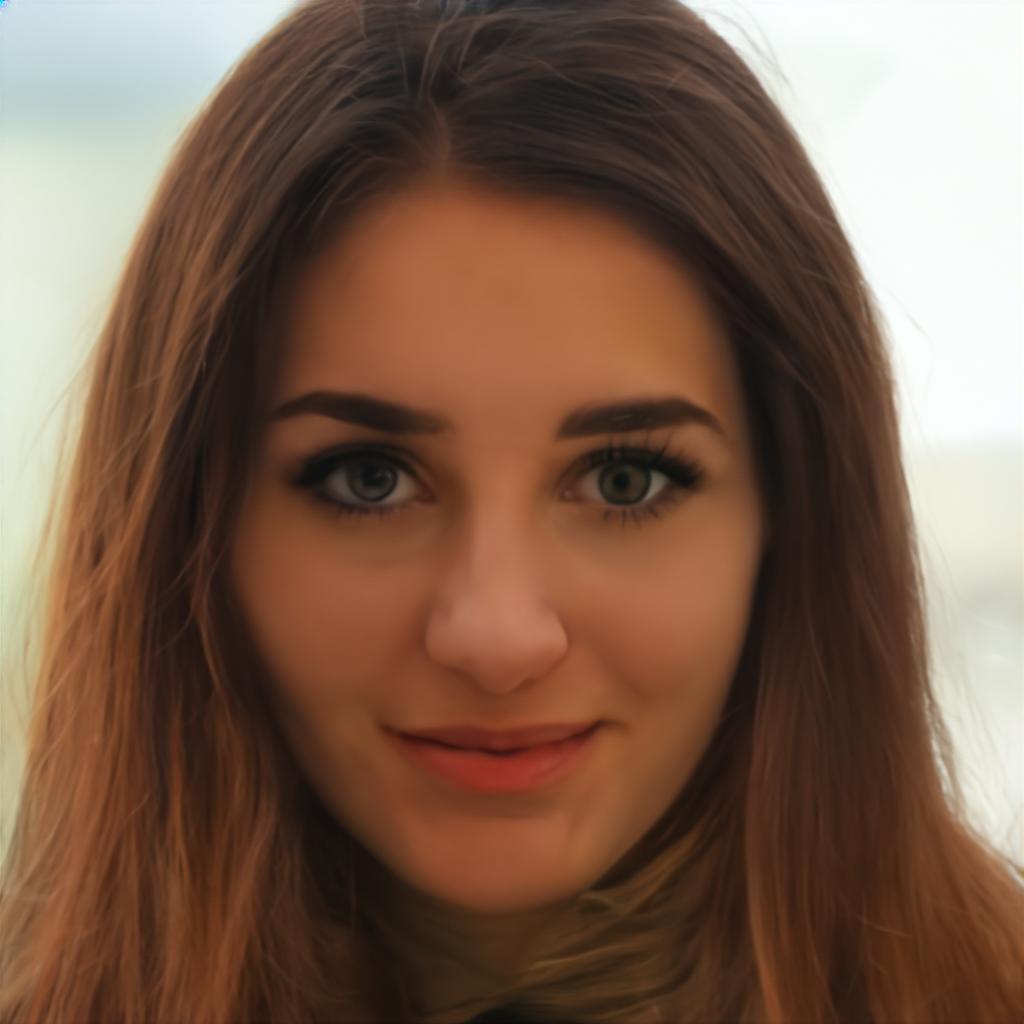} \\

\includegraphics[width=0.15\textwidth]{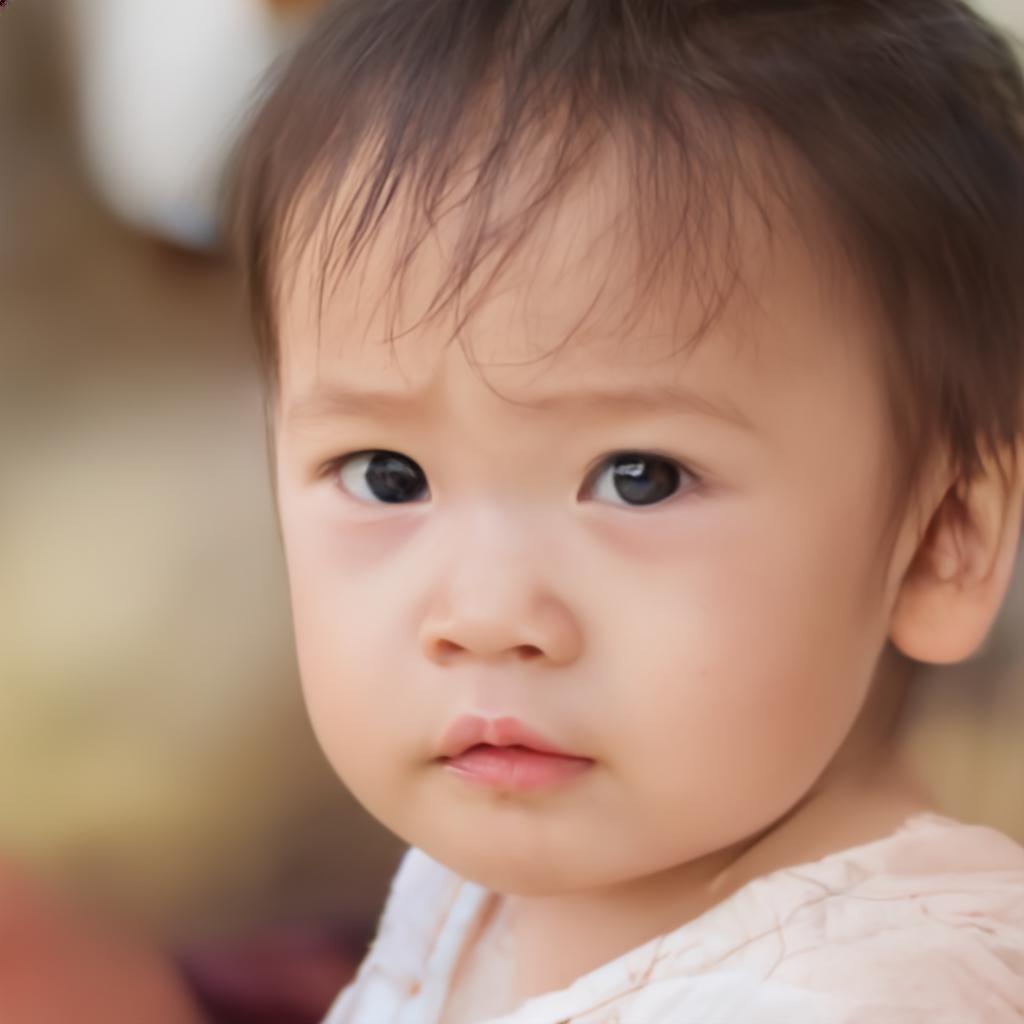} &
\includegraphics[width=0.15\textwidth]{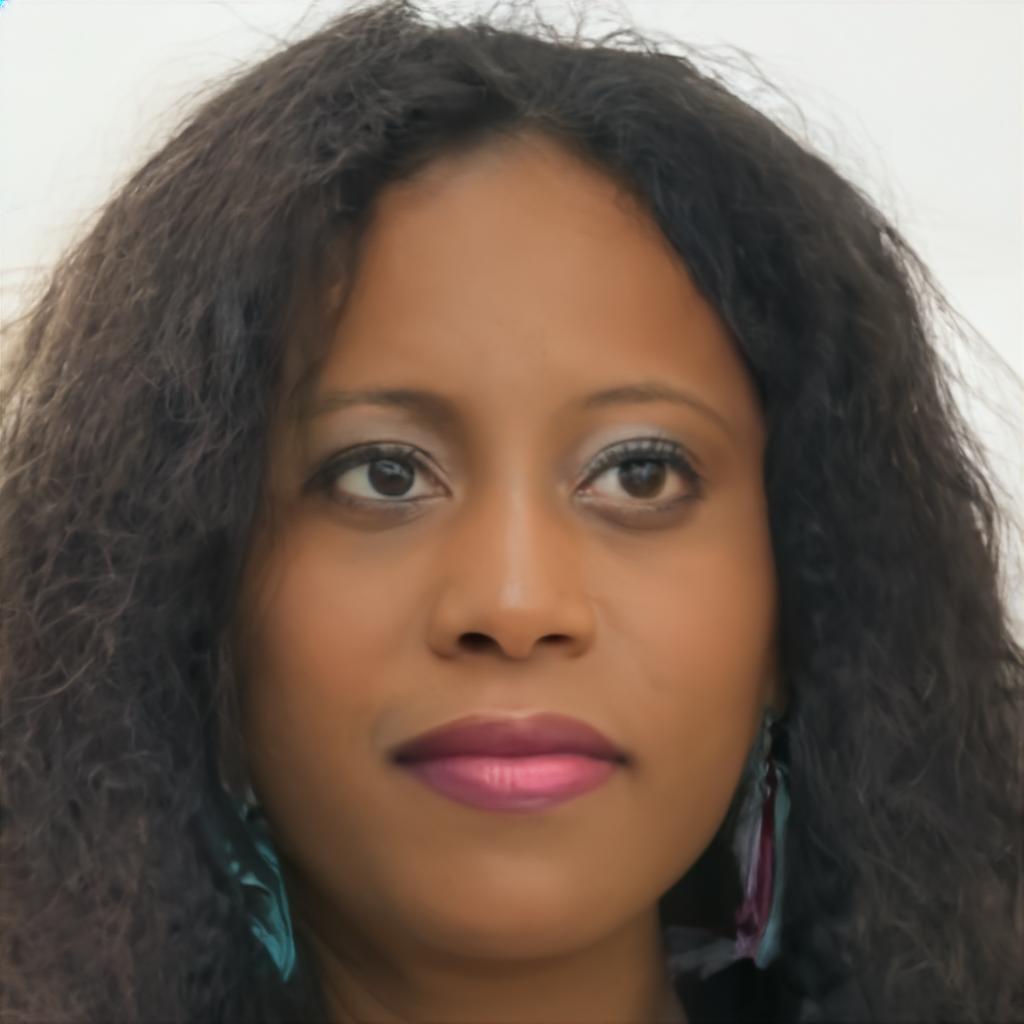} &
\includegraphics[width=0.15\textwidth]{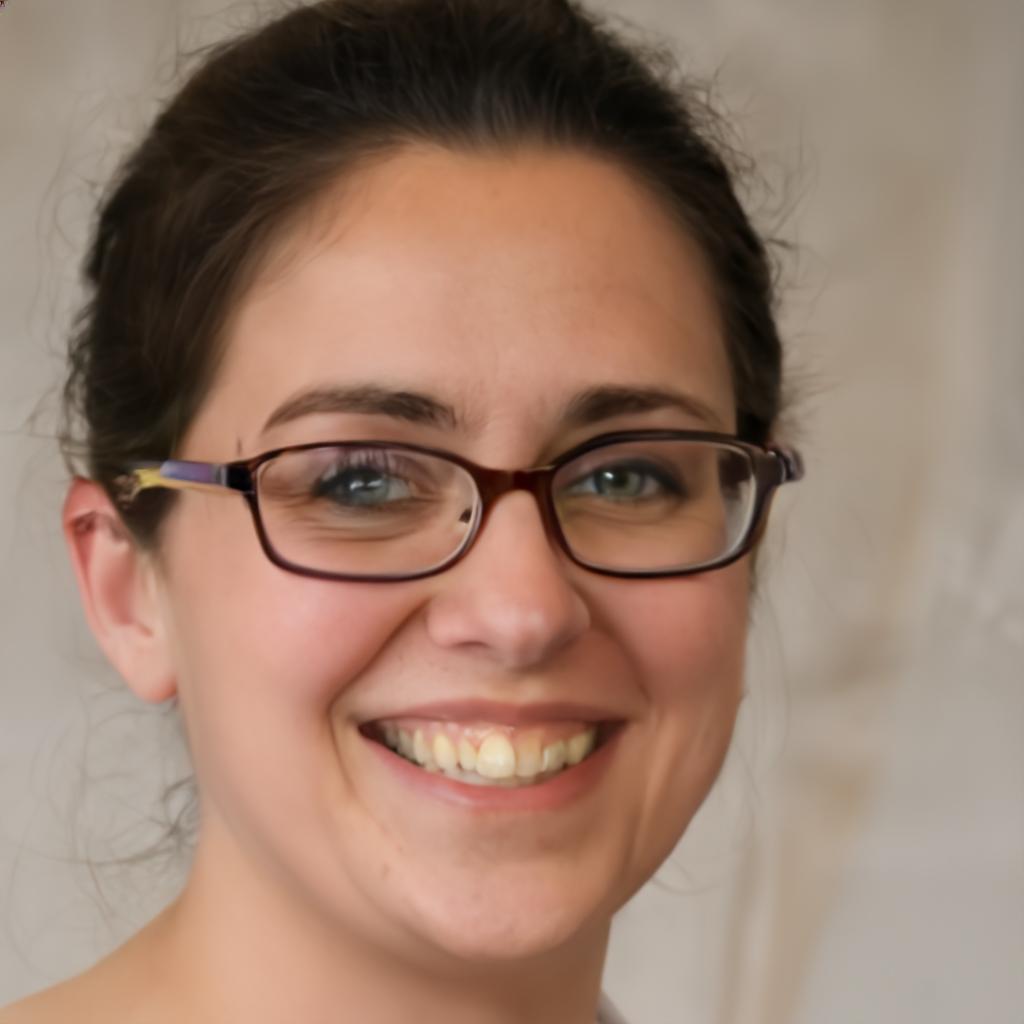} &
\includegraphics[width=0.15\textwidth]{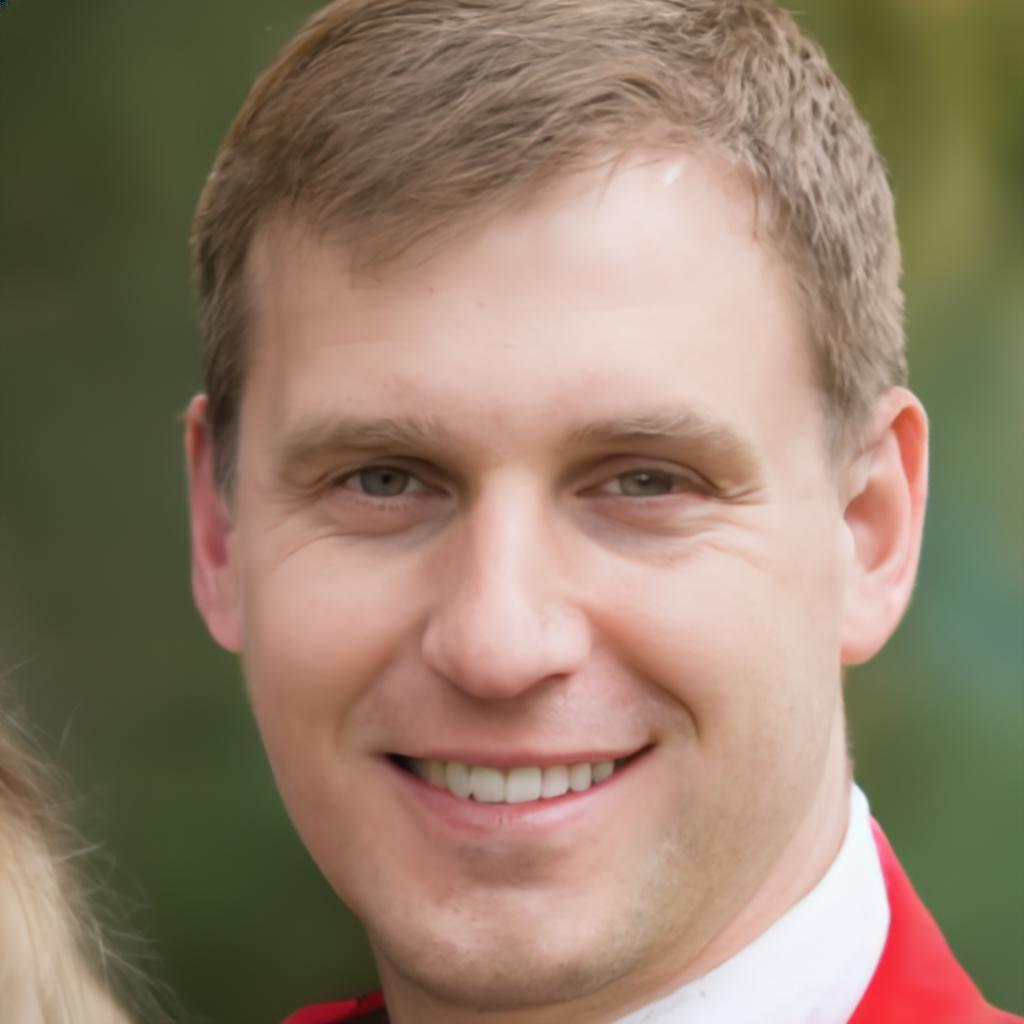} &
\includegraphics[width=0.15\textwidth]{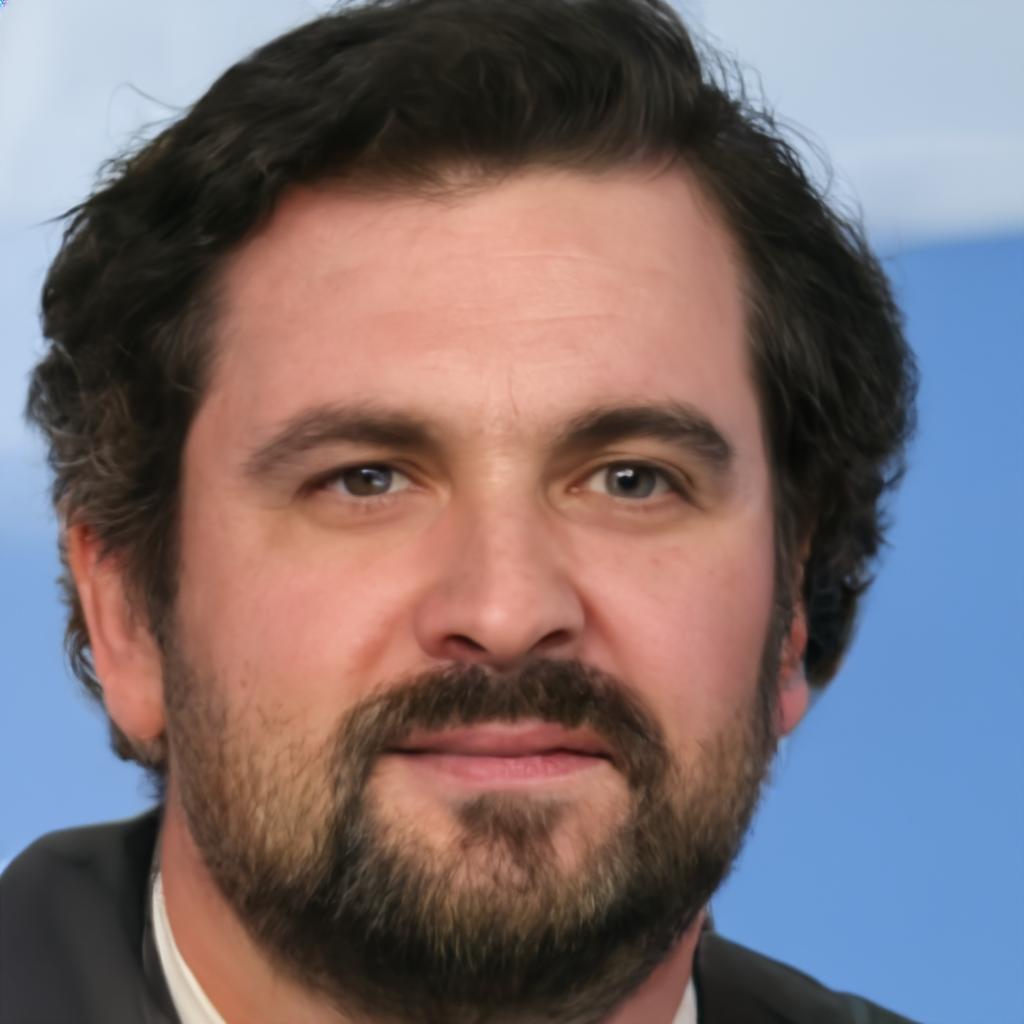} &
\includegraphics[width=0.15\textwidth]{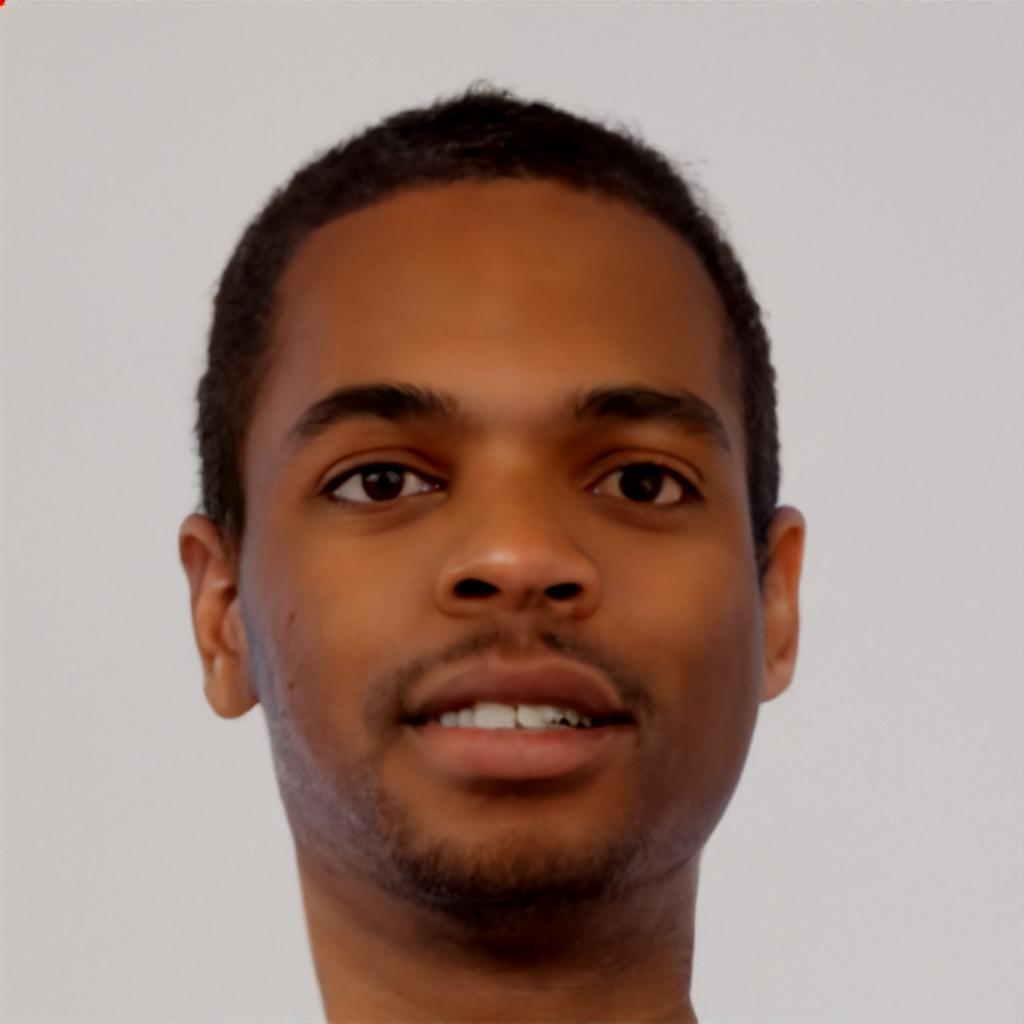} \\
\end{tabular}
\end{center}
\vspace*{-0.35cm}
\caption{Additional Synthetic 1024$\times$1024 faces images. We first sample from an unconditional 64$\times$64 diffusion model, then pass the samples through two 4$\times$ \modelname models, \ie~ 64$\times$64 $\rightarrow$ 256$\times$256 $\rightarrow$ 1024$\times$1024.
\vspace*{-.4cm}
}
\label{fig:1024x_cascade4}
\vspace*{-0.35cm}
\end{figure}



\begin{center}
{\large \bf Class Conditional ImageNet Samples \  256$\times$256}
\end{center}

\begin{figure}[H]
\vspace*{-0.2cm}
\setlength{\tabcolsep}{1.25pt}
\begin{center}
\begin{tabular}{cccccc}
\includegraphics[width=0.142\textwidth]{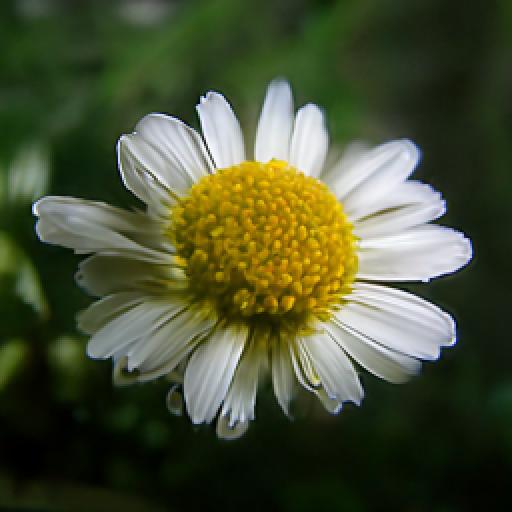} & 
 \includegraphics[width=0.142\textwidth]{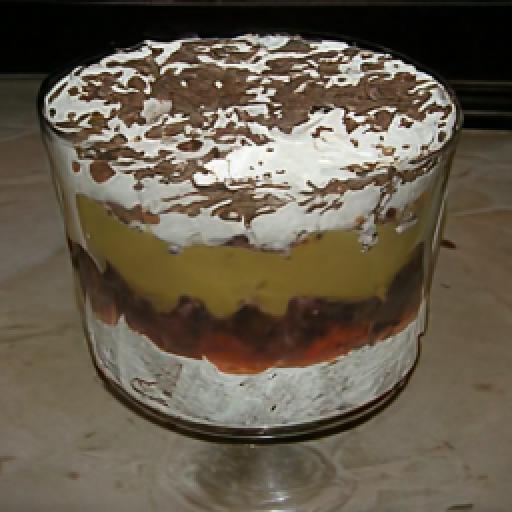} & 
 \includegraphics[width=0.142\textwidth]{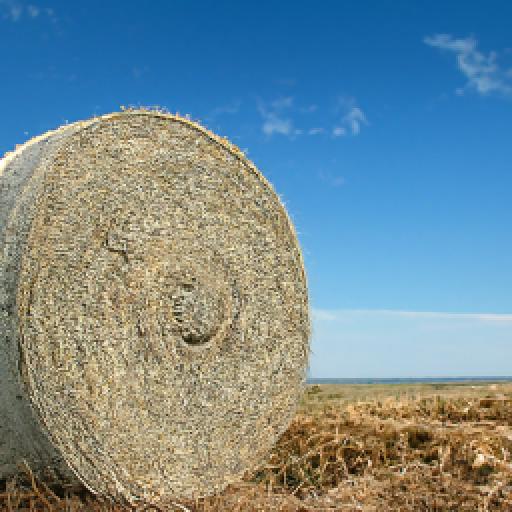} & 
 \includegraphics[width=0.142\textwidth]{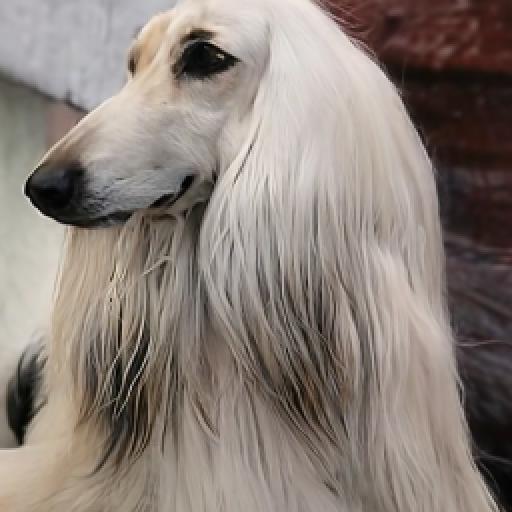} & 
 \includegraphics[width=0.142\textwidth]{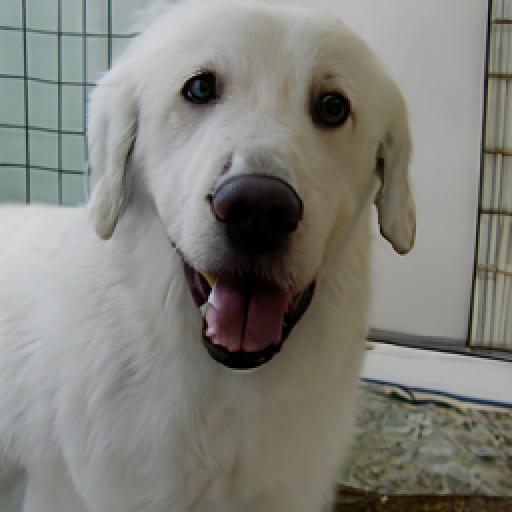} & 
 \includegraphics[width=0.142\textwidth]{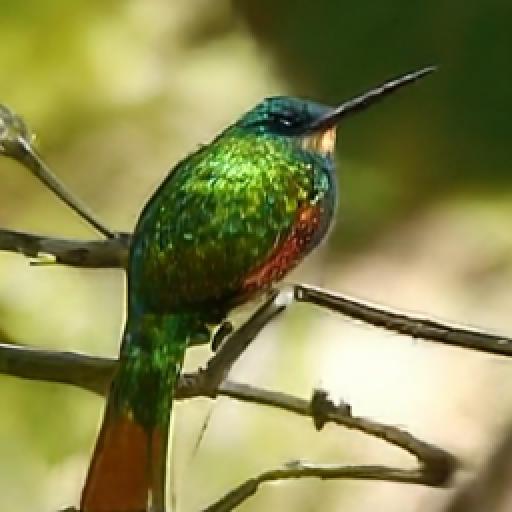}\\
 
 \includegraphics[width=0.142\textwidth]{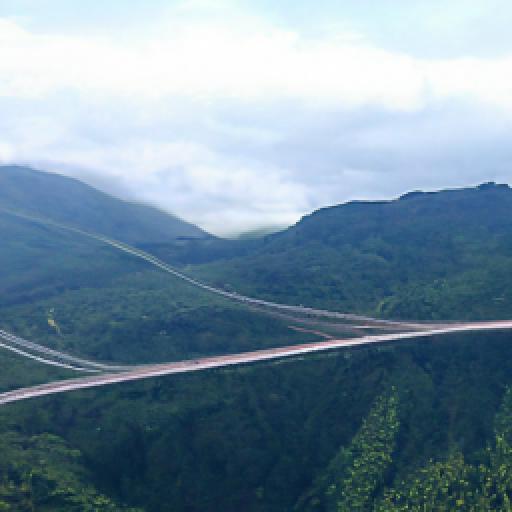} & 
 \includegraphics[width=0.142\textwidth]{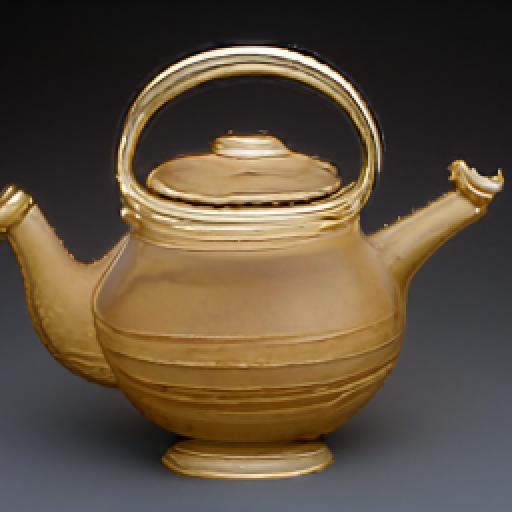} & 
 \includegraphics[width=0.142\textwidth]{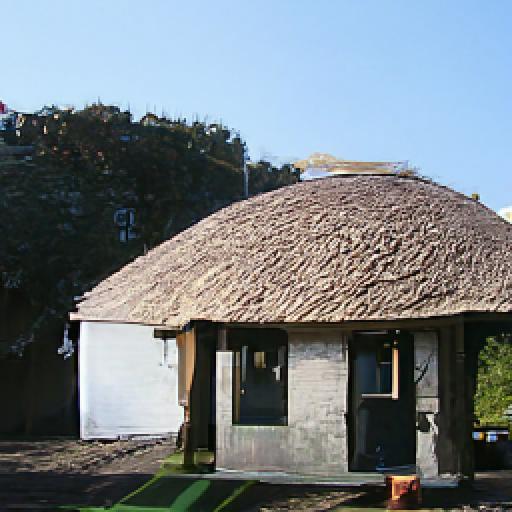} & 
 \includegraphics[width=0.142\textwidth]{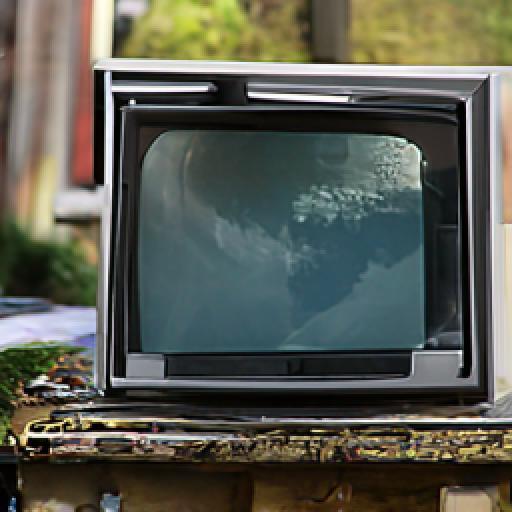} & 
 \includegraphics[width=0.142\textwidth]{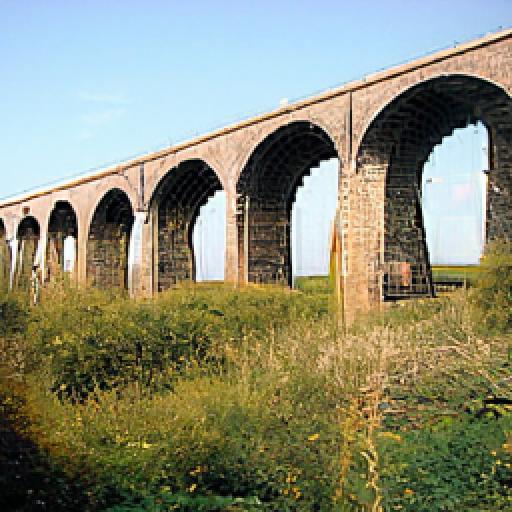} & 
 \includegraphics[width=0.142\textwidth]{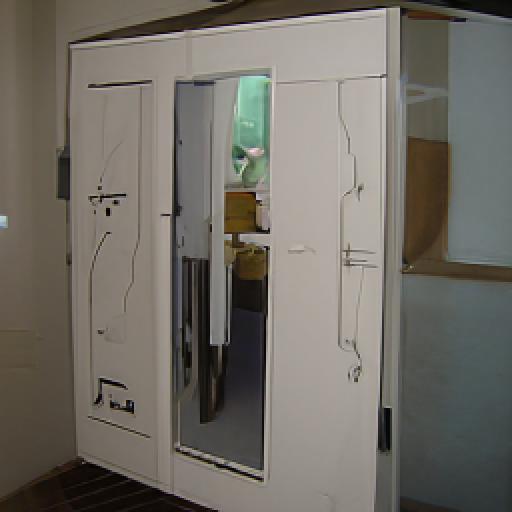}\\

\includegraphics[width=0.142\textwidth]{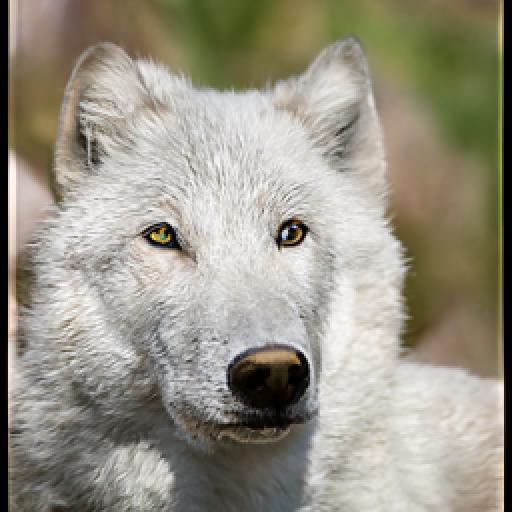} & 
\includegraphics[width=0.142\textwidth]{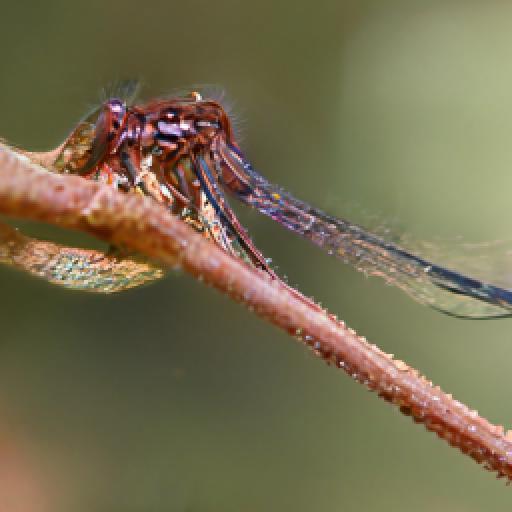} &
 \includegraphics[width=0.142\textwidth]{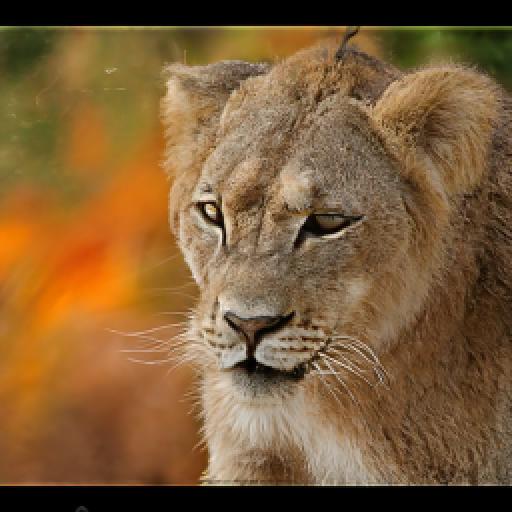} & 
 \includegraphics[width=0.142\textwidth]{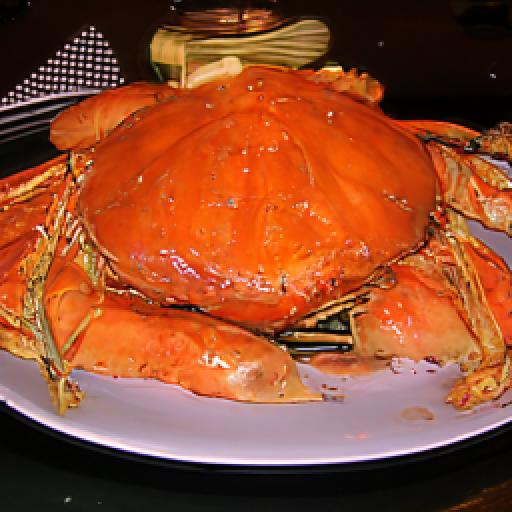} & 
  \includegraphics[width=0.142\textwidth]{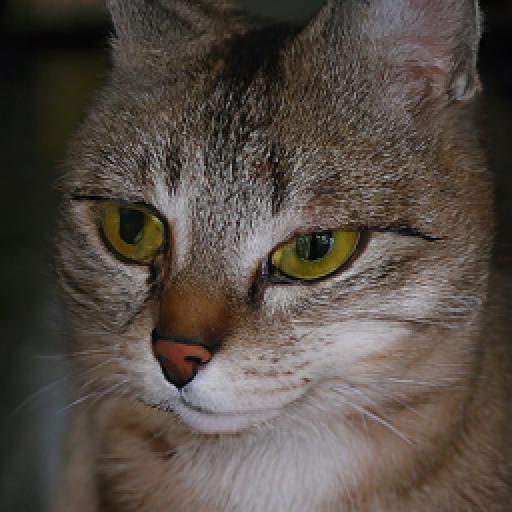} & 
 \includegraphics[width=0.142\textwidth]{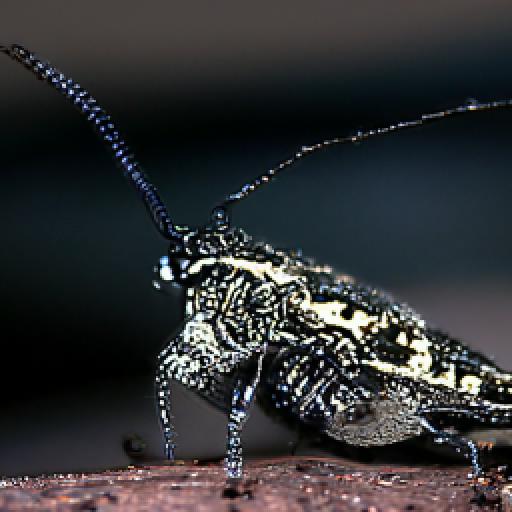}\\
 
 \includegraphics[width=0.142\textwidth]{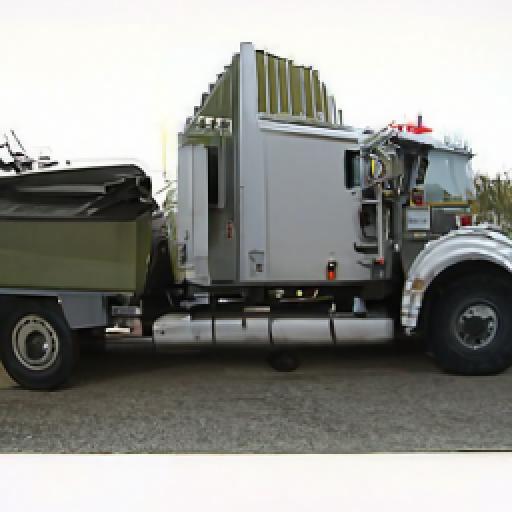} & 
 \includegraphics[width=0.142\textwidth]{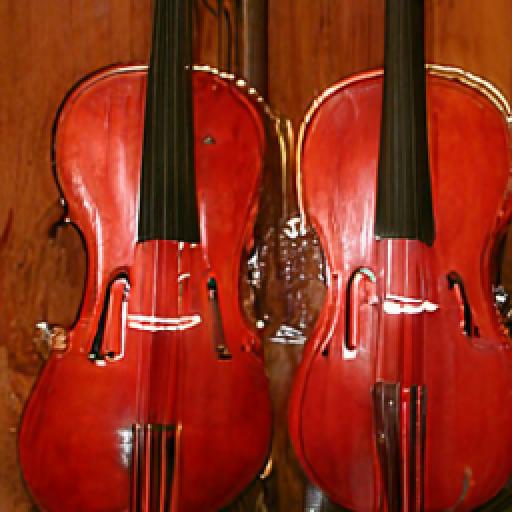} & 
 \includegraphics[width=0.142\textwidth]{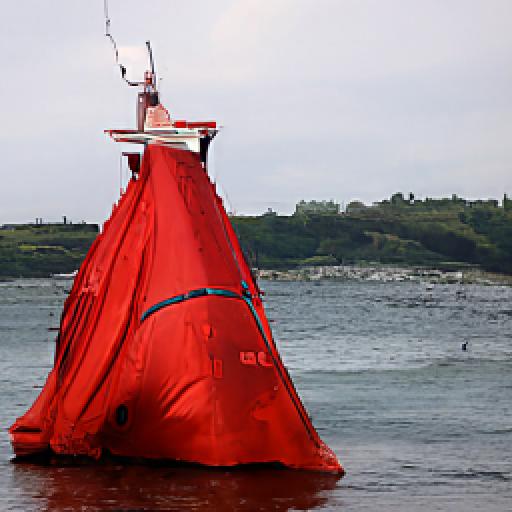} & 
 \includegraphics[width=0.142\textwidth]{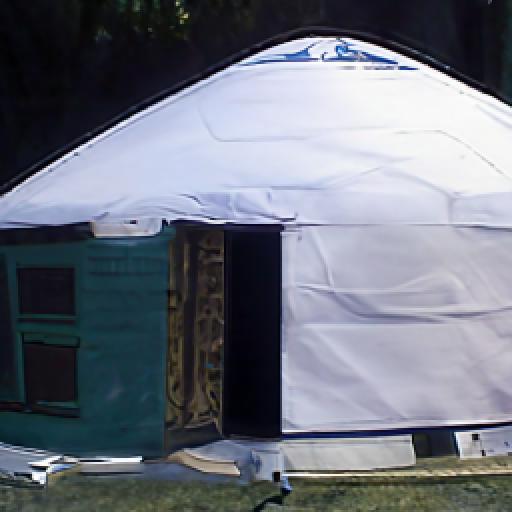} & 
 \includegraphics[width=0.142\textwidth]{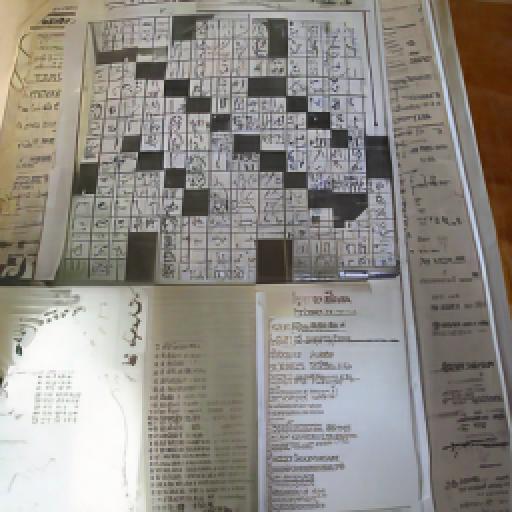} & 
 \includegraphics[width=0.142\textwidth]{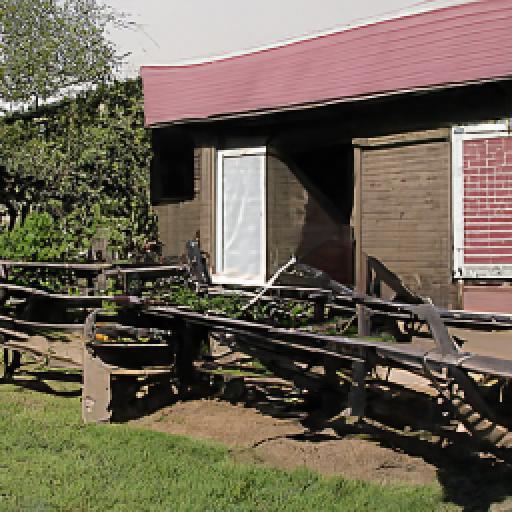}\\

\includegraphics[width=0.142\textwidth]{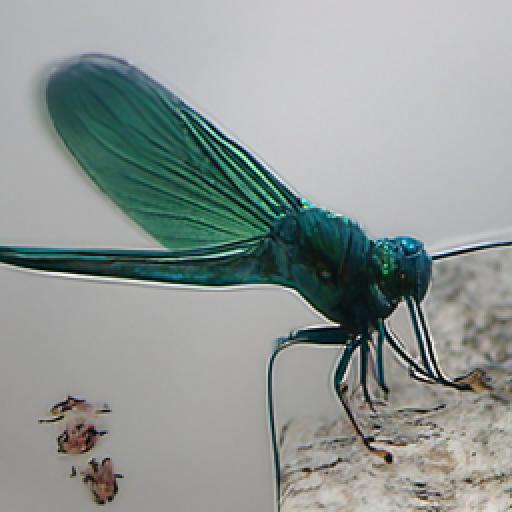} & 
 \includegraphics[width=0.142\textwidth]{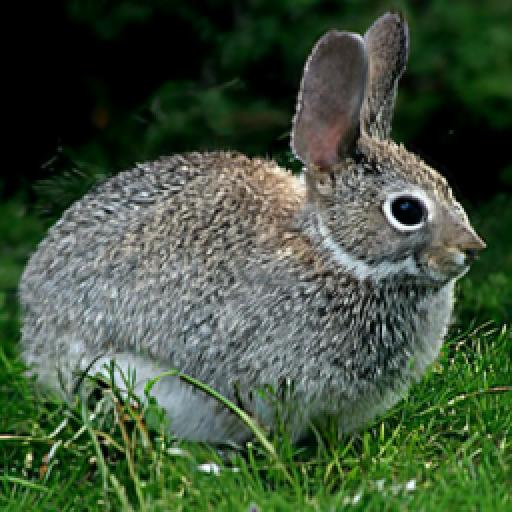} & 
 \includegraphics[width=0.142\textwidth]{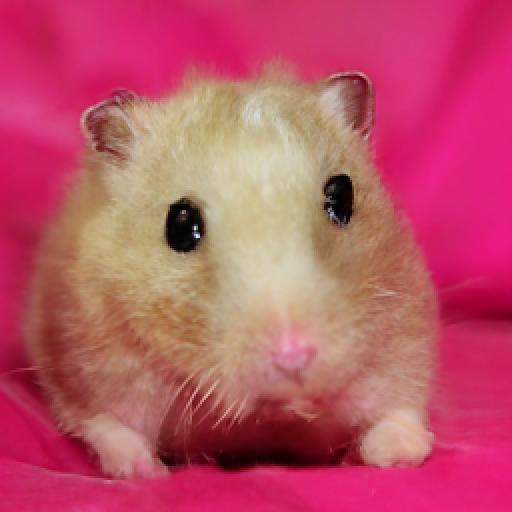} & 
 \includegraphics[width=0.142\textwidth]{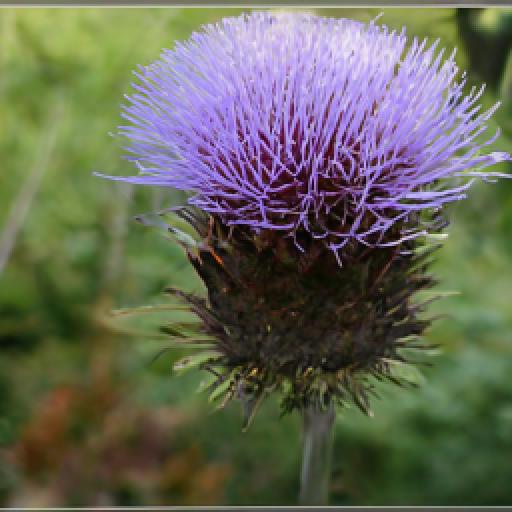} & 
 \includegraphics[width=0.142\textwidth]{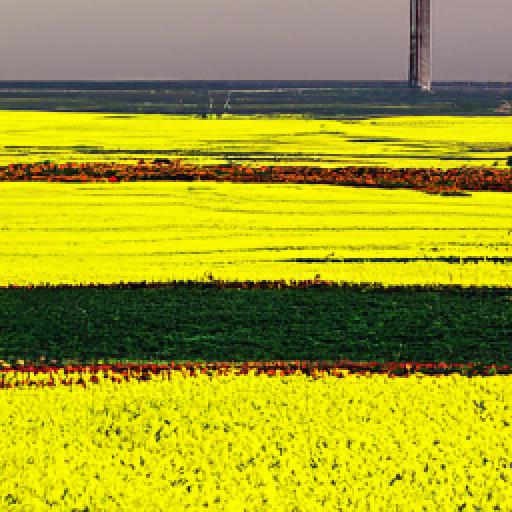} & 
 \includegraphics[width=0.142\textwidth]{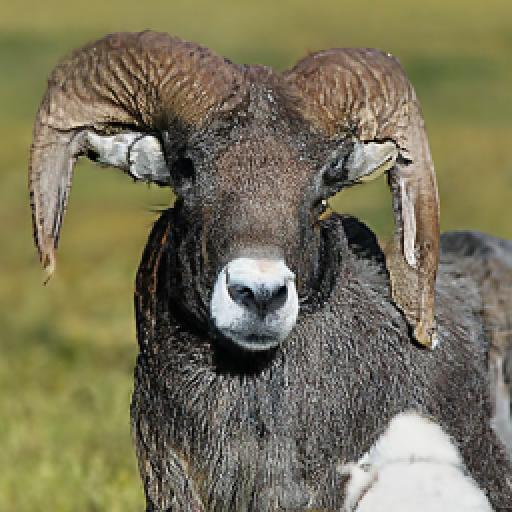}\\

\includegraphics[width=0.142\textwidth]{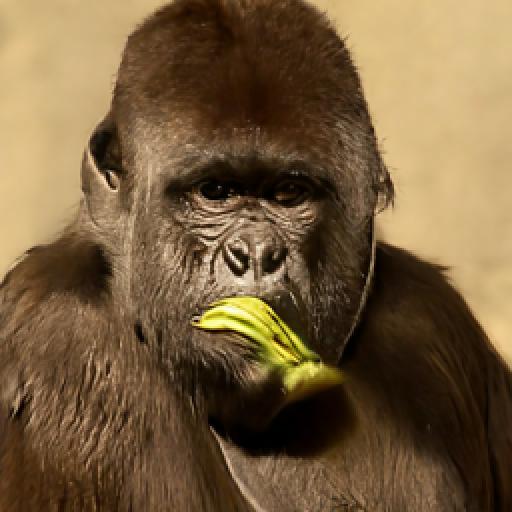} & 
 \includegraphics[width=0.142\textwidth]{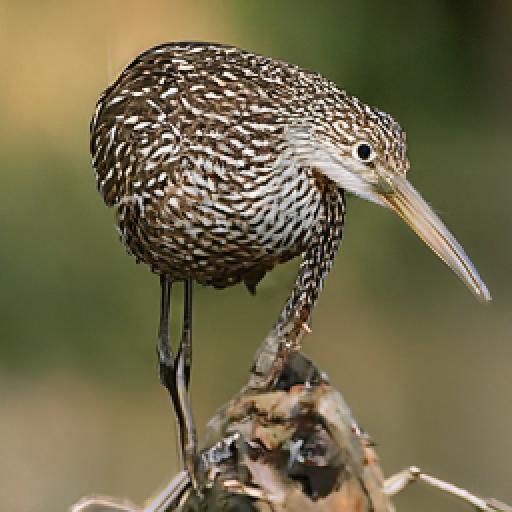} & 
 \includegraphics[width=0.142\textwidth]{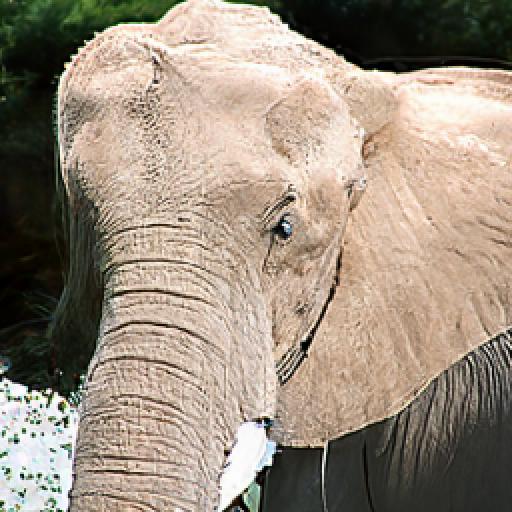} & 
 \includegraphics[width=0.142\textwidth]{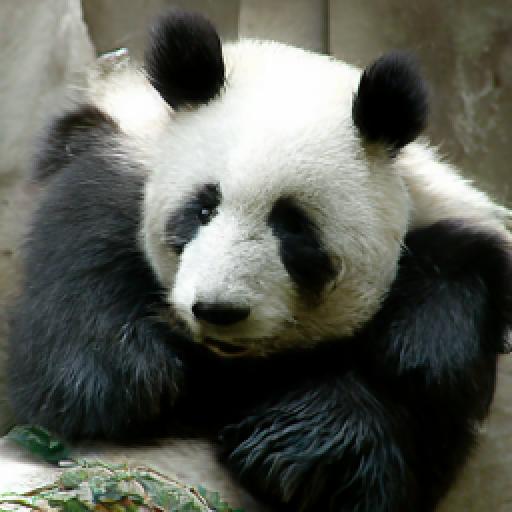} & 
 \includegraphics[width=0.142\textwidth]{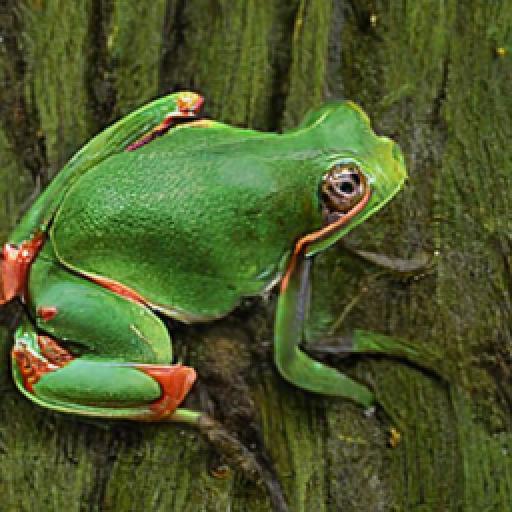} & 
 \includegraphics[width=0.142\textwidth]{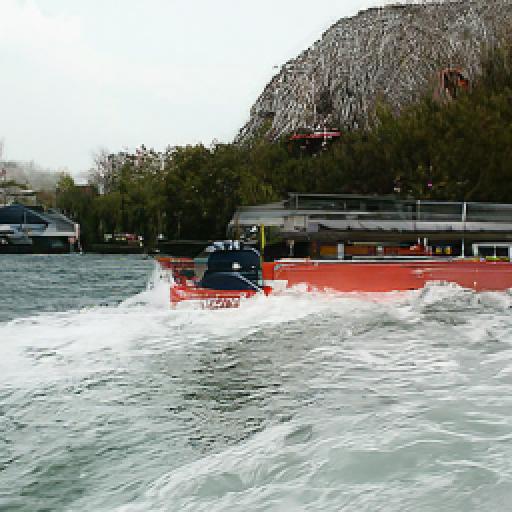}\\

\includegraphics[width=0.142\textwidth]{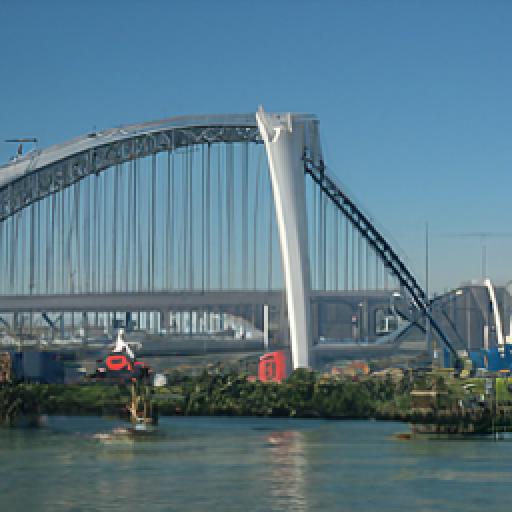} & 
 \includegraphics[width=0.142\textwidth]{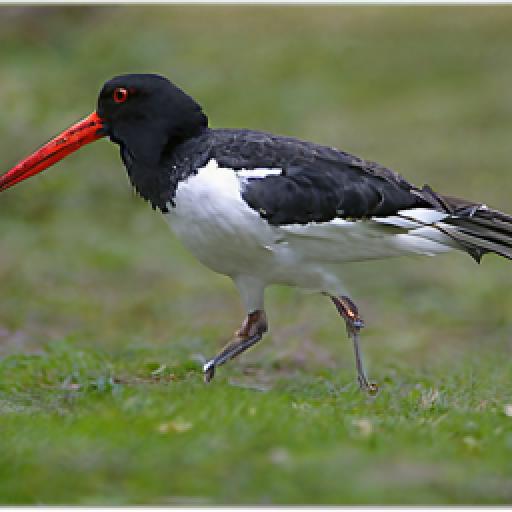} & 
 \includegraphics[width=0.142\textwidth]{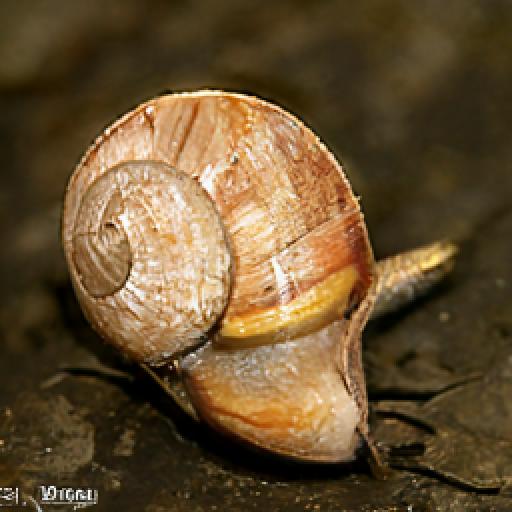} & 
 \includegraphics[width=0.142\textwidth]{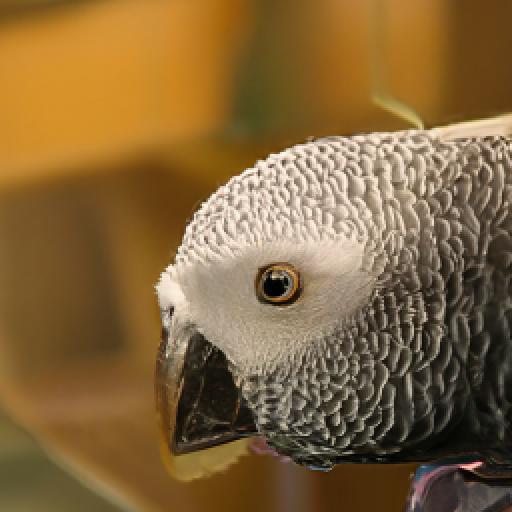} & 
 \includegraphics[width=0.142\textwidth]{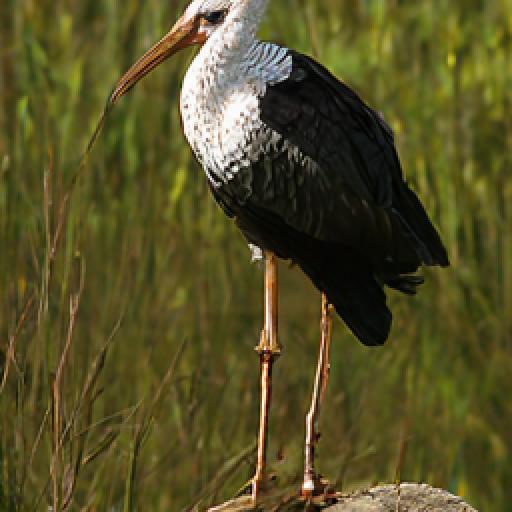} & 
 \includegraphics[width=0.142\textwidth]{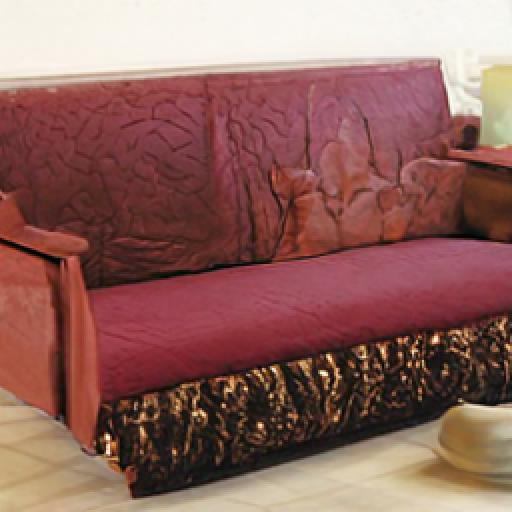}\\

\includegraphics[width=0.142\textwidth]{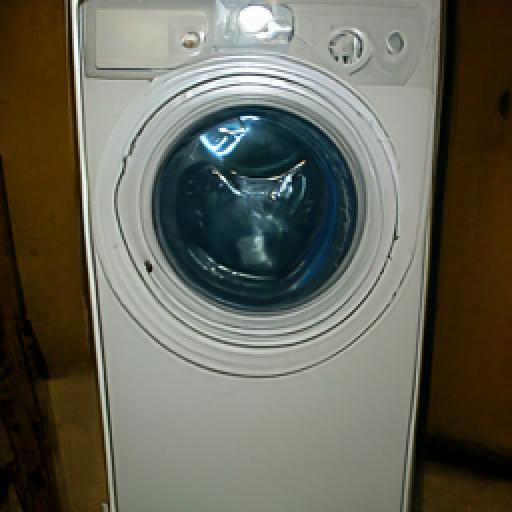} & 
 \includegraphics[width=0.142\textwidth]{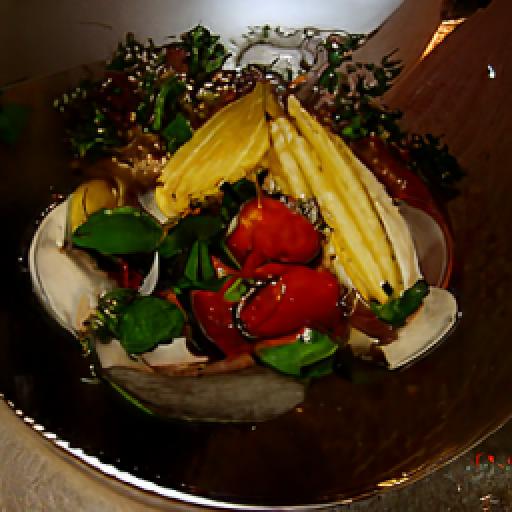} & 
 \includegraphics[width=0.142\textwidth]{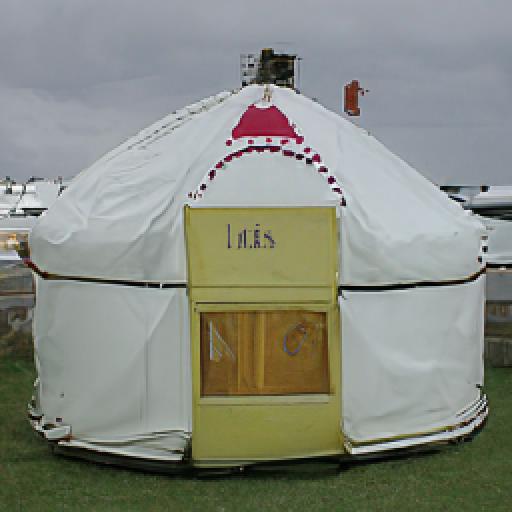} & 
 \includegraphics[width=0.142\textwidth]{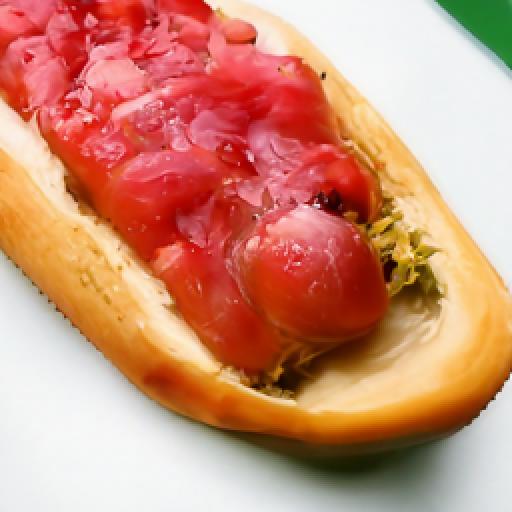} & 
 \includegraphics[width=0.142\textwidth]{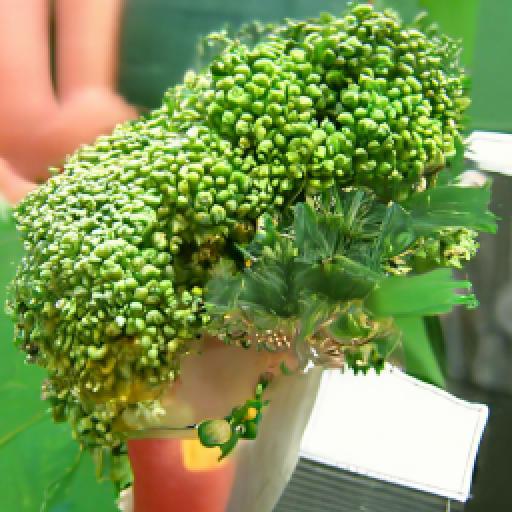} & 
 \includegraphics[width=0.142\textwidth]{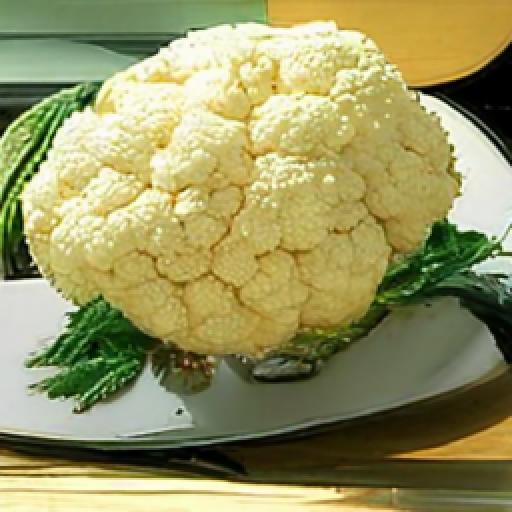}\\

\end{tabular}
\end{center}
\vspace*{-0.4cm}
\caption{Additional Synthetic 256$\times$256 ImageNet images. We first draw a random label, then sample a 64$\times$64 image from a class-conditional diffusion model, and apply a 4$\times$ \modelname model to obtain 256$\times$256 images.}
\vspace*{-0.2cm}
\label{fig:imagenet_256x_montage2}
\end{figure}

\begin{figure}[H]
\vspace*{-0.2cm}
\setlength{\tabcolsep}{1.25pt}
\begin{center}
\begin{tabular}{cccccc}
\includegraphics[width=0.142\textwidth]{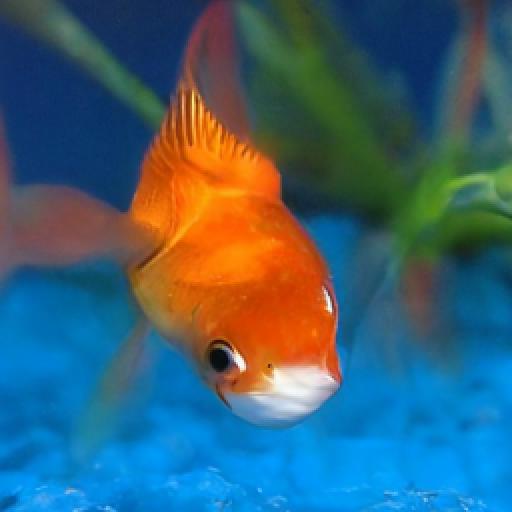} & 
 \includegraphics[width=0.142\textwidth]{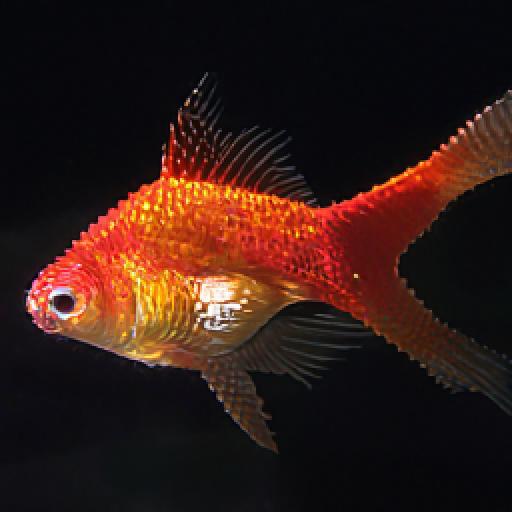} & 
 \includegraphics[width=0.142\textwidth]{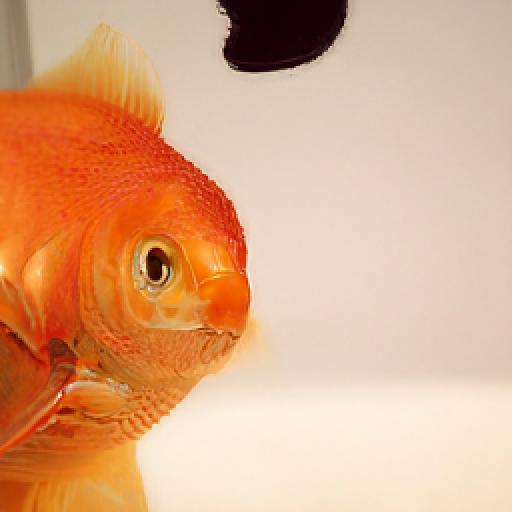} & 
 \includegraphics[width=0.142\textwidth]{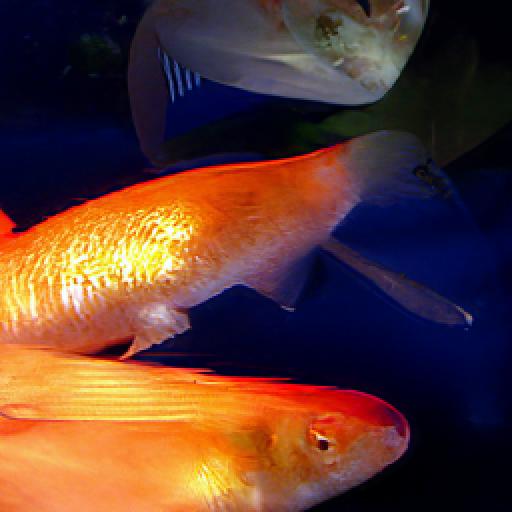} & 
 \includegraphics[width=0.142\textwidth]{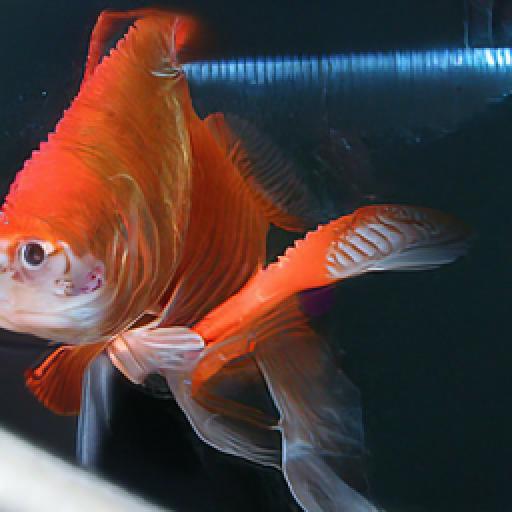} & 
 \includegraphics[width=0.142\textwidth]{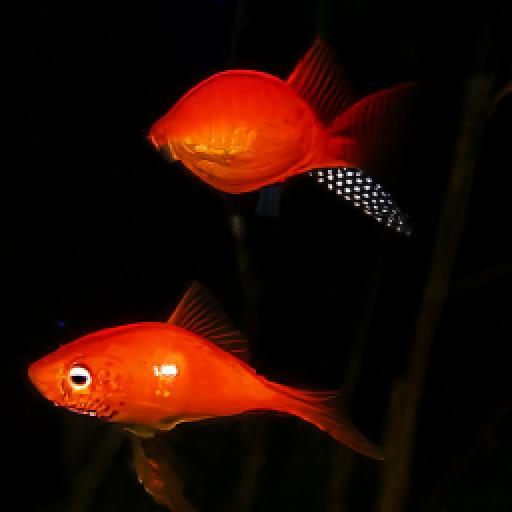}\\
 
\includegraphics[width=0.142\textwidth]{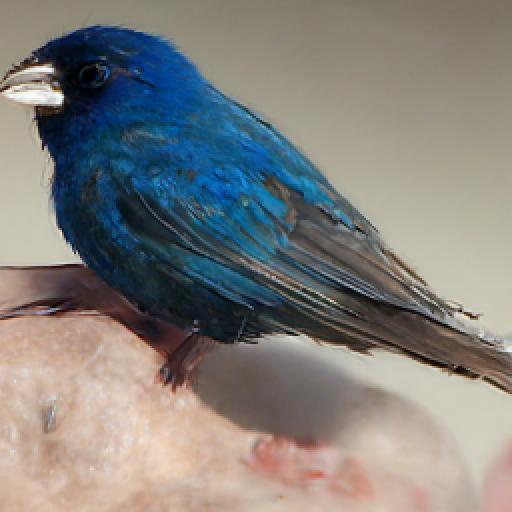} & 
 \includegraphics[width=0.142\textwidth]{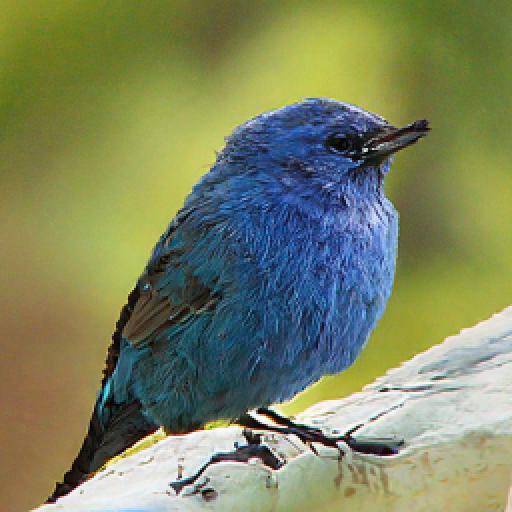} & 
 \includegraphics[width=0.142\textwidth]{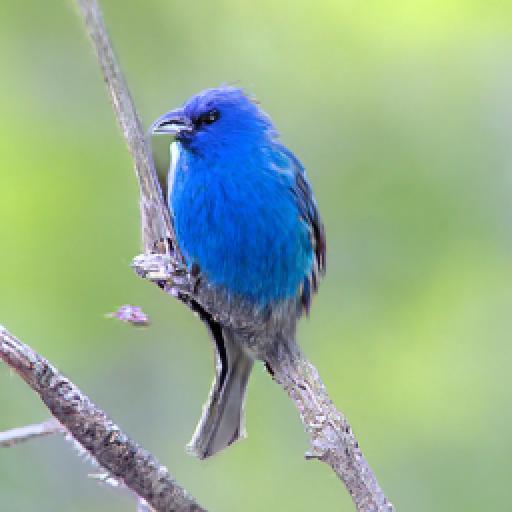} & 
 \includegraphics[width=0.142\textwidth]{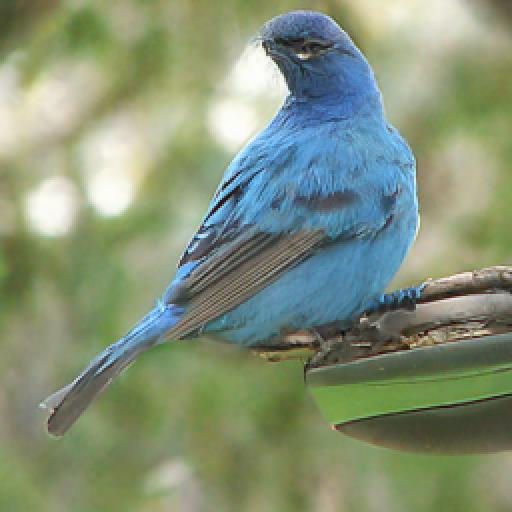} & 
 \includegraphics[width=0.142\textwidth]{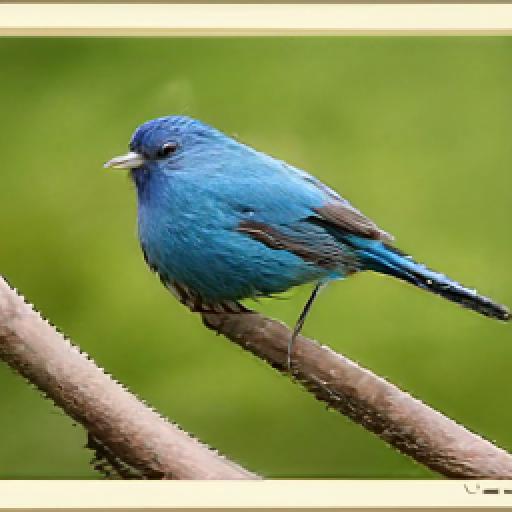} & 
 \includegraphics[width=0.142\textwidth]{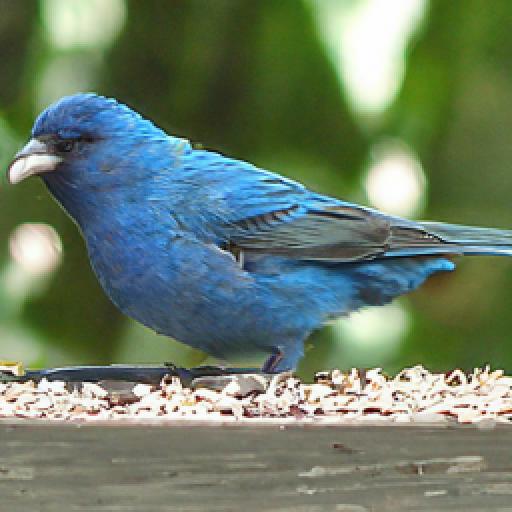}\\
 
 \includegraphics[width=0.142\textwidth]{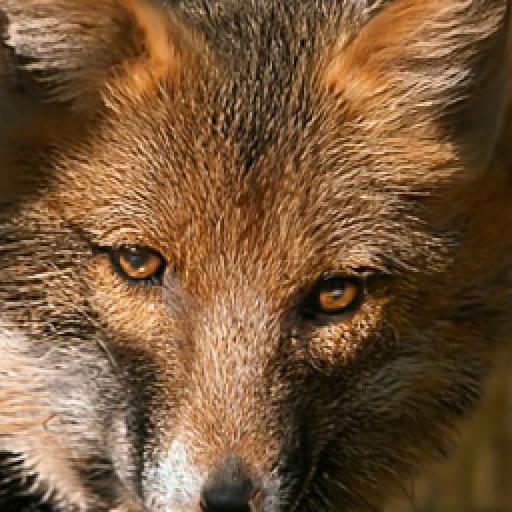} & 
 \includegraphics[width=0.142\textwidth]{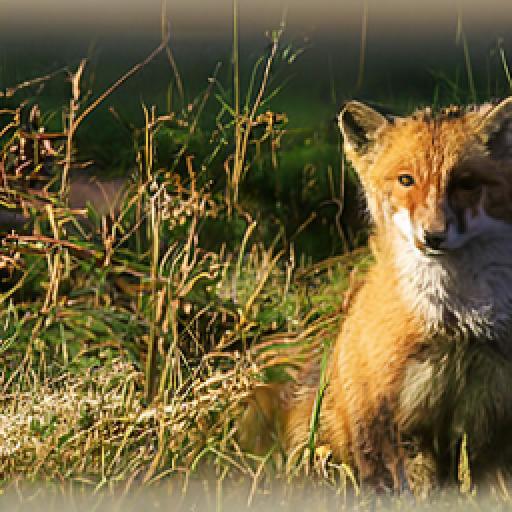} & 
 \includegraphics[width=0.142\textwidth]{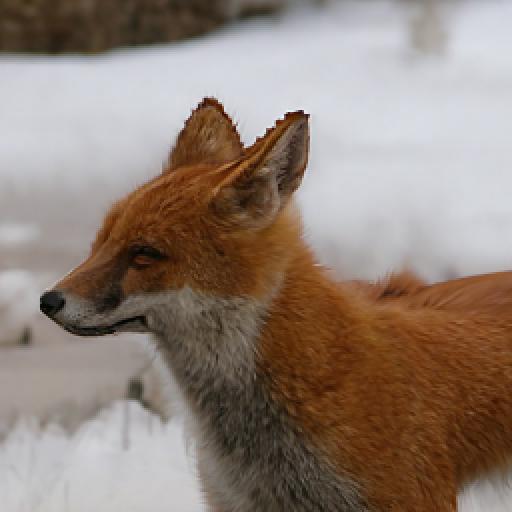} & 
 \includegraphics[width=0.142\textwidth]{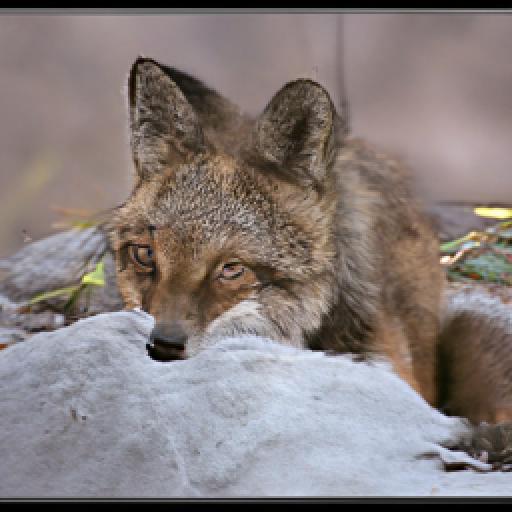} & 
 \includegraphics[width=0.142\textwidth]{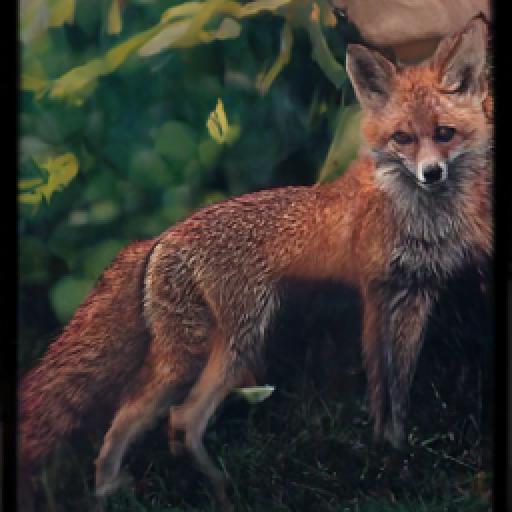} & 
 \includegraphics[width=0.142\textwidth]{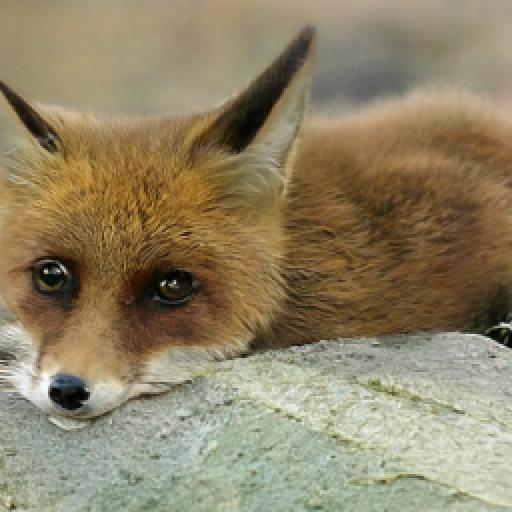}\\
 
 \includegraphics[width=0.142\textwidth]{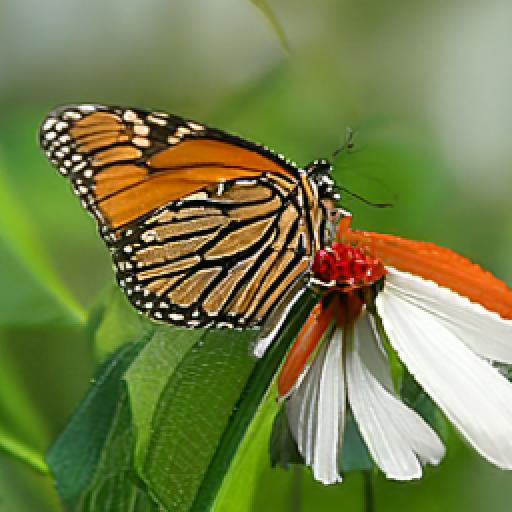} & 
 \includegraphics[width=0.142\textwidth]{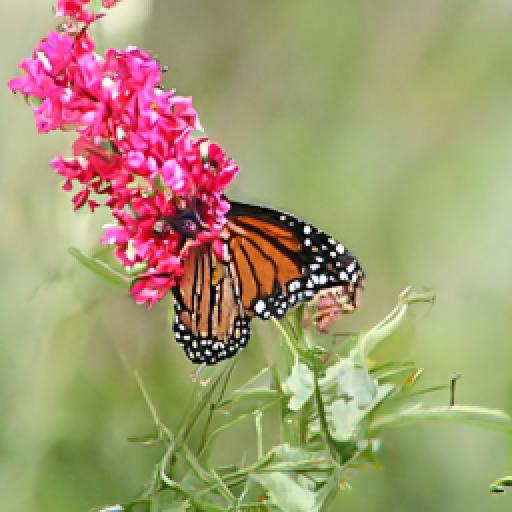} & 
 \includegraphics[width=0.142\textwidth]{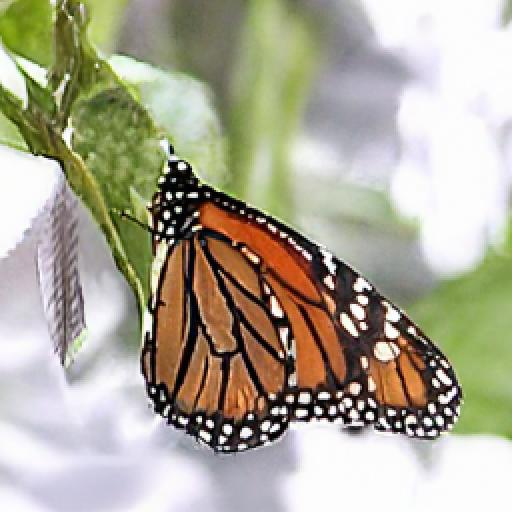} & 
 \includegraphics[width=0.142\textwidth]{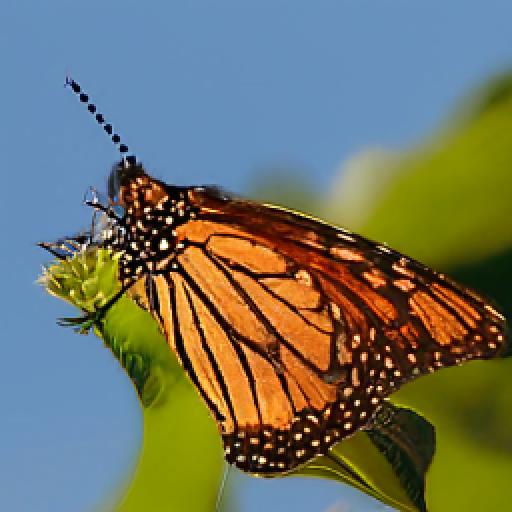} & 
 \includegraphics[width=0.142\textwidth]{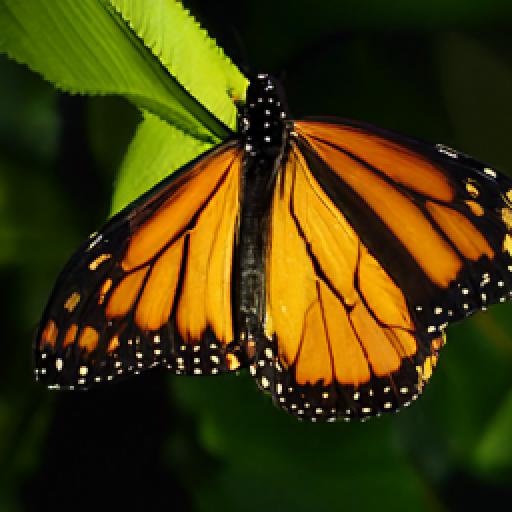} & 
 \includegraphics[width=0.142\textwidth]{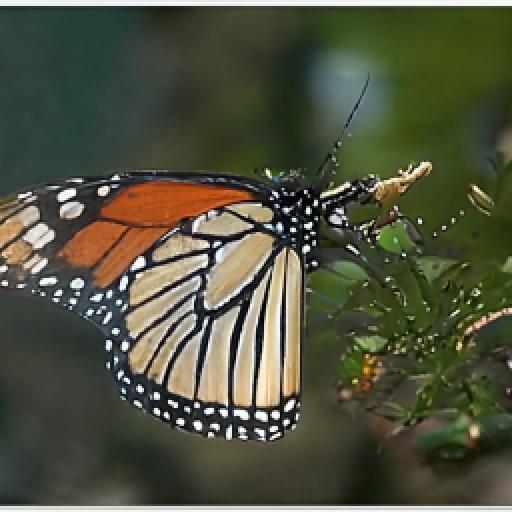}\\
 
 \includegraphics[width=0.142\textwidth]{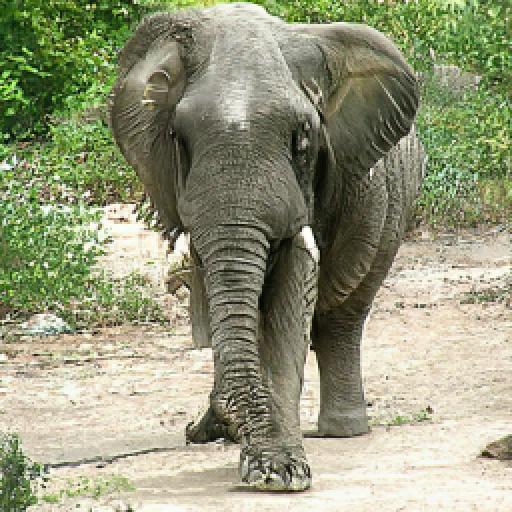} & 
 \includegraphics[width=0.142\textwidth]{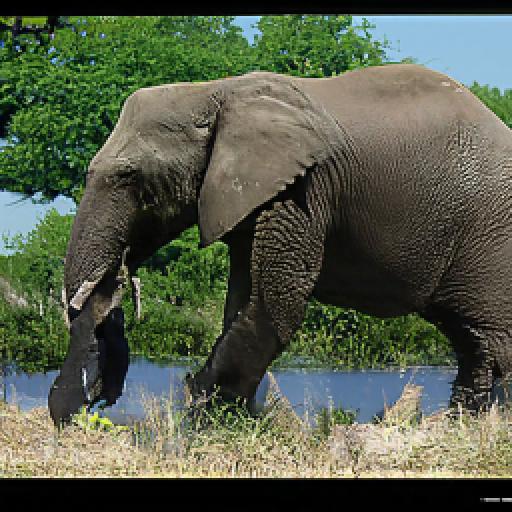} & 
 \includegraphics[width=0.142\textwidth]{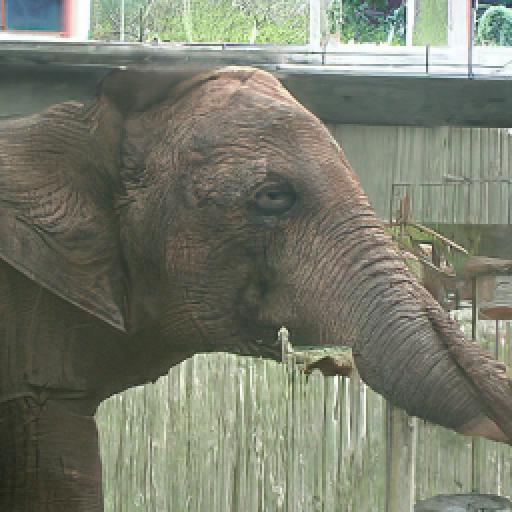} & 
 \includegraphics[width=0.142\textwidth]{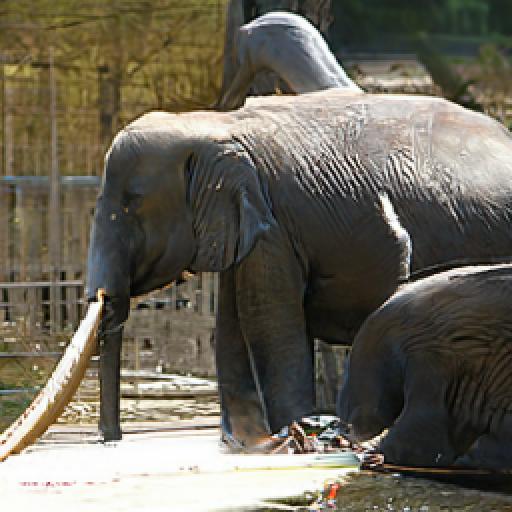} & 
 \includegraphics[width=0.142\textwidth]{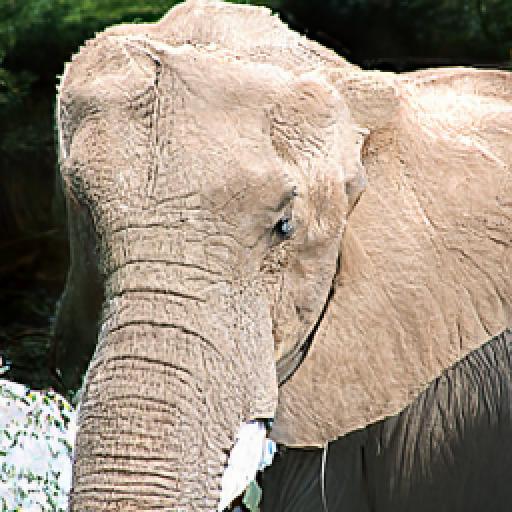} & 
 \includegraphics[width=0.142\textwidth]{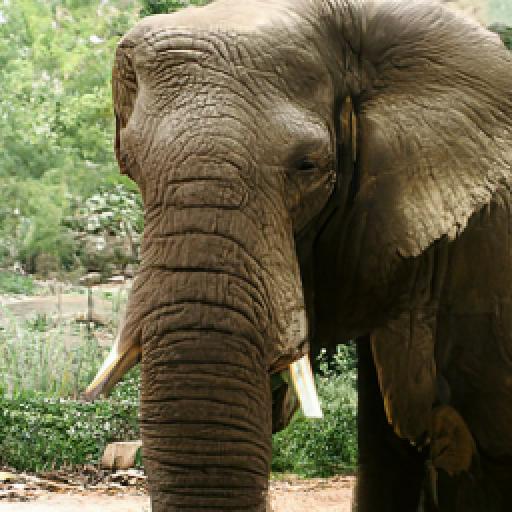}\\
 
 \includegraphics[width=0.142\textwidth]{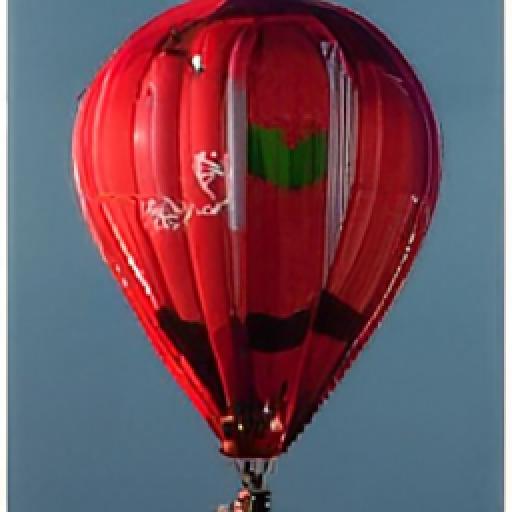} & 
 \includegraphics[width=0.142\textwidth]{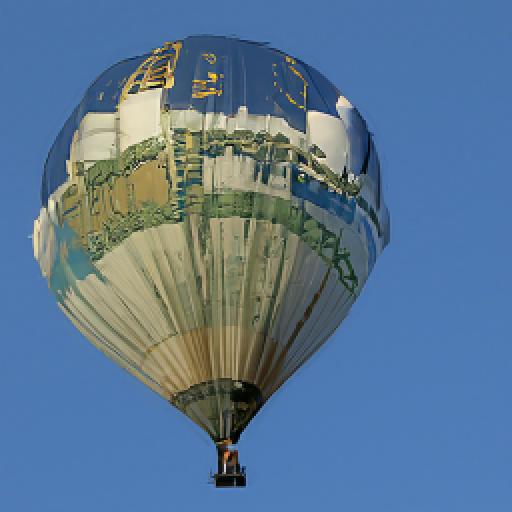} & 
 \includegraphics[width=0.142\textwidth]{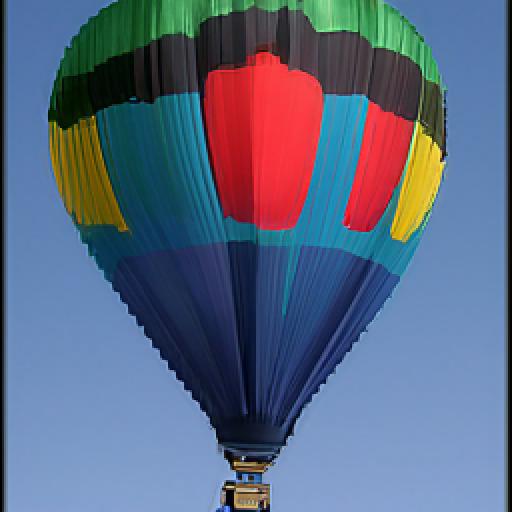} & 
 \includegraphics[width=0.142\textwidth]{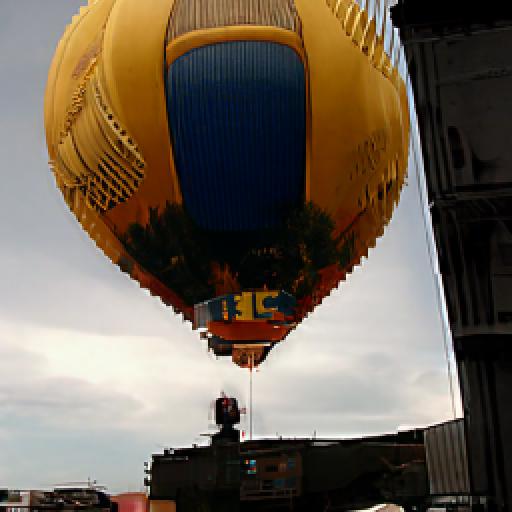} & 
 \includegraphics[width=0.142\textwidth]{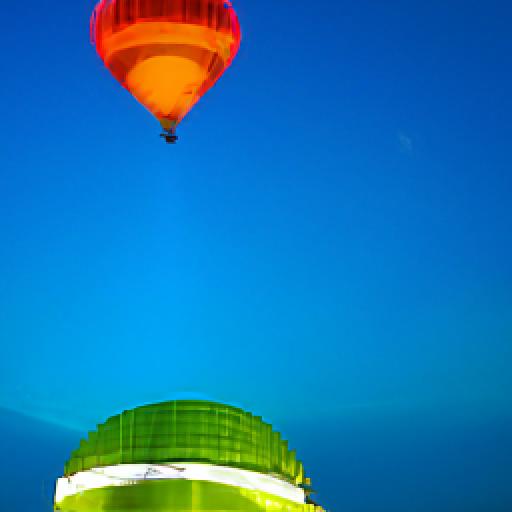} & 
 \includegraphics[width=0.142\textwidth]{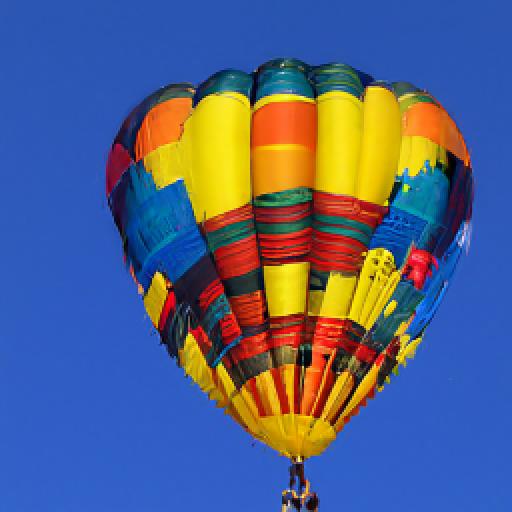}\\
 
 \includegraphics[width=0.142\textwidth]{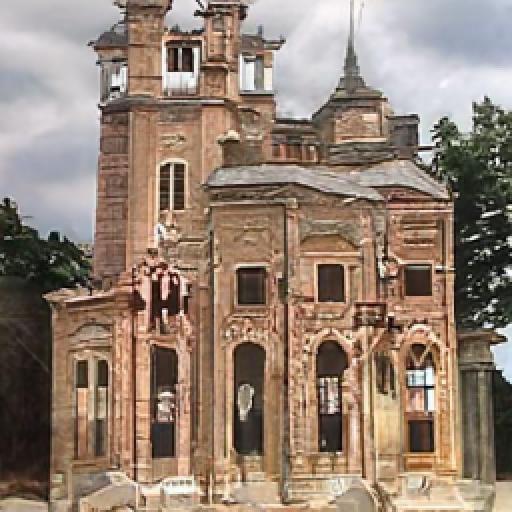} & 
 \includegraphics[width=0.142\textwidth]{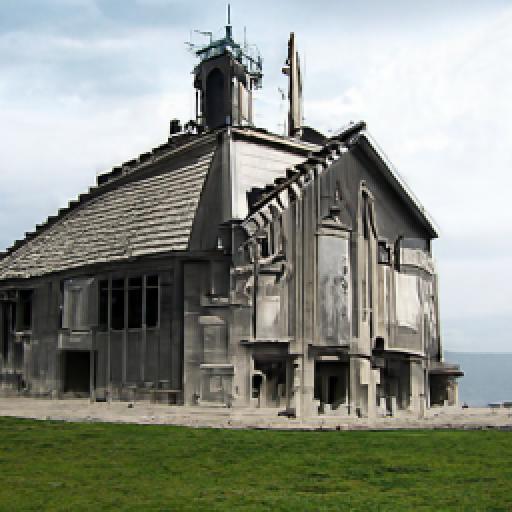} & 
 \includegraphics[width=0.142\textwidth]{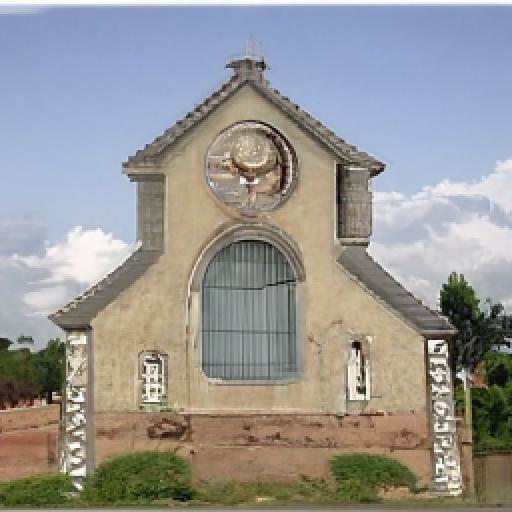} & 
 \includegraphics[width=0.142\textwidth]{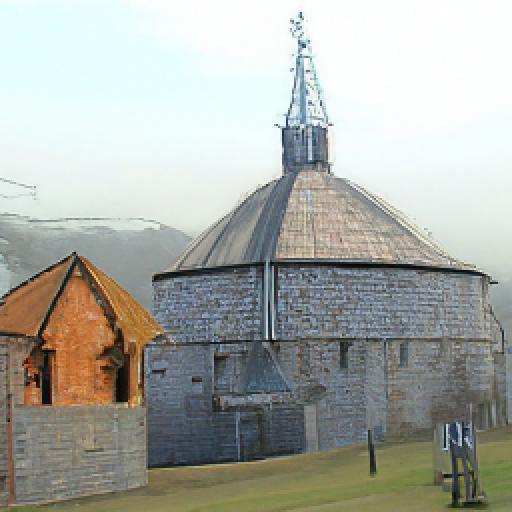} & 
 \includegraphics[width=0.142\textwidth]{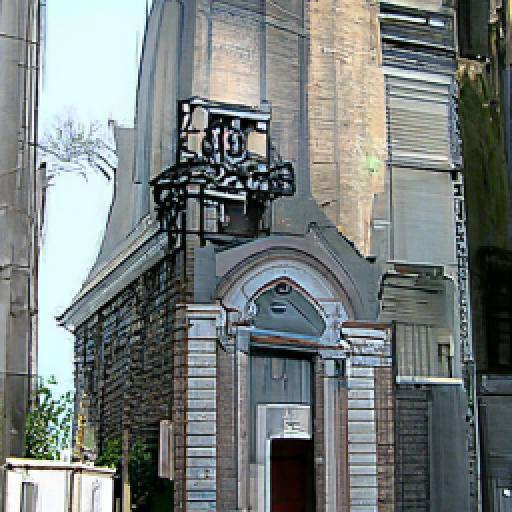} & 
 \includegraphics[width=0.142\textwidth]{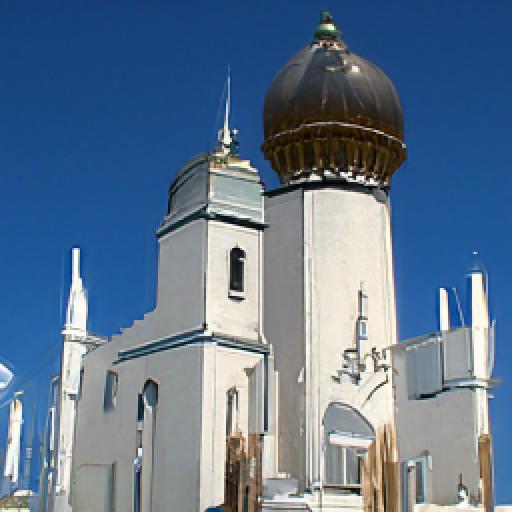}\\
 
 \includegraphics[width=0.142\textwidth]{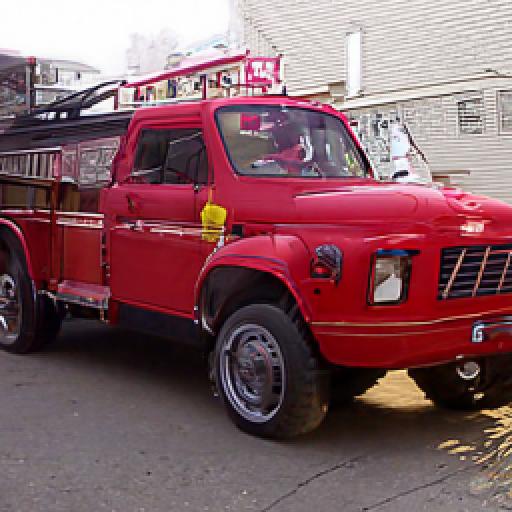} & 
 \includegraphics[width=0.142\textwidth]{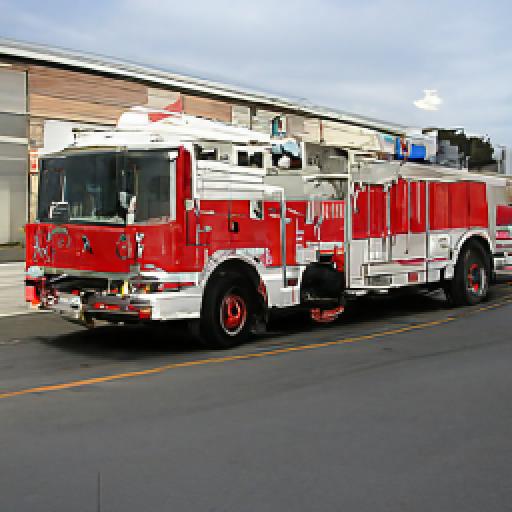} & 
 \includegraphics[width=0.142\textwidth]{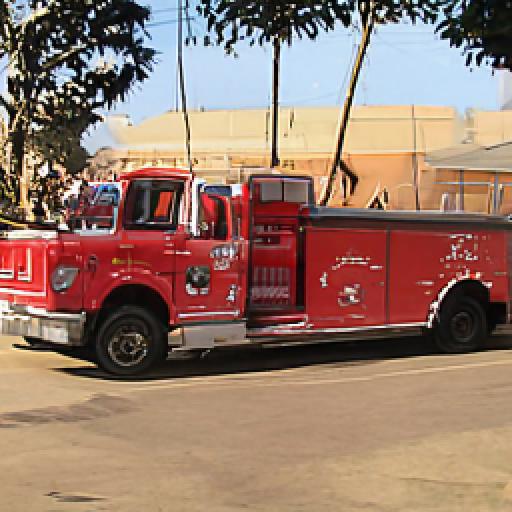} & 
 \includegraphics[width=0.142\textwidth]{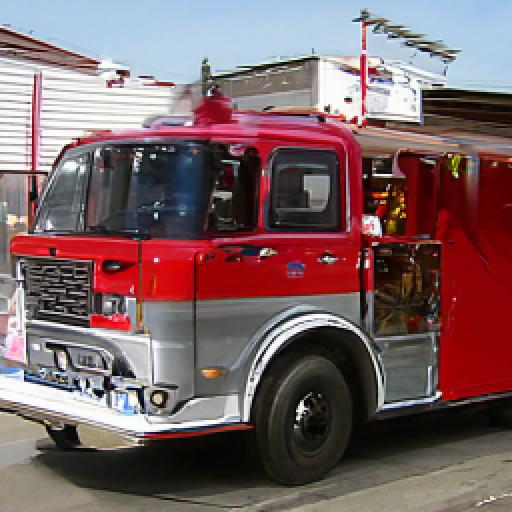} & 
 \includegraphics[width=0.142\textwidth]{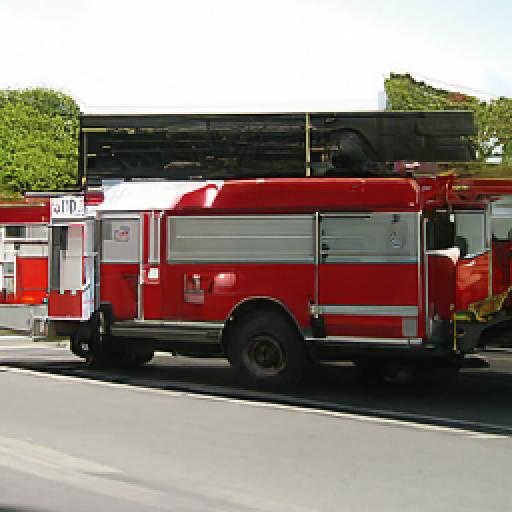} & 
 \includegraphics[width=0.142\textwidth]{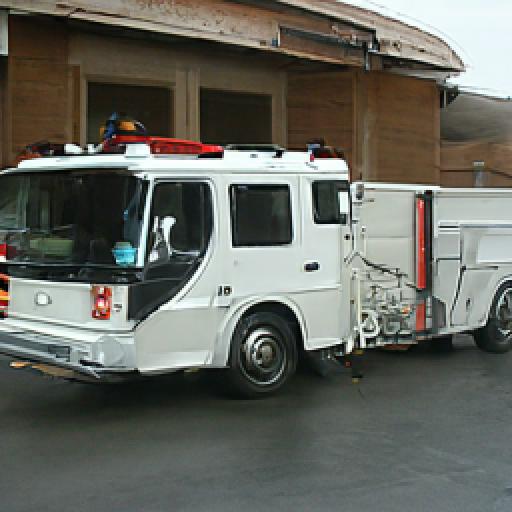}\\

\end{tabular}
\end{center}
\vspace*{-0.4cm}
\caption{Classwise Synthetic 256$\times$256 ImageNet images. Each row represents a specific ImageNet class. Classes from top to bottom - Goldfish, Indigo Bird, Red Fox, Monarch Butterfly, African Elephant, Balloon, Church, Fire Truck. For a given label, we sample a 64$\times$64 image from a class-conditional diffusion model, and apply a 4$\times$ \modelname model to obtain 256$\times$256 images.}
\vspace*{-0.2cm}
\label{fig:imagenet_256x_montage3}
\end{figure}

\newpage 
\subsection{Failure Cases of SR3}

While SR3 generates high quality super-resolution natural images as well as aligned face images, we do observe certain instances where the model falls short. In this section, we highlight some of these examples.

SR3 struggles with generating certain complex, regular hair patterns, an example of which is the finely braided hair in the top row of Figure \ref{fig:64x_512x_faces3_arxiv}. 
Since the FFHQ dataset used for training is relatively small, it is possible that the model is not exposed to enough examples of such structures. Long range correlations of fine details such as the consistency of highlights in eyes can also be challenging.
As shown in the 2nd row of Figure \ref{fig:64x_512x_faces3_arxiv},
the model also struggles with generating eye-glasses in certain cases, especially when the glasses are frameless.  In such cases 
the structure of the glasses is much harder to discern from the low resolution inputs.

SR3 also fails to generate natural looking text in certain ImageNet images. While it has learned to capture some properties of text, it has not learned common alphabets.  So while SR3 is able to generate much sharper characters compared to the regression baseline, the lack of meaningful structure of its generated text makes it easier for the human subjects to distinguish between real and generated images. The last row in Figure \ref{fig:64x_256x_natural_images2} shows one such instance.

The bottom row in Figure \ref{fig:64x_256x_natural_images3} shows an instance where the model has not correctly inferred the fine structure on the side of the building.  When the low resolution input does not reflect many of the fine details in the high resolution original, SR3 will infer structure.  In some cases it will introduce texture (e.g., the collar of the shirt 
in Figure \ref{fig1}).  In others, like the building here it may 
infer a more uniform texture.
As such, while the SR3 output in the bottom row of Figure \ref{fig:64x_256x_natural_images3}
is much sharper than the regression baseline, it also lacks many perceptually relevant details when compared with the reference image.

\vspace*{0.25cm}

\section{Images with the Lowest and Highest Fool Rates}

In interpreting the fool rate results in Figure \ref{fig:face_foolrate}, it is interesting to inspect those images that maximize the fool rates for a given technique, as well as those images that minimize the fool rate. This provides insight into the nature of the problems that models exhibit, as well as cases in which the model outputs are good enough to regularly fool people.  

\begin{figure}[H]
\small
    \begin{center}
    \setlength{\tabcolsep}{2pt}
    \begin{tabular}{ccccccc}
    \multicolumn{7}{c}{\large{\textbf{Task-1: Human Evaluation given low-resolution inputs}}} \\[.5cm]
    \multicolumn{3}{c}{\large{{Lowest Mean Fool Rates for PULSE}}} &\hspace*{1.0cm} & \multicolumn{3}{c}{\large{{Lowest Mean Fool Rates for SR3}}}\\ [0.2cm]
    { PULSE} & { Input} &  { SR3} & &
    { SR3} & { Input} &  { PULSE} \\[.1cm]
    {\includegraphics[width=.15\textwidth]{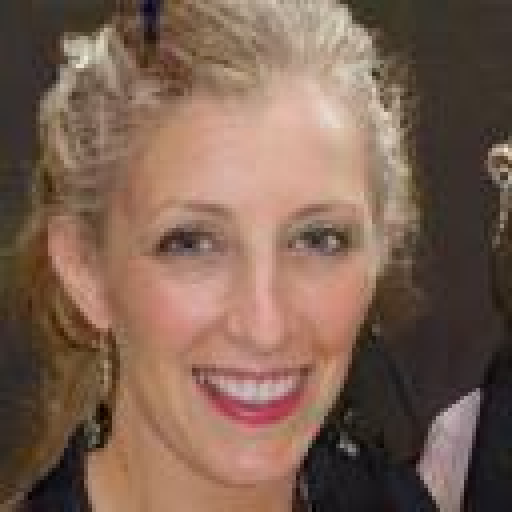}} &
    \raisebox{0.0375\textwidth}{\includegraphics[width=.075\textwidth]{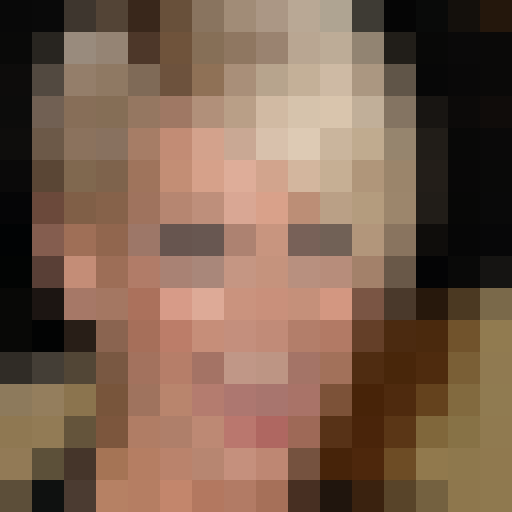}} &
    {\includegraphics[width=.15\textwidth]{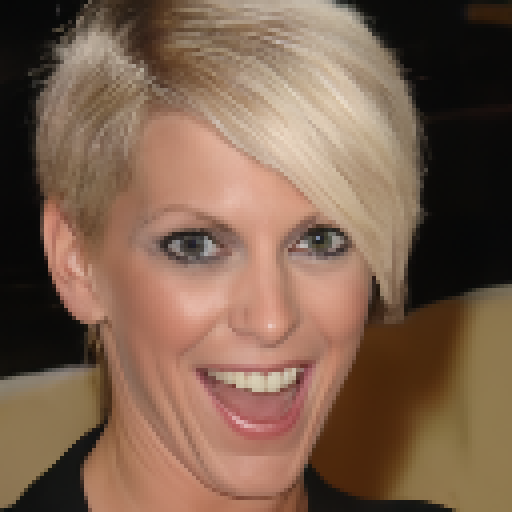}} &
    \hspace*{1.0cm} & 
    {\includegraphics[width=.15\textwidth]{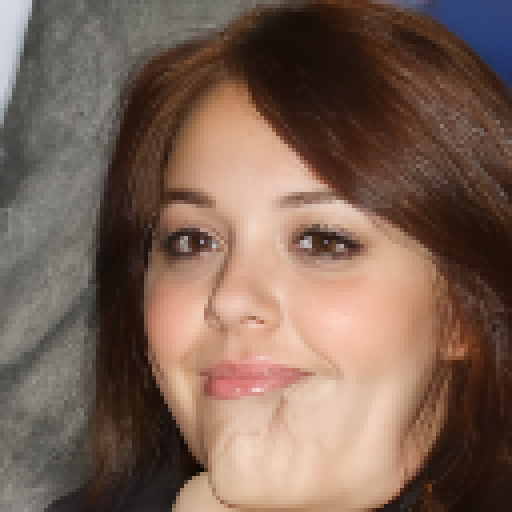}} &    
    \raisebox{0.0375\textwidth}{\includegraphics[width=.075\textwidth]{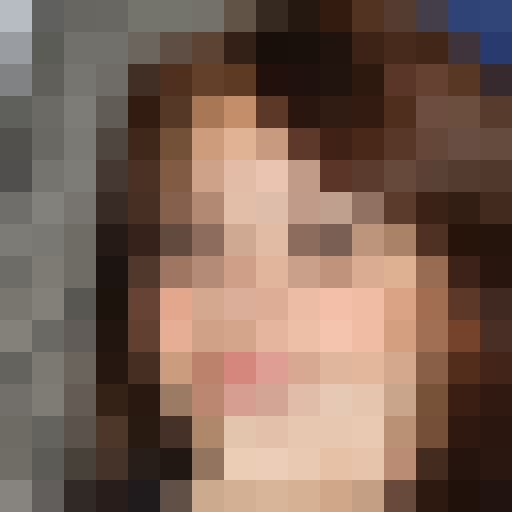}} &
    {\includegraphics[width=.15\textwidth]{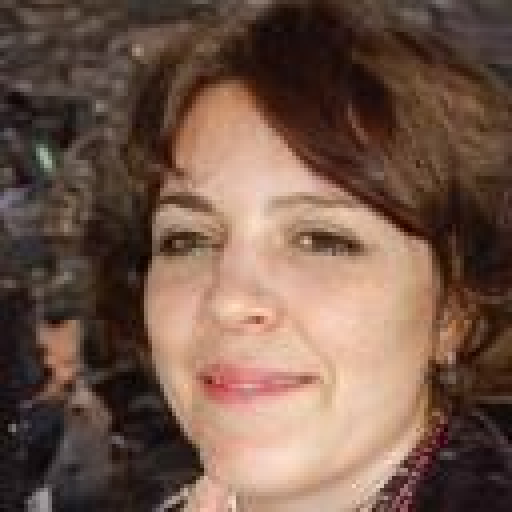}} \\
    Fool Rate: 0\% & & Fool Rate: 53.4\% & \hspace*{1.0cm} & Fool Rate: 14\% & & Fool Rate: 4.5\% \\ [0.35cm]
    
    {\includegraphics[width=.15\textwidth]{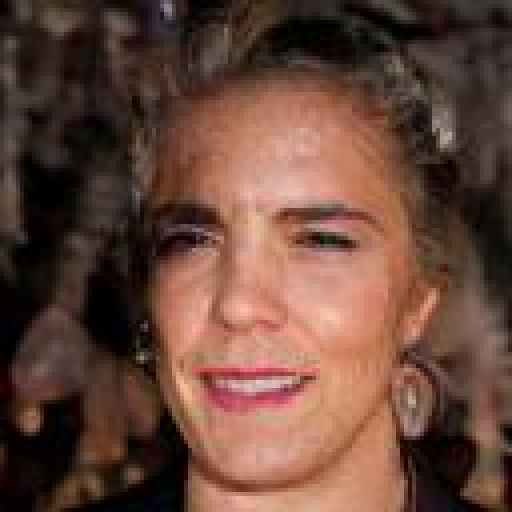}} &
    \raisebox{0.0375\textwidth}{\includegraphics[width=.075\textwidth]{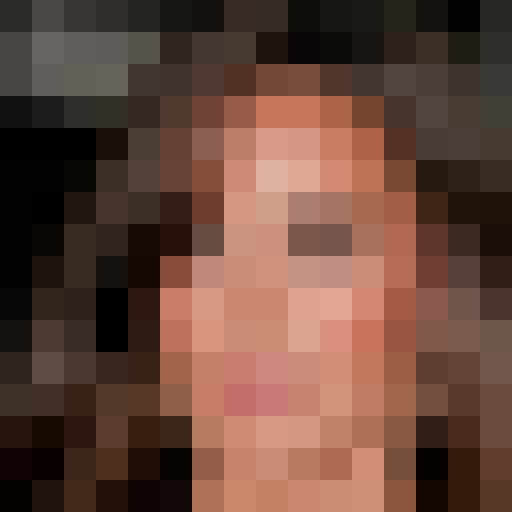}} &
    {\includegraphics[width=.15\textwidth]{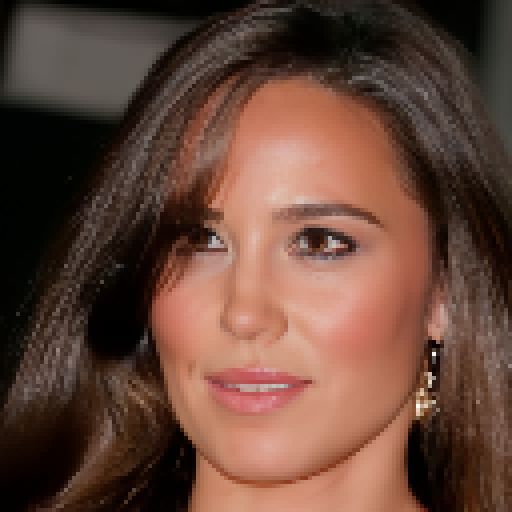}} &
     \hspace*{1.0cm} & 
    {\includegraphics[width=.15\textwidth]{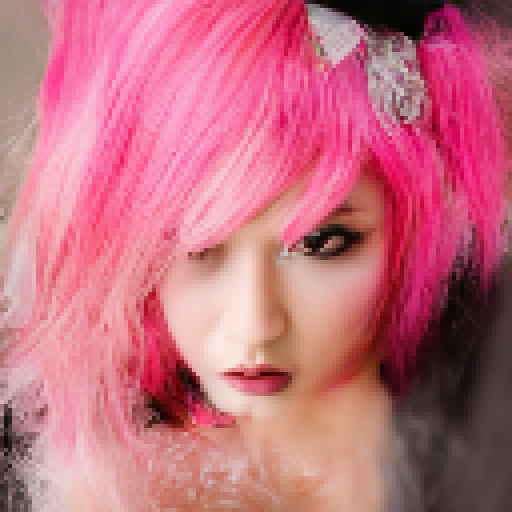}} &
    \raisebox{0.0375\textwidth}{\includegraphics[width=.075\textwidth]{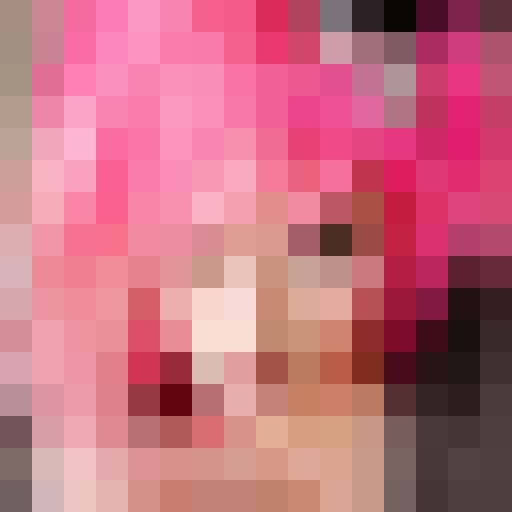}} &
    {\includegraphics[width=.15\textwidth]{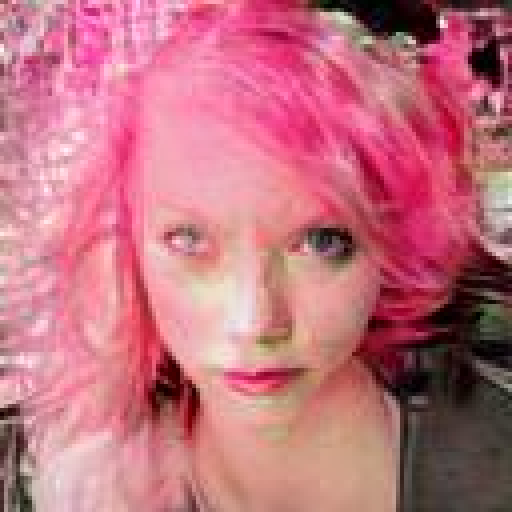}} \\
    Fool Rate: 2.2\% & & Fool Rate: 60.4\% & \hspace*{1.0cm} & Fool Rate: 18.6\% & & Fool Rate: 11.4\% \\ [0.2cm]

    \medskip\\

    \multicolumn{3}{c}{\large{{Highest Mean Fool Rates for PULSE}}} &\hspace*{1.0cm} & \multicolumn{3}{c}{\large{{Highest Mean Fool Rates for SR3}}}\\ [0.2cm]
    { PULSE} & { Input} &  { SR3} & &
    { SR3} & { Input} &  { PULSE} \\[.1cm]
    {\includegraphics[width=.15\textwidth]{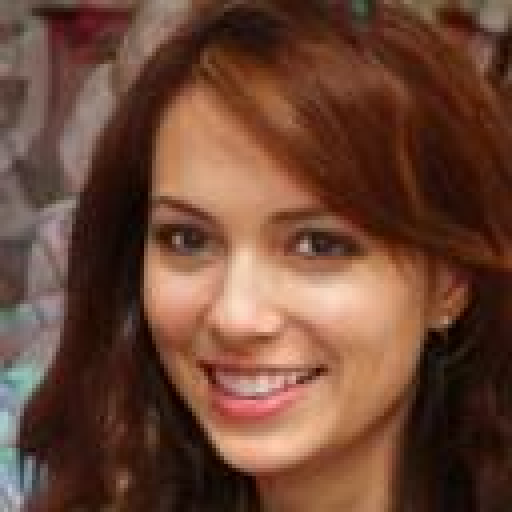}} &
    \raisebox{0.0375\textwidth}{\includegraphics[width=.075\textwidth]{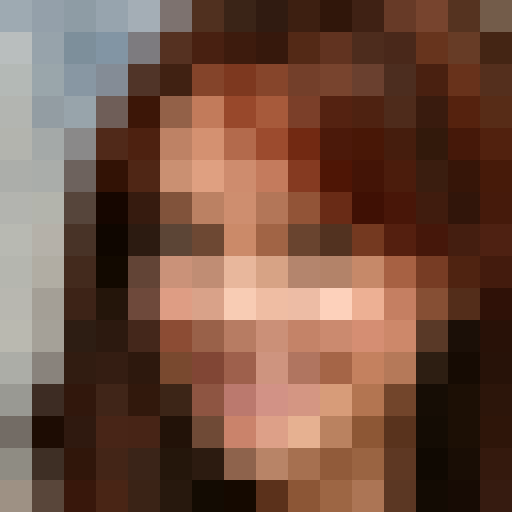}} &
    {\includegraphics[width=.15\textwidth]{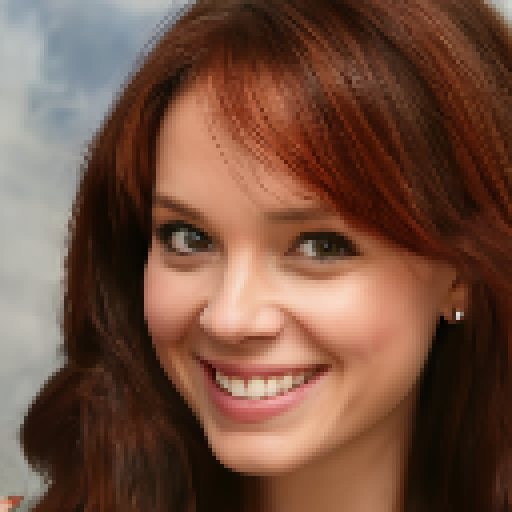}} & 
    \hspace*{1.0cm} & 
    {\includegraphics[width=.15\textwidth]{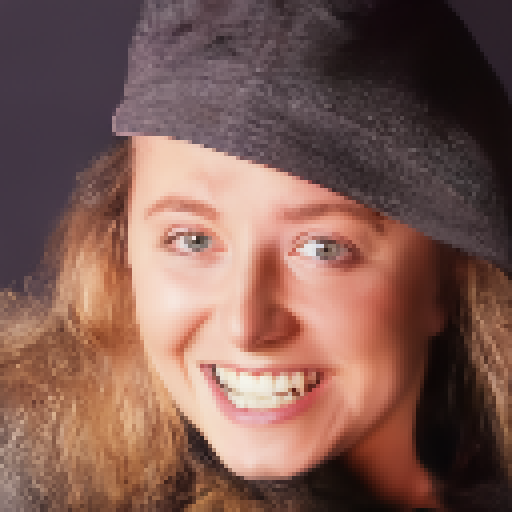}} &
    \raisebox{0.0375\textwidth}{\includegraphics[width=.075\textwidth]{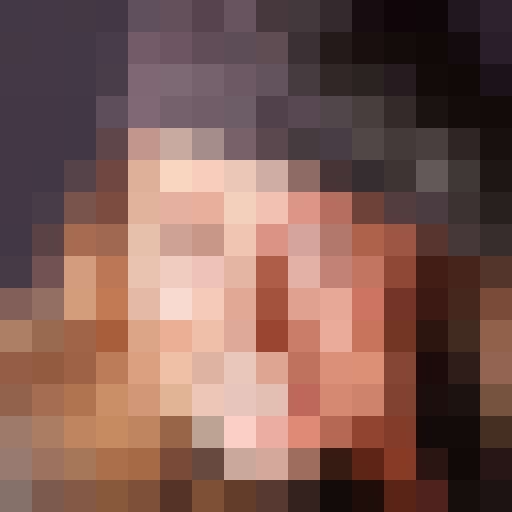}} &
    {\includegraphics[width=.15\textwidth]{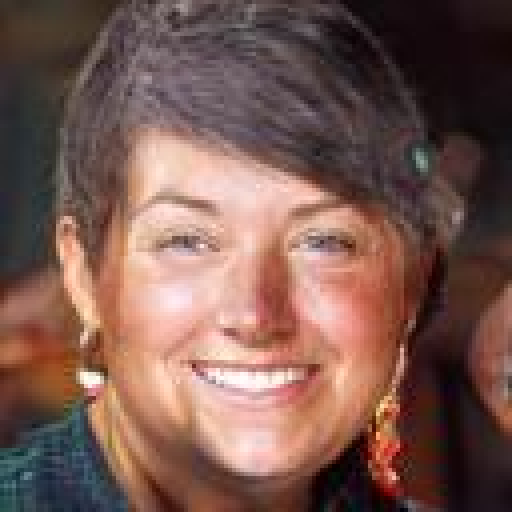}} \\
    Fool Rate: 63.4\% & & Fool Rate: 62.7\% & \hspace*{1.0cm} & Fool Rate: 88.3\% & & Fool Rate: 38.6\% \\ [0.35cm]
    
    {\includegraphics[width=.15\textwidth]{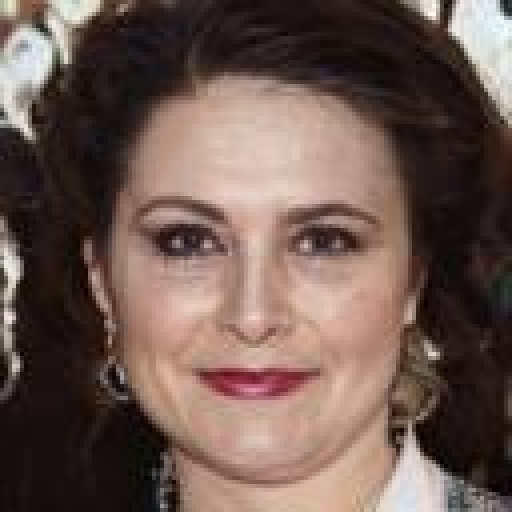}} &
    \raisebox{0.0375\textwidth}{\includegraphics[width=.075\textwidth]{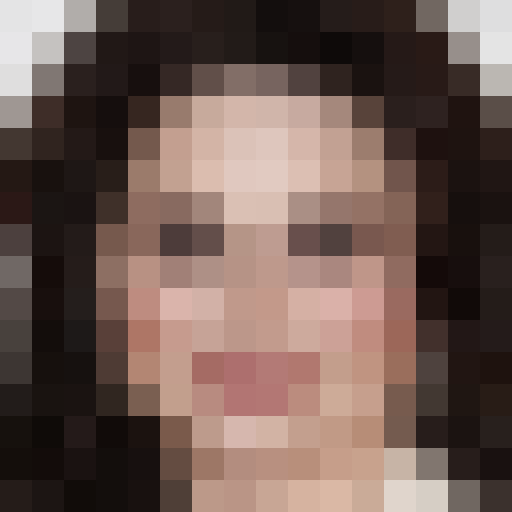}} &
    {\includegraphics[width=.15\textwidth]{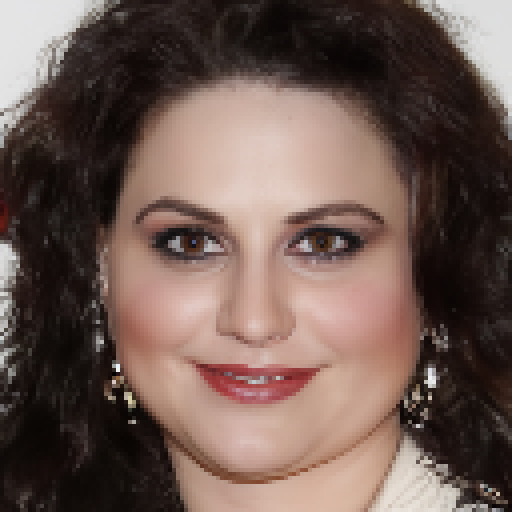}} &
    \hspace*{1.0cm} & 
    {\includegraphics[width=.15\textwidth]{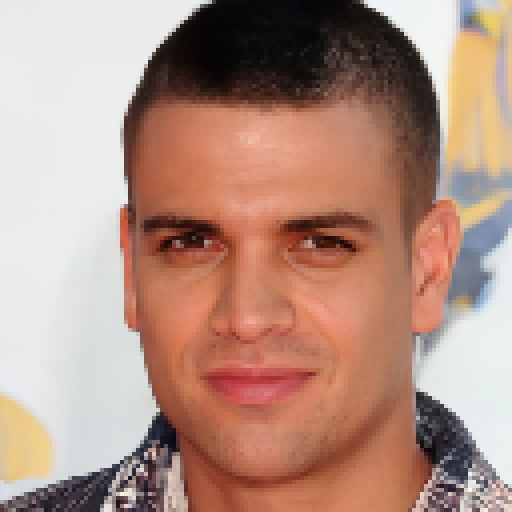}} &    
    \raisebox{0.0375\textwidth}{\includegraphics[width=.075\textwidth]{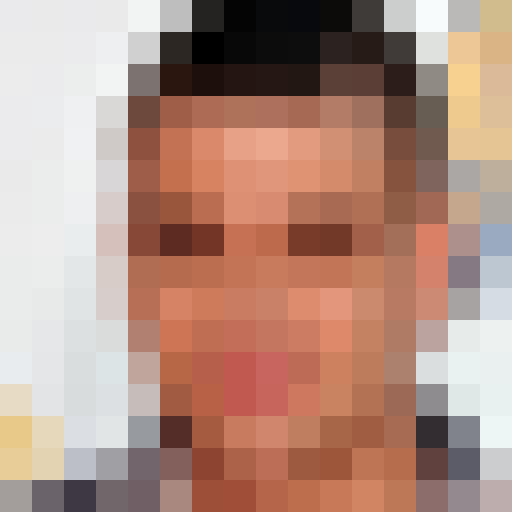}} &
    {\includegraphics[width=.15\textwidth]{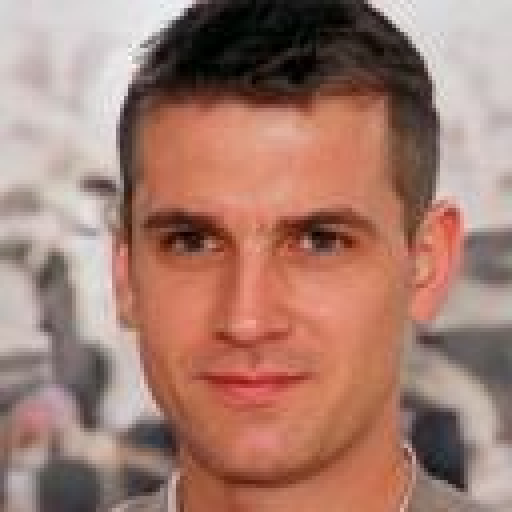}} \\
    Fool Rate: 63.4\% & & Fool Rate: 69.7\% & \hspace*{1.0cm} & Fool Rate: 83.7\% & & Fool Rate: 54.5\% \\ [0.2cm]

    \end{tabular}
    \end{center}
    \vspace*{-0.35cm}
    \caption{ Examples with lowest and highest fool rates for PULSE and SR3 based on Task-1. 
    Task-1 involves comparing the outputs of each algorithm with reference high-resolution images in the presence of low-resolution inputs,
    but for privacy reasons reference images are not included.
    Instead, we show the corresponding outputs from PULSE and SR3 for each input image and report the Mean Fool Rate for each image right below it.
    }
    \label{fig:worst_fool_rates}
    \end{figure}

In Figure \ref{fig:worst_fool_rates} we display the images with the lowest fool rates generated by PULSE \cite{menon2020pulse} and SR3 for both Task-1 (the conditional task), and Task-2, (the unconditional task). In order to be consistent with our human study interface, we show the corresponding low resolution image only for Task-1. 
Notice that images from PULSE for which the fool rate is low have obvious distortions, and the fool rates are
lower than 10\% for both tasks.  For SR3, by comparison, the images with the lowest fool rates are still reasonably
good, with much higher fool rates of 14\% and 19\% in Task-1, and 21\% and 26\% in Task-2.

\begin{figure}[H]
\small
    \begin{center}
    
    \setlength{\tabcolsep}{2pt}
    \begin{tabular}{ccccccc}
    \multicolumn{7}{c}{\large{\textbf{Task-2: Human Evaluation without low-resolution inputs}}} \\[.5cm]
    \multicolumn{3}{c}{\large{{Lowest Mean Fool Rates for PULSE}}} &\hspace*{1.0cm} & \multicolumn{3}{c}{\large{{Lowest Mean Fool  Rates for SR3}}}\\ [0.2cm]
    { PULSE} &  & { SR3} & & { SR3} &  & { PULSE} \\[.1cm]
    {\includegraphics[width=.15\textwidth]{figures/faces_16x_128x/pulse_out/28.png}} & &
    {\includegraphics[width=.15\textwidth]{figures/faces_16x_128x/diffusion_output/28.png}} & & 
    {\includegraphics[width=.15\textwidth]{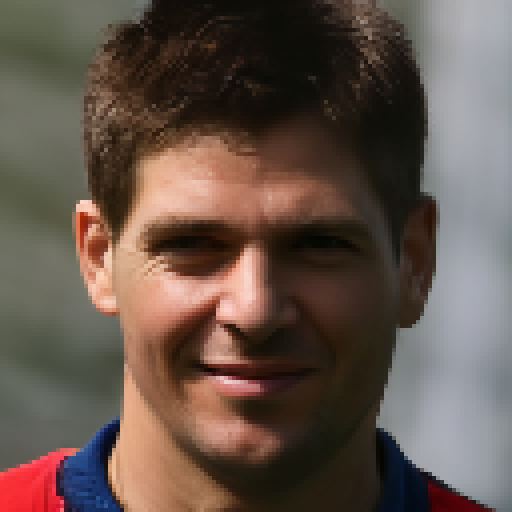}} & &
    {\includegraphics[width=.15\textwidth]{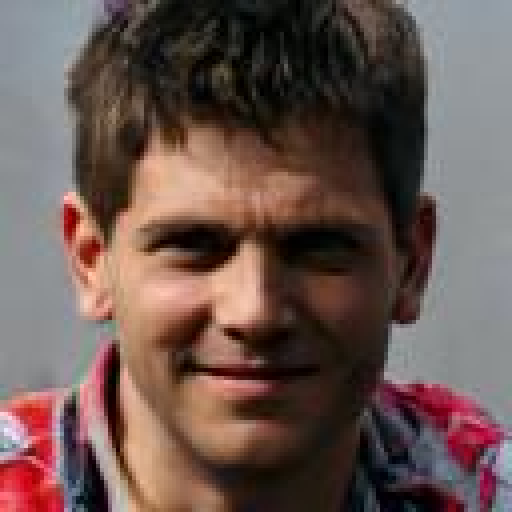}} \\
    Fool Rate: 8.9\% & & Fool Rate: 19\% & \hspace*{1.0cm} & Fool Rate: 21.4\% & & Fool Rate: 15.5\% \\[0.35cm]
    {\includegraphics[width=.15\textwidth]{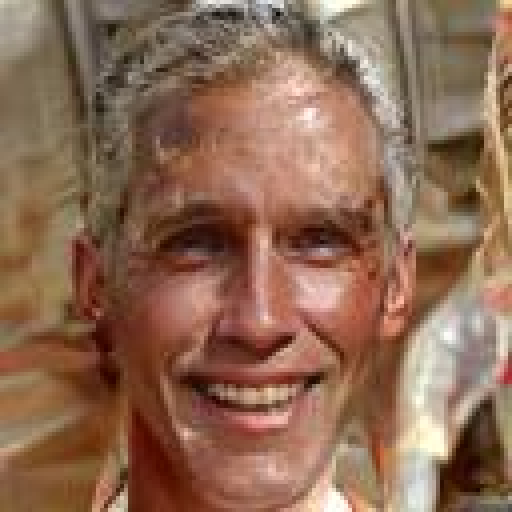}} & &
    {\includegraphics[width=.15\textwidth]{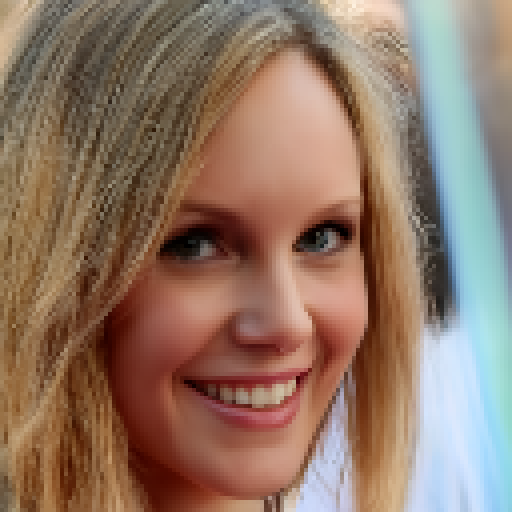}} & &
    {\includegraphics[width=.15\textwidth]{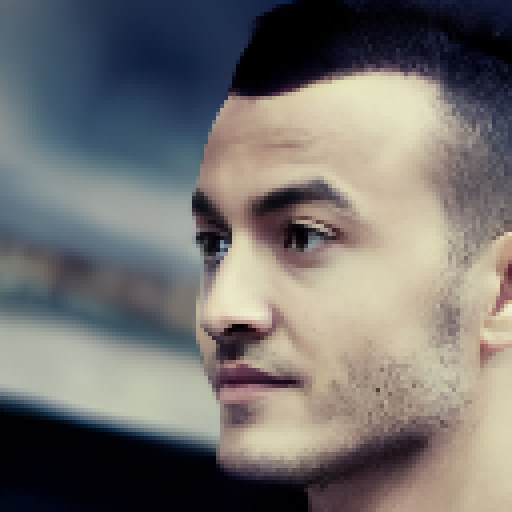}} & &
    {\includegraphics[width=.15\textwidth]{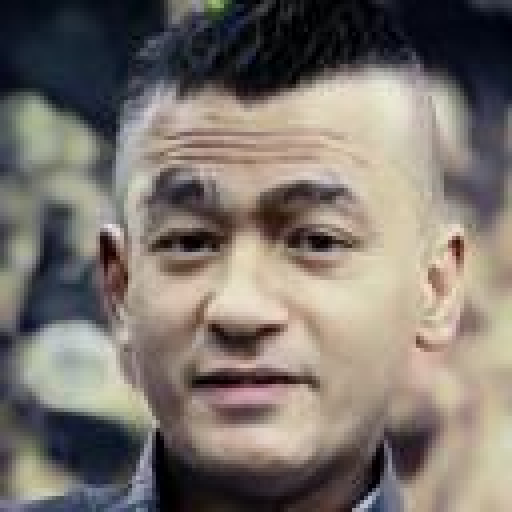}} \\
    Fool Rate: 8.9\% & & Fool Rate: 54.8\% & \hspace*{1.0cm} & Fool Rate: 26.2\% & & Fool Rate: 31.1\% \\

    \medskip\\

    \multicolumn{3}{c}{\large{{Highest Mean Fool Rates for PULSE}}} &\hspace*{1.0cm} & \multicolumn{3}{c}{\large{{Highest Mean Fool  Rates for SR3}}}\\ [0.2cm]
    { PULSE} &  & { SR3} & & { SR3} &  & { PULSE} \\[.1cm]
    {\includegraphics[width=.15\textwidth]{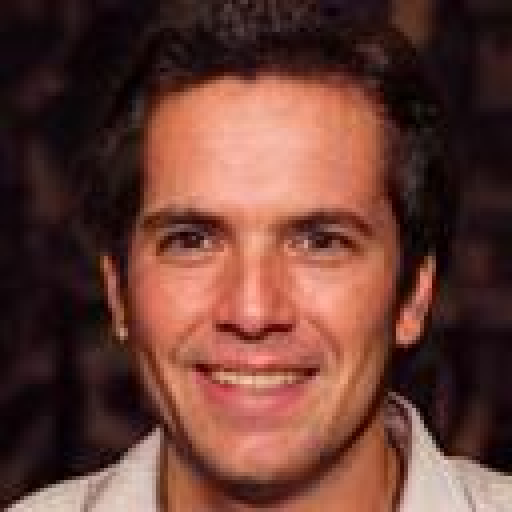}} & &
    {\includegraphics[width=.15\textwidth]{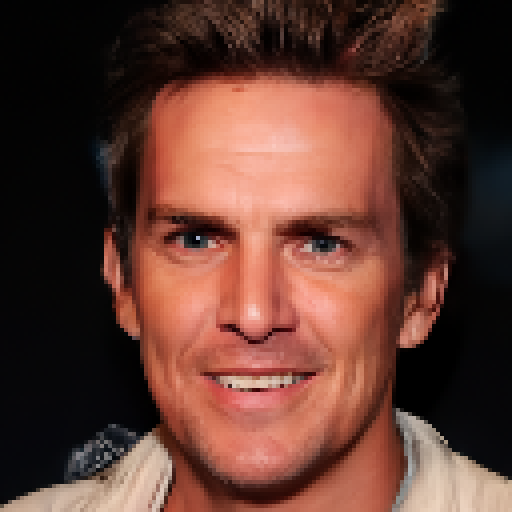}} & &
    {\includegraphics[width=.15\textwidth]{figures/faces_16x_128x/diffusion_output/37.png}} & &
    {\includegraphics[width=.15\textwidth]{figures/faces_16x_128x/pulse_out/37.png}} \\
    Fool Rate: 75.5\% & & Fool Rate: 61.9\% & \hspace*{1.0cm} & Fool Rate: 78.5\% & & Fool Rate: 35.5\% \\ [0.35cm]
    
    {\includegraphics[width=.15\textwidth]{figures/faces_16x_128x/pulse_out/34.png}} & &
    {\includegraphics[width=.15\textwidth]{figures/faces_16x_128x/diffusion_output/34.png}} & &
    {\includegraphics[width=.15\textwidth]{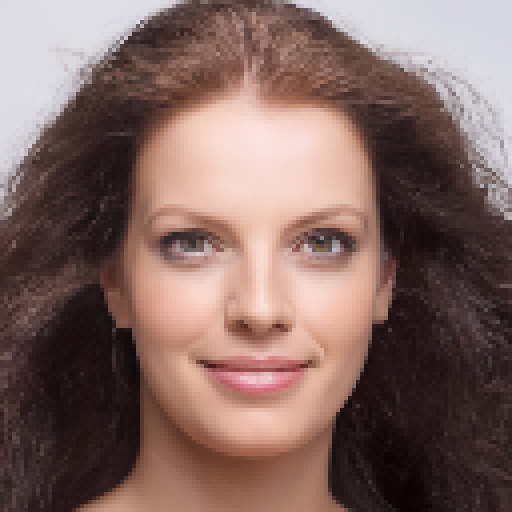}} & &
    {\includegraphics[width=.15\textwidth]{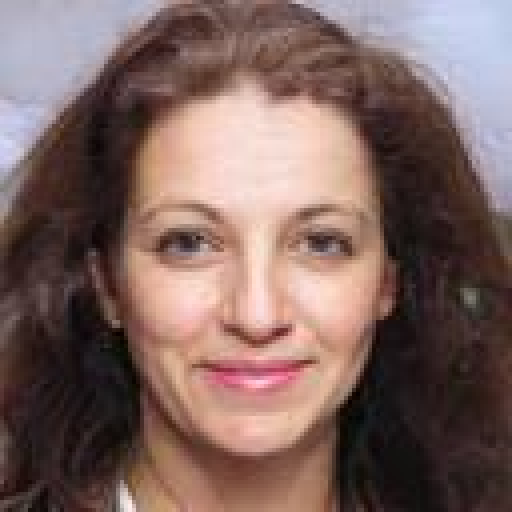}} \\
    Fool Rate: 66.7\% & & Fool Rate: 47.6\% & \hspace*{1.0cm} & Fool Rate: 66.7\% & & Fool Rate: 55.6\% \\

    \end{tabular}
    \end{center}
    
    \vspace*{-0.35cm}
    \caption{Examples with lowest and highest fool rates for PULSE and SR3 based on Task-2. 
    Task-2 involves comparing the outputs of each algorithm with reference high-resolution images in the absence of low-resolution inputs,
    but for privacy reasons reference images are not included.
    Instead, we show the corresponding outputs from PULSE and SR3 for each input image and report the Mean Fool Rate for each image right below it.}
    \label{fig:best_fool_rates}
    \end{figure}

Figure  \ref{fig:best_fool_rates} shows images that best fool human subjects.  In this case, it is interesting to note that the best fool rates for SR3 are 84\% and 88\%.
The corresponding original images are somewhat noisy, and as a consequence, many subjects refer
the SR3 outputs.

\vfill

\newpage


\vfill

\newpage

\vfill

\end{document}